\documentclass[aps,pra,reprint,nofootinbib]{revtex4-2}
\usepackage[T1]{fontenc}
\usepackage{lmodern,microtype}
\usepackage{amsmath,amssymb,amsthm,mathtools,booktabs,graphicx,comment}
\usepackage{tikz}
\usetikzlibrary{arrows.meta,positioning}
\usepackage{xcolor}

\usepackage[colorlinks=true,allcolors=blue!45!black]{hyperref}
\makeatletter
\AtBeginDocument{\renewcommand{\p@subsection}{\thesection.}}
\AtBeginDocument{\renewcommand{\p@subsubsection}{\thesection.\thesubsection.}}
\renewcommand{\frontmatter@title@below}{\vspace{-0.5em}}
\makeatother
\newtheorem{theorem}{Theorem}
\newtheorem{proposition}{Proposition}
\newtheorem{lemma}{Lemma}
\newcommand{\ket}[1]{\lvert #1\rangle}
\newcommand{\bra}[1]{\langle #1\rvert}
\newcommand{\norm}[1]{\lVert #1\rVert}
\newcommand{\dd}{\mathrm{d}}
\newcommand{\ii}{\mathrm{i}}
\newcommand{\diag}{\operatorname{diag}}
\newcommand{\T}{\top} 

\begin{document}
\title{Efficient quantum ultraspherical transform and\texorpdfstring{\\}{ }a well-conditioned quantum spectral method}
\author{Trong Duong, Norbert Hosters}
\affiliation{Chair for Computational Analysis of Technical Systems, RWTH Aachen University, Schinkelstrasse 2, 52062 Aachen, Germany}

\begin{abstract}
Solving differential equations is a promising application of quantum computing. Spectral methods, where the solution is represented by coefficients in a basis, are particularly well suited to quantum solvers because their structured discretizations can enable efficient linear-algebra operations while providing rapid convergence. Their quantum implementation, however, requires an efficient transform between amplitudes encoding function values and amplitudes encoding expansion coefficients. This work develops two key components of a quantum spectral method. First, we construct the quantum ultraspherical transform (QUT) for orthonormal ultraspherical polynomials, a basis that supports sparse and well-conditioned discretizations of differential equations. For the basis parameters $m=0$ and $1$, this transform reduces to the discrete cosine and sine transforms with known efficient circuits. Our contribution covers every fixed integer $m\geq2$: for dimensions $N$ with $N+m=2^n$, where $n$ is the number of qubits, the QUT achieves error at most $\varepsilon$ using a gate count polynomial in $n$ and $\log(1/\varepsilon)$. The construction builds on the quantum Fourier transform, augmented by approximations whose accuracy guarantees are independent of $N$. Second, we develop a quantum spectral solver for regular linear ordinary differential equations based on the QUT. Under suitable boundary conditions and data-access assumptions, preconditioning keeps condition numbers bounded independently of the basis size, allowing the solution state to be prepared at polylogarithmic cost in system size and inverse error. Together, these contributions provide a framework for quantum spectral methods with the potential to solve linear differential equations efficiently.
\end{abstract}

\maketitle
\raggedbottom

\section{Introduction}
\label{sec:intro}

A common workflow in quantum algorithms for differential equations~\cite{Clader2013,Berry2014DE,Montanaro2016,Berry2017,Childs2020,Childs2021,Liu2021,Krovi2023,Berry2024} involves discretizing the equation, embedding the discretized operator in a unitary circuit as a block encoding~\cite{Low2019,Gilyen2019,Sunderhauf2024}, and preparing a state proportional to the solution with a quantum linear-system algorithm~\cite{Harrow2009,Childs2017,Costa2022,Dalzell2024}. Its cost grows with the condition number of the discretized operator and with the cost of the block encoding. Spectral discretizations suit this workflow: they expand the solution in a basis of orthogonal functions and converge rapidly for smooth solutions~\cite{Boyd2001,Shen2011}, so small systems often suffice. The choice of basis affects this cost in two ways. First, it determines how differential operators act on the coefficients and therefore the structure and conditioning of the linear system. Second, it determines whether function values at the associated quadrature nodes and expansion coefficients can be converted into one another efficiently. On a quantum computer, this conversion must act coherently on the amplitudes of a state, as the quantum Fourier transform does for the Fourier basis~\cite{Coppersmith1994}.

Existing quantum spectral methods typically use Fourier, sinusoidal, or Chebyshev bases, for which the quantum Fourier transform and its sine and cosine variants provide efficient transforms~\cite{Klappenecker2001,Williams2023}. The Chebyshev basis, which accommodates general nonperiodic boundary conditions, is employed in linear-system solvers~\cite{Childs2020,Childs2021}, variational models~\cite{Kyriienko2021,Paine2023}, and effective-Hamiltonian methods~\cite{Wu2025}. However, differentiation matrices in the Chebyshev basis are dense. The resulting linear systems typically have condition numbers that grow polynomially with the basis size~\cite{Childs2021,Ghoreishi2004,Olver2013}. The classical ultraspherical spectral method avoids this growth by representing each derivative in the ultraspherical (Gegenbauer) basis whose parameter equals the order of the derivative~\cite{Olver2013,Townsend2015,Olver2020,Fortunato2021}. The resulting systems are sparse, and upon diagonal preconditioning, their condition numbers can remain bounded as the system size increases~\cite{Olver2013}. A quantum version of this method needs an efficient transform for each basis. However, only the first two bases\textemdash the Chebyshev polynomials of the first and second kinds\textemdash correspond to cosine and sine transforms, for which efficient quantum implementations are known.

\begin{figure}[t]
    \centering
    \includegraphics[width=\columnwidth]{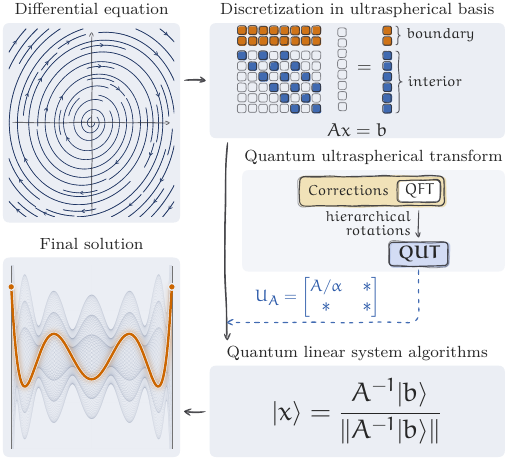}
    \caption{Discretizing a differential equation in the ultraspherical basis produces a linear system with a banded matrix, except for boundary rows. The QUT, built from a quantum Fourier transform (QFT) with corrections followed by hierarchical rotations, enables a block encoding of the matrix. It can be used by a quantum linear-system algorithm to prepare the solution state.}
    \label{fig:overview}
\end{figure}

Our first contribution is an efficient circuit for the quantum ultraspherical transform (QUT) with any fixed integer parameter $m$. The QUT maps function values at Gauss quadrature nodes to expansion coefficients in an orthonormal ultraspherical basis. For $m=0$ and $m=1$, the QUT reduces to the discrete cosine and sine transforms with efficient circuit implementations~\cite{Klappenecker2001,Williams2023}. For $m\geq 2$, Theorem~\ref{thm:efficient-qut} presents a circuit that implements the QUT to error $\varepsilon$ with a gate count polylogarithmic in the system size and $1/\varepsilon$ and a small ancilla overhead. The construction builds on the quantum Fourier transform, followed by corrections and rotational changes of basis that can be implemented efficiently. Together with the cosine and sine transforms, the QUT enables a quantum ultraspherical spectral method, as summarized in Fig.~\ref{fig:overview}.

Other efficient quantum polynomial transforms have recently been developed for Hermite polynomials~\cite{Jain2025}, integer-parameter Laguerre polynomials~\cite{Prakash2026}, and several families of hypergeometric orthogonal polynomials~\cite{Prakash2026,Guan2026}. The work~\cite{Prakash2026} also addresses the Jacobi family, which contains the ultraspherical polynomials, but its gate count scales linearly with the dimension, with additional polylogarithmic factors. To our knowledge, the QUT is the first efficient transform for a Jacobi family beyond the Chebyshev cases (Table~\ref{tab:transforms}). We also discuss how a key idea underlying the QUT extends to Jacobi polynomials with half-integer parameters.

\begin{table*}[t]
\caption{\label{tab:transforms}Quantum transforms for classical orthogonal polynomials with known circuits. Polylog denotes a gate count polynomial in $\log N$ and $\log(1/\varepsilon)$.}
\setlength{\tabcolsep}{6pt}
\renewcommand{\arraystretch}{1.2}
\begin{tabular}{@{}p{0.27\textwidth}p{0.39\textwidth}p{0.29\textwidth}@{}}
\toprule
Family & Mechanism & Gate complexity\\
\midrule
\raggedright Chebyshev, first and second kinds ($\lambda=0,1$) & \raggedright Cosine and sine transforms from the QFT~\cite{Klappenecker2001,Williams2023} & polylog\\
\raggedright Hermite & \raggedright Harmonic-oscillator evolution~\cite{Jain2025} & polylog\\
\raggedright Laguerre of integer parameter, Charlier, Meixner, Kravchuk & \raggedright Quantum optical operations~\cite{Prakash2026}; fast-forwarding in oscillator representation for Kravchuk~\cite{Guan2026} & polylog\\
\raggedright Jacobi, integer parameters $\alpha, \beta$ & \raggedright Discrete Hahn transform~\cite{Prakash2026} & linear, with extra polylog factors\\
\raggedright Ultraspherical, integer $m\geq2$ (Jacobi with $\alpha=\beta=m-\frac12$) & \raggedright 
QFT with corrections, followed by rotations (this work) & polylog (Theorem~\ref{thm:efficient-qut})\\
\bottomrule
\end{tabular}
\end{table*}

Our second contribution uses the QUT to construct a quantum ultraspherical spectral method for regular one-dimensional linear boundary-value problems. Within this method, QUTs convert functions to their expansion coefficients, enabling an efficient block encoding of the discretized system. For uniquely solvable problems with a broad class of boundary conditions, we show that its condition number, upon diagonal preconditioning, is bounded independently of the basis size (Proposition~\ref{prop:us-conditioning}). Given efficient quantum access to the problem data, the preconditioned solution state can therefore be prepared at a cost polylogarithmic in the basis size and the inverse error. Extracting useful quantities from the solution would add further costs, which determines whether the solver could provide a computational advantage in practice. We therefore do not claim an end-to-end quantum speedup for these one-dimensional problems.

The rest of the paper is organized as follows. Section~\ref{sec:qut} introduces the ultraspherical basis, defines the QUT, and states the main theorem. Section~\ref{sec:preliminaries} collects the main tools used throughout the paper. Section~\ref{sec:circuit} develops the QUT circuit and its complexity analysis. Readers primarily interested in the application may skip this section and proceed directly to Section~\ref{sec:us-solver}, where we apply the transform to a quantum ultraspherical spectral method and analyze its conditioning and complexity. Finally, Section~\ref{sec:discussion} discusses the scope of the results and outlook.

\section{Quantum ultraspherical transform}
\label{sec:qut}

This section defines the ultraspherical basis and the corresponding quantum transform before stating the main circuit guarantee and outlining its proof. Throughout the rest of the paper, we use $\|\cdot\|_2$ for the $\ell^2$ vector norm and the induced operator norm. For a scalar function $f$ on $[-1,1]$, we write $\|f\|_\infty=\sup_{x\in[-1,1]}|f(x)|$ for the supremum norm.

\subsection{Ultraspherical basis}
\label{sec:ultraspherical-basis}

The ultraspherical polynomials, also called Gegenbauer polynomials, are denoted by $C_k^{(\lambda)}(x)$. The index $k=0,1,2,\ldots$ is the polynomial degree, and $\lambda>-1/2$, $\lambda\neq0$, denotes the basis parameter. The polynomials are orthogonal on $[-1,1]$ with respect to the weight $w_\lambda(x)=(1-x^2)^{\lambda-1/2}$:
\begin{equation}
    \int_{-1}^{1} w_{\lambda}(x) C_k^{(\lambda)}(x) C_\ell^{(\lambda)}(x)\, \mathrm{d}x
    = h_k^{(\lambda)}\delta_{k\ell},
    \label{eq:gegenbauer-orthogonality}
\end{equation}
where $\delta_{k\ell}$ is the Kronecker delta and $h_k^{(\lambda)}$ is the squared norm of $C_k^{(\lambda)}$. In terms of Euler's Gamma function $\Gamma$,
\begin{equation}
    h_k^{(\lambda)} = \frac{\pi 2^{1-2\lambda}\Gamma(k+2\lambda)}
    {k!(k+\lambda)\Gamma(\lambda)^2}.
    \label{eq:gegenbauer-norm}
\end{equation}
For $\lambda>0$, the polynomials can be defined through the generating function
\begin{equation}
    (1-2xt+t^2)^{-\lambda}
    =\sum_{k=0}^{\infty}C_k^{(\lambda)}(x)t^k, \quad
    |x|, |t|<1.
    \label{eq:gegenbauer-generating}
\end{equation}

Ultraspherical polynomials constitute the symmetric subclass of the Jacobi polynomial family $P_k^{(\alpha,\beta)}$. In particular, $C_k^{(\lambda)}(x) = \frac{(2\lambda)_k}{(\lambda+1/2)_k}P_k^{(\lambda-1/2,\lambda-1/2)}(x)$, where $(\alpha)_k = \frac{\Gamma(\alpha+k)}{\Gamma(\alpha)}=\alpha(\alpha+1)\cdots(\alpha+k-1)$ denotes the rising factorial. Several classical polynomial families appear as special cases: $\lambda=\frac12$ yields the Legendre polynomials, while $\lambda=1$ corresponds to the Chebyshev polynomials of the second kind. The Chebyshev polynomials of the first kind, $T_k(x)$, arise as a limiting case:
\begin{equation}
    \lim_{\lambda\to0}\frac{k}{2\lambda}C_k^{(\lambda)}(x)=T_k(x), \quad k\geq1.
    \label{eq:chebyshev-limit}
\end{equation}

The ultraspherical basis is useful in spectral methods because differentiation maps each nonconstant basis polynomial to a scalar multiple of one polynomial in the next family. We begin by approximating the solution with a Chebyshev expansion,
\begin{equation}
    \tilde{u}_N(x) \coloneqq \sum_{k=0}^{N-1} u_k T_k(x),
    \label{eq:spectral-ansatz}
\end{equation}
where $u_k$ are the expansion coefficients. Differentiation then follows from
\begin{equation}
    \frac{\mathrm{d}T_k}{\mathrm{d}x} = \begin{cases}
        k C_{k-1}^{(1)}, &\quad k \geq 1 \\
        0, &\quad k = 0
    \end{cases}
    \label{eq:chebyshev-first-derivative}
\end{equation}
and, for $\lambda \geq 1$,
\begin{equation}
    \frac{\mathrm{d}C_k^{(\lambda)}}{\mathrm{d}x} = \begin{cases}
        2\lambda C_{k-1}^{(\lambda +1)}, &\quad k \geq 1 \\
        0, &\quad k = 0
    \end{cases}.
    \label{eq:gegenbauer-first-derivative}
\end{equation}
Iterating these identities gives the $q$th derivative of a Chebyshev polynomial,
\begin{equation}
    \frac{\mathrm{d}^q}{\mathrm{d}x^q} T_k(x) =2^{q-1}(q-1)!\,k\,C_{k-q}^{(q)}(x),
    \qquad k\geq q\geq1.
    \label{eq:solution-derivative}
\end{equation}
The derivative vanishes for $k<q$. The solution's derivatives can be written compactly as
\begin{equation}
    \frac{\mathrm{d}^{q} \tilde{u}_N }{\mathrm{d}x^{q}} =2^{q-1}(q-1)! \sum_{k=q}^{N-1} k u_k C_{k-q}^{(q)}(x).
\end{equation}
Therefore, for $1\leq q<N$, the $q$th-order differentiation matrix $D_q\in\mathbb{R}^{(N-q)\times N}$ maps Chebyshev coefficients $u_0,\dots,u_{N-1}$ to coefficients in the ultraspherical basis $C_0^{(q)},\dots,C_{N-q-1}^{(q)}$. Its first $q$ columns vanish, and the remaining nonzero entries lie on the $q$th superdiagonal,
\begin{equation}
    (D_q)_{jk}
    = 2^{q-1}(q-1)!\,k\,\delta_{j,k-q}.
    \label{eq:differentiation-matrix}
\end{equation}

A differential equation generally involves derivatives of several orders, whose coefficient representations must be expressed in a common basis. The initial conversion from the Chebyshev basis is given by $T_k(x)=\frac12 [C_k^{(1)}(x)-C_{k-2}^{(1)}(x)]$ for $k\geq1$, with $T_0=C_0^{(1)}=1$. For $\lambda\geq1$, consecutive ultraspherical families are related by
\begin{equation}
    C_k^{(\lambda)}(x)
    = \frac{\lambda}{k+\lambda}
    \left[
        C_k^{(\lambda+1)}(x)-C_{k-2}^{(\lambda+1)}(x)
    \right],
    \label{eq:gegenbauer-conversion}
\end{equation}
with the convention that $C_j^{(\lambda)}(x)=0$ for $j<0$. Accordingly, each conversion matrix contains only the main diagonal and the second superdiagonal. Composing these sparse conversion operators with the differentiation matrices yields a banded representation of the residual operator. Imposing boundary conditions introduces only a small number of dense rows~\cite{Olver2013}.

\subsection{Quantum transform}
\label{sec:quantum-transform}
The QUT uses normalized polynomials $p_k^{(\lambda)}=C_k^{(\lambda)}/\sqrt{h_k^{(\lambda)}}$. For $N\geq1$, let $x_0>x_1>\cdots>x_{N-1}$ be the zeros of $C_N^{(\lambda)}$ and let $w_0,\ldots,w_{N-1}>0$ be the corresponding Gaussian quadrature weights. These nodes and weights integrate every polynomial of degree at most $2N-1$ exactly against $w_\lambda$. We define the $N$-dimensional QUT operator $Q_N^{(\lambda)}$, with one row per polynomial degree and one column per quadrature node, by
\begin{equation}
    \bigl(Q_N^{(\lambda)}\bigr)_{kj}
    \coloneqq \sqrt{w_j}\,p_k^{(\lambda)}(x_j),
    \qquad 0\leq k,j<N.
    \label{eq:quantum-ultraspherical-transform}
\end{equation}
Since $p_k^{(\lambda)}p_\ell^{(\lambda)}$ has degree at most $2N-2$, Gaussian quadrature yields
\begin{equation}
    \sum_{j=0}^{N-1} w_j p_k^{(\lambda)}(x_j)p_\ell^{(\lambda)}(x_j)
    = \delta_{k\ell}.
    \label{eq:discrete-orthogonality}
\end{equation}
Thus $Q_N^{(\lambda)}Q_N^{(\lambda)\dagger}=I$, so $Q_N^{(\lambda)}$ is unitary on the $N$-dimensional space. In a quantum register, its input and output vectors are represented on the basis states $\ket{0},\ldots,\ket{N-1}$.

This work focuses on nonnegative integer parameters $\lambda=m$. For $m=0$, we use the Chebyshev limit $p_0^{(0)}=1/\sqrt{\pi}$ and $p_k^{(0)}=\sqrt{2/\pi}\,T_k$ for $k\geq1$, while $m=1$ corresponds to the Chebyshev polynomials of the second kind, $U_k = C_k^{(1)}$.

To interpret the QUT as a change of representation, let $f$ be a nonzero polynomial of degree at most $N-1$. It can be written in the ultraspherical polynomial basis as
\begin{equation}
    f(x) = \sum_{k=0}^{N-1} \hat f_k p_k^{(m)}(x),
    \label{eq:function-expansion}
\end{equation}
where $\widehat f_k$ is the coefficient of the $k$th basis polynomial. Collect those coefficients into the vector $\widehat{\boldsymbol f}=(\hat f_0,\ldots,\hat f_{N-1})^\T$. The discrete orthogonality property says that this coefficient vector and the vector of weighted samples, $(\sqrt{w_j}f(x_j))_{j=0}^{N-1}$, have the same Euclidean norm. Thus, after dividing by $\|\widehat{\boldsymbol f}\|_2$, each can be viewed as a normalized quantum state. One state encodes the values of $f$ at the sample points; the other encodes its expansion coefficients in the ultraspherical basis:
\begin{equation}
    \begin{aligned}
        \ket{\boldsymbol{f}^{(m)}}
        & \coloneqq \frac{1}{\lVert\widehat{\boldsymbol f}\rVert_2}
        \sum_{j=0}^{N-1}\sqrt{w_j}\,f(x_j)\ket{j}, \\
        \ket{\widehat{\boldsymbol{f}}^{(m)}}
        &\coloneqq\frac{1}{\lVert\widehat{\boldsymbol f}\rVert_2}
        \sum_{k=0}^{N-1} \hat f_k\ket{k}.
    \end{aligned}
    \label{eq:value-coefficient}
\end{equation}
The transform and its inverse then satisfy
\begin{equation}
    Q_N^{(m)}\ket{\boldsymbol{f}^{(m)}}
    =\ket{\widehat{\boldsymbol{f}}^{(m)}},
    \qquad
    Q_N^{(m)\dagger}\ket{\widehat{\boldsymbol{f}}^{(m)}}
    =\ket{\boldsymbol{f}^{(m)}}.
    \label{eq:transform-pair}
\end{equation}

The first two cases, $m=0$ and $m=1$, reduce to standard discrete cosine and sine transforms. For $m=0$,
\begin{equation}
    Q_{kj}^{(0)} =
    \begin{cases}
        1/\sqrt{N}, & k=0,\\
        \sqrt{2/N}\cos\!\left[\dfrac{k(j+1/2)\pi}{N}\right],
        & 1\leq k<N,
    \end{cases}
    \label{eq:dct-case}
\end{equation}
which is the discrete cosine transform of type II (DCT-II).  For $m=1$, the identities $C_k^{(1)}=U_k$ and $x_j=\cos[(j+1)\pi/(N+1)]$ yield
\begin{equation}
    Q_{kj}^{(1)}
    = \sqrt{\frac{2}{N+1}}
    \sin\! \left[ \frac{(k+1)(j+1)\pi}{N+1} \right],
    \label{eq:dst-case}
\end{equation}
which is the discrete sine transform of type I (DST-I). Efficient quantum circuits are known for both transforms~\cite{Klappenecker2001}, and the DCT-II has also been implemented as the quantum Chebyshev transform~\cite{Williams2023}.

Our main result is an efficient implementation of the QUT for every fixed integer $m\geq2$. The following theorem states the circuit guarantee, followed by a proof sketch. A complete proof is given in Section~\ref{sec:circuit-assembly}, where we assemble the component circuits and establish the error and resource bounds.

\subsection{Main result}
\label{sec:main-result}

For the circuit construction, we work in an enlarged space of dimension $K=N+m$. Taking $K=2^n$ allows this space to be encoded on $n$ qubits and manipulated using quantum circuits.

\begin{theorem}[Efficient quantum ultraspherical transform]
\label{thm:efficient-qut}

Fix an integer $m\geq2$ and a dimension $N\geq1$ with $K=N+m=2^n$. Let $0<\varepsilon<1/2$ be the target error. There is a unitary circuit $U=U_{m,N,\varepsilon}$ on $n$ system
qubits and $a$ ancillary qubits such that, for every normalized
input $\ket\psi\in\mathbb C^K$ supported on
$\ket0,\ldots,\ket{N-1}$,
\begin{equation}
\bigl\|U\ket{0^a}\ket\psi
-\ket{0^a}Q_N^{(m)}\ket\psi\bigr\|_2
\leq\varepsilon.
\label{eq:efficient-qut-guarantee}
\end{equation}
This uniform bound is an operator-norm guarantee on the valid
input subspace, including leakage into unused output labels or
nonzero ancillas. The adjoint circuit implements
$Q_N^{(m)\dagger}$ with the same error and resource bounds.

The circuit $U$ can be implemented using
\begin{equation}
    \begin{aligned}
    G_m(n,\varepsilon)
    &=O_m\!\Bigl(
    n^3\log\frac{n}{\varepsilon}
    +n\log^2\frac1\varepsilon
    \Bigr)
    \end{aligned}
\label{eq:efficient-qut-gates}
\end{equation}
single-qubit rotations and CNOT gates, together with
\begin{equation}
a=O_m\!\left(
\log n+\log\log\frac1\varepsilon
\right)
\label{eq:efficient-qut-ancillas}
\end{equation}
ancillary qubits, where $O_m$ allows constants that depend on the fixed parameter $m$. All ancillas start and return to $\ket{0^a}$ within the error in Eq.~\eqref{eq:efficient-qut-guarantee}. Circuit parameters can be computed and stored classically in time polynomial in $n$ and $\log(1/\varepsilon)$ before quantum execution.
\end{theorem}

\subsection{Proof sketch}
\label{sec:proof-sketch}
We construct the QUT from the quantum Fourier transform (QFT) together with additional circuits of polylogarithmic complexity. Writing $x=\cos\theta$, we use the fact that, for integer $m$, multiplying an ultraspherical polynomial by $\sin^m\theta$ yields a finite sine or cosine expansion. This relates the QUT to discrete trigonometric transforms evaluated on an equispaced grid in $\theta$, which can be implemented efficiently through the QFT. The remaining steps correct the sampling points and quadrature weights and convert the trigonometric coefficients into ultraspherical coefficients.

To correct the sampling points, we expand each sine or cosine about a nearby point on the equispaced \(\theta\) grid. The Taylor terms depend on the product $p\delta$, where $p$ is the Fourier-mode index and $\delta$ is the angular displacement. With $K=N+m$, we prove that $|\delta|=O_m(K^{-1})$, while $p<K$, so $p\delta$ remains bounded as $N$ grows. The factorial decay of the Taylor remainder therefore allows the truncation order to depend only on the target accuracy, not on $N$. The quadrature-weight corrections are uniformly bounded as well.

A direct table of the corrections would require $O(N)$ data, so we instead approximate them as functions of the node label. Because these correction functions vary most rapidly near the first and last few labels, which belong to the nodes closest to the endpoints $x = \pm 1$, we handle a fixed number of endpoint labels directly and partition the remaining labels into intervals whose lengths double away from each endpoint. After rescaling each interval to $[-1,1]$, we show that the corresponding correction functions remain analytic and uniformly bounded on a fixed complex neighborhood, independently of $N$. This yields polynomial approximations whose degree depends only on the Taylor order and target accuracy. We block encode the rescaled labels and use quantum eigenvalue transformation (QET) to apply these polynomial approximations in superposition. All polynomial coefficients and circuit angles are computed classically beforehand. The input-correction stage uses approximately $O(n^2\log n)$ gates, up to constants depending on $m$. We omit the dependence on error tolerance in this proof sketch for brevity.

For the trigonometric-to-ultraspherical basis change, we factor the transformation into $\lfloor m/2 \rfloor$ stages. Each stage acts separately on even and odd Fourier modes and uses a hierarchy of rotations in two passes. With each mode assigned a specific weight, the first pass forms weighted averages and contrasts of neighboring mode groups, while the second adjusts them into the combinations that map trigonometric coefficients to ultraspherical coefficients. Each pass therefore consists of $O(n)$ layers of two-dimensional rotations acting on selected pairs of amplitudes. The rotation angles depend smoothly on the mode labels and can be approximated by polynomials whose degree depends only on the target accuracy, and we again use QET to apply them coherently in superposition. Because the QET implementation introduces a fixed subnormalization factor, we amplify each layer before applying the next so that these factors do not accumulate through the hierarchy. For fixed $m$, only a constant number of amplification rounds is required per layer, and the full basis change uses approximately $O(n^3\log n)$ gates.

Finally, we combine the Taylor terms using the linear combination of unitaries (LCU) method. An auxiliary register stores the Taylor order $s$ and selects the corresponding input correction and Fourier block, while the basis change is shared across all orders. The resulting circuit block-encodes the approximate QUT with an overall subnormalization factor, which is known and bounded by a function of $m$ alone, so a constant number of final amplitude-amplification rounds suffices to obtain the full transform.

\section{Preliminaries}
\label{sec:preliminaries}

This section briefly reviews the quantum algorithmic tools used in the QUT construction, including block encodings, linear combinations of unitaries, quantum eigenvalue transformation, and oblivious amplitude amplification.

\subsection{Block encoding}
\label{sec:block-encoding}

A block encoding represents a possibly nonunitary operator as a block of a larger unitary circuit. Let $A$ act on the system register, and let $U$ act on the system together with $a$ ancillary qubits. The unitary $U$ is an $(\alpha,a,\varepsilon)$ block encoding of $A$ if $\alpha>0$, $\varepsilon\geq0$, and~\cite{Gilyen2019}
\begin{equation}
    \left\|A-\alpha(\bra{0^a}\otimes I)U
       (\ket{0^a}\otimes I)\right\|_2\leq\varepsilon,
    \label{eq:prelim-block-encoding}
\end{equation}
where $I$ is the identity on the system register. The parameter $\varepsilon$ bounds the error in $A$, so the corresponding error in the normalized block is $\varepsilon/\alpha$. For an exact encoding, $\varepsilon=0$ and the encoded block is $A/\alpha$, which requires $\alpha\geq\|A\|_2$. We refer to $\alpha$ as the subnormalization factor.

For a normalized input $\ket\psi$, an exact block encoding applied to $\ket{0^a}\ket\psi$ contains the component $\ket{0^a}A\ket\psi/\alpha$ in the zero-ancilla subspace. Measuring the ancillas therefore returns zero with probability $\|A\ket\psi\|_2^2/\alpha^2$. When $A$ is unitary and $\alpha$ is known, oblivious amplitude amplification can increase this success probability to one~\cite{Berry2014}.

The linear combination of unitaries (LCU) method combines encoded operators into a weighted sum~\cite{Berry2015,Childs2017}. Suppose each unitary $V_j$ has encoded block $B_j$, with all circuits using the same ancillary space. For weights $c_j\geq0$ and $\alpha=\sum_j c_j>0$, define $\operatorname{PREP}$ and $\operatorname{SELECT}$ by
\begin{equation}
    \operatorname{PREP}\ket0
    =\sum_j\sqrt{c_j/\alpha}\ket j
\end{equation}
and
\begin{equation}
    \operatorname{SELECT}
    =\sum_j\ket j\bra j\otimes V_j.
\end{equation}
Applying $\operatorname{PREP}$, then $\operatorname{SELECT}$, and finally $\operatorname{PREP}^\dagger$ produces the encoded block $\alpha^{-1}\sum_j c_j B_j$ when both the selector and component ancillas are selected in zero. Signs and phases in the coefficients can be absorbed into the controlled circuits $V_j$.

Products of block-encoded operators can be implemented using separate ancillary registers. If $U_1$ and $U_2$ encode $A_1/\alpha_1$ and $A_2/\alpha_2$, respectively, then applying $U_1$ followed by $U_2$ encodes $A_2A_1/(\alpha_2\alpha_1)$~\cite{Gilyen2019}. Each circuit acts on the system and its own ancillas, with all ancillary registers selected in zero at the input and output.

\subsection{Quantum eigenvalue transformation}
\label{sec:qet}
\label{sec:polynomial-approximation}

Quantum eigenvalue transformation (QET) applies a polynomial to the eigenvalues of a Hermitian operator while preserving its eigenvectors~\cite{Low2019,Gilyen2019}. Let $H=\sum_k\lambda_k\ket{u_k}\bra{u_k}$ be Hermitian with $\|H\|_2\leq1$, where the vectors $\ket{u_k}$ form an orthonormal eigenbasis. We call $H$ the \emph{signal operator}, since QET acts on its eigenvalues. Given a block encoding of $H$, QET constructs a block encoding of
\begin{equation}
    P(H)=\sum_k P(\lambda_k)\ket{u_k}\bra{u_k},
\end{equation}
up to a known subnormalization factor. We consider real polynomials $P$ satisfying $\|P\|_\infty\leq1$. QET is the Hermitian case of quantum singular value transformation (QSVT).

We describe the phase sequence using an exact $(1,a,0)$ block encoding $U_H$ that is also Hermitian, so $U_H^2=I$.
An arbitrary exact block encoding of $H$ can be put in this form using one additional ancilla and controlled calls to the encoding and its adjoint~\cite{Low2019,Gilyen2019}.
Let $\Pi=\ket{0^a}\bra{0^a}\otimes I$ project onto the zero-ancilla subspace.
For real phase angles $\Phi=(\phi_0,\ldots,\phi_d)$, the QET sequence is
\begin{equation}
    U_\Phi = e^{\ii\phi_0(2\Pi-I)}U_H
                 e^{\ii\phi_1(2\Pi-I)} \cdots U_H e^{\ii\phi_d(2\Pi-I)},
    \label{eq:qet-sequence}
\end{equation}
which uses $d$ calls to $U_H$ and $d+1$ phase rotations.
Its encoded block is a generally complex polynomial in $H$ of degree at most $d$.
The polynomial is even when $d$ is even and odd when $d$ is odd.

For a real polynomial $P$ of degree at most $d$ with this parity and $\|P\|_\infty\leq1$, the phases can be chosen so that the real part of the encoded polynomial equals $P$~\cite{Gilyen2019}.
On each eigenspace, replacing $\Phi$ by $-\Phi$ conjugates the encoded polynomial. A selector qubit coherently selects between the QET sequences with phases $\Phi$ and $-\Phi$. Applying a Hadamard before and after the selection extracts their average, whose encoded block is $P(H)$.

To implement a polynomial containing both parities, decompose it as $P=P_{\mathrm e}+P_{\mathrm o}$, where
\begin{equation}
    \begin{aligned}
        P_{\mathrm e}(x)&=\frac{P(x)+P(-x)}2,\\
        P_{\mathrm o}(x)&=\frac{P(x)-P(-x)}2.
    \end{aligned}
\end{equation}
Both parts are bounded by one on $[-1,1]$. After constructing their real blocks as above, a second selector qubit combines them with equal LCU weights. The resulting circuit $U_P$ satisfies
\begin{equation}
    (\bra0\otimes I)U_P(\ket0\otimes I)
       =\frac{P(H)}2,
    \label{eq:main-polynomial-block}
\end{equation}
where $\ket0$ denotes the joint all-zero state of the block-encoding ancillas and the two selector qubits.
The factor $1/2$ comes from averaging $P_{\mathrm e}(H)$ and $P_{\mathrm o}(H)$.
For a polynomial of definite parity, this average is unnecessary, and the encoded block is $P(H)$.
In either case, a degree-$d$ transformation uses $O(d)$ calls to $U_H$.

\paragraph*{Dyadic intervals and diagonal signals.}
For a function of a binary label, a useful choice of $H$ is a diagonal operator whose entries are rescaled labels. If the function varies on different scales across the label range, we can approximate it separately on dyadic intervals. For an $n$-qubit register, define
\begin{equation}
    I_\nu=\{2^\nu,\ldots,2^{\nu+1}-1\}, \qquad 1\leq\nu\leq n-1.
\end{equation}
For $b\geq2$, with bit positions numbered from zero, any $b\in I_\nu$ has its highest nonzero bit at position $\nu$, while the lower $\nu$ bits encode the offset $v=b-2^\nu$. Thus the binary representation identifies both the interval and the position within it. We rescale this offset to $[-1,1]$ by
\begin{equation}
    x_{\nu,v}=\frac{2v-2^\nu+1}{2^\nu-1},
    \qquad 0\leq v<2^\nu.
\end{equation}
For example, the labels in $I_2=\{4,5,6,7\}$ map to $-1,-1/3,1/3,1$. The corresponding diagonal signal operator is
\begin{equation}
    B_\nu=\diag(x_{\nu,v})
       =-\sum_{a=0}^{\nu-1}\frac{2^a}{2^\nu-1}Z_a,
    \label{eq:qet-local-signal}
\end{equation}
where $Z_a\ket v=(-1)^{v_a}\ket v$ and $v=\sum_a2^av_a$. Since the weights $2^a/(2^\nu-1)$ sum to one, LCU offers an exact block encoding with subnormalization one. We prepare a selector register with amplitudes $\sqrt{2^a/(2^\nu-1)}$, apply $-Z_a$ when the selector is in state $a$, and undo the preparation. The selector uses $\lceil\log_2\nu\rceil=O(\log n)$ qubits, with unused selector states assigned the identity. The resulting block encoding is Hermitian and diagonal in the label basis, leaving the label $b$ unchanged. Applying QET to this encoding implements $P(B_\nu)=\diag(P(x_{\nu,v}))$ with the subnormalization in Eq.~\eqref{eq:main-polynomial-block} when both parities are used.

The labels $b \in \{0,1\}$ are handled separately by operations controlled on the label. The same treatment applies to any additional labels excluded from the polynomial approximation.

\paragraph*{Bounded polynomial approximation.}
To implement a target function $f$, we need a polynomial that approximates $f$ and satisfies the QET bound on the whole interval $[-1,1]$.
Suppose $f$ is real, $\|f\|_\infty\leq1$, and a polynomial $p_d$ has uniform error $\|p_d-f\|_\infty\leq\delta$.
The rescaled polynomial $P=p_d/(1+\delta)$ then satisfies
$\|P\|_\infty\leq1$ and $\|P-f\|_\infty\leq\frac{2\delta}{1+\delta}\leq2\delta$. By the spectral decomposition of $H$, we have the error bound $\|P(H)-f(H)\|_2 \leq 2\delta$. If $f$ is needed only on a subinterval containing the spectrum of $H$, the approximation error need only be controlled there, but $P$ must still be bounded by one throughout $[-1,1]$.

For analytic functions, Chebyshev interpolation converges exponentially with the polynomial degree. Let $E_\varrho$, $\varrho>1$, be the closed Bernstein ellipse whose boundary is $z=(w+w^{-1})/2$ for $|w|=\varrho$. If $f$ is analytic in a neighborhood of $E_\varrho$ and bounded there by $M$, then its degree-$d$ interpolant at the Chebyshev nodes $x_j=\cos(j\pi/d)$, $j=0,\ldots,d$, satisfies~\cite[Theorem~8.2]{Trefethen2013}
\begin{equation}
    \|f-p_d\|_\infty \leq\frac{4M\varrho^{-d}}{\varrho-1}, \qquad d\geq1.
\end{equation}
Thus, for fixed $M$ and $\varrho$, an error at most $\delta$ requires $d=O(\log(1/\delta))$. If the rescaled functions on all dyadic intervals satisfy the same bounds, the same degree works on every interval.

\paragraph*{Coherent selection of phase sequences.}
The QET phase angles $\phi_0,\dots,\phi_d$ are computed classically from the polynomial coefficients~\cite{Haah2019,Dong2021}. Several polynomials can use the same signal block encoding and differ only in their phase angles. Consider a family of $L$ polynomials $P_j$, $j=0,\ldots,L-1$, with the same parity and phase sequences of equal length. At each QET step, an index register storing $j$ selects the phase angle associated with $P_j$, while all branches share the same signal call. Because the index register is preserved, the selected polynomial is applied coherently even when $j$ is in superposition.

The phase selection is implemented with uniformly controlled rotations, where each control value specifies one rotation angle~\cite{Mottonen2004}. With $p=\lceil\log_2 L\rceil$ control bits, a general uniformly controlled rotation requires $O(2^p)$ single-qubit rotations and CNOT gates. Selecting among $L$ phase sequences therefore adds $O(L)$ gates per phase step, in addition to the shared signal calls.

This coherent phase selection can be combined with the dyadic construction above to apply different polynomial approximations on different intervals. We apply QET to the dyadic intervals one at a time. For each interval, a flag qubit marks whether the current label belongs to that interval and controls the corresponding phase sequence. Because the label register is unchanged, the flag can be uncomputed and reused for the next interval. Each interval circuit acts as the identity outside its assigned label set. Since the intervals are disjoint, the resulting operator is a direct sum over the intervals, so its operator-norm error is the maximum interval error while the gate counts add. Within each interval, an additional control register can similarly select among several polynomials $P_j$ through their phase angles, giving a direct sum over both interval labels and polynomial indices.

\subsection{Oblivious amplitude amplification}
\label{sec:amplification-primitive}

The constructions above produce the desired operator with a known subnormalization factor. When the target operator is unitary, oblivious amplitude amplification (OAA) can remove this factor coherently~\cite{Berry2014,Berry2015}. OAA uses the encoding circuit, its inverse, and reflections about designated input and output subspaces. Because these reflections are independent of the particular input state, the procedure works for an arbitrary state in the input subspace.

Let $W$ be a unitary circuit, and let $\Pi_{\mathrm{in}}$ and $\Pi_{\mathrm{out}}$ project onto its input and successful output subspaces.
Suppose that $V$ maps these subspaces unitarily and that, for every normalized state $\ket\psi$ in the input subspace,
\begin{equation}
    \Pi_{\mathrm{out}} W\ket\psi = \frac1\alpha V\ket\psi, \qquad \alpha\geq1.
\end{equation}
The success probability is therefore $1/\alpha^2$, independently of $\ket\psi$.
For an ordinary block encoding with $a$ ancilla qubits, both projectors are $\ket{0^a}\bra{0^a}\otimes I$.

To amplify the successful component, define the reflections $S_{\mathrm{in}}=I-2\Pi_{\mathrm{in}}$ and $S_{\mathrm{out}}=I-2\Pi_{\mathrm{out}}$.
After $t\geq0$ amplification rounds, the circuit is
\begin{equation}
    \mathcal U_t=\bigl(-WS_{\mathrm{in}}W^\dagger S_{\mathrm{out}}\bigr)^tW.
\end{equation}
With $\theta=\arcsin(1/\alpha)$, we obtain
\begin{equation}
    \Pi_{\mathrm{out}}\mathcal U_t\ket\psi
       =\sin\bigl((2t+1)\theta\bigr)V\ket\psi.
\end{equation}
The same change in amplitude occurs for every input state. The circuit uses $2t+1$ calls to $W$ or $W^\dagger$ and $2t$ reflections.

For a known $\alpha$, we can adjust the initial amplitude so that an integer number of rounds achieves unit success probability.
Choose $t\geq1$ such that
\begin{equation}
    \alpha\leq\beta_t,
    \qquad \beta_t=\frac1{\sin\bigl(\pi/(4t+2)\bigr)}.
    \label{eq:amplifiable-amplitude}
\end{equation}
Prepare an additional ancilla with amplitude $\alpha/\beta_t$ on $\ket0$, and include its zero state in both the input and successful output projectors. The adjusted circuit $W_t$ then encodes $V/\beta_t$.
Applying the same amplification procedure to $W_t$ implements $V$ exactly, since $(2t+1)\arcsin(1/\beta_t)=\pi/2$. Choosing the smallest such $t$ requires $O(\alpha)$ circuit calls; a constant normalization factor therefore requires only a constant number of rounds.

The construction remains stable under small errors in the encoded operator~\cite{Berry2015,Gilyen2019}.
Suppose $W_t$ encodes $\widehat V/\beta_t$ with error $\|\widehat V-V\|_2\leq\Delta$. For fixed $t\geq1$ and exact circuit operations, the amplified output differs from the ideal output by $O(\Delta)$ in vector norm, uniformly over normalized inputs, including any leakage outside the successful output subspace. Errors in the encoding circuit and reflections contribute separately and add over their occurrences.

\section{Construction of the QUT}
\label{sec:circuit}

\subsection{Operator factorization}
\label{sec:factorization}
For $m\geq2$, the zeros $x_0,\dots,x_{N-1}$ of $C_N^{(m)}$, which we call the Gauss--ultraspherical nodes,\footnote{Also known as Gauss--Gegenbauer nodes.} do not coincide with the sampling points of a standard discrete sine or cosine transform. The key observation is that, after writing $x=\cos\theta$ and multiplying by an angular weight, each orthonormal ultraspherical polynomial becomes a finite trigonometric series. We can therefore evaluate these series on an auxiliary equispaced grid in $\theta$ using a discrete sine or cosine transform, and then shift the sampled values to the angles corresponding to $x_0,\dots,x_{N-1}$. A Taylor expansion performs this shift, while additional factors account for the quadrature weights. Truncating the expansion provides an approximate factorization into a trigonometric transform, a change from trigonometric to ultraspherical coefficients, and diagonal corrections for the sampling points and weights.

Throughout this section, we follow the assumptions of Theorem~\ref{thm:efficient-qut}, in particular $m\geq2$ and $K=N+m=2^n$. Writing $x=\cos\theta$, we define the angular functions
\begin{equation}
    G_k^{(m)}(\theta)
    \coloneqq
    \sin^m\!\theta\,p_k^{(m)}(\cos\theta).
    \label{eq:angular-functions}
\end{equation}
The factor $\sin^m\theta$ incorporates the integration weight, so the functions $G_k^{(m)}$ are orthonormal with respect to $\mathrm d\theta$ on $[0,\pi]$. For integer $m$, each $G_k^{(m)}$ is a trigonometric polynomial. Since $p_k^{(m)}(\cos\theta)$ is even in $\theta$, $G_k^{(m)}$ has a finite cosine expansion when $m$ is even and a finite sine expansion when $m$ is odd. For $0\leq k<N$, its highest frequency is $k+m<K$.

This finite frequency support allows exact evaluation on an equispaced angular grid. For even integers $m\geq2$, we use
\begin{equation}
    \widetilde\theta_\ell
    =\frac{(\ell+1/2)\pi}{K},
    \qquad 0\leq\ell\leq K-1,
\end{equation}
together with the cosine basis
\begin{equation}
    e_p(\theta)=
    \begin{cases}
        1/\sqrt{\pi}, & p=0,\\
        \sqrt{2/\pi}\cos(p\theta), & p=1,\dots,K-1,
    \end{cases}
    \label{eq:cosine-basis}
\end{equation}
which is orthonormal in $L^2([0,\pi],\mathrm d\theta)$. For odd integers $m\geq3$, we instead use
\begin{equation}
    \widetilde\theta_\ell=\frac{\ell\pi}{K},
    \qquad 1\leq\ell\leq K-1,
\end{equation}
with the orthonormal sine basis
\begin{equation}
    e_p(\theta)
    =\sqrt{2/\pi}\sin(p\theta),
    \qquad p=1,\dots,K-1.
    \label{eq:sine-basis}
\end{equation}
Sampling the trigonometric basis on these grids defines the sine or cosine transform matrix $\mathcal T=(\mathcal T_{p\ell})$,
\begin{equation}
    \mathcal T_{p\ell}
    =
    \sqrt{\frac{\pi}{K}}\,e_p(\widetilde\theta_\ell).
    \label{eq:sinusoidal-transform}
\end{equation}
Discrete orthogonality makes $\mathcal T$ unitary in dimension $K$ for even $m$ and $K-1$ for odd $m$. It maps grid samples to trigonometric coefficients.

To change from trigonometric coefficients to the angular basis $\{G_k^{(m)}\}$, define $L=(L_{kp})$ by
\begin{equation}
    L_{kp}
    =\int_0^\pi
    G_k^{(m)}(\theta)e_p(\theta)\,\mathrm d\theta,
    \qquad 0\leq k<N.
    \label{eq:frequency-to-degree}
\end{equation}
The entries $L_{kp}$ are the trigonometric expansion coefficients of $G_k^{(m)}$. Evaluating these expansions on the equispaced grid gives
\begin{equation}
    (L\mathcal T)_{k\ell} =
    \sqrt{\frac{\pi}{K}}\,
    G_k^{(m)}(\widetilde\theta_\ell).
    \label{eq:uniform-factorization}
\end{equation}
Since the orthonormal functions $G_k^{(m)}$ have exact expansions in the orthonormal basis $\{e_p\}$, their coefficient vectors, which form the rows of $L$, are orthonormal. Hence $LL^\dagger=I_N$. Since $L$ has fewer rows than columns, its circuit implementation requires a unitary completion. Section~\ref{sec:helmert-circuit} constructs this completion through successive rotations, while Appendix~\ref{app:weighted-contrasts} derives the corresponding coefficients.

It remains to shift the equispaced angles to the Gauss--ultraspherical nodes. Let $\theta_j\coloneqq\arccos x_j$ and pair each node with the grid index $\ell_j\coloneqq j+\lceil m/2\rceil$. The corresponding angular displacement and weight correction are
\begin{equation}
    \delta_j= \theta_j-\widetilde\theta_{\ell_j},
    \qquad
    \xi_j=
    \sqrt{\frac{K w_j}{\pi\sin^{2m}\theta_j}}.
    \label{eq:node-corrections}
\end{equation}

\begin{figure}[t]
    \centering
    \includegraphics[width=0.45\textwidth]{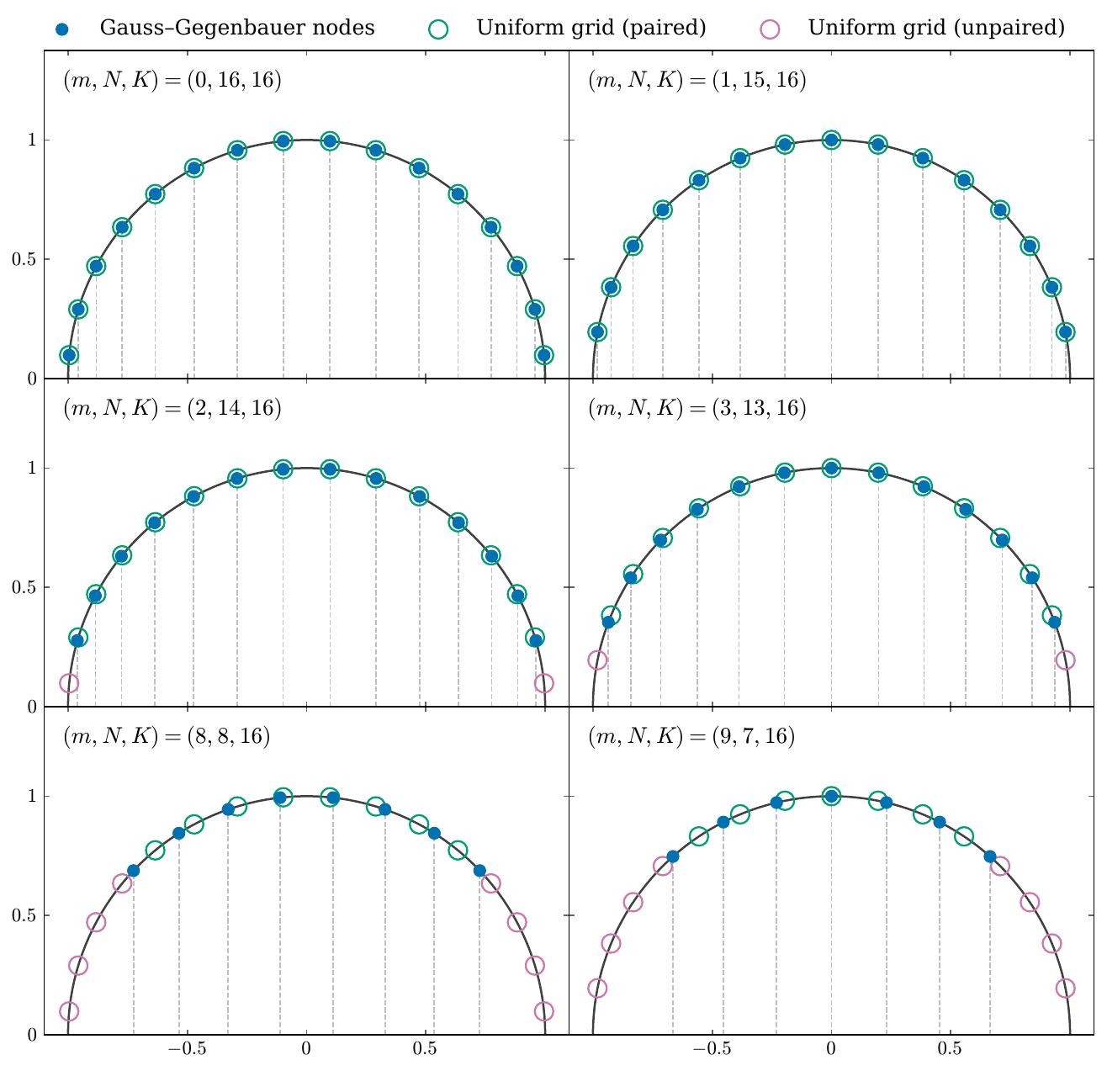}
    \caption{Gauss--ultraspherical nodes and paired uniform-grid points.  Each panel specifies $(m,N,K)$ with $K=N+m$ at fixed $K = 16$ and places its markers at $(\cos\theta,\sin\theta)$ on the upper unit semicircle.  Filled markers show the Gauss nodes; green rings show their paired uniform-grid points, and pink rings show unpaired grid points.  For $m=0$ and $m=1$, the Gauss nodes coincide with the uniform grid. The separation becomes visible as $m$ increases, with the largest mismatches occurring near the endpoints.}
    \label{fig:node-grid-pairing}
\end{figure}

This constant-shift pairing matches the leading bulk location of the Gauss nodes. For fixed $m$, Jacobi zero asymptotics imply $\theta_j=[j+(m+1)/2]\pi/K+O(K^{-2})$ when $j/N$ is bounded away from $0$ and $1$~\cite{DLMF}, while the parity-dependent grid places $\widetilde\theta_{j+\lceil m/2\rceil}$ exactly at this leading-order angle. By reflection symmetry, this choice minimizes the maximum angular displacement among constant-shift pairings $\ell_j=j+c$. Near the endpoints, where the bulk asymptotics are not uniform, a nonconsecutive nearest-grid pairing can achieve a smaller maximum displacement. The constant shift is nevertheless preferable for the circuit because it reduces the input relabeling to a fixed addition while keeping the bulk corrections small. Figure~\ref{fig:node-grid-pairing} illustrates the two angle sets and their pairing, with the largest angular mismatches $|\delta_j|$ occurring near the endpoints.

Both corrections in Eq.~\eqref{eq:node-corrections} are determined by a single phase function. At each node angle $\theta_j$, the phase is an integer multiple of $\pi$, while its derivative determines the corresponding quadrature weight. Appendix~\ref{app:phase-arithmetic} constructs this continuous phase function, denoted by $\Phi_K^{(m)}(\theta)$, and proves
\begin{equation}
    \Phi_K^{(m)}(\theta_j) =(j+m)\pi,
    \qquad
    \xi_j^2 =\frac{K}{\Phi_K^{(m)\prime}(\theta_j)}.
    \label{eq:phase-corrections}
\end{equation}
Thus, inverting the phase function gives the node angle $\theta_j$ and hence the displacement $\delta_j$, while its derivative determines the weight correction $\xi_j$.

Next, collect the correction quantities in the diagonal matrices $D_\delta \coloneqq \diag(\delta_j)$ and $\Xi \coloneqq \diag(\xi_j)$. For each derivative order $s\geq0$, define the differentiated sampling matrix $E_s$ by
\begin{equation}
    (E_s)_{pj} = \sqrt{\frac{\pi}{K}} \left. \frac{\mathrm{d}^se_p(\theta)}{\mathrm{d}\theta^s} \right\vert_{\theta = \widetilde{\theta}_{\ell_j}}.
    \label{eq:differentiated-sampling}
\end{equation}
Expanding $G_k^{(m)}(\theta)=\sum_p L_{kp}e_p(\theta)$ about $\widetilde\theta_{\ell_j}$ and using $\theta_j=\widetilde\theta_{\ell_j}+\delta_j$ yields
\begin{equation}
    \begin{aligned}
    (Q_N^{(m)})_{kj}
    &=\xi_j\sqrt{\frac{\pi}{K}}\,G_k^{(m)}(\theta_j)\\
    &=\xi_j\sqrt{\frac{\pi}{K}}
    \sum_{s=0}^{\infty}\frac{\delta_j^s}{s!}
    \sum_p L_{kp} \left. \frac{\mathrm{d}^se_p(\theta)}{\mathrm{d}\theta^s} \right\vert_{\theta = \widetilde{\theta}_{\ell_j}} \\
    &=\sum_{s=0}^{\infty}\frac{\xi_j \delta_j^s (LE_s)_{kj}}{s!}.
    \end{aligned}
\end{equation}
Right multiplication by $D_\delta^s\Xi$ will provide the factors $\delta_j^s\xi_j$. Truncating the Taylor series at order $r$ therefore results in
\begin{equation}
    \widetilde{Q}_{N,r}^{(m)} \coloneqq L\left(\sum_{s=0}^{r}\frac{E_sD_\delta^s}{s!}\right)\Xi.
    \label{eq:approximate-factorization}
\end{equation}

Reading Eq.~\eqref{eq:approximate-factorization} from right to left,
$\Xi$ corrects the quadrature weights, $D_\delta^s$ supplies the angular displacements, $E_s$ samples derivatives of the trigonometric basis, and $L$ maps trigonometric coefficients to ultraspherical coefficients. Each $E_s$ can be implemented from a Fourier transform together with frequency-dependent factors and phases. The retained Taylor terms are implemented in superposition and combined via LCU.

We next bound the truncation error in operator norm. Define
\begin{equation}
    \tau\coloneqq K\max_j|\delta_j|,
    \qquad
    \xi_{\max}\coloneqq\max_j\xi_j.
    \label{eq:factorization-parameters}
\end{equation}
Since the rows of $L$ are orthonormal, $\|L\|_2=1$. To bound $\|E_s\|_2$, note that taking $s$ derivatives of a sine or cosine mode multiplies it by a factor of magnitude $p^s$. Without these frequency factors, the corresponding sampling matrices have operator norm at most one. For even $m$, the cosine rows and the nonzero sine rows sampled at
$\widetilde\theta_\ell=(\ell+1/2)\pi/K$ are orthonormal. For odd $m$, the sine rows sampled at $\widetilde\theta_\ell=\ell\pi/K$ are orthonormal, while the cosine rows with $1\leq p<K$ become orthonormal after adding the endpoint samples at $\theta=0$ and $\theta=\pi$, each scaled by $1/\sqrt2$. Removing these endpoint columns, and then selecting only the paired grid columns, cannot increase the operator norm. Since every frequency satisfies $0\leq p<K$, $\|E_s\|_2\leq(K-1)^s\leq K^s$. Hence
\begin{equation}
    \begin{aligned}
        \left\lVert Q_N^{(m)}-\widetilde{Q}_{N,r}^{(m)}\right\rVert_2
        &\leq \xi_{\max}\sum_{s=r+1}^{\infty}\frac{\tau^s}{s!}\\
        &\leq \xi_{\max}e^\tau\frac{\tau^{r+1}}{(r+1)!}.
    \end{aligned}
    \label{eq:factorization-error}
\end{equation}
Thus, when the scaled displacements $K\delta_j$ and weight corrections $\xi_j$ are bounded, the error decreases factorially with $r$. Choosing $r$ so that the error in Eq.~\eqref{eq:factorization-error} is at most $\varepsilon$ yields
\begin{equation}
    r = O\!\left(\frac{\log(1/\varepsilon)}
    {\log\log(1/\varepsilon)}\right),
    \label{eq:taylor-order-scaling}
\end{equation}
independently of the transform dimension. Appendix~\ref{app:taylor-order} shows explicit sufficient choices. When $\tau=0$, the zeroth-order term is exact.

To normalize the factors for block encoding, choose bounds $T\geq\tau$ and $X\geq\xi_{\max}$ and define
\begin{equation}
    \Gamma_s = \diag(\gamma_{s,j}), \qquad
    \gamma_{s,j} =\frac{\xi_j}{X}\left(\frac{K\delta_j}{T}\right)^s.
    \label{eq:combined-correction-amplitude}
\end{equation}
By construction, $\|\Gamma_s\|_2\leq1$. Since $\|E_s\|_2\leq(K-1)^s$, we normalize the derivative block by
$(K-1)^s$ and absorb this scale into the Taylor coefficient. Define
\begin{equation}
    \lambda_{\mathrm{freq}}=\frac{K-1}{K},
    \qquad c_s=\frac{X(\lambda_{\mathrm{freq}}T)^s}{s!}.
    \label{eq:frequency-scaled-coefficients}
\end{equation}
Then
\begin{equation}
    \begin{aligned}
    \widetilde Q_{N,r}^{(m)}
    &=L\sum_{s=0}^r\frac{XT^s}{s!}\frac{E_s}{K^s}\Gamma_s\\
    &=L\sum_{s=0}^r c_s\frac{E_s}{(K-1)^s}\Gamma_s.
    \end{aligned}
    \label{eq:combined-correction}
\end{equation}
The two expressions agree term by term because $c_s/(K-1)^s=XT^s/(s!K^s)$.

The positive LCU coefficients sum to
\begin{equation}
    \begin{aligned}
    \alpha_r&=\sum_{s=0}^r c_s
        =X\sum_{s=0}^r\frac{(\lambda_{\mathrm{freq}}T)^s}{s!},
    \end{aligned}
    \label{eq:block-subnormalization}
\end{equation}
with $\alpha_r\leq Xe^{\lambda_{\mathrm{freq}}T}<Xe^T$. We construct a block encoding of $\Gamma_s/2$, so the assembled circuit block-encodes $\widetilde Q_{N,r}^{(m)}/(2\alpha_r)$. For a normalized input $\ket\psi$, the corresponding success probability is
\begin{equation}
    p_{\mathrm{succ}}(\psi)
    =\frac{\|\widetilde Q_{N,r}^{(m)}\ket\psi\|^2}{(2\alpha_r)^2}.
    \label{eq:pre-amplification-success}
\end{equation}

Because $\widetilde Q_{N,r}^{(m)}$ approximates the unitary QUT, this probability is close to $(2\alpha_r)^{-2}$. We use oblivious amplitude amplification to raise the selected amplitude to nearly one. The bounds on $T$ and $X$ therefore control both the required Taylor order and the number of amplification rounds.

\paragraph*{Normalization bounds.}
We now describe explicit choices for $T$ and $X$, with the full derivation given in Appendix~\ref{app:tighter-normalization}. Both corrections attain their largest magnitude at the endpoint nodes,
\begin{equation}
    \tau=\frac{(m+1)\pi}{2}-K\theta_0,
    \qquad \xi_{\max}=\xi_0.
    \label{eq:endpoint-correction-maxima}
\end{equation}
These maxima can be bounded without evaluating the node $x_0$. Let $z_m$ be the first positive zero of the Bessel function $J_{m-1/2}$~\cite{DLMF}, and define
\begin{equation}
\vartheta_{m,K}=\frac{z_m}{\sqrt{K^2-m(m-1)/3}}.
\end{equation}
Using the lower bound $\theta_0\geq\vartheta_{m,K}$ established in Appendix~\ref{app:tighter-normalization}, we choose
\begin{equation}
T=\frac{(m+1)\pi}{2}-K\vartheta_{m,K}.
\label{eq:tight-input-displacement}
\end{equation}
for the displacement bound. The same lower bound on $\theta_0$ gives the weight bound
\begin{equation}
X^2=\sum_{q=0}^{m-1}\sin^{-2q}\vartheta_{m,K}\prod_{a=1}^{q}\frac{(2a-1)(m-a)(m+a-1)}{2a(K^2-a^2)},
\label{eq:tight-input-weight}
\end{equation}
with the convention that an empty product equals one.

\begin{figure}[!t]
    \centering
    \includegraphics[width=\columnwidth]{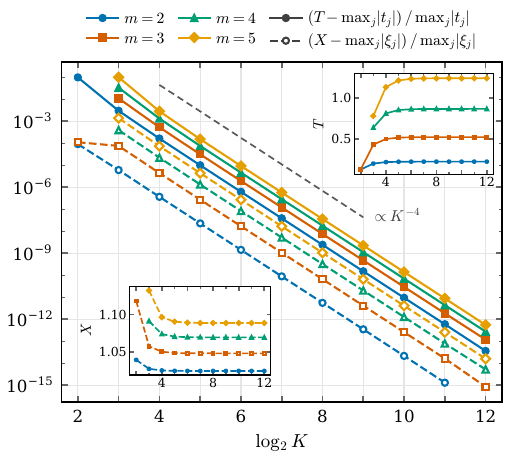}
    \caption{Tightness of the normalization bounds for $m=2,3,4,5$. Solid curves with filled markers show $(T-\tau)/\tau$, and dashed curves with open markers show $(X-\xi_{\max})/\xi_{\max}$. Here $\tau=\max_j|t_j|$ with $t_j=K\delta_j$. The gray dashed line indicates a $K^{-4}$ reference slope. Insets show $T$ (upper right) and $X$ (lower left) as functions of $\log_2K$.}
    \label{fig:normalization-tightness}
\end{figure}

These choices satisfy $T\geq\tau$ and $X\geq\xi_{\max}$ for all integers $m\geq2$ and $K\geq m+1$. Figure~\ref{fig:normalization-tightness} shows that the bounds rapidly approach the exact maxima as $K$ grows. Moreover, $T\leq(m+1)\pi/2-z_m$ and $X$ remains bounded independently of $K$.

The resulting subnormalization $\alpha_{\mathrm{QET}}=2\alpha_r\leq2Xe^{\lambda_{\mathrm{freq}}T}\leq2Xe^T$ is therefore bounded uniformly in $K$. Table~\ref{tab:amplification-rounds} lists the corresponding bounds and sufficient round counts for $2\leq m\leq8$.

\begin{table}[t]
\centering
\caption{
\label{tab:amplification-rounds}
Correction bounds, uniform subnormalization bounds and sufficient amplification rounds for $K=2^n\geq m+1$. Since $T$ is increasing and $X$ is decreasing with respect to $K$, we report $T$ at $K \rightarrow \infty$ and a representative value of $X$ at $K=64$. The upper bound $A$ of $\alpha_{\mathrm{QET}}$ and the round count are obtained from Eq.~\eqref{eq:uniform-tight-normalization} and Eq.~\eqref{eq:normalization-rounds}, respectively.}
\begin{tabular}{@{}rrrrr@{}}
\toprule
$m$ & $\left.T\right|_{K\to\infty}$ & $\left.X\right|_{K=64}$ & Bound $A$ & Rounds $t$\\
\midrule
2 & $0.2190$ & $1.0245$ & $2.551$ & $2$\\
3 & $0.5197$ & $1.0482$ & $3.526$ & $3$\\
4 & $0.8660$ & $1.0697$ & $5.087$ & $4$\\
5 & $1.2422$ & $1.0891$ & $7.545$ & $6$\\
6 & $1.6398$ & $1.1069$ & $11.410$ & $9$\\
7 & $2.0535$ & $1.1233$ & $17.513$ & $14$\\
8 & $2.4801$ & $1.1385$ & $27.194$ & $21$\\
\bottomrule
\end{tabular}
\end{table}

\subsection{Circuit overview}
\label{sec:circuit-overview}

According to Eq.~\eqref{eq:combined-correction}, each Taylor term has three main operations. The input correction $\Gamma_s$ accounts for the node displacement and quadrature weight, the Fourier block $E_s/(K-1)^s$ evaluates the corresponding derivative of the trigonometric expansion, and the basis change $L$ maps the resulting trigonometric coefficients to ultraspherical coefficients. We now construct quantum circuits for these components.

The input correction depends on the node label, while the basis change uses rotations whose angles depend on the frequency label. We implement these label-dependent functions using quantum eigenvalue transformation (QET) of block-encoded diagonal Hermitian signal operators~\cite{Martyn2021}. For each target function, we construct a bounded-degree polynomial approximation classically and synthesize the corresponding QET phase sequence. The Taylor terms are combined using LCU, and oblivious amplitude amplification removes the resulting subnormalization.

\begin{figure*}[t]
\centering
\begin{tikzpicture}[x=1cm,y=1cm,font=\small,
  gate/.style={draw,fill=white,minimum width=0.80cm,minimum height=0.48cm,inner sep=3pt,align=center},
  multi/.style={gate,minimum height=1.20cm},
  select/.style={draw,fill=black!60,rectangle,minimum size=4pt,inner sep=0pt},
  panel/.style={anchor=west,font=\small\bfseries}]
\path[use as bounding box] (-0.05,0.35) rectangle (17.35,-6.00);

\node[panel] at (0,0.12) {(a) Node-to-degree transform};
\foreach \y in {-0.72,-1.47} {\draw (1.70,\y)--(13.42,\y);}
\node[anchor=east] at (1.60,-0.72) {$\ket{0}_{\mathrm{aux}}$};
\node[anchor=east] at (1.60,-1.47) {$\ket{j}_{\mathrm{sys}}$};
\node[gate,minimum width=1.36cm] at (2.65,-1.47) {$+\lceil m/2\rceil$};
\node[multi] at (4.10,-1.095) {$W_t$};
\node[multi] at (5.70,-1.095) {$S_{\mathrm{out}}$};
\node[multi] at (7.30,-1.095) {$W_t^\dagger$};
\node[multi] at (8.90,-1.095) {$S_{\mathrm{in}}$};
\node[multi] at (10.50,-1.095) {$W_t$};
\draw[dashed,rounded corners=2pt] (4.94,-0.18) rectangle (11.10,-2.00);
\node[fill=white,inner sep=2pt,font=\footnotesize] at (8.02,-0.18) {$t$ rounds};
\node[gate] at (12.08,-1.47) {$-m$};
\node[anchor=west] at (13.52,-0.72) {$\ket{0}_{\mathrm{aux}}$};
\node[anchor=west] at (13.52,-1.47) {$Q_N^{(m)}\ket{j}$};

\node[panel] at (0,-2.52) {(b) One block $W_t$};
\draw[dashed,rounded corners=2pt] (5.25,-2.94) rectangle (12.33,-5.88);
\node[fill=white,inner sep=2pt,font=\footnotesize] at (8.79,-2.94)
  {$U_E=\sum_{s=0}^{r}\ket{s}\!\bra{s}_{\mathrm{ord}}\otimes U_E(s)$};
\foreach \y in {-3.28,-4.03,-4.78,-5.53} {\draw (1.70,\y)--(15.23,\y);}
\node[anchor=east] at (1.60,-3.28) {$\ket{0}_{\mathrm{ord}}$};
\node[anchor=east] at (1.60,-4.03) {$\ket{0}_{\mathrm{br}}$};
\node[anchor=east] at (1.60,-4.78) {$\ket{\ell_j}_{\mathrm{sys}}$};
\node[anchor=east] at (1.60,-5.53) {$\ket{0}_{\mathrm{work}}$};
\node[gate,minimum width=1.00cm,font=\footnotesize] at (2.57,-3.28) {$\mathrm{PREP}$};
\node[gate,minimum width=1.00cm] at (2.57,-5.53) {$R_{\mathrm{pad}}$};

\draw (4.26,-3.28)--(4.26,-4.555);
\node[select] at (4.26,-3.28) {};
\node[multi,minimum width=1.10cm] at (4.26,-5.155) {$U_{\mathrm{in}}$};
\draw (6.02,-3.28)--(6.02,-3.805);
\node[select] at (6.02,-3.28) {};
\node[multi,minimum width=1.28cm] at (6.02,-4.405) {$B_{\mathrm{in}}(s)$};
\node[multi,minimum width=1.54cm] at (7.95,-4.405) {$\mathrm{QFT}_{n+1}$};
\draw (9.87,-3.28)--(9.87,-3.805);
\node[select] at (9.87,-3.28) {};
\node[multi,minimum width=1.35cm] at (9.87,-4.405) {$B_{\mathrm{out}}(s)$};
\draw (11.64,-3.28)--(11.64,-4.555);
\node[select] at (11.64,-3.28) {};
\node[multi,minimum width=1.08cm] at (11.64,-5.155) {$R_p(s)$};

\node[multi,minimum width=1.08cm] at (13.18,-5.155) {$U_{\mathcal L}$};
\node[gate,minimum width=1.24cm,font=\footnotesize] at (14.27,-3.28) {$\mathrm{PREP}^{\dagger}$};

\node[anchor=west] at (15.33,-3.28) {$\bra{0}_{\mathrm{ord}}$};
\node[anchor=west] at (15.33,-4.03) {$\bra{0}_{\mathrm{br}}$};
\node[anchor=west] at (15.33,-4.78) {$k+m$};
\node[anchor=west] at (15.33,-5.53) {$\bra{0}_{\mathrm{work}}$};

\end{tikzpicture}
\caption{QUT circuit for $K=N+m=2^n$ and $m\geq2$.
\textbf{(a)} Amplified QUT circuit, including the fixed input and output label shifts. The dashed block is repeated $t$ times.
\textbf{(b)} One call to the padded block $W_t$. The input circuit encodes $\Gamma_s/2$, the Fourier circuit encodes $E_s/(K-1)^s$, and $U_{\mathcal L}$ implements the trigonometric-to-ultraspherical basis change. The work wires group the ancillary registers used by the input, Fourier, and basis-change circuits. The output bras select the encoded block without measurement, and valid output labels are $k+m$ for $0\leq k<N$.}
\label{fig:transform-circuit}
\end{figure*}
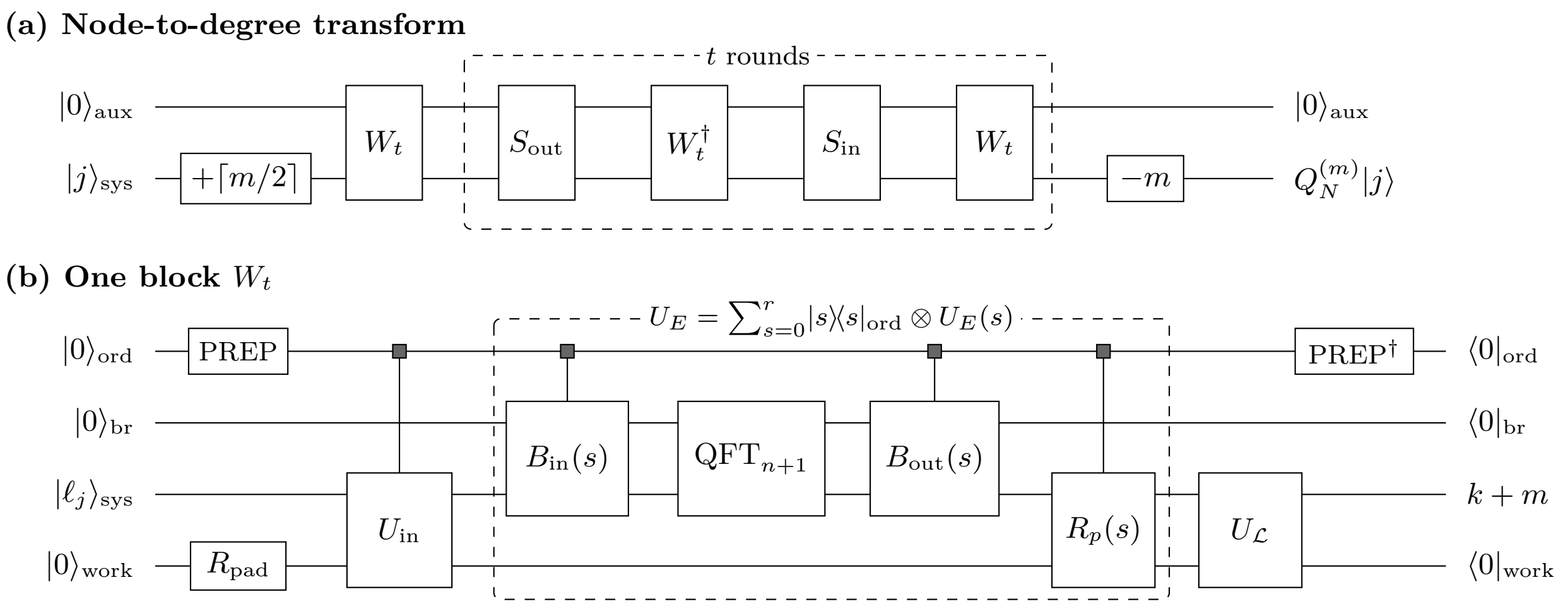

The system register contains $n=\log_2K$ qubits. Its label changes from the paired grid point $\ell_j=j+\lceil m/2\rceil$, to a frequency $p$, and finally to $k+m$ for ultraspherical degree $k$. Standard quantum adders implement the fixed input and output shifts. An order register of $\lceil\log_2(r+1)\rceil$ qubits stores the Taylor order $s$. A branch qubit is used within the Fourier block to combine the opposite Fourier labels $p$ and $2K-p$ into the required sine or cosine mode. Selector registers choose among controlled operations, while reusable flag qubits record conditions such as interval membership or successful block selection. Figure~\ref{fig:transform-circuit} shows these registers and the complete circuit. Here $U_{\mathrm{in}}$ denotes the input-correction circuit, $U_E$ the Fourier circuit, $U_{\mathcal L}$ the basis-change circuit, and $W_t$ the combined block used in $t$ amplification rounds.

\subsection{Input correction}
\label{sec:input-correction}
We construct the input circuit $U_{\mathrm{in}}$, which block-encodes $\Gamma_s/2$ for every fixed integer $m\geq2$. The correction combines the scaled displacement $t_j=K\delta_j$ and weight factor $\xi_j$. Using the bounds $T\geq|t_j|$ and $X\geq\xi_j$ from Eqs.~\eqref{eq:tight-input-displacement} and~\eqref{eq:tight-input-weight}, we extend these discrete corrections to continuous functions of the grid label and approximate them by polynomials. The continuous extension is obtained from a phase representation of the Gauss--ultraspherical nodes. Appendix~\ref{app:phase-arithmetic} derives this representation using Darboux transformations; we summarize the main construction here.

\begin{figure}[t]
    \centering
    \includegraphics[width=\columnwidth]{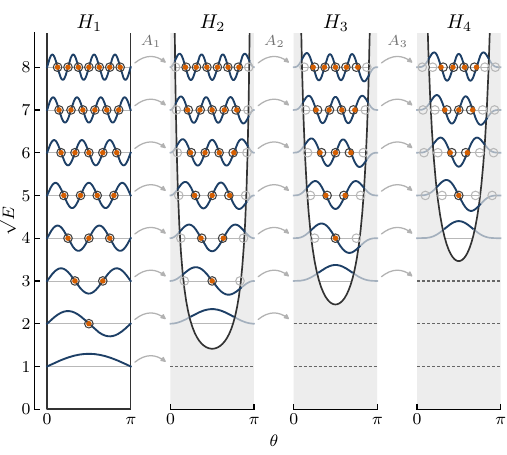}
    \caption{From Fourier modes to ultraspherical modes: Each panel $H_q$ shows the potential corresponding to $-\partial_\theta^2+q(q-1)\csc^2\theta$. We start with a particle in a box, whose modes are $\sin K\theta$ with energies $E=K^2$. Each Darboux transformation $A_q = \partial_\theta-q\cot\theta$ changes the potential and maps the modes to the next well without changing their energies, except for the lowest mode, which disappears. After $m-1$ steps, $\sin K\theta$ becomes proportional to $\sin^m\theta\,C_N^{(m)}(\cos\theta)$. Orange dots mark the Gauss--ultraspherical nodes, while circles mark the equispaced grid used by sine or cosine transforms. The two grids nearly coincide, but differ near the walls, where some grid points have no corresponding node (gray circles). The QUT uses a QFT on the equispaced grid and corrects for these small shifts.}
    \label{fig:darboux_ladder}
\end{figure}

For $1\leq q\leq m$, define the Schr\"odinger operator
\begin{equation}
    H_q\coloneqq -\partial_\theta^2+q(q-1)\csc^2\theta,
\end{equation}
and, for $1\leq q\leq m-1$, define
\begin{equation}
    A_q\coloneqq \partial_\theta-q\cot\theta.
\end{equation}
They satisfy the intertwining relation $H_{q+1}A_q=A_qH_q$. This is the standard Darboux transformation for this hierarchy of
Schr\"odinger operators~\cite{Crum1955}. Hence, if
$H_qZ_{q-1}=K^2Z_{q-1}$, then $Z_q\coloneqq K^{-1}A_qZ_{q-1}$
satisfies $H_{q+1}Z_q=K^2Z_q$. Thus each step changes the potential
from $q(q-1)\csc^2\theta$ to $q(q+1)\csc^2\theta$ while preserving
the eigenvalue $K^2$. We note that the angular function $G_N^{(m)}$ satisfies $H_m G_N^{(m)}=K^2G_N^{(m)}$. Figure~\ref{fig:darboux_ladder} illustrates this ladder of Darboux steps.

Starting from $Z_0(\theta)=e^{\ii K\theta}$ and applying $m-1$ steps produces $Z_{m-1}$ such that $\operatorname{Im}Z_{m-1}(\theta)\propto G_N^{(m)}(\theta)$. Set $u=\cot(\theta)/K$ and $h=1/K$. Expanding $Z_q = K^{-1}A_q Z_{q-1}$ yields the expression
\begin{equation}
    Z_q(\theta) = \mathsf P_q(u;h)e^{\ii K\theta},
\end{equation}
where $\mathsf P_q(u;h)$ is a polynomial in $u$ defined recursively by $\mathsf P_0=1$ and
\begin{equation}
    \mathsf P_q(u;h) =
    (\ii-qu)\mathsf P_{q-1}(u;h)
    -(u^2+h^2)\partial_u\mathsf P_{q-1}(u;h).
    \label{eq:main-phase-polynomial-recurrence}
\end{equation}
Note that $\operatorname{deg}(\mathsf P_q) = q$, independent of $K$. Using the continuous branch fixed in Appendix~\ref{app:phase-arithmetic}, we define the phase function
\begin{equation}
    \begin{aligned}
        \Phi_K^{(m)}(\theta)
        &= \arg Z_{m-1}(\theta) \\
        &= K\theta+\frac{(m-1)\pi}{2} +\psi_m(u;h),
    \end{aligned}
    \label{eq:main-phase-representation}
\end{equation}
where
\begin{equation}
    \psi_m(u;h) \coloneqq \arg\mathsf P_{m-1}(u;h) -\frac{m-1}{2}\pi.
\end{equation}
Thus $\psi_m$ measures the deviation of the phase from its linear
part $K\theta+(m-1)\pi/2$.

We use this phase function to extend the node corrections to a continuous grid label. Let $\zeta_m=1/2$ for even $m$ and $\zeta_m=0$ for odd $m$, so that the grid angles are $\widetilde\theta_\ell=\pi(\ell+\zeta_m)/K$. Treating $\ell$ as a real variable, define $\theta(\ell)$ as the solution of
\begin{equation}
    K\theta(\ell)
    +\psi_m\!\left(
        \frac{\cot\theta(\ell)}K;\frac1K
    \right)
    =
    \pi(\ell+\zeta_m).
    \label{eq:main-general-angle}
\end{equation}
At $\ell=\ell_j$, since $\ell_j+\zeta_m=j+(m+1)/2$, Eq.~\eqref{eq:main-general-angle} reduces to the node condition in Eq.~\eqref{eq:phase-corrections}, and hence $\theta(\ell_j)=\theta_j$. The corresponding scaled displacement is
\begin{equation}
    t(\ell) = K\theta(\ell)-\pi(\ell+\zeta_m) = K(\theta(\ell) - \widetilde \theta_\ell),
    \label{eq:main-general-displacement}
\end{equation}
so that $t(\ell_j)=K\delta_j$.  Appendix~\ref{app:phase-arithmetic} also establishes
\begin{equation}
    |\mathsf P_{m-1}(u;h)|^2
    \Phi_K^{(m)\prime}(\theta)
    =
    K D_{m-1}(h),
    \label{eq:main-phase-derivative}
\end{equation}
where $D_{m-1}(h) = \prod_{q=1}^{m-1}(1-q^2h^2)$. With $u(\ell)=\cot(\theta(\ell))/K$, we extend the weight correction by
\begin{equation}
    \xi(\ell)
    = \sqrt{\frac{K} {\Phi_K^{(m)\prime}(\theta(\ell))}} 
    = \frac{ |\mathsf P_{m-1}(u(\ell);h)| }{ \sqrt{D_{m-1}(h)} }.
    \label{eq:main-general-weight}
\end{equation}
Thus $\xi(\ell_j)=\xi_j$. We combine the two extensions into
\begin{equation}
    g_{s,m,K}(\ell)
    = \frac{\xi(\ell)}X \left(\frac{t(\ell)}T\right)^s,
    \label{eq:main-general-target}
\end{equation}
with $g_{s,m,K}(\ell_j)=\gamma_{s,j}$.

All evaluations of these functions are performed classically. For each interpolation label $\ell$, we solve Eq.~\eqref{eq:main-general-angle} for $\theta(\ell)$, or equivalently for $t(\ell)$. The phase correction $\psi_m$ is evaluated from the fixed-degree polynomial $\mathsf P_{m-1}$, after which Eqs.~\eqref{eq:main-general-weight} and \eqref{eq:main-general-target} determine $\xi(\ell)$ and $g_{s,m,K}(\ell)$.

We next exploit reflection symmetry to restrict the polynomial
approximation to one half of the grid. Let $\rho$ be the highest bit of the grid label $\ell$ and define the \emph{folded label} $b\coloneqq\min(\ell,K-1-\ell)$, which ranges over the left half $\{0,\ldots,K/2-1\}$ and is obtained by complementing the lower $n-1$ bits of $\ell$ when $\rho=1$. The reflection $\theta\mapsto\pi-\theta$ maps the grid angle $\widetilde\theta_\ell$ to $\widetilde\theta_{K-1-\ell}$ for even $m$ and to $\widetilde\theta_{K-\ell}$ for odd $m$; it reverses the displacement but preserves the weight. Hence the correction at $\ell$ is $(-1)^{\rho s}g_{s,m,K}(b)$ for even $m$ and $(-1)^{\rho s}g_{s,m,K}(b+\rho)$ for odd $m$, where the shift by $\rho$ appears because the reflected label $K-\ell$ exceeds $K-1-\ell$ by one.

The correction varies most rapidly near the grid endpoints, where the constant-shift pairing produces the largest mismatch, as shown in Fig.~\ref{fig:node-grid-pairing}. A single rescaling of the full label range to $[-1,1]$ would compress these variations into intervals of width $O(K^{-1})$. We therefore treat the first $B_m$ folded labels directly, where $B_m$ depends only on $m$ and is independent of $K$, and approximate the remaining labels on dyadic intervals. Labels with $b<B_m$ are implemented using fixed rotations, while the remaining labels are divided into intervals $I_\nu$ with $2^\nu\geq B_m$. Appendix~\ref{app:general-correction} specifies a sufficient choice of $B_m$ and proves uniform approximation bounds on these intervals. A sufficient choice is the smallest power of two larger than $m(m-1)/2$, for example $B_2=2$, $B_3=4$, and $B_4=8$.

Fix an approximation error $0<\eta_\Gamma<1$. On
$I_\nu=\{2^\nu,\ldots,2^{\nu+1}-1\}$, rescale the continuous folded label
$b\in[2^\nu,2^{\nu+1}-1]$ to $[-1,1]$ by
\begin{equation}
    x = \frac{2(b-2^\nu)}{2^\nu-1}-1,
\end{equation}
or equivalently,
\begin{equation}
    b = b_\nu(x)\coloneqq 2^\nu+\frac{2^\nu-1}{2}(x+1).
\end{equation}
For even $m$, we approximate $g_{s,m,K}(b_\nu(x))$ as a function of $x$. For odd $m$, we construct the approximation separately for $\rho=0$ and $\rho=1$, using $g_{s,m,K}(b_\nu(x)+\rho)$.

After the fold, the lower $n-1$ qubits of the system register hold $b$ (Appendix~\ref{app:input-circuit}). On $I_\nu$, the polynomial is applied through the dyadic signal $B_\nu$ of Eq.~\eqref{eq:qet-local-signal}, which acts on the $\nu$ least significant of these qubits. Its diagonal entry on the folded label $b=2^\nu+v\in I_\nu$ is $x_{\nu,v}=(2v-2^\nu+1)/(2^\nu-1)$, the rescaled coordinate $x$ of $b$, so that $b_\nu(x_{\nu,v})=b$. QET with a polynomial $P$ then encodes $P(B_\nu)/2$ as in Eq.~\eqref{eq:main-polynomial-block}, and $P(x_{\nu,v})$ approximates $g_{s,m,K}(b)$ for even $m$ and $g_{s,m,K}(b+\rho)$ for odd $m$. The same signal serves every order $s$ and, for odd $m$, both values of $\rho$; these enter only through the selected QET phase angles. Since the interval circuit acts as the identity outside $I_\nu$, the entries of $B_\nu$ on other folded labels do not enter the input block.

Appendix~\ref{app:general-correction} shows that these rescaled functions are analytic and uniformly bounded by the constant $M$ of Eq.~\eqref{eq:general-input-ellipse-constant} on a fixed complex
neighborhood of $[-1,1]$. Chebyshev interpolation yields
polynomial approximations of degree
\begin{equation}
    d_\Gamma
    =
    \left\lceil
        \log_2\frac{16M}{\eta_\Gamma}
    \right\rceil.
    \label{eq:main-input-degree}
\end{equation}
The same degree works for every Taylor order $0\leq s\leq r$ and,
when $m$ is odd, for both $\rho=0$ and $\rho=1$. At fixed $m$,
\begin{equation}
    d_\Gamma = O_m\!\left( r+\log\frac{1}{\eta_\Gamma} \right),
\end{equation}
independently of $K$.

We implement the interval polynomials using the QET circuits of
Section~\ref{sec:preliminaries}, together with fixed rotations for the
endpoint labels. After restoring the reflection sign and the original
grid labels, the resulting input circuit satisfies
\begin{equation}
    \bra0 U_{\mathrm{in}}\ket0
    =
    \sum_{s=0}^r
    \ket s\bra s\otimes
    \frac{\widehat\Gamma_s}{2},
    \label{eq:ideal-input-block}
\end{equation}
on the valid paired labels, with
$\|\widehat\Gamma_s-\Gamma_s\|_2\leq\eta_\Gamma$.
All interpolation data, rotation angles, and QET phase sequences are
computed classically before quantum execution.

\emph{Extension to other Jacobi parameters.}
For integer $m$, the QUT covers Jacobi polynomials with equal parameters $\alpha=\beta=m-1/2$, and both ingredients of the input correction extend to asymmetric pairs of half-integer parameters. First-order Darboux steps analogous to $A_q$ shift $(\alpha,\beta)$ by $(\pm1,\pm1)$, so a number of steps independent of $N$ connects the four Chebyshev families $(\pm1/2,\pm1/2)$ to every admissible half-integer pair~\cite{Grandati2014}. The resulting intertwiner has a fixed order and therefore maps each Fourier mode to the same mode multiplied by a polynomial of fixed degree in its frequency, whose coefficients depend only on the angle~\cite{Crum1955}, just as $\mathsf P_{m-1}(u;h)$ in $Z_{m-1}$ is a polynomial of fixed degree in $u=\cot\theta/K$ and $h=1/K$. For modes on which every step has a positive norm, the phase of the resulting normalized wave again determines the nodes and weights as in Eq.~\eqref{eq:phase-corrections}. We present these parameter shifts and phase--quadrature identities in Appendix~\ref{app:darboux-extensions}.

The exact construction stops at half-integer parameters. By Proposition~\ref{prop:darboux-obstruction}, a nonzero finite-order differential operator with analytic, frequency-independent coefficients can map Fourier modes to solutions of the Jacobi angular equation for infinitely many frequencies only if $\alpha,\beta\in1/2+\mathbb Z$, which is also exactly when the Jacobi angular functions are finite trigonometric polynomials, as the finite basis change $L$ requires. Legendre polynomials, with $\alpha=\beta=0$, therefore fall outside this construction. For such parameters, trigonometric asymptotic expansions~\cite{Hahn1980} and the nonoscillatory phase functions of classical fast Jacobi transforms~\cite{Bremer2020} still describe the angular functions as Fourier modes with amplitude and phase corrections, but these corrections are no longer finite sums of powers of the frequency. A fixed-degree correction does not automatically make a Jacobi transform efficient on a quantum computer. In the example discussed in Appendix~\ref{app:darboux-extensions}, one coefficient in the angular function diverges at an endpoint although the complete function is regular. Block-encoding this coefficient on the Gauss grid costs a subnormalization of order $N$. A Jacobi transform with the complexity of Theorem~\ref{thm:efficient-qut} would therefore need a discretization that preserves the regularity of the angular functions and an efficient circuit that implements it.

\subsection{Fourier block and frequency powers}
\label{sec:fourier-block}

We next implement $E_s/(K-1)^s$ for each Taylor order $s$. Recall
from Eq.~\eqref{eq:differentiated-sampling} that $E_s$ samples the
$s$th derivatives of the trigonometric basis functions at the paired
grid points. For $p>0$, these functions are
$e_p(\theta)=\sqrt{2/\pi}\cos(p\theta)$ for even $m$ and
$e_p(\theta)=\sqrt{2/\pi}\sin(p\theta)$ for odd $m$. Differentiating
$s$ times contributes a factor $p^s$, changes between sine and cosine
when $s$ is odd, and introduces a sign that depends on $m$ and $s$.
We therefore implement $E_s/(K-1)^s$ in two steps: a Fourier-based
circuit produces the required trigonometric samples, and a second
circuit applies the factor $[p/(K-1)]^s$.

The trigonometric function required after $s$ derivatives is determined
by
\begin{equation}
    \sigma_s=(m+s)\bmod 2,
\end{equation}
where $\sigma_s=0$ denotes cosine and $\sigma_s=1$ denotes sine. To
obtain these real transforms from the complex QFT, we first apply a
unitary $B_{\mathrm{in}}(s)$ that forms symmetric or antisymmetric
combinations of grid labels and their reflections in a $2K$-dimensional
register. Define
\begin{equation}
    \operatorname{ref}_m(\ell)=
    \begin{cases}
        2K-1-\ell,&m\text{ even},\\
        2K-\ell,&m\text{ odd}.
    \end{cases}
    \label{eq:grid-reflection}
\end{equation}
For a paired grid label $\ell_j=j+\lceil m/2\rceil$,
\begin{equation}
    B_{\mathrm{in}}(s)
    \ket0_{\mathrm{br}}\ket{\ell_j}_{\mathrm{sys}}
    =
    \frac{
        \ket{\ell_j}
        +(-1)^{\sigma_s}\ket{\operatorname{ref}_m(\ell_j)}
    }{\sqrt2}.
    \label{eq:Bin-action}
\end{equation}
The kets on the right denote labels of the joint
$\mathrm{br}\otimes\mathrm{sys}$ register, with the branch qubit as
the most significant bit. Thus the input is symmetric for the cosine
transform and antisymmetric for the sine transform. The QFT unitary, denoted by $\mathcal F$, then acts on these $n+1$
qubits. Its matrix entries are
\begin{equation}
    \mathcal F_{p\ell} = \frac{1}{\sqrt{2K}}e^{\ii\pi p\ell/K},
    \qquad 0\leq p,\ell<2K.
    \label{eq:qft-convention}
\end{equation}
The same QFT is shared by all Taylor orders.

After the QFT, a second unitary $B_{\mathrm{out}}(s)$ combines the
opposite frequencies $p$ and $2K-p$ and returns the branch qubit to
zero. For even $m$, this combination includes the phase associated
with the half-grid shift
$\widetilde\theta_\ell=(\ell+1/2)\pi/K$; for odd $m$, no such shift is
needed. Define
\begin{equation}
    \phi_{m,p}=
    \begin{cases}
        \pi p/(2K),&m\text{ even},\\
        0,&m\text{ odd},
    \end{cases}
\end{equation}
and let the derivative sign be
\begin{equation}
    \chi_{m,s}=
    \begin{cases}
        (-1)^{\lceil s/2\rceil},&m\text{ even},\\
        (-1)^{\lfloor s/2\rfloor},&m\text{ odd}.
    \end{cases}
    \label{eq:derivative-sign}
\end{equation}
For $1\leq p<K$, we choose $B_{\mathrm{out}}(s)$ so that
\begin{equation}
    \begin{aligned}
        &B_{\mathrm{out}}(s)\frac{\ii^{\sigma_s}}{\sqrt2}
        \left(
            e^{-\ii\phi_{m,p}}\ket p
            +(-1)^{\sigma_s}
             e^{\ii\phi_{m,p}}\ket{2K-p}
        \right)\\
        &\qquad=
        \chi_{m,s}\ket0_{\mathrm{br}}\ket p_{\mathrm{sys}}.
    \end{aligned}
    \label{eq:main-output-pairing}
\end{equation}
This adds the two frequency components for cosine and subtracts them,
with a factor of $\ii$, for sine. The orthogonal combination
is mapped to the branch state $\ket1_{\mathrm{br}}$. Appendix~\ref{app:fourier} presents an explicit construction of $B_{\mathrm{out}}(s)$ and treats the special Fourier labels $p=0$ and $p=K$ separately.

It remains to apply the frequency factor $[p/(K-1)]^s$. Since this
factor depends only on the frequency label, we implement it using QET. Writing $p=\sum_{a=0}^{n-1}2^ap_a$ with $p_a\in\{0,1\}$, we construct
\begin{equation}
    A_{\mathrm{freq}} = \diag(p/K)
    = \frac{K-1}{2K}I - \sum_{a=0}^{n-1} \frac{2^a}{2K}Z_a.
    \label{eq:qet-frequency-signal}
\end{equation}
The absolute values of these coefficients sum to
\begin{equation}
    \lambda_{\mathrm{freq}} = \frac{K-1}{2K} + \sum_{a=0}^{n-1} \frac{2^a}{2K} =\frac{K-1}{K}.
\end{equation}
Using these coefficients in an LCU construction therefore produces a
Hermitian unitary $U_{\mathrm{freq}}$ that block-encodes
\begin{equation}
    \overline A_{\mathrm{freq}}
    =
    \frac{A_{\mathrm{freq}}}{\lambda_{\mathrm{freq}}}
    =
    \diag\!\left(\frac{p}{K-1}\right).
    \label{eq:normalized-frequency-signal}
\end{equation}
The corresponding selector register requires $\lceil\log_2(n+1)\rceil$ qubits. Since the selected Pauli terms square to the identity, the resulting signal unitary satisfies $U_{\mathrm{freq}}^2=I$.

For $s\geq1$, the target polynomial $x^s$ is bounded by one on
$[-1,1]$ and has definite parity. A single QET parity sequence,
together with the real-part extraction described in
Section~\ref{sec:preliminaries}, therefore produces a circuit $R_p(s)$ such
that
\begin{equation}
    (\bra0\otimes I)R_p(s)\ket0\ket p
    =
    \left(\frac{p}{K-1}\right)^s\ket p.
    \label{eq:main-frequency-flag}
\end{equation}
Here $\ket0$ denotes the joint zero state of the frequency selector
and real-part ancilla. No even--odd combination is needed, so this
block introduces no additional subnormalization. For $s=0$, we define
$R_p(0)=I$.

To implement all orders $0\leq s\leq r$ coherently, we use a common
sequence with $r$ signal-query positions. The degree-$s$ sequence
requires only $s$ queries; the remaining positions are padded so that the extra signal calls cancel in pairs using
$U_{\mathrm{freq}}^2=I$. The order register selects the QET phase at each step. We show in Appendix~\ref{app:frequency-polynomials} that this shared construction reproduces the individual degree-$s$ sequences on every order branch. When $r=0$, the frequency circuit is omitted.

By combining the trigonometric transform and the frequency factor, we obtain
\begin{equation}
    U_E(s)
    = R_p(s) B_{\mathrm{out}}(s) \mathcal F B_{\mathrm{in}}(s),
    \label{eq:controlled-fourier-stage}
\end{equation}
which block-encodes $E_s/(K-1)^s$ on the paired grid inputs.
Composing it with $U_{\mathrm{in}}$ therefore results in, for each Taylor order $s$, the block $\frac{E_s}{(K-1)^s} \frac{\widehat\Gamma_s}{2}$. The next step is to apply the basis change $L$ to map these frequency amplitudes to ultraspherical coefficients.

\subsection{Weighted Helmert basis change}
\label{sec:helmert-circuit}

To convert frequency amplitudes to ultraspherical coefficients, we extend
the rectangular matrix $L$ from Eq.~\eqref{eq:frequency-to-degree} to
a unitary $\mathcal L$ on the system register. Its relevant output rows
satisfy
\begin{equation}
    \bra{k+m}\mathcal L\ket p=L_{kp},
    \qquad
    0\leq k<N,\quad 0\leq p<K.
    \label{eq:main-L-action}
\end{equation}

We construct $\mathcal L$ through successive changes between the angular
bases $\{G_k^{(\lambda)}\}$, starting from $\lambda=m\bmod2$ and increasing $\lambda$ by two until $\lambda=m$.
Using $\sin^2\theta=1-\cos^2\theta$, we expand $G_k^{(\lambda+2)}$ as a linear combination of $G_j^{(\lambda)}$ with $j\leq k+2$ and the same parity as $k$. After $\lfloor m/2\rfloor$ such steps, we obtain the target basis.

For a fixed intermediate parameter $\lambda$, assign each mode
$G_k^{(\lambda)}$ a positive weight $\mu_k$. Let
$\mathcal I_k=\{j:0\leq j<k,\ j\equiv k\pmod 2\}$ denote the preceding mode labels with the same parity as \(k\). Their normalized weighted mean is
\begin{equation}
    \ket{w_k}
    = \frac{1}{\sqrt{M_k}}
    \sum_{j\in\mathcal I_k}\sqrt{\mu_j}\ket j,
    \qquad
    M_k=\sum_{j\in\mathcal I_k}\mu_j .
    \label{eq:prefix-mass}
\end{equation}
A \emph{weighted contrast} compares the current mode $\ket k$ with this weighted mean. For $k\geq2$, define
\begin{equation}
    \ket{h_k}
    =
    \sqrt{\frac{M_k}{M_{k+2}}}\ket{k}
    -
    \sqrt{\frac{\mu_k}{M_{k+2}}}\ket{w_k}.
    \label{eq:main-weighted-contrast}
\end{equation}
Since $M_{k+2}=M_k+\mu_k$, this combination has unit norm and is
orthogonal to the weighted mean obtained after adding mode $k$.
Thus $\ket{h_k}$ represents the difference between mode $k$ and the weighted mean of all preceding modes with the same parity. For each parity group, the weighted mean of all its modes together
with these successive contrasts forms an orthonormal basis, known as
the weighted Helmert basis~\cite{Rayner2024}.

To reproduce the angular basis conversion, we choose
\begin{equation}
    \mu_k=
    \begin{cases}
        \dfrac{(k+\lambda)\Gamma(k+2\lambda)}{k!},
        &\lambda\geq1,\\[1ex]
        \dfrac12,&\lambda=0,\ k=0,\\
        1,&\lambda=0,\ k>0.
    \end{cases}
    \label{eq:contrast-weight}
\end{equation}
For $k\geq2$, the corresponding cumulative weight is
\begin{equation}
    M_k=
    \frac{1}{2(2\lambda+1)}
    \prod_{j=1}^{2\lambda+1}(k+j-2),
    \label{eq:contrast-prefix-mass}
\end{equation}
with $M_0=M_1=0$. Thus, for fixed $\lambda$, $M_k$ is a polynomial in $k$
of degree $2\lambda+1$, independent of the number of modes. With these weights, the contrast indexed by $k+2$ implements the basis conversion through
\begin{equation}
    G_k^{(\lambda+2)}(\theta) = -\sum_{\substack{0\leq j\leq k+2\\j\equiv k\;(\mathrm{mod}\,2)}}
    \langle h_{k+2}\mid j\rangle
    G_j^{(\lambda)}(\theta).
    \label{eq:main-angular-contrast}
\end{equation}
The derivation of the weights, the expression for $M_k$, and the overall sign in the conversion are given in Appendix~\ref{app:weighted-contrasts}.

Let $\mathcal H_\lambda$ denote a unitary whose relevant rows are these
weighted contrasts. Since degree $j$ at parameter $\lambda$ is stored
at system label $j+\lambda$, we require
\begin{equation}
    \bra{k+\lambda+2}\mathcal H_\lambda\ket{j+\lambda}
    =
    \langle h_{k+2}\mid j\rangle.
\end{equation}
The remaining rows can be chosen to complete the transformation to a
unitary; an explicit completion is provided in Appendix~\ref{app:weighted-contrasts}. The full basis change is therefore
\begin{equation}
    \mathcal L = (-1)^{\lfloor m/2\rfloor}
    \mathcal H_{m-2}\mathcal H_{m-4}\cdots
    \mathcal H_{m\bmod2},
    \label{eq:L-composition}
\end{equation}
where the prefactor accounts for the minus sign in each conversion.

\begin{figure*}[t]
\centering
\input{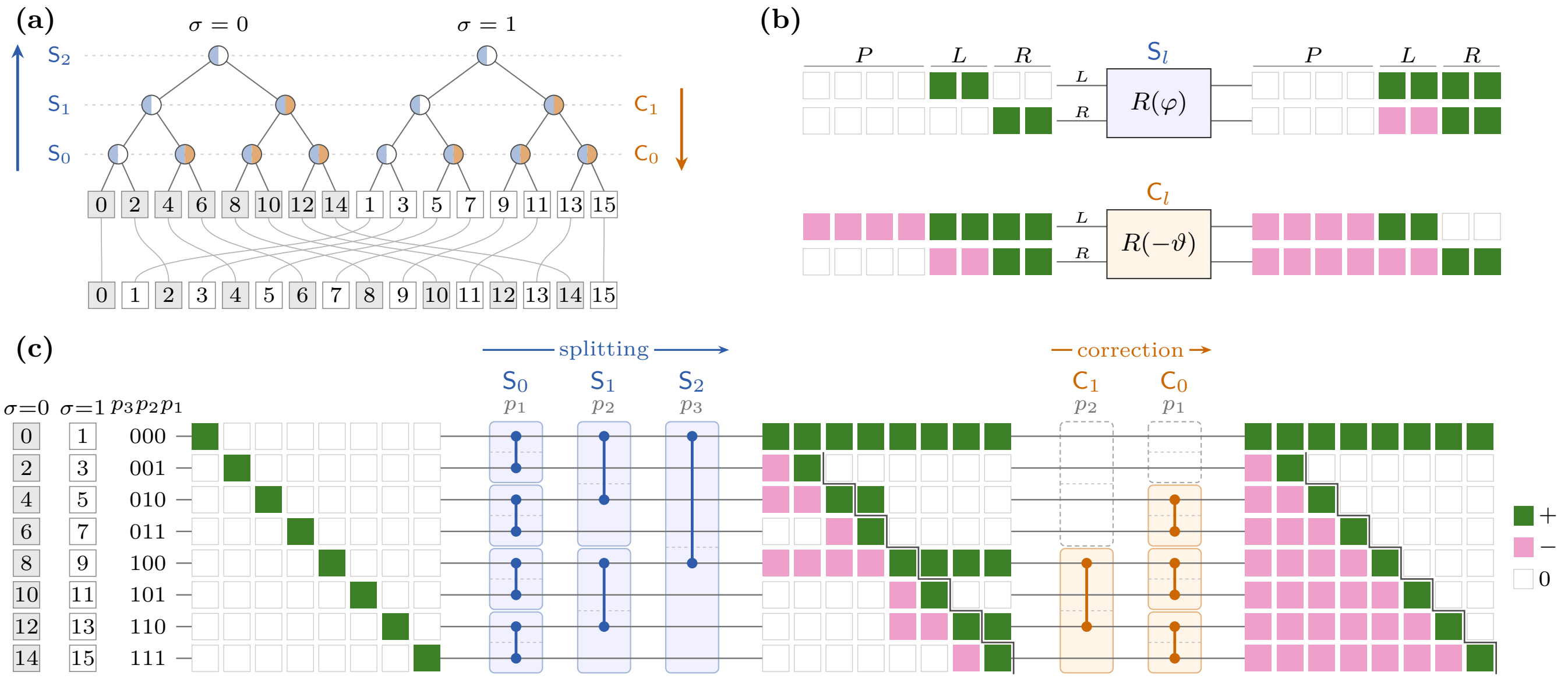}
\caption{Two-pass tree construction of $\mathcal H_\lambda$ for $K=16$. \textbf{(a)}~The parity bit $\sigma=p_0$ separates the labels into two parity groups with identical block trees.
\textbf{(b)}~For a block $A=L\cup R$ with preceding modes $P$, the splitting layer applies $R(\varphi)$ and the correction layer applies $R(-\vartheta)$ to the first labels of $L$ and $R$.
\textbf{(c)}~Sequence of splitting and correction layers, together with the resulting coefficient sign patterns. The bit shown below each layer is the qubit acted on by the rotations in that layer. White correction halves in \textbf{(a)} and dashed correction blocks in \textbf{(c)} correspond to $P=\varnothing$, for which the correction is the identity.}
\label{fig:helmert-tree}
\end{figure*}

We implement each $\mathcal H_\lambda$ separately on the even and odd
parity groups using two passes through a binary tree. We call the first
the \emph{splitting pass} and the second the \emph{correction pass}.
The splitting pass recursively forms weighted means and contrasts
between neighboring groups of modes. The correction pass then converts
these intermediate contrasts into the weighted Helmert contrasts of
Eq.~\eqref{eq:main-weighted-contrast}. Figure~\ref{fig:helmert-tree} illustrates the two-pass construction, which we describe below.

Consider a block $A=L\cup R$ of consecutive modes within one parity
group, divided into left and right halves. Let
\begin{equation}
    \mu_L=\sum_{j\in L}\mu_j,\qquad
    \mu_R=\sum_{j\in R}\mu_j,\qquad
    \mu_A=\mu_L+\mu_R.
\end{equation}
The splitting rotation acts on the two-dimensional subspace spanned by the first labels of $L$ and $R$. When $L$ and $R$ each contain a single mode, their amplitudes are the original mode amplitudes; when they contain several modes, they store the normalized weighted means of the two groups. We apply
\begin{equation}
    R(\varphi)=
    \begin{pmatrix}
        \cos\varphi&\sin\varphi\\
        -\sin\varphi&\cos\varphi
    \end{pmatrix},
    \label{eq:main-splitting-rotation}
\end{equation}
where $\varphi$ is determined by $\cos\varphi=\sqrt{\mu_L /\mu_A}$ and $\sin\varphi=\sqrt{\mu_R /\mu_A}$.

The first output is the weighted mean of $A$, while the second is a
contrast between $L$ and $R$. At the first layer, $L$ and $R$ each
contain one mode. Subsequent layers combine pairs into groups of four,
then eight, and so on. The splitting pass therefore produces the
weighted mean of the full parity group together with a hierarchy of
contrasts between neighboring blocks.

Consider $\lambda=0$ and $K=8$, and focus on the odd labels $1,3,5,7$, whose weights are all one. Let their input amplitudes be $c_1,c_3,c_5,c_7$. The first splitting layer acts on the pairs $(1,3)$ and $(5,7)$, and the second combines the two resulting means. After these two layers, the amplitudes are
\begin{equation}
    \begin{aligned}
        v_1&=\frac{c_1+c_3+c_5+c_7}{2},\\
        v_3&=\frac{-c_1+c_3}{\sqrt2},\\
        v_5&=\frac{-c_1-c_3+c_5+c_7}{2},\\
        v_7&=\frac{-c_5+c_7}{\sqrt2}.
    \end{aligned}
    \label{eq:helmert-splitting-example}
\end{equation}
Here $v_1$ is the mean of all four modes, while $v_3$, $v_5$, and
$v_7$ are contrasts produced at different levels of the tree. These contrasts are not yet all in the form required by Eq.~\eqref{eq:main-weighted-contrast}. For example, $v_5$ still contains $c_7$, whereas the contrast for mode $5$ should involve only $c_5$ and preceding modes $c_1$ and $c_3$. The correction pass can remove these later-mode contributions.

For the correction pass, consider again a block $A=L\cup R$. Let $P$ denote the modes of the same parity that precede $A$, with total weight $\mu_P$. The pass applies $R(-\vartheta)$ to the amplitudes at the first labels of $L$ and $R$, where
\begin{equation}
    \sin^2\vartheta = \frac{\mu_P\mu_R}{\mu_A(\mu_P+\mu_L)}.
    \label{eq:main-helmert-angles}
\end{equation}
This replaces the two intermediate contrasts by the contrast of $L$
with $P$ and a contrast of $R$ with $P\cup L$. When $P$ is empty,
$\vartheta=0$ and no correction is needed.

In the example above, only the block containing modes $5$ and $7$
requires a correction. Here $P=\{1,3\}$, $L=\{5\}$, and $R=\{7\}$, so
$\mu_P=2$ and $\mu_L=\mu_R=1$, which implies
$\sin^2\vartheta=1/3$. Applying $R(-\vartheta)$ to $v_5$ and $v_7$
yields
\begin{equation}
    \begin{aligned}
        \sqrt{\frac23}\,v_5-\frac{v_7}{\sqrt3}
        &=\frac{-c_1-c_3+2c_5}{\sqrt6},\\
        \frac{v_5}{\sqrt3}+\sqrt{\frac23}\,v_7
        &=\frac{-c_1-c_3-c_5+3c_7}{\sqrt{12}}.
    \end{aligned}
    \label{eq:helmert-correction-example}
\end{equation}
The first output is the contrast of mode $5$ with the preceding modes $1$ and $3$, while the second is the contrast of mode $7$ with modes $1$, $3$, and $5$. Together with the existing contrast at label $3$, these are the weighted Helmert contrasts for this parity group.

In general, the correction pass traverses the tree in the reverse
direction of the splitting pass. It begins with the two halves of the full parity group, then proceeds through quarters and successively smaller blocks until pairs are reached. The final amplitudes are the weighted mean and the contrasts in Eq.~\eqref{eq:main-weighted-contrast}.

On the system register, the least significant bit identifies the parity group, while the remaining $n-1$ bits specify the position within that group. For $0\leq l\leq n-2$, let $\mathsf S_l$ denote the splitting layer acting on position bit $l$, and for $0\leq l\leq n-3$, let $\mathsf C_l$ denote the corresponding correction layer. The splitting pass applies $\mathsf S_0,\ldots,\mathsf S_{n-2}$ from small to large blocks, while the correction pass applies
$\mathsf C_{n-3},\ldots,\mathsf C_0$ in the reverse direction. Thus,
\begin{equation}
    \mathcal H_\lambda
    =
    \mathsf C_0\mathsf C_1\cdots\mathsf C_{n-3}\,
    \mathsf S_{n-2}\mathsf S_{n-3}\cdots\mathsf S_0,
    \label{eq:helmert-tree-factorization}
\end{equation}
where the rightmost operator acts first. Each parameter conversion
$\lambda\to\lambda+2$ therefore uses $n-1$ splitting layers and $n-2$ correction layers, for a total of $2n-3$ rotation layers.

\begin{figure*}[t]
\centering
\begin{tikzpicture}[x=1cm,y=1cm,font=\footnotesize,
  hcpanel/.style={anchor=west,font=\small\bfseries},
  hcq/.style={black!80,line width=0.45pt},
  hcr/.style={black!80,line width=1.1pt},
  hcgate/.style={draw=black!80,line width=0.4pt,fill=white,minimum width=0.44cm,minimum height=0.38cm,inner sep=1pt},
  hcctrl/.style={circle,fill=black!85,inner sep=0pt,minimum size=4.2pt},
  hcsel/.style={rectangle,fill=black!65,inner sep=0pt,minimum size=4.6pt},
  hcdash/.style={line width=0.45pt,dash pattern=on 2pt off 1.5pt,rounded corners=1.5pt},
  hcmode/.style={draw=black!45,line width=0.3pt,minimum size=0.33cm,inner sep=0pt,font=\scriptsize},
  hcoff/.style={draw=black!30,line width=0.3pt,dash pattern=on 1.2pt off 1pt,minimum size=0.33cm,inner sep=0pt},
  ]
\definecolor{hwSplit}{RGB}{46,92,170}
\colorlet{hwSplitFill}{blue!6}
\newcommand\hcket[1]{\lvert #1\rangle}
\newcommand\hcbra[1]{\langle #1\rvert}
\newcommand\hcoplus[2]{
  \draw[black!80,line width=0.45pt,fill=white] (#1,#2) circle (0.105);
  \draw[black!80,line width=0.45pt] ({#1-0.105},#2) -- ({#1+0.105},#2) ({#1},{#2-0.105}) -- ({#1},{#2+0.105});}
\path[use as bounding box] (-0.05,0.0) rectangle (17.80,5.12);

\begin{scope}[shift={(0,4.40)}]
\node[hcpanel] at (-0.05,0.55) {(a)};
\foreach \y/\lab/\st in {0/{$b,\sigma,p_l{\cdots}p_1$}/hcr,-0.45/{$p_{l+1}$}/hcq,-0.90/{$\hcket{0}_{\mathrm{cs}}$}/hcq,
                         -1.35/{$\hcket{0}_{g,\mathrm{re},\mathrm{par}}$}/hcr,-1.80/{$\hcket{0}_{\mathrm{pr}}$}/hcq,-2.25/{$\hcket{0}_{\mathrm{pad}}$}/hcq}{
  \draw[\st] (1.70,\y) -- (8.78,\y);
  \node[anchor=east] at (1.64,\y) {\lab};}
\foreach \o/\c in {0.14/22,0.07/35}{
  \draw[hcdash,black!\c] ({2.68+\o},{0.25+\o}) rectangle ({5.82+\o},{-2.03+\o});}
\path[hcdash,fill=white,draw=black!75] (2.68,0.25) rectangle (5.82,-2.03);
\foreach \y in {-0.45,-0.90,-1.80}{\draw[hcq] (2.68,\y) -- (5.82,\y);}
\draw[hcr] (2.68,0) -- (5.82,0);
\draw[hcr] (2.68,-1.35) -- (5.82,-1.35);
\node[fill=white,inner sep=1.5pt,font=\scriptsize] at (3.65,0.30) {$\prod_{\mathcal R}$};
\foreach \x in {3.05,5.45}{
  \draw[black!80,line width=0.45pt] (\x,0) -- (\x,-1.80);
  \node[hcsel] at (\x,0) {};
  \hcoplus{\x}{-1.80}}
\draw[black!80,line width=0.45pt] (4.25,0) -- (4.25,-1.80);
\node[hcsel] at (4.25,0) {};
\node[hcsel] at (4.25,-0.90) {};
\node[hcctrl] at (4.25,-1.80) {};
\node[hcgate,fill=hwSplitFill,minimum width=0.72cm] at (4.25,-1.35) {$U_P$};
\foreach \x/\g in {6.35/X,7.00/Z}{
  \draw[black!80,line width=0.45pt] (\x,-0.45) -- (\x,-0.90);
  \node[hcctrl] at (\x,-0.90) {};
  \node[hcgate,fill=hwSplitFill] at (\x,-0.45) {$\g$};}
\node[hcgate] at (2.20,-0.90) {$H$};
\node[hcgate] at (7.80,-0.90) {$H$};
\node[hcgate] at (2.20,-2.25) {$R_y$};
\foreach \y/\t in {-0.90/{$\hcbra{0}$},-1.35/{$\hcbra{0}$},-1.80/{$\hcket{0}$},-2.25/{$\hcbra{0}$}}{
  \node[anchor=west,inner sep=1pt] at (8.80,\y) {\t};}
\foreach \y/\t in {-0.90/{$1/2$},-1.35/{$1/2$},-2.25/{$4/\beta_3$}}{
  \node[anchor=west,inner sep=1pt,text=black!55,font=\scriptsize] at (9.32,\y) {\t};}
\end{scope}

\begin{scope}[shift={(0,0.98)}]
\node[hcpanel] at (-0.05,0.50) {(b)};
\draw[hcr] (1.33,0) -- (8.08,0);
\draw[hcq] (1.33,-0.55) -- (8.08,-0.55);
\node[anchor=east] at (1.27,0) {$\hcket{0}_{\mathrm{work}}$};
\node[anchor=east] at (1.27,-0.55) {$\hcket{\psi}_{\mathrm{sys}}$};
\draw[hcdash,draw=black!75] (2.50,0.38) rectangle (7.90,-0.93);
\node[fill=white,inner sep=1.5pt] at (5.20,0.38) {$\times3$};
\foreach \x/\t in {1.95/{$W_l$},4.50/{$W_l^\dagger$},6.82/{$W_l$}}{
  \path[fill=hwSplitFill,draw=black!80,line width=0.4pt] ({\x-0.31},0.22) rectangle ({\x+0.31},-0.77);
  \node at (\x,-0.275) {\t};}
\foreach \x in {3.32,5.68}{\node[hcgate,minimum width=1.0cm,font=\scriptsize] at (\x,0) {$I{-}2\hcket{0}\hcbra{0}$};}
\node[hcgate] at (7.47,-0.55) {$-I$};
\node[anchor=west] at (8.12,0) {$\hcket{0}$};
\node[anchor=west] at (8.12,-0.55) {$\mathsf S_l\hcket{\psi}$};
\end{scope}

\begin{scope}[shift={(10.25,4.40)}]
\node[hcpanel] at (-0.05,0.55) {(c)};
\foreach \p in {0,...,15}{\node[font=\tiny,text=black!50] at ({1.25+0.40*\p},0.33) {\p};}
\draw[black!35,line width=0.5pt,dash pattern=on 1pt off 1pt] (1.25,0) -- (2.05,-1.22) -- (2.85,-2.44) -- (3.65,-3.66);
\foreach \lam/\y in {0/0,2/-1.22,4/-2.44,6/-3.66}{
  \node[anchor=east] at (0.98,\y) {$G^{(\lam)}$};
  \foreach \p in {0,...,15}{
    \pgfmathtruncatemacro\j{\p-\lam}
    \ifnum\j<0 \node[hcoff] at ({1.25+0.40*\p},\y) {};
    \else\node[hcmode,fill=white] at ({1.25+0.40*\p},\y) {\j};\fi}}
\foreach \y/\lam in {0/0,-1.22/2,-2.44/4}{
  \draw[-{Stealth[length=4pt,width=3.5pt]},black!60,line width=0.6pt] (0.66,{\y-0.24}) -- (0.66,{\y-0.98});
  \node[anchor=east] at (0.60,{\y-0.61}) {$\mathcal H_{\lam}$};}
\end{scope}

\end{tikzpicture}%
\caption{Helmert basis-change circuits.
\textbf{(a)}~Padded block $W_l$ for a splitting layer $\mathsf S_l$.
The product runs over disjoint address regions $\mathcal R$, selected by the predicate qubit $\ket{\cdot}_{\operatorname{pr}}$: addresses $b<C_m$ use fixed rotations, dyadic intervals $I_\nu$ use QET, and positions with nonzero lower bits are idle. The small filled squares indicate the register values that activate the corresponding operation.
The block encoding $U_P$ implements the region-dependent cosine or sine factor, either by a fixed rotation or a QET block. Here, $W_l$ is an approximate block encoding of $R(\varphi)/\beta_3$. For a correction layer $\mathsf C_l$, the same circuit replaces $\varphi$ by $\vartheta$ and reverses the controlled Pauli gates to $Z$ then $X$, yielding $R(-\vartheta)/\beta_3$.
\textbf{(b)}~Three rounds of amplitude amplification achieve unit success amplitude in the ideal case.
\textbf{(c)}~Successive parameter conversions for $K=16$ and $m=6$, with
$\mathcal L=(-1)^3\mathcal H_4\mathcal H_2\mathcal H_0$.
At parameter $\lambda$, system label $p$ represents degree $p-\lambda$ in the basis $\{G_k^{(\lambda)}\}$.}
\label{fig:helmert-circuit}
\end{figure*}
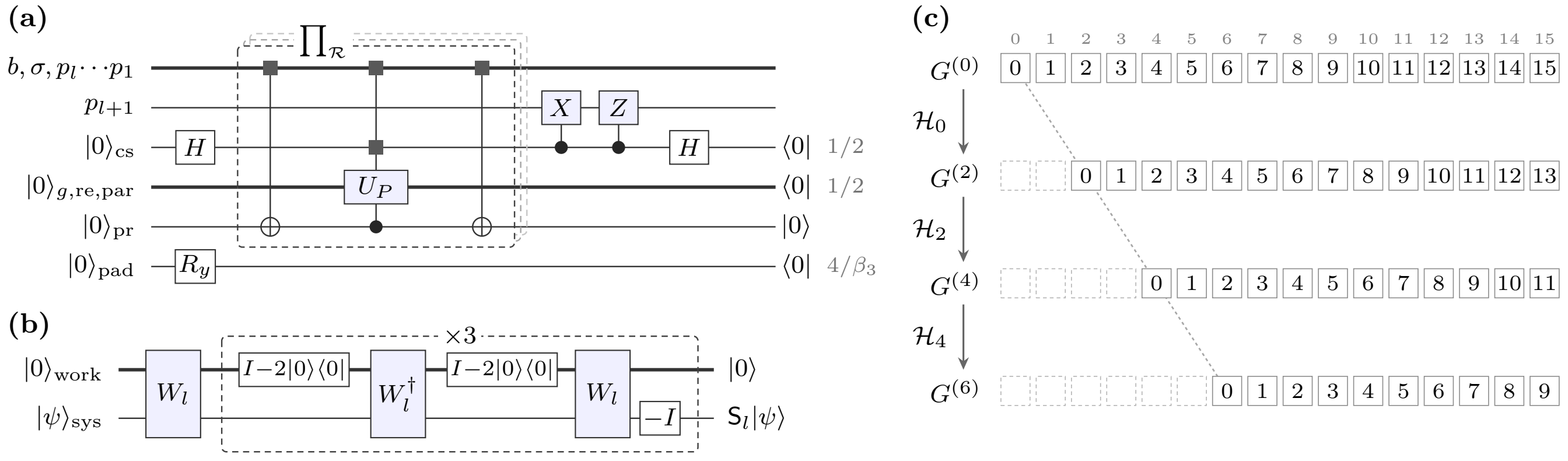

We next describe how the rotation angles are implemented within each layer. Write the system label as $(p_{n-1}\cdots p_1p_0)_2$, where $p_0$ identifies the parity group. In layer $l$, the rotation acts on $p_{l+1}$. Requiring the less significant position bits $p_1,\ldots,p_l$ to be zero selects the first labels of the two halves $L$ and $R$. The more significant bits encode the \emph{block address}
\begin{equation}
    b=(p_{n-1}\cdots p_{l+2})_2,
\end{equation}
which identifies the block acted on by that layer.

For each layer and parity group, the weights $\mu_k$ and cumulative
weights $M_k$ determine the rotation angle as a function of $b$:
$\varphi=\varphi_l(b)$ for a splitting layer $\mathsf S_l$ and
$\vartheta=\vartheta_l(b)$ for a correction layer $\mathsf C_l$.
We therefore implement the cosine and sine of the rotation angle as
functions of the block address. These functions vary most rapidly at the first few block addresses. We handle the first $C_m$ addresses directly using fixed rotations, where $C_m$ depends only on $m$ and is independent of $K$. Any power of two $C_m\geq\max\{2,\tfrac23(m-1)\}$ suffices for the approximation bounds (Appendix~\ref{app:weighted-contrasts}, Lemma~\ref{lem:helmert-approximation}). Since fixed rotations are cheaper than QET sequences on short intervals, we use the smallest power of two no smaller than $16(m+1)$.

On each interval, we encode the block address in a diagonal Hermitian signal operator $B_\nu^{\mathrm H}$. Its eigenvalues lie in a fixed interval independent of the layer and of $K$. We approximate the cosine and sine of the layer's rotation angle by real polynomials $P_C$ and $P_S$, respectively. We describe the signal construction in Appendix~\ref{app:helmert-polynomials} and prove that, for fixed $m$, degree
\begin{equation}
    d_L=O_m\!\left(\log\frac{1}{\eta_L}\right)
    \label{eq:helmert-polynomial-degree}
\end{equation}
suffices to approximate the angle functions to accuracy $\eta_L$, uniformly over all layers and block addresses.

Using QET, we construct block encodings of
$P_C(B_\nu^{\mathrm H})/2$ and $P_S(B_\nu^{\mathrm H})/2$.
Since
\begin{equation}
    R(\alpha)=\cos\alpha\,I+\ii\sin\alpha\,Y,
\end{equation}
we combine the two scalar blocks with an additional selector qubit. For a splitting layer, the sine branch applies $\ii Y$ to the target qubit, while for a correction layer it applies $-\ii Y$. The selected block is therefore
\begin{equation}
    \frac{
        P_C(B_\nu^{\mathrm H})I
        \pm \ii P_S(B_\nu^{\mathrm H})Y
    }{4}
    =
    \frac{\widehat R}{4},
    \label{eq:qet-helmert-rotation-block}
\end{equation}
where the plus sign approximates the splitting rotation $R(\varphi)$ and the minus sign approximates the correction rotation $R(-\vartheta)$. One factor of $1/2$ comes from combining the even and odd parts of each QET polynomial, and the second comes from combining the cosine and sine branches.

The fixed rotations for $b<C_m$ are encoded with the same factor $1/4$. Positions where a layer should not act receive $I/4$. Thus every block address within a given layer has the same subnormalization, allowing the whole layer to be amplified coherently.

We amplify each layer before applying the next so that the factor $1/4$ does not accumulate through the tree. We first pad the block encoding to the subnormalization $\beta_3=\csc\frac{\pi}{14}$. Three rounds of oblivious amplitude amplification then raise the success amplitude to one for an exact rotation. Including the initial block application, an amplified layer uses seven calls to the padded block encoding or its inverse. The amplification also returns the layer workspace to zero, so the same workspace can be reused by the next layer. However, approximation and gate-synthesis errors can leave a small residual component outside the zero-workspace subspace. Figure~\ref{fig:helmert-circuit} summarizes the complete basis-change construction.

\subsection{Circuit assembly and analysis}
\label{sec:circuit-assembly}

We assemble the input correction, Fourier block, and weighted Helmert basis change into the QUT circuit shown in
Fig.~\ref{fig:transform-circuit}. The following error and resource analysis completes the proof of Theorem~\ref{thm:efficient-qut}.

\subsubsection{Full circuit implementation}
\label{sec:full-circuit}

We first combine the component circuits to implement the Taylor sum in Eq.~\eqref{eq:combined-correction}. Using the coefficients $c_s$ from Eq.~\eqref{eq:frequency-scaled-coefficients} and the normalization $\alpha_r$ from Eq.~\eqref{eq:block-subnormalization}, we prepare the order register as
\begin{equation}
    \operatorname{PREP}\ket0_{\mathrm{ord}}
    = \sum_{s=0}^r \sqrt{\frac{c_s}{\alpha_r}} \ket s_{\mathrm{ord}}.
    \label{eq:order-amplitudes}
\end{equation}

The input and Fourier stages in Eqs.~\eqref{eq:ideal-input-block}
and~\eqref{eq:controlled-fourier-stage} depend on the Taylor order
$s$, while the basis change is shared by all orders. Let $U_E$ denote the controlled circuit that applies $U_E(s)$ on the order state $\ket s$ for $0\leq s\leq r$ and acts as the identity on unused order labels. Since the Fourier block encodes $E_s/(K-1)^s$, the coefficients $c_s$ restore the corresponding Taylor weights. Applying the three components and then undoing the order preparation creates
\begin{equation}
    W = \operatorname{PREP}^{\dagger}
    U_{\mathcal L}U_EU_{\mathrm{in}}
    \operatorname{PREP}.
    \label{eq:raw-block-composition}
\end{equation}

Let $\ket{\mathrm{in}_j} \coloneqq \ket{0}_{\mathrm{aux}}\ket{\ell_j}_{\mathrm{sys}}$ and $\ket{\mathrm{out}_k} \coloneqq \ket{0}_{\mathrm{aux}}\ket{k+m}_{\mathrm{sys}}$, where $\ket{0}_{\mathrm{aux}}$ denotes the state with all auxiliary registers zero. In the ideal implementation, the encoded block is
\begin{equation}
\bra{\mathrm{out}_k}W\ket{\mathrm{in}_j} = \frac{(\widetilde Q_{N,r}^{(m)})_{kj}} {\alpha_{\mathrm{QET}}}, \qquad \alpha_{\mathrm{QET}} = 2\alpha_r.
\label{eq:selected-transform-block}
\end{equation}
The basis-change unitary also contains output labels outside $k+m$, $0\leq k<N$. These labels are part of its unitary completion but are not valid QUT outputs.

We next amplify the encoded transform. Choose the smallest $t\geq1$ such that
\begin{equation}
    \alpha_{\mathrm{QET}}\leq\beta_t,
    \qquad
    \beta_t=
    \frac{1}{\sin[\pi/(4t+2)]},
\end{equation}
as in Eq.~\eqref{eq:amplifiable-amplitude}. Following
Section~\ref{sec:amplification-primitive}, we add a padding flag whose amplitude on $\ket0$ is $\alpha_{\mathrm{QET}}/\beta_t$. The resulting
circuit $W_t$ block-encodes $\widetilde Q_{N,r}^{(m)}/\beta_t$.

Let $\Pi_{\mathrm{in}}$ project onto states for which all work registers and the padding flag are zero and the system label is one of the paired labels $\ell_j$, $0\leq j<N$. Similarly, let $\Pi_{\mathrm{out}}$ project onto states with zero auxiliaries and a valid output label
$k+m$, $0\leq k<N$. Define
\begin{equation}
    S_{\mathrm{in}}=I-2\Pi_{\mathrm{in}},
    \qquad
    S_{\mathrm{out}}=I-2\Pi_{\mathrm{out}}.
    \label{eq:amplification-reflections}
\end{equation}
The amplified circuit is
\begin{equation}
    \mathcal U_t
    =(-W_tS_{\mathrm{in}}W_t^\dagger S_{\mathrm{out}})^tW_t,
    \label{eq:main-amplified-circuit}
\end{equation}
which uses $2t+1$ calls to $W_t$ and $W_t^\dagger$. For a unitary target, the amplification raises the success amplitude to one. At $m=2$, the bound $\alpha_{\mathrm{QET}}\leq2.551<\beta_2=1+\sqrt5$ indicates that two amplification rounds and five block calls suffice.

Finally, we include the fixed input and output label shifts needed to obtain the circuit $U$ of Theorem~\ref{thm:efficient-qut}. Let
$\operatorname{ADD}_c$ add the integer $c$ to the system label modulo $K$. Then
\begin{equation}
U =(I_{\mathrm{aux}} \otimes \operatorname{ADD}_{-m}) \, \mathcal U_t \,
(I_{\mathrm{aux}}\otimes\operatorname{ADD}_{\lceil m/2\rceil}),
\label{eq:assembled-qut}
\end{equation}
where $I_{\mathrm{aux}}$ acts on all ancillary registers. The input shift sends $j$ to its paired grid label $\ell_j=j+\lceil m/2\rceil$, while the output shift maps the label $k+m$ back to the ultraspherical degree $k$.

\subsubsection{Complexity analysis}
\label{sec:complexity-analysis}

There are two sources of error: approximation of the transform
components and implementation of the resulting circuit. We bound these contributions separately, combine them to meet the target precision, and then count the required gates and qubits. We first analyze the circuit with separate work registers for its components.

\begin{proof}[Proof of Theorem~\ref{thm:efficient-qut}]
Fix $m\geq2$ and write $K=2^n$. The displacement and weight
bounds $T$ and $X$ remain bounded as $K$ grows, hence $\alpha_{\mathrm{QET}} = 2\alpha_r\leq2Xe^T = O_m(1)$, so the outer amplification requires only $t=O_m(1)$ rounds.

\emph{Component approximation errors.}
We first consider the errors that arise from truncating the Taylor series and approximating the input correction, frequency block, and basis change. From Eq.~\eqref{eq:factorization-error}, a truncation at
order $r$ has the error bound
\begin{equation}
    \varepsilon_T(r)=Xe^T\frac{T^{r+1}}{(r+1)!}.
\end{equation}

Let $\eta_\Gamma$ bound the operator error in $\Gamma_s$ for all
$0\leq s\leq r$. Since the circuit encodes $\Gamma_s/2$, the
corresponding block error is at most $\eta_\Gamma/2$. Likewise, let
$\eta_E$ bound the error in $E_s/(K-1)^s$. The polynomial $x^s$ can be implemented exactly by QET, so $\eta_E=0$ for exact phase angles; we retain $\eta_E$ to account for finite-precision classical phase computation. For the basis change, let $\varepsilon_L$ satisfy
\begin{equation}
    \bigl\|U_{\mathcal L}(\ket0\otimes I)
       -\ket0\otimes\mathcal L\bigr\|_2
    \leq\varepsilon_L,
    \label{eq:helmert-basis-error}
\end{equation}
where $\ket0$ is the joint zero state of the basis-change work qubits. This bound includes both accumulated layer errors and any amplitude
remaining in nonzero work states. With separate work registers, the encoded component blocks multiply, giving an error at most $\eta_\Gamma+\eta_E+\varepsilon_L$ for each Taylor term. Let
$\widehat Q$ denote the resulting unnormalized transform. Summing over the Taylor orders and adding the truncation remainder yields the overall approximation error
\begin{equation}
    \begin{aligned}
    \|\widehat Q-Q_N^{(m)}\|_2
    &\leq\varepsilon_T(r)+\sum_{s=0}^r c_s(\eta_\Gamma+\eta_E+\varepsilon_L) \\
    &= \varepsilon_T(r)+ \alpha_r(\eta_\Gamma+\eta_E+\varepsilon_L)
    \eqqcolon\Delta.
    \end{aligned}
    \label{eq:qet-transform-error}
\end{equation}

The quantity $\Delta$ is an error bound of the encoded transform.
Appendix~\ref{app:amplification} shows that, with exact elementary
gates, amplification changes this into an output-state error of at most
$4\Delta$. More precisely, for every normalized
$\ket\psi\in\mathcal H_{\mathrm{valid}}$, where $\mathcal H_{\mathrm{valid}} = \operatorname{span}\{\ket0,\ldots,\ket{N-1}\}$, the amplified output differs from
$\ket{0^a}\otimes Q_N^{(m)}\ket\psi$ by at most $4\Delta$ in vector
norm, provided $t\geq1$ and $\Delta\leq1/10$.

\emph{Circuit implementation errors.}
Next, we fix these component approximations and analyze errors in the circuit implementation. Let $\varepsilon_W$ bound the
operator-norm difference between an implemented padded block $W_t$ and
its reference circuit with exact elementary gates; the same bound
applies to $W_t^\dagger$. Let $\varepsilon_R$ bound the error in either
reflection $S_{\mathrm{in}}$ or $S_{\mathrm{out}}$ from
Eq.~\eqref{eq:amplification-reflections}. Since the amplification
circuit uses $2t+1$ calls to $W_t$ or $W_t^\dagger$ and $2t$
reflections, these errors contribute at most $(2t+1)\varepsilon_W+2t\varepsilon_R$. Errors in the input and output label shifts would add to this bound, but we treat these shifts as exact for simplicity.

\emph{Total error and parameter choices.} For the implemented circuit $U$, the total output error satisfies
\begin{equation}
\begin{aligned}
    \varepsilon_{\mathrm{out}}
    &\coloneqq\sup_{\substack{\ket\psi\in\mathcal H_{\mathrm{valid}}\\
                              \|\ket\psi\|_2=1}}
       \bigl\|U\ket{0^a}\ket\psi
              -\ket{0^a}Q_N^{(m)}\ket\psi\bigr\|_2\\
    &\leq4\Delta+(2t+1)\varepsilon_W+2t\varepsilon_R.
\end{aligned}
    \label{eq:qet-general-total-error}
\end{equation}
This norm includes the system and all ancillary registers, including
amplitude on unused system labels.

It is sufficient to choose
\begin{equation}
    \varepsilon_T(r) \leq\frac{\varepsilon}{32}, \qquad
    \eta_\Gamma,\eta_E,\varepsilon_L\leq\frac{\varepsilon}{32\alpha_r},
    \label{eq:approximation-error-allocation}
\end{equation}
for the approximation errors and 
\begin{equation}
    \varepsilon_W \leq\frac{\varepsilon}{4(2t+1)}, \qquad
    \varepsilon_R\leq\frac{\varepsilon}{8t}
    \label{eq:implementation-error-allocation}
\end{equation}
for the implementation errors. Then $\Delta\leq\varepsilon/8<1/10$ and $\varepsilon_{\mathrm{out}}\leq\varepsilon$.

We now choose the approximation orders to meet these tolerances. The factorially decreasing Taylor remainder shows that it is sufficient to take
\begin{equation}
    r = O_m\!\left( \frac{\log(1/\varepsilon)}{\log(\log(1/\varepsilon)+c)} \right)
    \label{eq:efficient-qut-order}
\end{equation}
to achieve $\varepsilon_T(r)\leq\varepsilon/32$. Any fixed constant $c\geq 1-\log 2$ keeps the denominator positive for all $0<\varepsilon<1/2$ and does not affect the asymptotic scaling as $\varepsilon\to0$.

Once $r$ is fixed, the input correction must approximate all $\Gamma_s$, $0\leq s\leq r$, to accuracy $\eta_\Gamma$. From
Section~\ref{sec:input-correction}, a sufficient polynomial degree is $d_\Gamma=O_m(r+\log(1/\eta_\Gamma))$. Since $\alpha_r=O_m(1)$, Eq.~\eqref{eq:approximation-error-allocation} allows $\eta_\Gamma=\Theta_m(\varepsilon)$, and hence
\begin{equation}
    d_\Gamma=O_m(\log(1/\varepsilon)).
    \label{eq:input-degree-complexity}
\end{equation}

For the basis change, we divide the error budget $\varepsilon_L$ among
its successive layers. Let $\eta_L$ bound the approximation error in
each of $\cos\varphi$, $\sin\varphi$, $\cos\vartheta$, and $\sin\vartheta$. Appendix~\ref{app:helmert-amplification} shows that, provided $\eta_L\leq10^{-2}$, each amplified layer has error at most $10\eta_L$ on the system and work registers together. Since there are $L_{\mathrm{tot}} = \left\lfloor\frac m2\right\rfloor(2n-3)$ layers, we choose
\begin{equation}
    \eta_L=\min\left\{ 10^{-2}, \frac{\varepsilon_L}{10L_{\mathrm{tot}}} \right\}.
\end{equation}
Since $\varepsilon_L=\Theta_m(\varepsilon)$ and
$L_{\mathrm{tot}}=\Theta_m(n)$, this implies
$\eta_L=\Theta_m(\varepsilon/n)$. On the block addresses treated by QET, the rotation functions can be approximated with degree
\begin{equation}
    d_L = O_m\!\left(\log\frac{1}{\eta_L}\right)
    = O_m\!\left( \log \left( \frac{n}{\varepsilon} \right) \right).
    \label{eq:helmert-degree-complexity}
\end{equation}

\emph{Resource counts.} We count single-qubit rotations and CNOT gates. For fixed $m$, the
input correction, frequency powers, and basis change require
\begin{equation}
    \begin{aligned}
        C_\Gamma &= O_m\!\left( (d_\Gamma+1) [n^2\log n+n(r+1)] \right),\\
        C_{\mathrm{freq}} &= O\!\left( rn\log n+r^2 \right),\\
        C_{\mathcal L} &= O_m\!\left( (d_L+1)n^3 \right).
    \end{aligned}
\label{eq:component-gate-costs}
\end{equation}
The input and frequency circuits share signal calls across Taylor orders, while their phase rotations are selected by the order register. The basis-change cost includes all splitting and correction layers and their amplification. Appendices~\ref{app:input-circuit}, \ref{app:frequency-polynomials}, and~\ref{app:helmert-resources} establish these counts.

Each padded transform block also contains the QFT and the preparation and unpreparation of the order register, which require $O(n^2+r)$ gates. Since the outer amplification uses
$t=O_m(1)$ rounds, the block calls, reflections, and fixed label shifts change the total cost only by an $m$-dependent constant factor. Hence
\begin{equation}
    C_{\mathrm{tot}} = O_m\!\left( C_\Gamma+C_{\mathrm{freq}}+C_{\mathcal L} +n^2+r \right).
    \label{eq:circuit-cost}
\end{equation}
Using Eqs.~\eqref{eq:efficient-qut-order},~\eqref{eq:input-degree-complexity}, and~\eqref{eq:helmert-degree-complexity}, we obtain the bound on $G_m(n,\varepsilon)=C_{\mathrm{tot}}$ stated in Theorem~\ref{thm:efficient-qut},
\begin{equation}
    C_{\mathrm{tot}} = O_m\!\left( n^3\log\frac n\varepsilon + n\log^2\frac1\varepsilon \right).
    \label{eq:total-gate-complexity}
\end{equation}
In particular, the basis-change cost $C_{\mathcal L}$ controls the leading dependence on $n$, $O_m(n^3\log(n/\varepsilon))$, while the input correction $C_\Gamma$ accounts for the leading quadratic dependence on $\log(1/\varepsilon)$.

The order register requires $\lceil\log_2(r+1)\rceil$ qubits. The input and frequency selector registers together require $O(\log n)$ qubits. All remaining auxiliary registers require only a constant number of qubits. From Eq.~\eqref{eq:efficient-qut-order}, $\log(r+1) = O_m(\log\log(1/\varepsilon))$, so the total number of auxiliary qubits is
\begin{equation}
    a = O_m\!\left( \log n+\log\log\frac1\varepsilon \right).
    \label{eq:total-ancilla-complexity}
\end{equation}

The error analysis above used separate work registers. Reusing workspace does not affect asymptotic resource bounds, while its additional error can be absorbed into the circuit implementation budget. All polynomial coefficients, QET phase sequences, fixed rotations, and order-state amplitudes are computed classically before quantum execution. For fixed $m$, only polynomially many such quantities are required, and their computation takes time polynomial in $n$ and
$\log(1/\varepsilon)$.

The same error and resource bounds hold for the inverse transform
implemented by $U^\dagger$. For any normalized valid output
$\ket\phi$, apply Eq.~\eqref{eq:efficient-qut-guarantee} to
$Q_N^{(m)\dagger}\ket\phi$ and use the unitarity of $U^\dagger$ and
$Q_N^{(m)}$ to obtain the corresponding inverse bound.

Thus, $U$ satisfies Eq.~\eqref{eq:efficient-qut-guarantee} with the resource bounds above, completing the proof of Theorem~\ref{thm:efficient-qut}. These bounds assume fixed $m$, count arbitrary single-qubit rotations and CNOT gates, and exclude input-state preparation and synthesis into a discrete gate set.
\end{proof}

\subsubsection{Compiled circuits and empirical errors}
\label{sec:compiled-circuits}

\begin{figure}[t]
    \centering
    \includegraphics[width=\columnwidth]{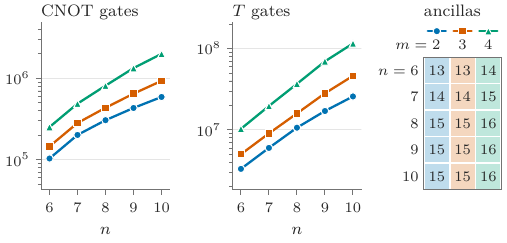}
    \caption{A compilation of resources for the full QUT circuit for $\varepsilon=10^{-5}$ and $m=2,3,4$. Left: number of CNOT gates in the gate set of single-qubit unitaries and CNOT gates. Center: number of $T$ gates in the Clifford+$T$ gate set. Right: number of ancillary qubits.}
    \label{fig:gate-counts}
\end{figure}

We compiled the full circuit $U$ of Eq.~\eqref{eq:assembled-qut} for $m=2,3,4$ and $6\leq n\leq10$ at the target error $\varepsilon=10^{-5}$. The compilation uses the cutoffs $C_m$ of Section~\ref{sec:helmert-circuit} and the error allocation of Eqs.~\eqref{eq:approximation-error-allocation} and~\eqref{eq:implementation-error-allocation}. The resulting counts include all $2t+1$ calls of $W_t$ and $W_t^\dagger$, the $2t$ reflections, and the two label shifts. Figure~\ref{fig:gate-counts} reports the number of CNOT gates in the gate set of single-qubit unitaries and CNOT gates. It also reports the number of $T$ gates in the Clifford+$T$ gate set, in which every rotation is synthesized with the algorithm of Ross and Selinger~\cite{RossSelinger2016} to precision $\varepsilon_W/G$, where $G$ is the number of rotations in one block call. For $n=10$ and $m=2$, which corresponds to $N=1022$, the circuit acts on $25$ qubits, $15$ of which are ancillas. It uses $4.9\times10^5$ CNOT gates or, after Clifford+$T$ synthesis, $2.2\times10^7$ $T$ gates. The basis change accounts for $55\%$ of the CNOT gates and $87\%$ of the synthesized rotations in one block call. For $n\leq8$ when $m=2,3$, and for $n\leq9$ when $m=4$, every block address of the basis change lies below $C_m$. In these cases, the basis change consists of fixed rotations only. The asymptotic scaling of Theorem~\ref{thm:efficient-qut} is therefore not expected at these sizes.

\begin{figure}[t]
    \centering
    \includegraphics[width=\columnwidth]{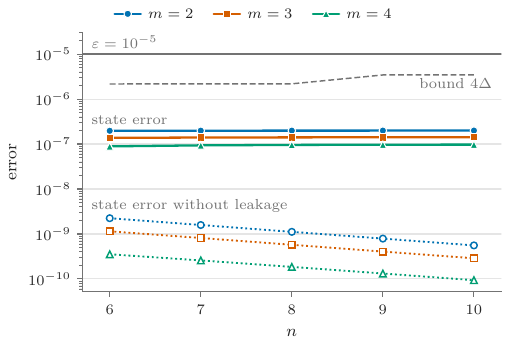}
    \caption{Simulation errors of the full QUT circuit constructed for $\varepsilon=10^{-5}$, applied to the uniform superposition over $\ket0,\ldots,\ket{N-1}$. Filled markers show the output error of Eq.~\eqref{eq:efficient-qut-guarantee}, including leakage into unused labels and nonzero ancilla states. Open markers show the error of the component with all ancillas in $\ket0$ and a valid output label. The dashed line shows the state error bound $4\Delta$ for $m=3$, the largest of the three for $6\leq n \leq 10$; the bounds for $m\in\{2,4\}$ lie slightly below it.}
    \label{fig:uniform-errors}
\end{figure}

To test the error analysis, we applied the same circuits to the uniform superposition $\ket\psi=N^{-1/2}\sum_{j=0}^{N-1}\ket j$ in a state-vector simulation. Because the simulation uses exact single-qubit rotations, only the approximation term $4\Delta$ of Eq.~\eqref{eq:qet-general-total-error} applies. Figure~\ref{fig:uniform-errors} compares the resulting output error with this bound and with the target error $\varepsilon=10^{-5}$. The output error, which includes amplitude in unused labels and in nonzero ancilla states, remains below $3\times10^{-7}$ for all $m$ and $n$ and varies little with $n$. It lies roughly one to two orders of magnitude below the bound $4\Delta$. The error of the component with all ancillas in $\ket0$ and a valid output label is about two to three orders of magnitude smaller than the output error. This error also decreases with $n$. The output error is therefore dominated by leakage out of the valid output subspace.

\section{Quantum ultraspherical spectral method}
\label{sec:us-solver}

We now use the QUT to construct a quantum version of the ultraspherical spectral method for linear boundary-value problems, following the classical framework of Olver and Townsend~\cite{Olver2013}. We suppose the solution is expanded in Chebyshev polynomials, so that its $\lambda$th derivative is represented in the ultraspherical basis with parameter $\lambda$. Sparse conversion operators map these terms into the ultraspherical basis with parameter equal to the differential order $q$, so they can be combined in a single linear system. For uniquely solvable problems with suitable boundary conditions, diagonal preconditioning can make this system well conditioned as the number of coefficients increases.

The QUT enables three operations within this discretization: preparing the forcing term directly in the parameter-$q$ basis, implementing multiplication by the variable coefficients $a_\lambda$ through their values at the Gauss--ultraspherical nodes of a single basis, and returning derivatives of the solution as weighted samples. With $K=2^n$ Chebyshev coefficients, the $\lambda$th derivative contains exactly $N_\lambda\coloneqq K-\lambda$ ultraspherical coefficients. These steps therefore require transforms $Q^{(\lambda)}_{N_\lambda}$ for $\lambda=0,1,\ldots,q$ on the same $n$-qubit register, which satisfy the dimension requirement $N+m=2^n$ of Theorem~\ref{thm:efficient-qut} with $N=N_\lambda$ and $m=\lambda$. The integer QUTs, together with the cosine and sine transforms for $\lambda=0$ and $1$, therefore provide the complete family of transforms needed for an order-$q$ ultraspherical discretization.

We first define the boundary-value problem and its discretization in orthonormal coefficient bases. We then show that diagonal preconditioning keeps the condition number bounded independently of the resolution $K$ for a broad class of boundary conditions. Using the resulting structure, we construct a block encoding of the preconditioned operator from QUTs and simple diagonal operations, and examine the cost of preparing the solution and estimating relevant outputs. Finally, we evaluate the discretization error and identify necessary conditions under which the quantum method might outperform classical solvers. Figure~\ref{fig:us-flow} provides an overview of the construction, with supporting proofs given in Appendix~\ref{app:us-proofs}.

\begin{figure*}[t]
\centering
\begin{tikzpicture}[x=1cm,y=1cm,font=\footnotesize,
  svpanel/.style={anchor=west,font=\small\bfseries},
  svreg/.style={black!80,line width=1.1pt},
  svgate/.style={draw=black!80,line width=0.4pt,fill=white,minimum width=0.58cm,minimum height=0.58cm,inner xsep=2.5pt,inner ysep=1.5pt},
  svqut/.style={svgate,fill=svQutFill},
  svnote/.style={text=black!55,font=\scriptsize},
  svdash/.style={line width=0.45pt,dash pattern=on 2pt off 1.5pt,rounded corners=1.5pt},
  ]
\definecolor{svBlue}{RGB}{46,92,170}
\definecolor{svOrange}{RGB}{204,102,0}
\colorlet{svQutFill}{blue!6}
\newcommand\svket[1]{\lvert #1\rangle}
\def\svnone{none}
\newcommand\svmat[3]{%
  \foreach \r in {0,...,7}{\foreach \c in {0,...,7}{%
    \pgfmathtruncatemacro\svd{abs(\r-\c)}%
    \pgfmathtruncatemacro\svb{round(85/(1+0.42*\c))}%
    \pgfmathtruncatemacro\svl{round((58-17*\svd)/(1+0.16*\c))}%
    \def\svfill{none}%
    \if#3A\ifnum\r<2 \edef\svfill{svOrange!\svb}%
          \else\ifnum\svd=0 \def\svfill{svBlue}\else\ifnum\svd<3 \edef\svfill{svBlue!\svl}\fi\fi\fi\fi
    \if#3I\ifnum\r>1 \ifnum\svd=0 \def\svfill{svBlue}\fi\fi\fi
    \if#3B\ifnum\r<2 \edef\svfill{svOrange!\svb}\fi\fi
    \if#3C\ifnum\r>1 \ifnum\svd<3 \edef\svfill{svBlue!\svl}\fi\fi\fi
    \ifx\svfill\svnone
      \path[fill=white,draw=black!20,line width=0.25pt] ({#1+0.14*\c},{#2+0.14*(7-\r)}) rectangle ++(0.115,0.115);
    \else
      \path[fill=\svfill] ({#1+0.14*\c},{#2+0.14*(7-\r)}) rectangle ++(0.115,0.115);
    \fi}}}
\path[use as bounding box] (-0.05,0.0) rectangle (17.80,4.55);

\begin{scope}[shift={(0,3.39)}]
\node[svpanel] at (-0.05,0.98) {(a)};
\draw[svreg] (0.66,0) -- (13.45,0);
\foreach \s in {1,-1}{\draw[svreg,rounded corners=3pt] (9.00,0) -- (9.40,0) -- (9.40,0.80*\s) -- (13.45,0.80*\s);}
\node[anchor=east] at (0.62,0) {$\svket F$};
\node[svqut] at (1.85,0) {$Q^{(q)}_{K-q}$};
\node[svgate] at (3.40,0) {$+q$};
\node[anchor=south,inner sep=2pt] at (4.65,0.05) {$\svket b$};
\node[svgate,minimum width=1.9cm,minimum height=0.72cm] at (6.55,0) {$\tfrac{\alpha}{2\bar\kappa}\,\mathcal A_K^{-1}$};
\node[anchor=south,inner sep=2pt] at (8.25,0.05) {$\svket{\boldsymbol v}$};
\node[svgate,minimum width=0.66cm] at (10.60,0.80) {$-q$};
\node[svqut,minimum width=1.14cm] at (12.25,0.80) {$Q^{(q)\dagger}_{K-q}$};
\node[svgate,minimum width=0.66cm] at (10.60,0) {$R$};
\node[svnote,anchor=north] at (10.60,-0.30) {$p_{\mathrm{rec}}$};
\node[svqut,minimum width=1.14cm] at (12.25,0) {$Q^{(0)\dagger}_{K}$};
\node[anchor=west] at (13.50,0.80) {$u_K^{(q)}(x^{(q)}_j)$};
\node[anchor=west] at (13.50,0) {$u_K(x^{(0)}_j)$};
\node[anchor=west] at (13.50,-0.80) {$\langle g\vert\boldsymbol v\rangle\propto\mathcal O[u_K]$};
\end{scope}

\draw[svdash,draw=black!55] (0.0,0.0) rectangle (17.80,2.12);
\node[fill=white,inner sep=1.5pt] (calls) at (6.55,2.12) {$\times\,O(\bar\kappa\log(\bar\kappa/\eta))$};
\draw[-{Stealth[length=4pt,width=3.5pt]},black!60,line width=0.6pt] (calls.north) -- (6.55,3.01);

\node[svpanel] at (0.05,1.86) {(b)};
\svmat{0.62}{0.44}{A}
\draw[black!45,line width=0.4pt] (0.56,1.28) -- (0.51,1.28) -- (0.51,1.535) -- (0.56,1.535);
\node[svnote,anchor=east,inner sep=1pt] at (0.50,1.41) {$q$};
\node at (2.07,1.00) {$=$};
\svmat{2.40}{0.44}{I}
\node at (3.85,1.00) {$+$};
\svmat{4.18}{0.44}{B}
\node at (5.60,1.00) {$+$};
\node at (6.05,1.00) {$\sum_\lambda$};
\svmat{6.45}{0.44}{C}
\node at (1.18,0.22) {$\mathcal A_K$};
\node[svnote] at (2.96,0.22) {$1$};
\node[svnote] at (4.74,0.22) {$\lVert\boldsymbol r_i\rVert_2$};
\node[svnote] at (7.01,0.22) {$(\tfrac32)^{q-\lambda}\rho_\lambda\bar a_\lambda$};

\draw[black!30,line width=0.35pt] (7.95,0.25) -- (7.95,1.90);
\node[svpanel] at (8.05,1.86) {(c)};
\draw[svreg] (8.35,1.00) -- (17.65,1.00);


\node[svgate,anchor=west] (c1) at (8.50,1.00) {$\widehat D_\lambda R$};
\node[svgate,anchor=west] (c2) at ([xshift=0.10cm]c1.east) {$\widehat S_{\lambda\to\mu^\star}$};
\node[svqut,anchor=west]  (c3) at ([xshift=0.10cm]c2.east) {$Q^{(\mu^\star)\dagger}_{N_{\mu^\star}}$};
\node[svgate,anchor=west] (c4) at ([xshift=0.10cm]c3.east) {$a_\lambda(x^{(\mu^\star)}_j)$};
\node[svqut,anchor=west]  (c5) at ([xshift=0.10cm]c4.east) {$Q^{(\mu^\star)}_{N_{\mu^\star}}$};
\node[svgate,anchor=west] (c6) at ([xshift=0.10cm]c5.east) {$\widehat S_{\mu^\star\to q}$};
\node[svgate,anchor=west] (c7) at ([xshift=0.10cm]c6.east) {$\Pi_{<K-q}$};
\node[svgate,anchor=west] (c8) at ([xshift=0.10cm]c7.east) {$+q$};
\draw[black!45,line width=0.4pt] ([yshift=-0.08cm]c3.south west) -- ++(0,-0.07) -| ([yshift=-0.08cm]c5.south east);
\node[anchor=north,inner sep=2pt] at ([yshift=-0.16cm]c4.south) {$M_{\mu^\star}[a_\lambda]$};

\end{tikzpicture}%
\caption{Quantum ultraspherical solver.
\textbf{(a)} Solver pipeline: preparation of the forcing state, QSVT inversion, and readout, from top to bottom, of the $q$th derivative samples, the solution samples, or a linear functional. The shift $+q$ reserves the first $q$ labels for boundary equations.
\textbf{(b)}~LCU decomposition of the preconditioned operator $\mathcal A_K$. Gray labels show the normalization weights, which sum to $\alpha$.
\textbf{(c)}~One lower-order contribution in Eq.~\eqref{eq:us-operator}, with basis conversions before and after multiplication in the common parameter-$\mu^\star$ basis. All lower-order terms share the same QUT pair.}
\label{fig:us-flow}
\end{figure*}
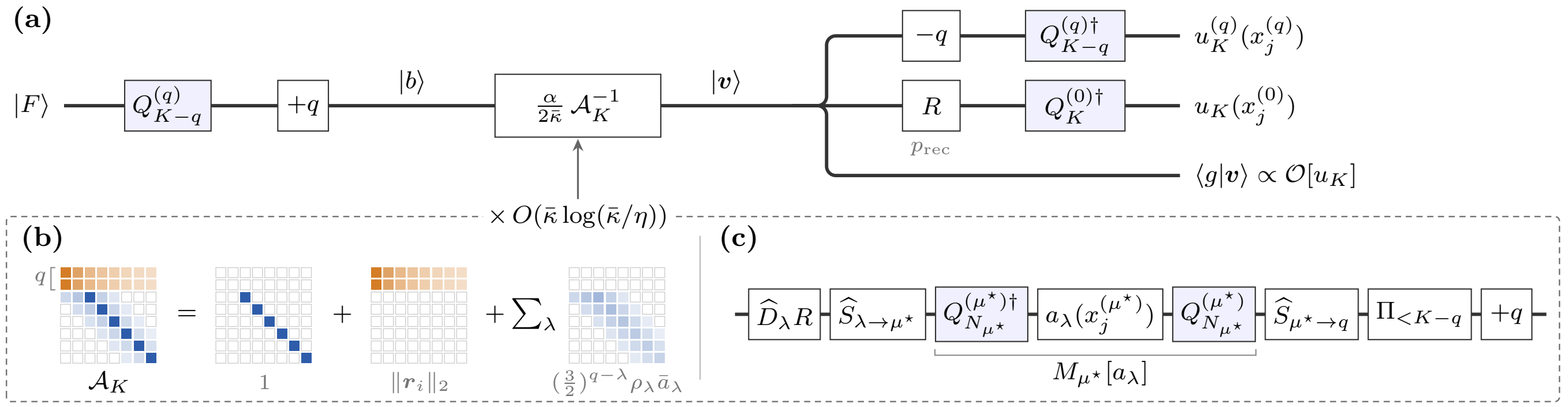

\subsection{Boundary-value problems of order \texorpdfstring{$q$}{q}}
\label{sub:us-problem}
Fix an integer $q\geq1$. We consider the linear differential equation
\begin{equation}
\sum_{\lambda=0}^{q}a_\lambda(x)\,u^{(\lambda)}(x)=f(x),
 \label{eq:us-ode}
\end{equation}
on $(-1,1)$, subject to $q$ linear boundary conditions
\begin{align}
 \mathcal B_iu&=c_i,\qquad i=0,\ldots,q-1.
 \label{eq:us-bc}
\end{align}
We assume that the equation is regular, with $a_q(x)\neq0$ on $[-1,1]$, and, without loss of generality, take $a_q\equiv 1$. We further assume that $a_0,\ldots,a_{q-1}$ are continuous on $[-1,1]$ and define $\bar a_\lambda\coloneqq\max_{[-1,1]}|a_\lambda|$.

Each $\mathcal B_i$ is a bounded linear functional on $C^q([-1,1])$. Examples include endpoint values $u^{(j)}(\pm1)$ and interior values $u^{(j)}(x_0)$ for $j<q$, weighted integrals $\int_{-1}^{1}\varpi(x)u(x)\,\dd x$ with integrable $\varpi$, and linear combinations of these. We assume that the corresponding homogeneous boundary-value problem has only the trivial solution. Equivalently, if $y_0,\ldots,y_{q-1}$ form a basis of solutions of the homogeneous differential equation, the $q\times q$ matrix with entries $\mathcal B_i y_j$ must be invertible. This is the standard uniqueness condition for a linear boundary-value problem.

The problem class includes first-order initial-value problems with $q=1$ and $\mathcal B_0u=u(-1)$, second-order advection--reaction--diffusion equations with Dirichlet conditions, and fourth-order beam equations with clamped ends. We exclude equations whose leading coefficient vanishes on $[-1,1]$, such as the ultraspherical operator $-(1-x^2)u''+(2m+1)xu'$, whose eigenfunctions are the polynomials $p_k^{(m)}$. However, for equations of the form $\mu u-(1-x^2)u''+(2m+1)xu'=f$ with $\mu>0$, the QUT directly diagonalizes the differential operator, allowing fast inversion~\cite{Tong2021} in its eigenbasis.

\begin{figure*}[t]
\centering
\includegraphics[width=\textwidth]{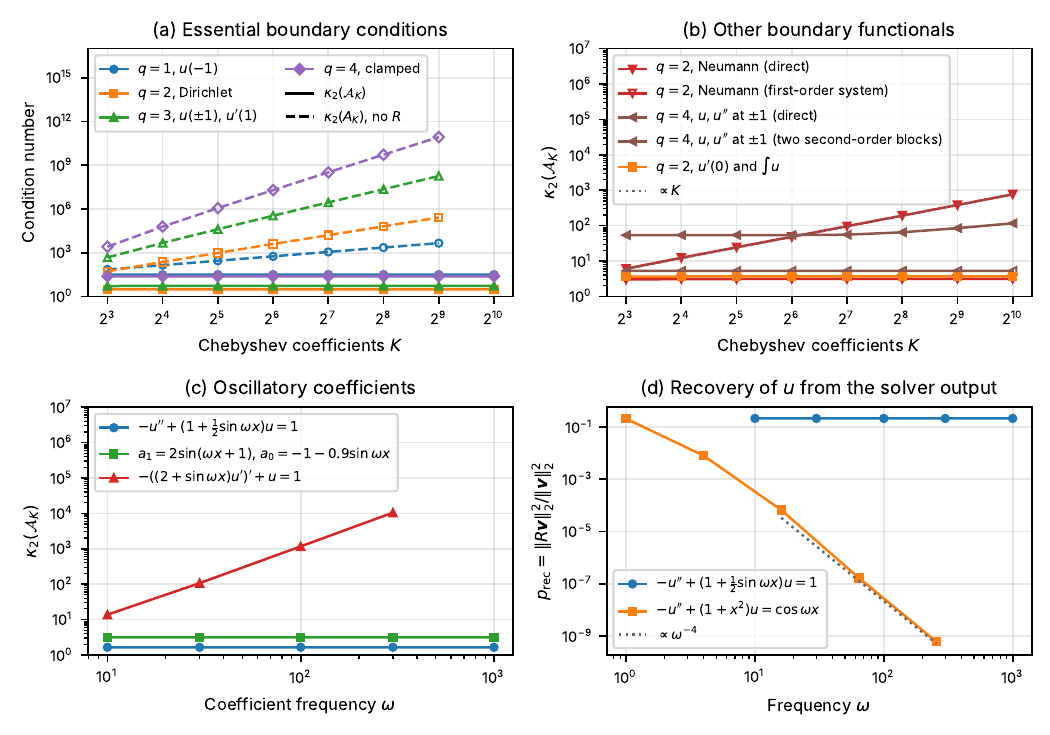}
\caption{Preconditioned ultraspherical systems. All equations are written as Eq.~\eqref{eq:us-ode} with $a_q=1$ and discretized by Eq.~\eqref{eq:us-system}. \textbf{(a)}~Condition numbers $\kappa_2(\mathcal A_K)$ (solid) and $\kappa_2(A_K)$ without rescaling (dashed) for $u'+4xu=f$ with known $u(-1)$ ($q=1$); $u''+\sin(2x)u'-(2+\cos3x)u=f$ with Dirichlet data ($q=2$); $u'''+xu''-(1+x^2)u'+(\cos x)u=f$ with $u(\pm1)$ and $u'(1)$ ($q=3$); and $u^{(4)}+(1+x^2)u''-xu'+(2+\sin x)u=f$ with clamped ends ($q=4$). 
\textbf{(b)}~Direct (filled markers) and split (open markers) discretizations of two problems: $u''-(1+x^2)u=f$ with Neumann data $u'(\pm1)$ (red), split into a first-order system, and the $q=4$ equation with $u(\pm1)$ and $u''(\pm1)$ (brown), split into two coupled second-order Dirichlet blocks for $u$ and $u''$. For comparison, the $q=2$ equation of panel~(a) with the conditions $u'(0)$ and $\int_{-1}^{1}u\,\dd x$ is shown in orange.
\textbf{(c)}~Coefficients of frequency $\omega$ with Dirichlet conditions. We take $K$ as the smallest power of two with $K\geq1.25\,\omega+64$, or $K\geq6\,\omega+64$ for the divergence form (red). Doubling $K$ changes each plotted value by at most $0.01\%$.
\textbf{(d)}~Recovery probability for an oscillatory coefficient with smooth forcing and for an oscillatory forcing, both with $u(\pm1)=0$.}
\label{fig:us-checks}
\end{figure*}

\subsection{Orthonormal ultraspherical discretization}
\label{sub:us-discretization}

Let $K=2^n>q$. We approximate the solution by a polynomial of degree less than $K$,
\begin{equation}
    u_K(x) = \sum_{k=0}^{K-1}u_k\,p_k^{(0)}(x),
 \label{eq:us-expansion}
\end{equation}
with coefficient vector $\boldsymbol u=(u_0,\ldots,u_{K-1})^\T$. Since the QUT acts on orthonormal coefficients, we express all operators in the orthonormal bases $p_k^{(\lambda)}$. Differentiation is particularly simple in these bases: each derivative maps a basis polynomial to a single polynomial in the next ultraspherical basis. Combining Eq.~\eqref{eq:solution-derivative} with the norms in Eq.~\eqref{eq:gegenbauer-norm} yields, for $k\geq\lambda\geq0$,
\begin{equation}
 \frac{\dd^\lambda p_k^{(0)}}{\dd x^\lambda}
 =d_k^{(\lambda)}\,p_{k-\lambda}^{(\lambda)},
 \qquad
 d_k^{(\lambda)}\coloneqq\prod_{j=0}^{\lambda-1}\sqrt{k^2-j^2},
 \label{eq:us-derivative}
\end{equation}
while the derivative vanishes for $k<\lambda$. Each factor arises from one differentiation, which maps $p_{k-j}^{(j)}$ to $\sqrt{k^2-j^2}\,p_{k-j-1}^{(j+1)}$. Thus, the differentiation matrix $\widehat D_\lambda\in\mathbb R^{N_\lambda\times K}$ has only one nonzero entry per column $k\geq\lambda$, given by $(\widehat D_\lambda)_{k-\lambda,k}=d_k^{(\lambda)}$. For example, $(\widehat D_2)_{k-2,k}=k\sqrt{k^2-1}$ for $k\geq2$.

The differentiation matrix $\widehat D_\lambda$ produces coefficients directly in the parameter-$\lambda$ ultraspherical basis. One can apply $Q^{(\lambda)\dagger}_{N_\lambda}$ to convert these coefficients into weighted samples of $u_K^{(\lambda)}$ at the corresponding Gauss--ultraspherical nodes. All these transforms use the same $n$-qubit register, with the first $N_\lambda$ labels holding the valid coefficients. For $\lambda=0$ and $\lambda=1$, the corresponding transforms are the DCT-II and DST-I from Eqs.~\eqref{eq:dct-case} and~\eqref{eq:dst-case}.

To combine derivatives of different orders, we use sparse conversion operators that express each polynomial in the next ultraspherical basis. Rewriting Eq.~\eqref{eq:gegenbauer-conversion} in the normalized basis gives
\begin{equation}
p_k^{(\lambda)}
=s_k^{(\lambda)}\,p_k^{(\lambda+1)}
-t_k^{(\lambda)}\,p_{k-2}^{(\lambda+1)},
\label{eq:us-conversion}
\end{equation}
where, for $\lambda\geq1$,
\begin{equation}
    \begin{aligned}
    s_k^{(\lambda)} &= \frac12\sqrt{\frac{(k+2\lambda)(k+2\lambda+1)}{(k+\lambda)(k+\lambda+1)}},\\
    t_k^{(\lambda)} &= \frac12\sqrt{\frac{k(k-1)}{(k+\lambda)(k+\lambda-1)}}.
    \end{aligned}
    \label{eq:us-conversion-entries}
\end{equation}
For $\lambda=0$, these formulas extend to $s_k^{(0)}$ for $k\geq1$ and to $t_k^{(0)}$ for $k\geq2$, with $s_0^{(0)}=1/\sqrt2$. We set $t_0^{(\lambda)}=t_1^{(\lambda)}=0$ for every $\lambda$. Applying Eq.~\eqref{eq:us-conversion} $\mu-\lambda$ times allows rewriting $p_k^{(\lambda)}$ as a linear combination of $p_k^{(\mu)},p_{k-2}^{(\mu)},\ldots,p_{k-2(\mu-\lambda)}^{(\mu)}$ for any $\mu\geq\lambda$. For $\lambda\leq\mu\leq q$, we therefore define the conversion matrix $\widehat S_{\lambda\to\mu}\in\mathbb R^{N_\mu\times N_\lambda}$, which maps the coefficients of a polynomial of degree less than $N_\lambda$ in the basis $p^{(\lambda)}$ to its first $N_\mu$ coefficients in the basis $p^{(\mu)}$. Its matrix representation is upper triangular: column $k$ of $\widehat S_{\lambda\to\mu}$ has nonzero entries only in the rows $k,k-2,\ldots,k-2(\mu-\lambda)$. For a single step, these entries are $(\widehat S_{\lambda\to\lambda+1})_{k,k}=s_k^{(\lambda)}$ and $(\widehat S_{\lambda\to\lambda+1})_{k-2,k}=-t_k^{(\lambda)}$. For $\mu>\lambda+1$, $\widehat S_{\lambda\to\mu}$ applies these single steps to all $N_\lambda$ coefficients and keeps only the first $N_\mu$ coefficients at the end. Since the weight functions $w_\lambda(x)=(1-x^2)^{\lambda-1/2}$ satisfy $w_\mu=(1-x^2)^{\mu-\lambda}w_\lambda\leq w_\lambda$ for $\mu\geq\lambda$, every conversion is contractive, $\|\widehat S_{\lambda\to\mu}\|_2\leq1$.

The QUT also provides a convenient representation of multiplication operators. Let $J^{(\lambda)}_{N_\lambda}$ be the $N_\lambda\times N_\lambda$ Jacobi matrix representing multiplication by $x$ in the basis $p_0^{(\lambda)},\ldots,p_{N_\lambda-1}^{(\lambda)}$, truncated to degree less than $N_\lambda$. If $x_j$ is a zero of $p_{N_\lambda}^{(\lambda)}$, the three-term recurrence implies that $(p_0^{(\lambda)}(x_j),\ldots,p_{N_\lambda-1}^{(\lambda)}(x_j))^\T$ is an eigenvector of $J^{(\lambda)}_{N_\lambda}$ with eigenvalue $x_j$. After weighting these vectors by $\sqrt{w_j}$, Eq.~\eqref{eq:discrete-orthogonality} yields an orthonormal eigenbasis~\cite{Golub1969}. Hence,
\begin{equation}
 Q^{(\lambda)}_{N_\lambda}\diag(x_0,\ldots,x_{N_\lambda-1})\,Q^{(\lambda)\dagger}_{N_\lambda}=J^{(\lambda)}_{N_\lambda}.
 \label{eq:us-golub-welsch}
\end{equation}
For a function $a(x)$ on $[-1,1]$, we therefore define the $N_\lambda\times N_\lambda$ Gauss multiplication matrix
\begin{equation}
    \begin{aligned}
     M_\lambda[a] &\coloneqq Q^{(\lambda)}_{N_\lambda}\diag\bigl(a(x_0),\ldots,a(x_{N_\lambda-1})\bigr)\,Q^{(\lambda)\dagger}_{N_\lambda} \\
     &= a\bigl(J^{(\lambda)}_{N_\lambda}\bigr).
    \end{aligned}
    \label{eq:us-multiplication}
\end{equation}
Applied to the coefficients of a polynomial $g$ of degree less than $N_\lambda$, $M_\lambda[a]$ returns the coefficients of the polynomial that interpolates $ag$ at the $N_\lambda$ Gauss nodes. If $a$ is a polynomial, Gaussian quadrature implies
\begin{align}
 (M_\lambda[a])_{ik}
 &=\int_{-1}^{1}w_\lambda(x)\,p_i^{(\lambda)}(x)a(x)p_k^{(\lambda)}(x) \,\dd x
\end{align}
whenever $i+k+\deg a\leq2N_\lambda-1$. Thus, except for entries involving the highest degrees, $M_\lambda[a]$ is exactly the matrix obtained by multiplying by $a$ and projecting the result onto the first $N_\lambda$ ultraspherical basis functions. The possible discrepancy is confined to entries with $i+k+\deg a>2N_\lambda-1$, where the product exceeds the degree for which the $N_\lambda$-point Gauss rule is exact. Moreover, $\norm{M_\lambda[a]}_2=\max_j|a(x_j)|\leq\max_{[-1,1]}|a|$.

We now apply these multiplication operators to the differential equation. Let $\mu^\star$ be the largest $\lambda<q$ with $a_\lambda\not\equiv0$. For each lower-order term $a_\lambda u^{(\lambda)}$, the differentiation matrix $\widehat D_\lambda$ first produces the coefficients of $u^{(\lambda)}$ in the parameter-$\lambda$ ultraspherical basis. We convert them to the parameter-$\mu^\star$ basis, apply $M_{\mu^\star}[a_\lambda]$ there, and convert the result to the common parameter-$q$ basis. With exact polynomial representations, multiplication by $a_\lambda$ commutes with the basis conversions: either order represents the same function $a_\lambda u^{(\lambda)}$. Differences can arise only from finite truncation near the highest degrees. This choice therefore allows all lower-order terms to use the same multiplication basis, and hence the same transform. The interior equations are therefore represented by the $N_q\times K$ matrix
\begin{equation}
 A_K^{\mathrm{int}}\coloneqq\widehat D_q+\sum_{\lambda=0}^{q-1}\widehat S_{\mu^\star\to q}\,M_{\mu^\star}[a_\lambda]\,\widehat S_{\lambda\to\mu^\star}\,\widehat D_\lambda.
 \label{eq:us-operator}
\end{equation}
Since a conversion only changes how a polynomial is written, any common basis at or above the highest derivative order that appears represents the same products. We use the lowest such basis, $\mu^\star$, because transforms become more expensive as the parameter grows, from the $O(n^2)$ DCT-II and DST-I for $\mu^\star\leq1$ to QUTs whose cost increases with the parameter.

The forcing is represented by its interpolant at the $N_q$ Gauss nodes of $C^{(q)}_{N_q}$. From the weighted samples $F_j\coloneqq\sqrt{w_j}f(x_j)$, Eq.~\eqref{eq:transform-pair} yields $\widehat{\boldsymbol f}\coloneqq Q^{(q)}_{N_q}\boldsymbol F\in\mathbb R^{N_q}$, the orthonormal coefficient vector of this interpolant. The boundary functionals act on the Chebyshev coefficients through $B\in\mathbb R^{q\times K}$, with entries $B_{ik}\coloneqq\mathcal B_i\bigl(p_k^{(0)}\bigr)$. With $\boldsymbol c=(c_0,\ldots,c_{q-1})^\T$, the boundary and interior equations form the square system
\begin{equation}
 A_K\boldsymbol u = \begin{bmatrix}\boldsymbol c\\ \widehat{\boldsymbol f}\end{bmatrix},
 \qquad A_K \coloneqq \begin{bmatrix}B \\ A_K^{\mathrm{int}}\end{bmatrix}.
 \label{eq:us-system}
\end{equation}
Since $A_K$ is $K\times K$ with $K=2^n$, the full system still stays on the same $n$-qubit register. The solution coefficient $u_k$ occupies label $k$, while the output labels $0,\ldots,q-1$ represent the boundary equations and labels $q,\ldots,K-1$ represent the interior equations. The transform $Q^{(q)}_{N_q}$ initially places the $N_q$ forcing coefficients on labels $0,\ldots,N_q-1$. We apply an addition by $q$ to shift them to labels $q,\ldots,K-1$, leaving the first $q$ labels for the boundary data.

\subsection{Numerical conditioning}
\label{sub:us-conditioning}

The expression in Eq.~\eqref{eq:us-derivative} implies $d_k^{(q)}=\Theta(k^q)$, causing $\kappa_2(A_K)$ to grow like $K^q$, as shown in Fig.~\ref{fig:us-checks}(a). We can rescale the unknowns with the $K\times K$ diagonal matrix $R \coloneqq \operatorname{diag}(R_{kk})$, where
\begin{equation}
 R_{kk} \coloneqq \begin{cases}1,&k<q,\\1/d_k^{(q)},&k\geq q.\end{cases}
 \label{eq:us-R}
\end{equation}
We then solve $\mathcal A_K\boldsymbol v=(\boldsymbol c,\widehat{\boldsymbol f})^\T$ with $\mathcal A_K\coloneqq A_KR$ and recover $\boldsymbol u=R\boldsymbol v$. The rescaling cancels the factor $d_k^{(q)}$ in the highest derivative: $\widehat D_qR$ maps column $k\geq q$ to row $k-q$ with unit coefficient. It therefore removes the resolution-dependent growth caused by differentiation. This is analogous to the fast-inversion strategy of~\cite{Tong2021}, but here the known differentiation factors are inverted explicitly.

In the register layout above, $\widehat D_qR$ acts as the identity on labels $q,\ldots,K-1$, so
\begin{equation}
 \mathcal A_K = I-\Pi_{<q}+\sum_{i=0}^{q-1}\ket{i}\bra{\Pi_{<K}\boldsymbol r_i}+C_K.
 \label{eq:us-structure}
\end{equation}
Here $\Pi_{<M}$ denotes the projection onto the first $M$ labels. The rescaled boundary row $\boldsymbol r_i$ is an infinite sequence with entries $r_{ik}\coloneqq R_{kk}\mathcal B_i(p_k^{(0)})$ for all $k\geq0$, with $R_{kk}$ from Eq.~\eqref{eq:us-R}, so $\Pi_{<K}\boldsymbol r_i$ is the $i$th row of $BR$. The lower-order terms are collected in the $K\times K$ matrix $C_K$ whose first $q$ rows vanish and whose last $N_q$ rows are $(A_K^{\mathrm{int}}-\widehat D_q)R$, with $A_K^{\mathrm{int}}$ from Eq.~\eqref{eq:us-operator}.

The rescaling also suppresses the lower-order terms at high polynomial degree.  For $k\geq q$, the contribution of the term of order $\lambda$ to column $k$ has norm at most
$\bar a_\lambda d_k^{(\lambda)}/d_k^{(q)}=O(k^{\lambda-q})$. Thus, every term with $\lambda<q$ decays as $k$ increases.

It remains to control the rescaled boundary rows uniformly in $K$. For endpoint derivatives, evaluating Eq.~\eqref{eq:solution-derivative} at $x=\pm1$ yields, for $k\geq1$, $\partial_x^jp_k^{(0)}(\pm1) =(\pm1)^{k+j}\sqrt{2/\pi}\,\prod_{l<j}(k^2-l^2)/(2l+1)$, and therefore $|r_{ik}|=\Theta(k^{2j-q})$. More generally, suppose that $\mathcal B_i$ is a bounded linear functional on $C^j([-1,1])$. Applying the higher-order Markov brothers' inequality~\cite{Markoff1916} to the degree-$k$ polynomial $p_k^{(0)}$ gives
\begin{equation}
    \| \partial_{x}^{\ell} p_k^{(0)}\|_{\infty} \leq \frac{
    k^2 (k^2-1^2)\dots (k^2 - (\ell-1)^2)
    }{(2\ell-1)!!} \|p_k^{(0)}\|_{\infty}.
\end{equation}
Since the normalized Chebyshev basis has bounded norm $\| p^{(0)}_k\|_{\infty} = O(1)$, it follows that $\| \partial_{x}^{\ell} p_k^{(0)}\|_{\infty} = O(k^{2\ell})$ for $\ell \leq j$. Boundedness of $\mathcal B_i$ on $C^j([-1,1])$ then implies $|\mathcal B_i(p_k^{(0)})|=O(k^{2j})$, and hence $|r_{ik}|=O(k^{2j-q})$. These estimates demonstrate the sufficient condition $j<q/2$ for square summability of the rescaled boundary rows. Appendix~\ref{app:us-bc} proves the following proposition, which also admits interior derivative conditions of every order below $q$.

\begin{proposition}[Bounded boundary rows]\label{prop:us-bc}
Let $\boldsymbol r=\bigl(R_{kk}\mathcal B(p_k^{(0)})\bigr)_{k\geq0}$ be the rescaled row associated with a boundary functional $\mathcal B$. Then $\boldsymbol r\in\ell^2(\mathbb N)$, and hence $\norm{\Pi_{<K}\boldsymbol r}_2$ is bounded uniformly in $K$, in either of the following cases:
\begin{enumerate}
\item[(i)] $\mathcal B$ is bounded on $C^j([-1,1])$ for some $j<q/2$.
This includes endpoint conditions $\mathcal Bu=u^{(j)}(\pm1)$ and, for $j=0$, weighted integrals $\mathcal Bu=\int_{-1}^{1}\varpi u\,\dd x$ with integrable $\varpi$;

\item[(ii)] $\mathcal Bu=u^{(j)}(x_0)$ with $|x_0|<1$ and $j<q$.
\end{enumerate}
If $\mathcal Bu=u^{(j)}(\pm1)$ with $q/2\leq j<q$, then
$\norm{\Pi_{<K}\boldsymbol r}_2=\Theta(K^{2j-q+1/2})$.
\end{proposition}

For second-order equations, this includes Dirichlet conditions but excludes Neumann and Robin conditions. For fourth-order equations, it also includes first-derivative conditions such as those at clamped boundaries, although higher-order endpoint derivatives may still fall outside the admissible range. Such cases can be handled by introducing derivatives as additional unknowns and rewriting the equation as a first-order system, as described at the end of this subsection.

For these admissible boundary conditions, the boundary contribution in Eq.~\eqref{eq:us-structure} consists of $q$ rows whose norms are bounded independently of $K$. The lower-order terms are also controlled: within $C_K$, the contribution of derivative order $\lambda<q$ has a $k$th column of norm $O(k^{\lambda-q})$, which decays with $k$. Together with uniqueness of the continuous boundary-value problem, these properties lead to uniform conditioning as the number of coefficients increases.

\begin{proposition}[Uniform conditioning]\label{prop:us-conditioning}
Assume that $a_0,\ldots,a_{q-1}$ are continuous on $[-1,1]$, that every boundary functional is bounded on $C^q([-1,1])$ and has a rescaled row in $\ell^2(\mathbb{N})$, and that the homogeneous problem has only the trivial solution. Extend $\mathcal A_K$ by the identity on labels $k\geq K$. Then $\mathcal A_K$ converges in operator norm to an invertible operator
\begin{equation}
    \mathcal A_\infty=I+\mathcal K
\end{equation}
on $\ell^2(\mathbb{N})$, where $\mathcal K$ is compact, i.e., a norm limit of finite-rank operators. Consequently, for every $\gamma>1$, there exists $K_0$ such that, for all $K\geq K_0$,
\begin{equation}
 \kappa_2(\mathcal A_K)\leq\gamma \, \kappa_2(\mathcal A_\infty).
 \label{eq:us-uniform}
\end{equation}
\end{proposition}
We prove this proposition by adapting Lemmas~4.3 and~4.4 of~\cite{Olver2013} to the orthonormal bases used here. The details are given in Appendix~\ref{app:us-conditioning}. The limiting condition number depends on the differential coefficients and boundary conditions. The proof does not provide an explicit threshold $K_0$, but Lemma~\ref{lem:us-certificate} in Appendix~\ref{app:us-certificate} specifies a computable one together with an explicit bound on $\norm{\mathcal A_K^{-1}}_2$.

Figure~\ref{fig:us-checks}(a) illustrates this behavior at finite resolution. We consider four linear ODEs of orders one through four, with variable coefficients and boundary conditions satisfying Proposition~\ref{prop:us-bc}. Over the range $8\leq K\leq1024$, $\kappa_2(\mathcal A_K)$ remains nearly constant and below $50$ in all four examples. Without the rescaling, by contrast, the condition number grows approximately as $K^q$. The singular values of the preconditioned matrices also cluster near one, consistent with the correction to the identity. For the $q=2$ example at $K=1024$, only $14$ of the $1024$ singular values lie outside $[0.99,1.01]$.

Rapid oscillation of the coefficients alone does not cause poor conditioning. Figure~\ref{fig:us-checks}(c) considers $-u''+(1+\tfrac12\sin\omega x)u=1$ with $u(\pm 1)=0$. As $\omega$ increases from $10$ to $1000$, the required resolution grows from $K=128$ to $K=2048$, while $\kappa_2(\mathcal A_K)$ remains small. Adding the term $2\sin(\omega x+1)u'$ and increasing the oscillation amplitude in the coefficient of $u$ from $1/2$ to $0.9$ raises the condition number only moderately over the same frequency range.

In these examples, the coefficient amplitudes are bounded independently of $\omega$. Conditioning can deteriorate more strongly when normalization by the leading coefficient introduces terms whose amplitudes grow with frequency. For $-((2+\sin\omega x)u')'+u=1$, dividing by the leading coefficient produces $a_1(x)=\omega\cos(\omega x)/(2+\sin\omega x)$, whose maximum magnitude is $\omega/\sqrt3$. Correspondingly, $\kappa_2(\mathcal A_K)$ increases from about $10^1$ at $\omega=10$ to about $10^4$ at $\omega=300$.

Higher-order endpoint conditions provide another source of growth with $K$. For $u''-(1+x^2)u=f$ with prescribed Neumann data $u'(\pm1)$, Proposition~\ref{prop:us-bc} predicts boundary-row norms of order $K^{1/2}$. In this example, $\kappa_2(\mathcal A_K)$ grows approximately linearly with $K$, as shown in Fig.~\ref{fig:us-checks}(b).

This growth can be avoided by introducing the derivatives as separate unknowns, $U_j\coloneqq u^{(j)}$ for $0\leq j<q$. Equation~\eqref{eq:us-ode} then becomes the first-order system
\begin{equation}
\begin{cases}
U_j' = U_{j+1}, & \text{for } 0\leq j<q-1,\\
U_{q-1}'+\sum_{\lambda=0}^{q-1}a_\lambda U_\lambda = f. &
\end{cases}
\end{equation}
Endpoint conditions on $u,u',\ldots,u^{(q-1)}$ become value conditions on the corresponding components. Applying the $q=1$ discretization and rescaling to each equation produces bounded boundary rows by Proposition~\ref{prop:us-bc}. For the Neumann example above, the condition number of the reformulated system remains close to $3$ throughout the tested range of $K$. This first-order formulation also covers initial-value problems of arbitrary order and requires only the transforms for $\lambda=0$ and $1$.

When $p>1$ divides $q$, we can instead rewrite the problem as coupled equations of order $p$. Each component equation then uses ultraspherical transforms up to parameter $p$, including QUTs with $\lambda\geq2$, while the boundary conditions involve lower-order derivatives of the new unknowns. To satisfy Proposition~\ref{prop:us-bc}, each endpoint condition must involve derivatives of order below $p/2$ of the corresponding unknown. For example, a simply supported beam with prescribed $u(\pm1)$ and $u''(\pm1)$ can be written as two coupled second-order equations with Dirichlet conditions on $u$ and $u''$. This formulation uses $Q^{(2)}_{N_2}$ and keeps $\kappa_2\approx5.28$ across the tested resolutions, whereas the direct fourth-order discretization has a condition number one to two orders of magnitude larger and continues to grow with $K$. Appendix~\ref{app:us-natural} provides the details of both reformulations.

\subsection{Block encoding with QUT}
\label{sub:us-encoding}

We construct a block encoding of $\mathcal A_K/\alpha$ by combining the terms in Eq.~\eqref{eq:us-structure} through the LCU method. Figure~\ref{fig:us-flow} shows the resulting circuit and its role in the solver. The identity contribution $I-\Pi_{<q}$ has normalization one and is implemented with a comparator that selects labels $k\geq q$.

For the boundary contribution, let $V_i$ prepare from $\ket0$ a state whose component with a flag qubit in $\ket0$ is $\Pi_{<K}\boldsymbol r_i/\norm{\boldsymbol r_i}_2$. Applying $V_i^\dagger$, selecting the zero label, and relabeling $0\mapsto i$ produces a block encoding of $\ket i\bra{\Pi_{<K}\boldsymbol r_i}/\norm{\boldsymbol r_i}_2$. If every rescaled boundary row is square summable, an LCU over $i$ with weights $\norm{\boldsymbol r_i}_2$ then represents the full boundary contribution with subnormalization $\sum_{i=0}^{q-1} \| \boldsymbol{r}_i\|_2$, which is independent of $K$.

For endpoint conditions $\mathcal B_i u=u^{(j)}(\pm1)$ with $j<q/2$, $V_i$ can be constructed using an envelope that is constant on dyadic intervals, as detailed in Appendix~\ref{app:us-encoding}. Beyond a fixed set of low degrees, the envelope has height $C\,2^{-s(q-2j)}$ on $2^s\leq k<2^{s+1}$, where $C$ depends only on $q$ and $j$. A normalized envelope state is first prepared over the degree labels. Then, a controlled rotation by the ratio of each row entry to the corresponding envelope value produces the desired amplitudes on a flagged branch. Since the unnormalized envelope has bounded $\ell^2$-norm, the success probability is bounded below independently of $K$. A constant number of amplification rounds followed by a padding rotation then prepares the flagged component $\Pi_{<K}\boldsymbol r_i/\norm{\boldsymbol r_i}_2$. For other boundary functionals, we assume efficient preparation of the rescaled row and denote its cost $T_{\boldsymbol r_i}$ below.

Each lower-order term contains the rescaled derivative $\widehat D_\lambda R$, the multiplication operator $M_{\mu^\star}[a_\lambda]$, and the conversions $\widehat S_{\lambda\to\mu^\star}$ and $\widehat S_{\mu^\star\to q}$ before and after it. We implement them from right to left in Eq.~\eqref{eq:us-operator}.

First, $\widehat D_\lambda R$ is block-encoded with subnormalization $\rho_\lambda\coloneqq\max_k d_k^{(\lambda)}R_{kk}=d_{q-1}^{(\lambda)}$. A rotation controlled by the label $k$ sets the selected ancilla amplitude to $d_k^{(\lambda)}R_{kk}/\rho_\lambda$. We then apply the adder $\mathrm{ADD}_{-\lambda}$ of Eq.~\eqref{eq:assembled-qut} to shift the degree label from $k$ to $k-\lambda$, accounting for the decrease in polynomial degree after $\lambda$ derivatives. Labels $k<\lambda$ have zero selected amplitude.

Second, we convert the coefficients to the parameter-$\mu^\star$ basis with $\widehat S_{\lambda\to\mu^\star}$, which is described in the next paragraph. A value oracle followed by a controlled rotation and uncomputation of the oracle workspace realizes $\diag\bigl(a_\lambda(x_j)/\bar a_\lambda\bigr)$ at the nodes $x_j$ of $C^{(\mu^\star)}_{N_{\mu^\star}}$. Conjugating this diagonal block by $Q^{(\mu^\star)}_{N_{\mu^\star}}$ implements $M_{\mu^\star}[a_\lambda]$ with subnormalization $\bar a_\lambda$, as in Eq.~\eqref{eq:us-multiplication}. Since this transform is the same for every $\lambda$, one pair, applied before and after the diagonals selected by $\lambda$, serves all lower-order terms.

Third, we convert the product coefficients to the parameter-$q$ basis using $\widehat S_{\mu^\star\to q}$. Both $\widehat S_{\lambda\to\mu^\star}$ and $\widehat S_{\mu^\star\to q}$ act on the register one step at a time and end with a comparator that keeps the first $N_{\mu^\star}$ or $N_q$ labels. For $\lambda\leq\nu<q$, the step from parameter $\nu$ to $\nu+1$ acts on the labels as
\begin{equation}
 \diag(s_k^{(\nu)})-\tfrac12\mathrm{ADD}_{-2}\diag(2t_k^{(\nu)}),
\end{equation}
which we implement by LCU. Since $s_k^{(\nu)}\leq1$ and $t_k^{(\nu)}\leq1/2$, both diagonal factors have subnormalization one. The corresponding LCU weights $1$ and $1/2$ yield subnormalization $3/2$ for each step and therefore $(3/2)^{q-\lambda}$ for $\widehat S_{\lambda\to\mu^\star}$ and $\widehat S_{\mu^\star\to q}$ together.

After the conversion, we apply $\mathrm{ADD}_q$ to place the $N_q$ coefficients in rows $q,\ldots,K-1$, below the $q$ boundary rows. Comparators around the QUT pair restrict its input and output to the valid labels $k<N_{\mu^\star}$, so the selected block vanishes on all other labels. This restriction allows the error guarantee of Theorem~\ref{thm:efficient-qut} to be applied directly to the QUT components.

Combining the identity, boundary, and lower-order terms creates a block encoding of $\mathcal A_K/\alpha$ with
\begin{equation}
 \alpha\coloneqq1+\sum_{i=0}^{q-1}\norm{\boldsymbol r_i}_2+\sum_{\lambda=0}^{q-1}\Bigl(\frac32\Bigr)^{q-\lambda}\rho_\lambda\,\bar a_\lambda.
 \label{eq:us-alpha}
\end{equation}
Under the conditions of Proposition~\ref{prop:us-bc}, $\alpha$ is independent of $K$. For the four problems in Fig.~\ref{fig:us-checks}(a), $\alpha$ ranges from $8.2$ to $82$, while the inversion parameter $\alpha\norm{\mathcal A_K^{-1}}_2$ ranges from $15$ to $166$ at $K=1024$. The value of $\alpha$ exceeds $\norm{\mathcal A_K}_2$ by factors between $3.4$ and $6.9$. We next apply QSVT to this block encoding to approximate $\mathcal A_K^{-1}$ and discuss the resulting error and gate complexity.

\begin{proposition}[QUT implementation of the preconditioned inverse]\label{prop:us-inverse}
Under the assumptions of Proposition~\ref{prop:us-conditioning}, let $K\geq K_0$ and suppose that a bound $\bar\kappa \geq \alpha\norm{\mathcal A_K^{-1}}_2$ is known.
Then there exists a circuit $U_{\mathrm{inv}}$ with $a_{\mathrm{inv}}$ ancillas such that
\begin{align}
 \Bigl\lVert(\bra{0^{a_{\mathrm{inv}}}}\otimes I)\,U_{\mathrm{inv}}\,(\ket{0^{a_{\mathrm{inv}}}}\otimes I)-\frac{\alpha}{2\bar\kappa}\,\mathcal A_K^{-1}\Bigr\rVert_2\leq\eta,
 \label{eq:us-inverse}
\end{align}
using $L_\eta=O(\bar\kappa\log(\bar\kappa/\eta))$ calls to the block encoding of $\mathcal A_K/\alpha$ and its adjoint. It suffices for each call to have error $\delta_\eta = O(\eta^2/L_\eta^2)$. Each call has gate cost
\begin{equation}
\begin{aligned}
     C_{\mathrm{blk}}(\delta)={}&2G_{\mu^\star}\bigl(n,O(\delta)\bigr)+\sum_{\lambda=0}^{q-1}T_{a_\lambda}\bigl(O(\log\tfrac1\delta)\bigr)\\
     &+\sum_{i=0}^{q-1}T_{\boldsymbol r_i}\bigl(O(\delta)\bigr)+\mathrm{poly}_q\bigl(n,\log\tfrac1\delta\bigr).
\end{aligned}
 \label{eq:us-blockcost}
\end{equation}
Here, $G_{\mu^\star}$ denotes the QUT gate complexity from Eq.~\eqref{eq:efficient-qut-gates} with $m=\mu^\star$ if $\mu^\star\geq2$, and $G_{\mu^\star}=O(n^2)$ if $\mu^\star\in\{0,1\}$~\cite{Klappenecker2001}. The quantity $T_{a_\lambda}(b)$ is the cost of evaluating $a_\lambda/\bar a_\lambda$ at a Gauss node of $C^{(\mu^\star)}_{N_{\mu^\star}}$ to $b$ bits, including computation of the node from its label, while $T_{\boldsymbol r_i}$ is the cost of implementing $V_i$. The $a_{\mathrm{inv}}$ ancillas comprise those used by the QUTs in Eq.~\eqref{eq:efficient-qut-ancillas}, together with the LCU selectors, flags, and oracle workspace.
\end{proposition}

The construction and error allocation are detailed in Appendix~\ref{app:us-encoding}. Briefly, we apply QSVT to implement the inverse by approximating $(2\bar\kappa x)^{-1}$ on $[1/\bar\kappa,1]$ with an odd polynomial and applying this polynomial to the singular values of the block-encoded operator~\cite{Gilyen2019,Childs2017}. If the value oracles and row preparations have cost polynomial in $n$ and $\log(1/\delta)$, then the inverse block has complexity polynomial in $n$ and $\log(1/\eta)$, with constants depending on $q$, $\alpha$, and $\bar\kappa$, but not on $K$.

The QSVT polynomial depends on $\bar\kappa$, so $\bar\kappa$ must be known before the circuit is built. We obtain a rigorous value from one small classical computation. To do so, we pick an integer $K_1\geq q$ and split the labels into the first $K_1$ labels and the rest. Let $K_3$ exceed $K_1$ by an amount that depends only on $q$ and on the degree of a polynomial approximation of the coefficients. For every $K\geq K_3$, the entries of $\mathcal A_K$ in the first $K_1$ rows or columns then equal those of $\mathcal A_\infty$. The only exceptions are the boundary rows, which are known explicitly, and a small, bounded error when the coefficients are not polynomials. We therefore compute these entries once, from $\mathcal A_{K_3}$. The remaining block differs from the identity by an explicitly bounded amount. We use a Schur-complement argument, which eliminates this block, to bound $\|\mathcal A_K^{-1}\|_2$ for all $K\geq K_3$ at a classical cost that depends on $K_3$ but not on $K$ (Lemma~\ref{lem:us-certificate} in Appendix~\ref{app:us-certificate}). For the four problems in Fig.~\ref{fig:us-checks}(a), with $K_1=32$, $\bar\kappa$ is at most $2.1\%$ above the largest computed $\alpha\|\mathcal A_K^{-1}\|_2$ for $K_3\leq K\leq4096$.

\subsection{Solution recovery and output extraction}\label{sub:us-outputs}

We now use the inverse circuit to prepare states encoding the solution and its derivatives, or to estimate scalar responses directly. We first prepare the forcing and boundary data as the input state.

The forcing is encoded through its weighted samples at the Gauss nodes of $C^{(q)}_{N_q}$. We assume a circuit that prepares
\begin{equation}
 \ket F=\frac1\beta\sum_{j=0}^{N_q-1}\sqrt{w_j}\,f(x_j)\ket j
 \label{eq:us-forcing}
\end{equation}
with cost $T_F(s)$ for error $s$, where $\beta^2\coloneqq\sum_jw_j\lvert f(x_j)\rvert^2>0$.
By Eq.~\eqref{eq:transform-pair}, $Q^{(q)}_{N_q}\ket F=\widehat{\boldsymbol f}/\beta$, and $\mathrm{ADD}_q$ shifts these coefficients to the residual labels. For homogeneous boundary data, $\ket b\coloneqq\mathrm{ADD}_qQ^{(q)}_{N_q}\ket F$ is therefore the normalized right-hand side of Eq.~\eqref{eq:us-system}, with preparation cost $P_b(s)=T_F(O(s))+G_q(n,O(s))+O(n)$. For inhomogeneous boundary data, one controlled rotation combines the boundary vector $\boldsymbol c$ on the first $q$ labels with the forcing state, using amplitudes proportional to $\norm{\boldsymbol c}_2$ and $\beta$. This requires the value of $\beta$, which must be supplied or estimated. For brevity, we take $\boldsymbol c=0$ below.

Applying $U_{\mathrm{inv}}$ to $\ket b$ produces on the successful branch a state proportional to $(\alpha/2\bar\kappa)\mathcal A_K^{-1}\ket b$.
Its ideal success amplitude is at least $1/(2\bar\kappa)$, since
\begin{equation}
    \norm{\mathcal A_K^{-1}\ket b}_2
\geq \frac{1}{\norm{\mathcal A_K}_2}
\geq \frac1\alpha.
\end{equation}
Therefore, preparing the normalized state $\ket{\boldsymbol v}$ to error $\epsilon$ through amplitude amplification~\cite{Brassard2002} requires
\begin{align}
 C_{\boldsymbol v}(\epsilon)=\widetilde O\Bigl(\bar\kappa\,\bigl[P_b(\epsilon/\bar\kappa)+C_{\mathrm{inv}}(\epsilon/\bar\kappa)\bigr]\Bigr)
 \label{eq:us-state-cost}
\end{align}
gates, where $C_{\mathrm{inv}}(\eta)=O(L_\eta C_{\mathrm{blk}}(\delta_\eta))$ and $\widetilde O$ suppresses logarithmic factors in the failure probability.
This success amplitude is bounded below by $1/(2\bar\kappa)$ independently of the forcing, which affects only the recovery step below. We note that the cost in Eq.~\eqref{eq:us-state-cost} involves $O(\bar\kappa^2\log(\bar\kappa/\epsilon))$ block calls. Linear-system solvers based on the discrete adiabatic theorem can reduce this to $O(\bar\kappa\log(1/\epsilon))$~\cite{Costa2022}.

\emph{Extracting the highest derivative.}
The state $\ket{\boldsymbol v}$ is itself a useful output. Its labels $q,\ldots,K-1$ contain the orthonormal coefficients of $\partial^q_x u_K$ in the parameter-$q$ basis, while the first $q$ labels contain $u_0,\ldots,u_{q-1}$. We can apply a comparator to select the derivative coefficients with probability $\norm{u_K^{(q)}}^2_{L^2(w_q)} /  \norm{\boldsymbol v}_2^2$. On this branch, applying $\mathrm{ADD}_{-q}$ followed by $Q^{(q)\dagger}_{N_q}$ converts the coefficients into amplitudes proportional to $\sqrt{w_j}\,u_K^{(q)}(x_j)$ at the Gauss nodes of $C^{(q)}_{N_q}$. Thus, weighted samples of the highest derivative can be extracted from $\ket{\boldsymbol v}$ with the cost associated with selecting this branch.

\emph{Solution recovery.}
Recovering the solution coefficients requires applying the rescaling $R$, which is a contraction and can be implemented by a label-controlled rotation. Applied to $\ket{\boldsymbol v}$, this step succeeds with \emph{recovery probability}
\begin{equation}
 p_{\mathrm{rec}}
 \coloneqq
 \frac{\norm{R\boldsymbol v}_2^2}{\norm{\boldsymbol v}_2^2}
 =
 \frac{\norm{u_K}_{L^2(w_0)}^2}
 {\sum_{k<q}u_k^2+\norm{\partial_x^q u_K}_{L^2(w_q)}^2}.
 \label{eq:us-prec}
\end{equation}
Applying the inverse quantum Chebyshev transform $Q^{(0)\dagger}_K$~\cite{Williams2023} to the recovered coefficients produces amplitudes proportional to $\sqrt{w_j}\,u_K(x_j)$ at the Chebyshev nodes. More generally, applying $\widehat D_\lambda R/\rho_\lambda$ to $\ket{\boldsymbol v}$ and then $Q^{(\lambda)\dagger}_{N_\lambda}$ produces weighted samples of $\partial_x^\lambda u_K$ at the Gauss nodes of $C^{(\lambda)}_{N_\lambda}$, with success probability $\norm{\widehat D_\lambda R\boldsymbol v}_2^2 / \rho_\lambda^2\norm{\boldsymbol v}_2^2$. Figure~\ref{fig:us-checks}(d) illustrates the behavior of $p_{\mathrm{rec}}$ in two different scenarios. One of them is the equation $-u''+(1+x^2)u=\cos\omega x$. Its solution satisfies $u\approx\cos(\omega x)/\omega^2$, so $u''\approx-\cos(\omega x)$ and $p_{\mathrm{rec}}$ decreases as $\omega^{-4}$. Amplitude amplification of the recovery step therefore incurs an additional factor $p_{\mathrm{rec}}^{-1/2}$.

\emph{Estimating scalar responses.}
If only a linear functional $\mathcal O[u]=\sum_k\ell_k u_k$ of the solution is required, we can estimate it directly without recovering the state. Since $\mathcal O[u_K]=\boldsymbol\ell^\T R\boldsymbol v$, we prepare $\ket g\coloneqq\frac{R\boldsymbol\ell}{\norm{R\boldsymbol\ell}_2}$ and define $\mathcal Y = \frac{\alpha}{2\bar\kappa}
 \bra g\mathcal A_K^{-1}\ket b$. Then
\begin{equation}
 \mathcal O[u_K] = \frac{2\bar\kappa}{\alpha}\,
 \beta\,\norm{R\boldsymbol\ell}_2\,\mathcal Y.
 \label{eq:us-response}
\end{equation}
A Hadamard test combining $U_{\mathrm{inv}}$ with the preparations of $\ket b$ and $\ket g$ can provide access to the overlap $\mathcal Y$. Estimating it to additive error $\tau$ using amplitude estimation needs
\begin{equation}
 C_{\mathrm{resp}}(\tau) = \widetilde O\!\left(
 \frac{P_b(\tau)+P_g(\tau) + C_{\mathrm{inv}}(\tau)
 }{\tau} \right)
 \label{eq:us-response-cost}
\end{equation}
gates, where $P_g$ is the cost of preparing $\ket g$.

\subsection{Accuracy and comparison with classical solvers}
\label{sub:us-scope}

We relate the solution of the finite system in Eq.~\eqref{eq:us-system} to the boundary-value problem through a discretization bound. Let $u$ be the exact solution, let $\mathcal T_Ku$ be its Chebyshev truncation to degree less than $K$, with coefficient vector $\boldsymbol u^\star$, and define the residual $\boldsymbol\zeta_K \coloneqq A_K\boldsymbol u^\star - (\boldsymbol c,\widehat{\boldsymbol f})^\T$. Since
$\mathcal A_K(\boldsymbol v-R^{-1}\boldsymbol u^\star)
=-\boldsymbol\zeta_K$
and $\norm R_2\leq1$, we show
\begin{equation}
\begin{aligned}
 &\norm{u_K-\mathcal T_Ku}_{L^2(w_0)}
 +\norm{\partial_x^q u_K-\partial_x^q\mathcal T_Ku}_{L^2(w_q)} \\
 &\qquad \leq
 2\norm{\mathcal A_K^{-1}}_2\norm{\boldsymbol\zeta_K}_2.
\end{aligned}
\label{eq:us-discretization}
\end{equation}
The residual $\boldsymbol\zeta_K$ can be bounded by  the tails of the rescaled boundary rows and of the expansions of $u^{(\lambda)}$, $a_\lambda u^{(\lambda)}$, and $f$, together with the Gaussian interpolation errors. In particular, when these functions are analytic in a Bernstein ellipse, $\boldsymbol\zeta_K$ decays geometrically with $K$, while $\mathcal A_K^{-1}$ is bounded by Proposition~\ref{prop:us-conditioning}. Details are presented in Appendix~\ref{app:us-error}.

When the discretization error is controlled and $\norm{\mathcal A_K^{-1}}_2$ is bounded, the quantum cost depends on the resolution only through $n=\log_2K$. For fixed $q$, $\ket{\boldsymbol v}$ can be prepared with gate complexity polynomial in $n$ and $\log(1/\epsilon)$ (Eq.~\eqref{eq:us-state-cost}), while recovering the solution adds the factor $p_{\mathrm{rec}}^{-1/2}$. By comparison, the adaptive QR solver for the classical ultraspherical spectral method costs $O(K)$~\cite{Olver2013}, and fast polynomial transforms cost $O(K\log^2K)$~\cite{Townsend2018}. Compared to the classical solver, the quantum dependence on $K$ is polylogarithmic, but the output is a quantum state rather than an explicit coefficient vector.

This reduction is useful only under three additional conditions. First, increasing the resolution should not cause $\kappa_2(\mathcal A_\infty)$ or $\alpha$ to grow substantially. Fine scales in bounded coefficients can increase $K$ without degrading conditioning, as in Fig.~\ref{fig:us-checks}(c), whereas large normalized coefficients may increase both. Second, the coefficients, forcing, and boundary rows must have efficient access; for example, a forcing specified by a formula can satisfy this requirement, while an arbitrary table of $K$ samples does not. Third, the desired output should be a quantum state, samples from it, or a scalar at moderate precision. Recovering the solution incurs the factor $p_{\mathrm{rec}}^{-1/2}$, while estimating a scalar response to additive error $\tau$ costs $O(1/\tau)$.

These conditions are necessary but not sufficient for an end-to-end quantum advantage, since a classical method may also compute the target output directly without reconstructing the full solution. For example, classical methods can compute scalar responses without reconstructing all $K$ coefficients. Analyses of quantum finite-element and PDE algorithms similarly show that, once output precision is included, the quantum speedup for smooth problems at fixed spatial dimension is typically only polynomial~\cite{Montanaro2016,Linden2022}. In particular, the problems considered here are one-dimensional and well preconditioned, so the $1/\tau$ dependence in Eq.~\eqref{eq:us-response-cost} can become the dominant cost in scalar estimation. We therefore do not claim an end-to-end quantum advantage for one-dimensional boundary-value problems.

Our contribution in this section is a QUT-based block encoding of the preconditioned ultraspherical operator whose subnormalization and condition number remain bounded independently of the resolution, together with the corresponding solution-state preparation. Such a construction is useful when the solution state is consumed coherently by a later quantum subroutine and may also serve as a building block for higher-dimensional solvers, where the relative advantage over classical methods can increase with dimension~\cite{Montanaro2016}. Our resolution-independent conditioning also improves on earlier quantum spectral solvers for nonperiodic problems. In particular, previous work on quantum Chebyshev spectral methods can lead to linear systems or operators that become ill-conditioned as the basis size grows~\cite{Childs2021, Wu2025}. By contrast, periodic wavelet methods can achieve resolution-independent conditioning through diagonal preconditioning~\cite{Bagherimehrab2023}. The ultraspherical formulation offers the same uniform conditioning to nonperiodic boundary conditions while preserving spectral accuracy.

\section{Discussion}
\label{sec:discussion}

Quantum linear-system algorithms provide a theoretically feasible way to solve linear differential equations efficiently on quantum computers. However, their rigorous efficiency guarantees require the discretized system to have a small condition number. This work makes progress toward this requirement for spectral discretizations of nonperiodic problems through two contributions. First, we have constructed an efficient circuit for the QUT, which maps function values at Gauss nodes to orthonormal ultraspherical coefficients. For fixed $m\geq2$ and $N+m=2^n$, it achieves error $\varepsilon$ with $O_m(n^3\log(n/\varepsilon)+n\log^2(1/\varepsilon))$ gates (Theorem~\ref{thm:efficient-qut}). Second, we have used QUTs to construct a quantum version of the ultraspherical spectral method. For regular, uniquely solvable linear boundary-value problems in one dimension with a broad class of boundary conditions, the preconditioned system has a condition number that remains bounded as the system size increases (Proposition~\ref{prop:us-conditioning}). Given efficient access to the problem data, this makes it possible to prepare a quantum state encoding the preconditioned solution at a cost polylogarithmic in the system size and inverse error. In what follows, we discuss the underlying mechanism of the QUT construction and its possible extension to other polynomial families and offer an outlook on extending the spectral method beyond one dimension.

Several mechanisms for constructing efficient quantum transforms have been proposed (Table~\ref{tab:transforms}). Our work contributes a new technique, in which a known efficient transform, here the QFT, is combined with an exact basis change and with corrections that are uniformly smooth and therefore can be implemented efficiently. This structure arises because a chain of $m-1$ Darboux transformations maps Fourier modes to ultraspherical modes. Darboux transformations, a common tool in supersymmetric quantum mechanics, relate partner Hamiltonians whose spectra coincide except possibly for finitely many levels~\cite{Crum1955,Fernandez2010}. Orthogonal-polynomial families beyond the ultraspherical case might admit efficient transforms of a similar form if their associated Schr\"odinger operators are connected by a finite Darboux chain to an operator whose eigenbasis already has an efficient transform. For such families, the complexity of the transform would depend on how the finite Darboux chain is implemented in the circuit.

The practical utility of the quantum ultraspherical spectral method remains an open question, particularly for higher-dimensional problems. Possible directions include tensor-product and multi-element discretizations~\cite{Townsend2015, Fortunato2021}, but extending the one-dimensional conditioning result to multiple dimensions is not straightforward. In one dimension, conversion operators appear only in lower-order terms, which are suppressed by the diagonal rescaling. In a tensor-product discretization, however, the highest-order terms can also require conversions, namely in each coordinate in which another term is differentiated to higher order. When these conversions appear in the highest-order terms, a tensor product of one-dimensional diagonal rescalings need not suppress them at high polynomial degree, so uniform conditioning is no longer guaranteed. Constructing a preconditioner with this guarantee, together with a quantum workflow for efficiently loading problem data and extracting quantities of interest, is therefore an important step toward assessing the method's practicality.

\section*{LLM assistance}
We made extensive use of frontier large language models (LLMs), including ChatGPT and Claude, throughout the development of this work. All programming and numerical implementations, together with the first draft, were generated by LLMs and subsequently refined. Based on our initial observation about the Gauss--ultraspherical nodes, we used LLMs to help design and optimize the QUT component circuits. We also used them to adapt the classical ultraspherical spectral method for a quantum algorithm based on the QUT. We have verified LLM-generated arguments and constructions, both theoretically and empirically, to the best of our knowledge.

\appendix

\section*{Guide to the appendices}

Appendices~\ref{app:taylor-order}--\ref{app:amplification} complete the proof of Theorem~\ref{thm:efficient-qut} in the order of Section~\ref{sec:circuit}: the Taylor order, the node and weight corrections, the Fourier block, the weighted Helmert basis change, and the amplification errors. Appendix~\ref{app:us-proofs} contains the proofs for the spectral solver of Section~\ref{sec:us-solver}, and Appendix~\ref{app:darboux-extensions}, on which no other result depends, discusses extensions to Jacobi transforms. Each appendix uses the notation of the main text and relies on no other appendix; its title names the section it supports.

\section{\texorpdfstring{Taylor expansion order (Section~\ref{sec:factorization})}{Taylor expansion order}}
\label{app:taylor-order}

The truncation bound in Eq.~\eqref{eq:factorization-error} decreases factorially with the Taylor order $r$ once the scaled displacements and weight corrections are bounded. With the circuit constants $T\geq\tau$ and $X\geq\xi_{\max}$ of Eqs.~\eqref{eq:tight-input-displacement} and~\eqref{eq:tight-input-weight}, the remainder is at most $Xe^TT^{r+1}/(r+1)!$, and we now determine how large $r$ must be for this bound to fall below a target error $0<\varepsilon<Xe^T$. Since $(r+1)!\geq[(r+1)/e]^{r+1}$, it suffices that
\begin{equation}
    (r+1)\log\frac{r+1}{eT}\geq\Lambda_\varepsilon,
    \qquad
    \Lambda_\varepsilon\coloneqq\log\frac{Xe^T}{\varepsilon}.
    \label{eq:taylor-order-condition}
\end{equation}
Writing $r+1=eT\,e^{y}$ turns this condition into $ye^{y}\geq\Lambda_\varepsilon/(eT)$, which holds exactly when $y\geq W_0(\Lambda_\varepsilon/(eT))$, where $W_0$ is the principal branch of the Lambert function. A sufficient order is therefore
\begin{equation}
    r+1=\left\lceil\frac{\Lambda_\varepsilon}{W_0\bigl(\Lambda_\varepsilon/(eT)\bigr)}\right\rceil,
    \label{eq:taylor-order-lambert}
\end{equation}
and the asymptotic form $W_0(z)=\log z-\log\log z+o(1)$ as $z\to\infty$ yields Eq.~\eqref{eq:taylor-order-scaling} whenever $T$ and $X$ stay bounded.

A choice that avoids the Lambert function follows from the same estimate. If $r+1\geq2eT$, then $[eT/(r+1)]^{r+1}\leq2^{-(r+1)}$, so that
\begin{equation}
    r+1=\max\Bigl\{\lceil2eT\rceil,\Bigl\lceil\log_2\frac{Xe^T}{\varepsilon}\Bigr\rceil\Bigr\}
    \label{eq:taylor-order-simple}
\end{equation}
also suffices. This order exceeds Eq.~\eqref{eq:taylor-order-lambert} by a factor of order $\log\log(1/\varepsilon)$ as $\varepsilon\to0$.

\section{\texorpdfstring{Node and weight corrections (Sections~\ref{sec:factorization} and~\ref{sec:input-correction})}{Node and weight corrections}}
\label{app:arithmetic}

This appendix analyzes the corrections applied by the input circuit. We derive the phase representation of the Gauss--ultraspherical nodes and weights stated in Eqs.~\eqref{eq:phase-corrections} and~\eqref{eq:main-phase-derivative}, prove the normalization bounds of Section~\ref{sec:factorization}, establish the $K$-independent polynomial degree of Eq.~\eqref{eq:main-input-degree}, and finally describe the input circuit and count its gates. Throughout, $m\geq2$ is an integer, $K=N+m\geq m+1$, and $u=\cot\theta/K$ and $h=1/K$ are the scaled variables of Section~\ref{sec:input-correction}; the circuit results assume in addition that $K=2^n$.

\subsection{Phase function}
\label{app:phase-arithmetic}

The phase function is built from the Darboux ladder of Section~\ref{sec:input-correction}. Starting from $Z_0=e^{\ii K\theta}$, the steps $Z_q=K^{-1}A_qZ_{q-1}$ with $A_q=\partial_\theta-q\cot\theta$ produce $Z_q=\mathsf P_q(u;h)e^{\ii K\theta}$, where $\mathsf P_q$ obeys the recurrence~\eqref{eq:main-phase-polynomial-recurrence}, and $Z_q$ solves $H_{q+1}Z_q=K^2Z_q$. The quantity that controls the phase of these solutions is the Wronskian of their real and imaginary parts,
\begin{equation}
    J[Z]\coloneqq\operatorname{Im}\bigl(\overline{Z}\,Z'\bigr),
\end{equation}
which equals $|Z|^2\Phi'$ for $Z=|Z|e^{\ii\Phi}$. A closed form for $J[Z_{m-1}]$ is therefore a closed form for the derivative of the phase.

\begin{lemma}[Phase function]
\label{lem:phase-function}
For $1\leq q\leq m-1$ and real $u$,
\begin{equation}
    \operatorname{Im}\bigl[\mathsf P_q(u;h)\overline{\mathsf P_{q-1}(u;h)}\bigr]=D_{q-1}(h)>0,
    \label{eq:phase-polynomial-wronskian}
\end{equation}
where $D_q(h)=\prod_{a=1}^{q}(1-a^2h^2)$. Hence each phase increment $\arg(\mathsf P_q/\mathsf P_{q-1})$ lies in $(0,\pi)$, and
\begin{equation}
    \psi_m(u;h)=-\sum_{q=1}^{m-1}\arctan\frac{\operatorname{Re}\bigl[\mathsf P_q\overline{\mathsf P_{q-1}}\bigr]}{D_{q-1}(h)}
    \label{eq:phase-deviation}
\end{equation}
is a continuous branch of $\arg\mathsf P_{m-1}-(m-1)\pi/2$ with $\psi_m(0;h)=0$. With this branch, the phase function $\Phi_K^{(m)}$ of Eq.~\eqref{eq:main-phase-representation} satisfies Eq.~\eqref{eq:main-phase-derivative}, and for $0\leq j<N$,
\begin{equation}
    \Phi_K^{(m)}(\theta_j)=(j+m)\pi,
    \qquad
    \xi_j^2=\frac{K}{\Phi_K^{(m)\prime}(\theta_j)}.
    \label{eq:phase-identities}
\end{equation}
\end{lemma}

\begin{proof}
\emph{Wronskian.} Because $Z_{q-1}$ solves $Z''+[K^2-q(q-1)\csc^2\theta]Z=0$, the pair $(A_qZ_{q-1},(A_qZ_{q-1})')$ follows from $(Z_{q-1},Z_{q-1}')$ by the real matrix
\begin{equation*}
    \begin{pmatrix}
        -q\cot\theta & 1\\
        q^2\csc^2\theta-K^2 & -q\cot\theta
    \end{pmatrix},
\end{equation*}
whose determinant is $K^2-q^2$. The Wronskian transforms with this determinant, so $J[Z_q]=(1-q^2h^2)J[Z_{q-1}]$. Then $J[Z_0]=K$ implies $J[Z_q]=KD_q(h)$. Since $|Z_q|=|\mathsf P_q|$, the case $q=m-1$ is Eq.~\eqref{eq:main-phase-derivative}. Moreover, $Z_q\overline{Z_{q-1}}=\mathsf P_q\overline{\mathsf P_{q-1}}$ has imaginary part $K^{-1}J[Z_{q-1}]=D_{q-1}(h)$, which proves Eq.~\eqref{eq:phase-polynomial-wronskian}; the bound $K\geq m+1$ makes it positive.

\emph{Branch.} A straightforward induction on the recurrence relation in Eq.~\eqref{eq:main-phase-polynomial-recurrence}
shows $\mathsf P_q(-u;h)=(-1)^q\overline{\mathsf P_q(u;h)}$ for real $u$ and $h$. Thus $\mathsf P_q(0;h)$ is real for even $q$ and imaginary for odd $q$, every real part in Eq.~\eqref{eq:phase-deviation} vanishes at $u=0$, and $\psi_m(0;h)=0$.

\emph{Nodes.} The operators $A_q$ are real. Using Eq.~\eqref{eq:gegenbauer-first-derivative}, we can compute $A_q[\sin^q\theta\,C_k^{(q)}(\cos\theta)]=-2q\sin^{q+1}\theta\,C_{k-1}^{(q+1)}(\cos\theta)$. Starting from $\operatorname{Im}Z_0=\sin K\theta=\sin\theta\,C_{K-1}^{(1)}(\cos\theta)$, the ladder shows that $\operatorname{Im}Z_{m-1}=|\mathsf P_{m-1}|\sin\Phi_K^{(m)}$ is a nonzero multiple of $\sin^m\theta\,C_N^{(m)}(\cos\theta)$. Since $\mathsf P_{m-1}$ has no real zeros by Eq.~\eqref{eq:phase-polynomial-wronskian}, the nodes $\theta_j$ are the zeros of $\sin\Phi_K^{(m)}$ in $(0,\pi)$. The phase increases strictly by Eq.~\eqref{eq:main-phase-derivative}, and its limits follow from the leading term $(-1)^q(2q-1)!!\,u^q$ of $\mathsf P_q$: every increment tends to $\pi$ as $\theta\to0^+$, where $u\to+\infty$, and to zero as $\theta\to\pi^-$. Hence $\Phi_K^{(m)}$ increases from $(m-1)\pi$ to $K\pi$ and takes each value $m\pi,\ldots,(K-1)\pi$ exactly once, at the $N$ nodes in increasing order, which is the node identity in Eq.~\eqref{eq:phase-identities}.

\emph{Weights.} For eigenfunctions of $H_q$ with eigenvalue $K^2$ that vanish at both endpoints, integration by parts gives $\|A_qf\|^2=(K^2-q^2)\|f\|^2$ in $L^2([0,\pi])$. Starting from $\|\sin K\theta\|^2=\pi/2$, the orthonormal angular function of degree $N$ is therefore $G_N^{(m)}=\pm[2/(\pi D_{m-1}(h))]^{1/2}|Z_{m-1}|\sin\Phi_K^{(m)}$. At a node, where $\sin\Phi_K^{(m)}=0$, the identity $|Z_{m-1}|^2\Phi_K^{(m)\prime}=KD_{m-1}(h)$ yields
\begin{equation}
    G_N^{(m)\prime}(\theta_j)^2=\frac{2K}{\pi}\,\Phi_K^{(m)\prime}(\theta_j).
    \label{eq:node-derivative}
\end{equation}
On the other hand, the Christoffel--Darboux formula $w_j^{-1}=\sum_{k<N}p_k^{(m)}(x_j)^2=\beta_Np_N^{(m)\prime}(x_j)p_{N-1}^{(m)}(x_j)$, where $\beta_N$ is the coefficient of $p_N^{(m)}$ in $xp_{N-1}^{(m)}$, combines with the derivative identity $(1-x^2)p_N^{(m)\prime}=2K\beta_Np_{N-1}^{(m)}$ at the zeros of $p_N^{(m)}$~\cite{Szego1975,DLMF} into $w_j=2K/[(1-x_j^2)\,p_N^{(m)\prime}(x_j)^2]$. Since $G_N^{(m)\prime}(\theta_j)=-\sin^{m+1}\theta_j\,p_N^{(m)\prime}(x_j)$, this reads $w_j=2K\sin^{2m}\theta_j/G_N^{(m)\prime}(\theta_j)^2$, and Eq.~\eqref{eq:node-derivative} turns the definition of $\xi_j$ in Eq.~\eqref{eq:node-corrections} into the weight identity in Eq.~\eqref{eq:phase-identities}.
\end{proof}

\subsection{Normalization bounds}
\label{app:tighter-normalization}

By Lemma~\ref{lem:phase-function}, the phase derivative is $\Phi_K^{(m)\prime}=K/F_{m,K}$ with
\begin{equation}
    F_{m,K}(\theta)\coloneqq\frac{|\mathsf P_{m-1}(\cot\theta/K;1/K)|^2}{D_{m-1}(1/K)}.
    \label{eq:normalization-weight-function}
\end{equation}
Both corrections are therefore determined by $F_{m,K}$ and by the phase deviation, which we write as $\psi_m(\theta)$ in place of $\psi_m(\cot\theta/K;1/K)$: the continuous corrections of Section~\ref{sec:input-correction} are $t=-\psi_m$ and $\xi=F_{m,K}^{1/2}$ at the angle $\theta(\ell)$ of Eq.~\eqref{eq:main-general-angle}, and at the nodes they reduce to $t_j=K\delta_j$ and $\xi_j$.

\begin{lemma}[Normalization bounds]
\label{lem:tight-normalization}
Let $m\geq2$ and $K\geq m+1$. Then
\begin{equation}
    F_{m,K}(\theta)=\sum_{q=0}^{m-1}\frac{b_q}{\sin^{2q}\theta}
    \label{eq:normalization-positive-expansion}
\end{equation}
with the positive coefficients
\begin{equation}
    b_q=\frac{(m-1+q)!}{(m-1-q)!}\,\frac{\binom{2q}{q}}{4^q\prod_{a=1}^{q}(K^2-a^2)}.
    \label{eq:normalization-coefficients}
\end{equation}
On $[\theta_0,\pi-\theta_0]$, the continuous corrections satisfy $|\psi_m|\leq\tau$ and $F_{m,K}\leq\xi_0^2$, with equality at the endpoints, which proves Eq.~\eqref{eq:endpoint-correction-maxima}. Moreover, $\theta_0\geq\vartheta_{m,K}$, so the constants of Eqs.~\eqref{eq:tight-input-displacement} and~\eqref{eq:tight-input-weight} satisfy $T\geq\tau$ and $X\geq\xi_{\max}$. As $K$ grows, $T$ increases from a positive value toward $(m+1)\pi/2-z_m$, $X$ decreases, and both bounds become tight.
\end{lemma}

\begin{proof}
\emph{Positive expansion.} Both the real and imaginary parts of $Z_{m-1}$ satisfy $y''+Qy=0$, where $Q=K^2-m(m-1)\csc^2\theta$. The product of any two solutions of this equation satisfies $w'''+4Qw'+2Q'w=0$. Since $|Z_{m-1}|^2$ is a sum of such products, it satisfies the same equation, as does $F_{m,K}=|Z_{m-1}|^2/D_{m-1}$. The symmetry $\mathsf P_q(-u;h)=(-1)^q\overline{\mathsf P_q(u;h)}$ makes $F_{m,K}$ even in $u$, hence a polynomial of degree $m-1$ in $u^2=(\csc^2\theta-1)/K^2$, that is, in $\csc^2\theta$. Its leading coefficient $b_{m-1}=[(2m-3)!!]^2/\prod_{a=1}^{m-1}(K^2-a^2)$ follows from the leading term of $\mathsf P_{m-1}$, and substituting $F=\sum_qb_q\csc^{2q}\theta$ into the third-order governing equation yields
\begin{equation}
    b_{q+1}=\frac{(2q+1)(m+q)(m-1-q)}{2(q+1)[K^2-(q+1)^2]}\,b_q.
    \label{eq:normalization-coefficient-recurrence}
\end{equation}
This recurrence determines all coefficients from $b_{m-1}$, and Eq.~\eqref{eq:normalization-coefficients} satisfies it.

\emph{Endpoint maxima.} All $b_q$ are positive for $K\geq m+1$, so $F_{m,K}\geq b_0=1$ is symmetric about $\pi/2$ and decreases on $(0,\pi/2]$. By Lemma~\ref{lem:phase-function}, $\psi_m'=K(F_{m,K}^{-1}-1)\leq0$, and $\psi_m(\pi-\theta)=-\psi_m(\theta)$ because the real parts in Eq.~\eqref{eq:phase-deviation} are odd in $u$. Both $|\psi_m|$ and $F_{m,K}$ therefore attain their maxima over $[\theta_0,\pi-\theta_0]$ at the endpoints. At $\theta_0$, the node identity $\Phi_K^{(m)}(\theta_0)=m\pi$ implies $\psi_m(\theta_0)=(m+1)\pi/2-K\theta_0=\tau$. Since $t_j=-\psi_m(\theta_j)$ and $\xi_j^2=F_{m,K}(\theta_j)$, this proves Eq.~\eqref{eq:endpoint-correction-maxima}.

\emph{Lower bound on $\theta_0$.} Removing the term $\theta^{-2}$ from the partial-fraction expansion $\csc^2\theta=\sum_{k\in\mathbb Z}(\theta-k\pi)^{-2}$~\cite{DLMF} leaves terms $(k\pi-\theta)^{-2}+(k\pi+\theta)^{-2}$ that increase with $\theta$, so that
\begin{equation}
    \frac13\leq\csc^2\theta-\frac1{\theta^2}\leq1-\frac4{\pi^2},
    \qquad 0<\theta\leq\frac\pi2.
    \label{eq:normalization-cosecant-bounds}
\end{equation}
The function $y=\sin^m\theta\,C_N^{(m)}(\cos\theta)$ solves $y''+[K^2-m(m-1)\csc^2\theta]y=0$, while $v=\sqrt\theta\,J_{m-1/2}(\kappa\theta)$ with $\kappa^2=K^2-m(m-1)/3$ solves $v''+[\kappa^2-m(m-1)\theta^{-2}]v=0$ and has its first zero at $\vartheta_{m,K}=z_m/\kappa$. Both behave as positive multiples of $\theta^m$ near zero, and by the lower bound in Eq.~\eqref{eq:normalization-cosecant-bounds}, the coefficient in the equation for $v$ is at least that for $y$. Hence, wherever $y,v>0$, the Wronskian-like quantity $y'v-yv'$ has a nonnegative derivative and is therefore nondecreasing. It vanishes at $0^+$, so $y/v$ is nondecreasing and $y$ cannot vanish before $v$. In other words, the first zero of $y$ occurs no earlier than the first zero of $v$. This proves $\theta_0\geq\vartheta_{m,K}$ and thus $T\geq\tau$. Since $F_{m,K}$ decreases on $(0,\pi/2]$ and Eq.~\eqref{eq:tight-input-weight} is $X^2=F_{m,K}(\vartheta_{m,K})$, it follows that $X\geq\xi_0$.

\emph{Dependence on $K$.} The upper bound in Eq.~\eqref{eq:normalization-cosecant-bounds} provides the reverse comparison, with $\kappa^2$ replaced by $K^2-m(m-1)(1-4/\pi^2)$, once the corresponding Bessel zero lies below $\pi/2$. Hence $K\theta_0\to z_m$, and by Eq.~\eqref{eq:normalization-positive-expansion} both bounds become tight as $K\to\infty$. Since $K\vartheta_{m,K}=z_m[1-m(m-1)/(3K^2)]^{-1/2}$ decreases with $K$, $T$ increases toward $(m+1)\pi/2-z_m$; it is already positive at $K=m+1$, where $N=1$, $\theta_0=\pi/2$, and the comparison is strict. For $X$, put $c=m(m-1)/3$ and abbreviate $\vartheta=\vartheta_{m,K}$. Each nonconstant term of Eq.~\eqref{eq:tight-input-weight} is a constant divided by $(K\sin\vartheta)^{2q}\prod_{a\leq q}(1-a^2/K^2)$, and
\begin{equation}
    \frac{\dd}{\dd K}\bigl(K\sin\vartheta\bigr)=\cos\vartheta\Bigl(\tan\vartheta-\vartheta-\frac{c}{z_m^2}\,\vartheta^3\Bigr)>0,
    \label{eq:normalization-X-monotone}
\end{equation}
because $\tan\vartheta\geq\vartheta+\vartheta^3/3$ and $z_m^2>3c$; the latter holds because $\sqrt\theta\,J_{m-1/2}(\theta)$ is convex and increasing while $\theta^2<m(m-1)$ and therefore cannot vanish there. Hence $X$ decreases with $K$.
\end{proof}

The monotonicity in Lemma~\ref{lem:tight-normalization} reduces the uniform subnormalization bound in Table~\ref{tab:amplification-rounds} to finitely many evaluations. For $K\geq64$, we bound $X$ by its value $X_{64}$ at $K=64$ and $T$ by its limit $T_\infty=(m+1)\pi/2-z_m$, while the power-of-two values $m+1\leq K<64$ are evaluated directly. With $\alpha_{\mathrm{QET}}=2\alpha_r\leq2Xe^T$, this means
\begin{equation}
    \alpha_{\mathrm{QET}}\leq A\coloneqq\max\Bigl\{2X_{64}\,e^{T_\infty},\ \max_{K<64}2Xe^T\Bigr\},
    \label{eq:uniform-tight-normalization}
\end{equation}
where the inner maximum runs over $K=2^n\geq m+1$. For $2\leq m\leq8$, the first term is the larger one, and Table~\ref{tab:amplification-rounds} lists $A$ rounded upward. The padding condition $\alpha_{\mathrm{QET}}\leq\beta_t$ of Eq.~\eqref{eq:amplifiable-amplitude} then holds for the smallest integer $t$ with $A\leq\beta_t$,
\begin{equation}
    t=\Bigl\lceil\frac{\pi}{4\arcsin(1/A)}-\frac12\Bigr\rceil,
    \label{eq:normalization-rounds}
\end{equation}
which gives the round counts in the table.

\subsection{Uniform polynomial approximation}
\label{app:general-correction}

We now show that the rescaled corrections of Section~\ref{sec:input-correction} are analytic and uniformly bounded on a fixed Bernstein ellipse around $[-1,1]$, which underlies the $K$-independent degree in Eq.~\eqref{eq:main-input-degree}. As in Section~\ref{sec:qet}, $E_\varrho$ denotes the closed region bounded by the ellipse $\{(w+w^{-1})/2:|w|=\varrho\}$; we use $\varrho=2$. Fix a dyadic interval $I_\nu$ in the range $B_m\leq2^\nu\leq K/4$, and let $b_\nu(x)=2^\nu+(2^\nu-1)(x+1)/2$ be the affine coordinate of Section~\ref{sec:input-correction}, which maps $[-1,1]$ onto $[2^\nu,2^{\nu+1}-1]$. For a folded label $b$, the corrections are evaluated at the grid label $b+\zeta-\zeta_m$, where $\zeta=\zeta_m=1/2$ for even $m$ and $\zeta=\rho\in\{0,1\}$ for odd $m$; this label is $b$ for even $m$ and $b+\rho$ for odd $m$, as required by the reflection. For real $b$, we write $t_\zeta(b)\coloneqq t(b+\zeta-\zeta_m)$ and $\xi_\zeta(b)\coloneqq\xi(b+\zeta-\zeta_m)$ for the displacement and weight as functions of the folded label. Since $\ell+\zeta_m=b+\zeta$ at this label, Eqs.~\eqref{eq:main-general-angle} and~\eqref{eq:main-general-displacement} show that $t_\zeta(b)$ is a fixed point of
\begin{equation}
    F_b(t)\coloneqq-\psi_m(u;h),
    \label{eq:general-input-fixed-point}
\end{equation}
where $u=h\cot z$ is the scaled cotangent at the angle $z=h[\pi(b+\zeta)+t]$. To extend $t_\zeta(b)$ to complex-valued labels, we first extend $\psi_m$ to complex $u$. The next lemma expresses its bounds through 
\begin{equation}
    \Lambda_m\coloneqq\binom m2 = \frac{m(m-1)}{2}
\end{equation}
and coefficientwise domination: $f\ll g$ means that every Taylor coefficient of $f$ at the origin is bounded in modulus by the corresponding coefficient of $g$.

\begin{lemma}[Phase bounds on a disk]
\label{lem:phase-disk}
Let $q=m-1$ and $0<h\leq h_0$ with $\epsilon\coloneqq h_0^2(q^3-q)/6<1/2$, and set $\gamma=(1-2\epsilon)^{-1}$ and
\begin{equation}
    \begin{aligned}
        \widehat D(R)&=\frac{\gamma\Lambda_mR}{1-\Lambda_mR/2},\\
        \Psi(R)&=\frac12\log\frac{1-\widehat D(-R)}{1-\widehat D(R)}.
    \end{aligned}
    \label{eq:general-input-simple-majorant}
\end{equation}
If $\Lambda_mR<2$ and $\widehat D(R)<1$, then $\psi_m(\cdot\,;h)$ extends analytically to a neighborhood of the disk $|u|\leq R$, where $|\psi_m|\leq\Psi(R)$ and $|\partial_u\psi_m|\leq\Psi'(R)$.
\end{lemma}

\begin{proof}
Multiplying the recurrence~\eqref{eq:main-phase-polynomial-recurrence} by $\ii^{-q}$ shows that $\ii^{-q}\mathsf P_q(u;h)=Y_q(\ii u,-h^2)$, where $Y_0=1$ and
\begin{equation}
    Y_k=(1+kx)\,Y_{k-1}+(x^2+\sigma)\,\partial_xY_{k-1}.
    \label{eq:general-input-bessel-recurrence}
\end{equation}
By induction, $Y_q$ has nonnegative coefficients in $x$ and $\sigma$, and at $\sigma=0$ it is the Bessel polynomial $y_q(x)=\sum_k\beta_kx^k$ with $\beta_k=(q+k)!/[k!\,(q-k)!\,2^k]$~\cite[Section~18.34]{DLMF}. In these variables, the equation $H_{q+1}Z_q=K^2Z_q$ becomes $(x^2+\sigma)Y_q''+2(x+1)Y_q'-q(q+1)Y_q=0$, so the coefficients of $Y_q=\sum_ka_k(\sigma)x^k$ obey the downward recurrence
\begin{equation}
    (q-k)(q+k+1)\,a_k=2(k+1)\,a_{k+1}+\sigma(k+1)(k+2)\,a_{k+2},
\end{equation}
starting from $a_q=(2q-1)!!$ and $a_{q+1}=0$. Dividing this recurrence at $\sigma=h_0^2$ by its form at $\sigma=0$ shows that the ratios $c_k=a_k(h_0^2)/\beta_k$ satisfy $c_q=c_{q-1}=1$ and $c_k=c_{k+1}+\tfrac14h_0^2(q-k-1)(q+k+2)\,c_{k+2}$. They increase as $k$ decreases, and since $\sum_{k=0}^{q-2}(q-k-1)(q+k+2)=\tfrac23(q^3-q)$, summing the increments gives $c_0\leq1+\epsilon c_0$, so that $a_k(h_0^2)\leq\beta_k/(1-\epsilon)$. Since $Y_q$ has nonnegative coefficients in $x$ and $\sigma$, each $a_k(\sigma)$ is a polynomial in $\sigma$ with nonnegative coefficients, so that $|a_k(\sigma)|\leq a_k(|\sigma|)\leq a_k(h_0^2)$ for $|\sigma|\leq h_0^2$. Similarly, because $a_0(0)=\beta_0=1$ and $a_0-1$ also has nonnegative coefficients, $|a_0(\sigma)-1|\leq a_0(h_0^2)-1$. Then, $a_0(\sigma)\geq2-a_0(h_0^2)\geq(1-2\epsilon)/(1-\epsilon)$ for real $\sigma$ with $|\sigma|\leq h_0^2$.

\emph{Majorant of the normalized polynomial.} Set $\widetilde Y(u)\coloneqq Y_q(\ii u,-h^2)/a_0(-h^2)$. By the identity $\ii^{-q}\mathsf P_q(u;h)=Y_q(\ii u,-h^2)$, this is $\widetilde Y(u)=\ii^{-q}\mathsf P_q(u;h)/a_0(-h^2)$, and $\widetilde Y(0)=1$. Its coefficient of $u^k$ is $\ii^ka_k(-h^2)/a_0(-h^2)$. Since $\sigma=-h^2$ satisfies $|\sigma|\leq h_0^2$, we apply the bounds just derived above to obtain $|a_k(-h^2)|\leq\beta_k/(1-\epsilon)$ and $a_0(-h^2)\geq(1-2\epsilon)/(1-\epsilon)>0$, so this coefficient has modulus at most $\beta_k/(1-2\epsilon)=\gamma\beta_k$. The Bessel coefficients satisfy $\beta_1=q(q+1)/2=\Lambda_m$ and, for $k\geq1$, $\beta_{k+1}/\beta_k=[q(q+1)-k(k+1)]/(2k+2)\leq q(q+1)/4=\Lambda_m/2$, so that $\beta_k\leq\Lambda_m(\Lambda_m/2)^{k-1}=2(\Lambda_m/2)^k$. For $\Lambda_mR<2$, we expand $\widehat D$, defined in Eq.~\eqref{eq:general-input-simple-majorant}, in a geometric series as $\widehat D(R)=\gamma\Lambda_mR\sum_{j\geq0}(\Lambda_mR/2)^j=\sum_{k\geq1}2\gamma(\Lambda_m/2)^kR^k$. Every coefficient of $\widetilde Y-1$ is therefore bounded in modulus by the corresponding coefficient of $\widehat D$, both constant terms being zero; that is, $\widetilde Y-1\ll\widehat D$.

\emph{Properties of $\ll$.} The remaining steps use four elementary properties of domination by a power series $g$ with nonnegative coefficients. (i) If $f\ll g$ and $g$ converges at some $R>0$, then the series of $f$ converges absolutely for $|u|\leq R$, and $|f(u)|\leq g(|u|)\leq g(R)$ there. (ii) If $f_1\ll g_1$ and $f_2\ll g_2$, then $f_1f_2\ll g_1g_2$, because each coefficient of a product is a finite sum of products of coefficients; in particular, $f^j\ll g^j$ for every $j\geq1$. (iii) The odd part $f_{\mathrm{odd}}(u)\coloneqq[f(u)-f(-u)]/2$ retains exactly the odd-degree coefficients of $f$, so $f\ll g$ implies $f_{\mathrm{odd}}\ll g_{\mathrm{odd}}$; multiplication by a constant of modulus one also preserves $\ll$. (iv) Differentiation multiplies the $k$th coefficient by $k$, so $f\ll g$ implies $f'\ll g'$.

\emph{Logarithm.} By (i), $|\widetilde Y(u)-1|\leq\widehat D(|u|)\leq\widehat D(R)<1$ for $|u|\leq R$. Hence $\widetilde Y$ stays within distance less than one of $1$, so $\operatorname{Re}\widetilde Y>0$ and the principal logarithm $\log\widetilde Y$ is analytic on the disk $|u| \leq R$. With $w\coloneqq\widetilde Y-1$, its Taylor series at the origin follows by substituting $w$ into $\log(1+w)=\sum_{j\geq1}(-1)^{j+1}w^j/j$. Since $w(0)=0$, the coefficient of $u^k$ receives contributions only from the finitely many terms with $j\leq k$. By (ii), $w^j\ll\widehat D^j$, and the coefficients $(-1)^{j+1}/j$ have the moduli $1/j$ of the coefficients of $-\log(1-x)=\sum_{j\geq1}x^j/j$. Hence $\log\widetilde Y\ll-\log(1-\widehat D)$, where the right side has nonnegative coefficients and converges at $R$ because $\widehat D(R)<1$.

\emph{Identification with $\psi_m$.} Define $S(u)\coloneqq[\log\widetilde Y(u)-\log\widetilde Y(-u)]/(2\ii)$, which equals $-\ii(\log\widetilde Y)_{\mathrm{odd}}(u)$ and is analytic on the radius-$R$ disk, since the disk is symmetric under $u\mapsto-u$. For real $u$, the real coefficients of $Y_q$ imply that $\widetilde Y(-u)=\overline{\widetilde Y(u)}$. Since $\widetilde Y(u)$ lies in the half-plane $\operatorname{Re}z>0$, the principal logarithm is well-defined there and satisfies $\log \overline{z} = \overline{\log z}$. Hence $S(u)=\operatorname{Im}\log\widetilde Y(u)$ is the principal argument of $\widetilde Y(u)$: on $[-R,R]$ it is real, continuous, valued in $(-\pi/2,\pi/2)$, and $S(0)=0$. Because $a_0(-h^2)>0$ and $q=m-1$, $\arg\widetilde Y=\arg\mathsf P_{m-1}-(m-1)\pi/2$ modulo $2\pi$, so $S$ is a continuous determination of $\arg\mathsf P_{m-1}-(m-1)\pi/2$ on $[-R,R]$, as is the branch $\psi_m$ of Lemma~\ref{lem:phase-function}. Two such determinations differ by a continuous function with values in $2\pi\mathbb Z$, which is constant on the interval and here vanishes at $u=0$, where $S(0)=\psi_m(0;h)=0$. Thus $S=\psi_m(\cdot\,;h)$ on $[-R,R]$, and $S$ is the analytic continuation of $\psi_m$ to the disk.

\emph{Bounds.} The hypotheses $\Lambda_mR<2$ and $\widehat D(R)<1$ are strict. Since $\widehat D$ is continuous, they remain true for some slightly larger radius $R'>R$. The preceding argument therefore applies on the disk $|u|\leq R'$, so $\psi_m$ is analytic on the open disk $|u|<R'$. By property (iii), $\psi_m=S\ll[-\log(1-\widehat D)]_{\mathrm{odd}}$. Evaluating this odd part at $R$ gives $\tfrac12\{\log[1-\widehat D(-R)]-\log[1-\widehat D(R)]\}=\Psi(R)$. Thus $\Psi$ provides a power-series bound for $\psi_m$. Because this bounding series has nonnegative coefficients and converges at $R'$, property (i) implies $|\psi_m(u)|\leq\Psi(|u|)\leq\Psi(R)$ for $|u|\leq R$. Finally, property (iv) gives the corresponding derivative bound $|\partial_u\psi_m(u)|\leq\Psi'(R)$ for $|u|\leq R$.
\end{proof}

\begin{lemma}[Analytic extension of the corrections]
\label{lem:input-approximation}
Let $B_m\leq2^\nu\leq K/4$ and $\zeta\in\{0,1/2,1\}$. For every $b$ in a neighborhood of $b_\nu(E_2)$, the map $F_b$ has a unique fixed point $t_\zeta(b)$ in the closed unit disk. It is holomorphic in $b$, satisfies $|t_\zeta(b)|\leq1/2$, and coincides with the displacement $t(b+\zeta-\zeta_m)$ of Eq.~\eqref{eq:main-general-displacement} for real $b$. The function
\begin{equation}
    \xi_\zeta(b)=\bigl[1-\partial_tF_b\bigl(t_\zeta(b)\bigr)\bigr]^{-1/2}
    \label{eq:general-input-complex-weight}
\end{equation}
is holomorphic, positive for real $b$, bounded in modulus by $\sqrt2$, and equal to the weight correction $\xi(b+\zeta-\zeta_m)$ of Eq.~\eqref{eq:main-general-weight} for real $b$. Consequently, for $0\leq s\leq r$, the local targets
\begin{equation}
    f_s(x)=\frac{\xi_\zeta(b_\nu(x))}{X}\Bigl(\frac{t_\zeta(b_\nu(x))}{T}\Bigr)^s
    \label{eq:general-input-local-target}
\end{equation}
are analytic on $E_2$, bounded by one on $[-1,1]$, and bounded on $E_2$ by
\begin{equation}
    M=\max\Bigl\{1,\frac{\sqrt2}{X}\max\bigl\{1,(2T)^{-r}\bigr\}\Bigr\}.
    \label{eq:general-input-ellipse-constant}
\end{equation}
\end{lemma}

\begin{proof}
\emph{Cotangent bounds.} The real semiaxis of $E_2$ is $5/4$, so $\operatorname{Re}b$ lies between $(7\cdot2^\nu+1)/8$ and $(17\cdot2^\nu-9)/8$ on $b_\nu(E_2)$. For $|t|\leq1$, $0\leq\zeta\leq1$, and $K\geq2^{\nu+2}$, the angle $z=h[\pi(b+\zeta)+t]$ therefore satisfies
\begin{equation}
    \operatorname{Re}z \geq h\bigl[\pi(7\cdot2^\nu+1)/8-1\bigr]
\end{equation}
and 
\begin{equation}
    \pi-\operatorname{Re}z\geq h\bigl[\pi(15\cdot2^\nu+1)/8-1\bigr].
\end{equation}
Since the cotangent satisfies $|\cot z|\leq\max\{1,|\cot\operatorname{Re}z|\}$ for complex $z$ and $|\cot y|\leq1/\min\{y,\pi-y\}$ for $0<y<\pi$, the bound $h\leq2^{-\nu-2}$, which also implies $h^2<0.48\,R_\nu^2$, yields
\begin{equation}
    \begin{aligned}
        |u|&\leq R_\nu\coloneqq\frac{8}{7\pi\,2^\nu+\pi-8},\\
        |\partial_tu|&=|u^2+h^2|\leq1.48\,R_\nu^2.
    \end{aligned}
    \label{eq:general-input-cotangent-bounds}
\end{equation}

\emph{Contraction.} Since $2^\nu\geq B_m>\Lambda_m$ and both $2^\nu$ and $\Lambda_m$ are integers, we have $\Lambda_m\leq2^\nu-1$. It follows that 
\begin{equation}
    \Lambda_mR_\nu\leq8(2^\nu-1)/(7\pi\,2^\nu+\pi-8)<8/(7\pi)<0.364.
\end{equation}
Also, since $h\leq h_0=2^{-\nu-2}$, the quantity $\epsilon$ in Lemma~\ref{lem:phase-disk} satisfies $\epsilon\leq(q-1)/[24q(q+1)]\leq1/144$, which implies $\gamma\leq72/71$. 

The quantities $\widehat D(R_\nu)$, $\Psi(R_\nu)$, and $\Psi'(R_\nu)/\Lambda_m$ increase with both $\gamma$ and $\Lambda_mR_\nu$. Evaluating them at the upper bounds $\gamma=72/71$ and $\Lambda_mR_\nu=8/(7\pi)$ gives $\widehat D(R_\nu)<0.46$, $\Psi(R_\nu)<0.44$, and $\Psi'(R_\nu)<1.66\,\Lambda_m$. Therefore, Lemma~\ref{lem:phase-disk} and Eq.~\eqref{eq:general-input-cotangent-bounds} imply that, for every $|t|\leq1$,
\begin{equation}
    \begin{aligned}
        |F_b(t)|&<0.44,\\
        |\partial_tF_b(t)|&\leq\Psi'(R_\nu)\,|\partial_tu|<0.9\,R_\nu.
    \end{aligned}
    \label{eq:general-input-contraction}
\end{equation}
In particular, $R_\nu\leq R_1<0.21$ makes the second bound smaller than $1/2$. Thus $F_b$ maps the closed unit disk into the disk of radius $0.44$ and is a contraction there. These strict bounds have enough margin to remain true in a small neighborhood of $b_\nu(E_2)$. 

Starting from $t^{(0)}=0$, define successive approximations by
\begin{equation}
    t^{(k+1)}=F_b(t^{(k)}).
\end{equation}
Because $F_b$ is a contraction, these iterates converge uniformly to one unique fixed point $t_{\zeta}(b)$. The iterates depend holomorphically on $b$, so their uniform limit does too. The fixed point also satisfies $|t_\zeta(b)|<0.44$. For real $b$, the iterates are real, so the limit is real and lies on the branch of the phase function specified in Lemma~\ref{lem:phase-function}. It solves Eq.~\eqref{eq:main-general-angle} at the grid label $b+\zeta-\zeta_m$; this solution is unique because the left-hand side of Eq.~\eqref{eq:main-general-angle} is strictly increasing.

\emph{Weight.} By differentiating Eq.~\eqref{eq:main-phase-representation}, with $\partial_\theta u=-K(u^2+h^2)$, we obtain $\Phi_K^{(m)\prime}/K=1-(u^2+h^2)\partial_u\psi_m$, which equals $1-\partial_tF_b(t)$ at the fixed point. By Eq.~\eqref{eq:general-input-contraction}, this lies in the disk of radius $1/2$ around one, so its inverse square root has a holomorphic branch that is positive for real $b$ and bounded in modulus by $\sqrt2$. For real $b$, Eq.~\eqref{eq:main-general-weight} identifies it with the weight correction $\xi(b+\zeta-\zeta_m)$.

\emph{Targets.} On the ellipse $E_2$, the bounds $|t|\leq1/2$ and $|\xi|\leq\sqrt2$ imply $|f_s|\leq(\sqrt2/X)(2T)^{-s}\leq M$ for every $s\leq r$. Specifically, for $x\in[-1,1]$, the continuous label $b_\nu(x)+\zeta-\zeta_m$ lies between two paired grid labels. Since $\theta(\ell)$ increases with $\ell$, the corresponding angles lie in $[\theta_0,\pi-\theta_0]$. Lemma~\ref{lem:tight-normalization} then gives $|t|\leq\tau\leq T$ and $\xi\leq\xi_0\leq X$, hence $|f_s(x)|\leq1$ on $[-1,1]$.
\end{proof}

\begin{figure}[t]
    \centering
    \includegraphics[width=\columnwidth]{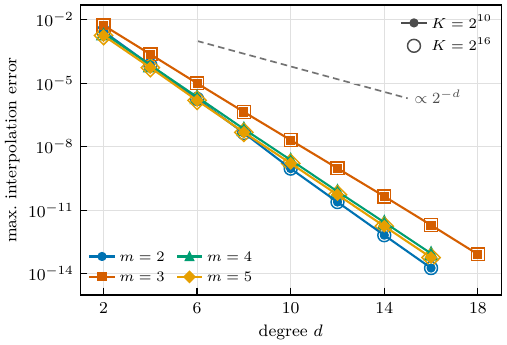}
    \caption{Chebyshev interpolation error of the input-correction targets $f_s$ in Eq.~\eqref{eq:general-input-local-target}, maximized over all dyadic intervals with $B_m\leq2^\nu\leq K/4$, reflection sectors, and orders $s\leq8$. The gray dashed line indicates the rate $2^{-d}$ guaranteed by Lemma~\ref{lem:input-approximation}.}
    \label{fig:input-degree}
\end{figure}

\emph{Approximation degree.} The degree bound follows from the bounded approximation of Section~\ref{sec:qet}. Because the target functions are analytic on the Bernstein ellipse $E_2$ and bounded there by $M$, Chebyshev interpolation with degree $d$ approximates them on $[-1,1]$ with uniform error at most $4M\,2^{-d}$. Choosing the degree $d_\Gamma$ as in Eq.~\eqref{eq:main-input-degree} makes this error at most $\eta_\Gamma/4$. Allowing a further error of $\eta_\Gamma/4$ in computing the Chebyshev coefficients, then rescaling the resulting polynomial by $1+\eta_\Gamma/2$, gives a real polynomial $P$ such that
\begin{equation}
    \|P\|_{[-1,1]}\leq1 \quad \text{and} \quad \|P-f_s\|_{[-1,1]}\leq\eta_\Gamma.
\end{equation}
These bounds hold uniformly over the interval, reflection sector, order $s\le r$, and dimension $K$. Since $X\ge 1$ and Lemma~\ref{lem:tight-normalization} shows that $T$ is bounded away from zero, $\log M=O_m(r+1)$. Consequently, the required degree satisfies 
\begin{equation}
    d_\Gamma=O_m\!\left(r+\log\frac{1}{\eta_\Gamma}\right).
\end{equation}
The proof uses the cutoff only through the condition $2^\nu > \Lambda_m$. Thus, we choose $B_m$ to be the smallest power of two greater than $\Lambda_m$, namely $B_m = 2^{\lfloor \log_2 \Lambda_m \rfloor + 1}$, as stated in Section~\ref{sec:input-correction}. In practice, the interpolation error decreases much faster than the guaranteed rate $2^{-d}$, and it is independent of $K$, as shown in Fig.~\ref{fig:input-degree}.

\emph{Tighter bound for $m=2$.} Lemma~\ref{lem:input-approximation} uses the bounds $|t|\leq1/2$ and $|\xi|\leq\sqrt2$ on $E_2$ for every $m$. For $m=2$ these bounds are loose. In this case, $z_2\approx4.4934$ is the first positive zero of $J_{3/2}$, that is, the smallest positive solution of $\tan z_2=z_2$. Also, Eq.~\eqref{eq:tight-input-displacement} reads $T=3\pi/2-Kz_2/\sqrt{K^2-2/3}$. By Lemma~\ref{lem:tight-normalization}, $T$ increases with $K$ toward $3\pi/2-z_2\approx0.219$, which is the value listed in Table~\ref{tab:amplification-rounds}. Since $T<0.22<1/2$ for every $K$, the factor $(t/T)^s$ may reach $(2T)^{-s}>2.28^s$ on the ellipse, and thus the constant $M$ of Eq.~\eqref{eq:general-input-ellipse-constant} grows exponentially with $r$. At $m=2$, however, $\mathsf P_1(u;h)=\ii-u$, so the phase correction is a single arctangent, $\psi_2(u;h)=\arctan u$. The displacement $t=-\arctan u$ is small wherever the scaled cotangent $u=h\cot z$ is small, that is, wherever the angle $z$ stays away from the endpoints $0$ and $\pi$. On every interval with $2\leq2^\nu\leq K/4$, the region $b_\nu(E_2)$ has $\operatorname{Re}b>7/4$, and since $|t|\leq1/2$ there by Lemma~\ref{lem:input-approximation}, the angle $z$ keeps a distance of at least $(2\pi-1/2)/K$ from both endpoints. Hence $|u|\leq\max\{1/8,1/(2\pi-1/2)\}<0.173$. The power series for the complex inverse tangent gives
\begin{equation}
    |t| = |\arctan u | \leq \operatorname{arctanh}|u| < 0.175.
\end{equation}
For $m=2$, the weight correction is $\xi=\sec t/\sqrt{1-K^{-2}}$. Therefore,
\begin{equation}
    |\xi|\leq\frac{\sec0.175}{\sqrt{1-K^{-2}}}.
\end{equation}

These two bounds control both factors of the target $f_s=(\xi/X)(t/T)^s$ of Eq.~\eqref{eq:general-input-local-target} on $E_2$. We take $K\geq8$, since the dyadic intervals with $2\leq2^\nu\leq K/4$ exist only then. Since $T\geq0.195$ for $K\geq8$ by the values above, we have $|t/T|<0.175/0.195<0.9$ and therefore $|t/T|^s\leq1$ for every $s\geq0$. For the prefactor, we use Eq.~\eqref{eq:tight-input-weight} at $m=2$, which reads $X^2=1+[(K^2-1)\sin^2\vartheta_{2,K}]^{-1}$, and the fact that $X\sqrt{1-K^{-2}}$ increases with $K$ from $1.0193$ at $K=8$ toward $\sqrt{1+z_2^{-2}}\approx1.0245$. Hence $|\xi|/X\leq\sec0.175/(X\sqrt{1-K^{-2}})<1.0156/1.019<1$. Together, $|f_s|<1$ on $E_2$ for every order $s$, every interval, and every $K\geq8$. The degree in Eq.~\eqref{eq:main-input-degree} uses $M$ only as a bound on the targets over $E_2$, through the Chebyshev error bound $4M\,2^{-d}$, so for $m=2$ we may take $M=1$ in place of Eq.~\eqref{eq:general-input-ellipse-constant}. The degree then becomes $d_\Gamma=\lceil\log_2(16/\eta_\Gamma)\rceil$, which no longer depends on the Taylor order $r$. For example, for $\eta_\Gamma=10^{-7}$ and $r=6$, Eq.~\eqref{eq:general-input-ellipse-constant} gives $M\approx2\times10^2$ for every $K\geq32$ and hence $d_\Gamma=35$ (and $36$ for $K=8,16$), whereas $M=1$ yields a lower degree $d_\Gamma=\lceil\log_2(1.6\times10^8)\rceil=28$.

\subsection{Input circuit and gate count}
\label{app:input-circuit}

\emph{Registers and label folding.} The input circuit $U_{\mathrm{in}}$ acts on the system register, which holds the paired grid label $\ell$, on the order register, and on a small workspace: the $O(\log n)$ selector qubits of the dyadic signal $B_\nu$ in Eq.~\eqref{eq:qet-local-signal}, the two selector qubits of Eq.~\eqref{eq:main-polynomial-block}, which extract the real part and average the even and odd parts, an interval predicate, and a flag. The circuit first folds the label: controlled on its highest bit $\rho$, it complements the lower $n-1$ bits, which then hold the folded label $b$. After the corrections described next, the phase $(-1)^{\rho s}$ restores the sign of the displacement on the reflected half, and the fold is undone.

\emph{Piecewise implementation.} The corrections are applied region by region. On each dyadic interval with $B_m\leq2^\nu\leq K/4$, the predicate marks $b\in I_\nu$, the QET sequence of Section~\ref{sec:qet} applies the bounded polynomial approximation of Appendix~\ref{app:general-correction} to the signal $B_\nu$ on the low $\nu$ bits, with phase angles selected by the order $s$, the even or odd part of the polynomial, and, for odd $m$, the bit $\rho$, and the predicate is then uncomputed. The even- and odd-part sequences share their signal calls, with one additional call controlled by the selector qubit of the even or odd part, so that each part is applied by a sequence of its own parity with at most $d_\Gamma+1$ calls. Only the select stage of the signal needs this control, because the preparation of its LCU selector and the inverse preparation cancel when the select stage is skipped. The remaining folded labels $b<B_m$ receive one fixed rotation of the flag qubit, whose angle is selected by $(b,s,\rho)$ and whose selected amplitude is $g_{s,m,K}(b)/2$ for even $m$ and $g_{s,m,K}(b+\rho)/2$ for odd $m$ on paired grid labels and zero on all other labels. This region contains every label outside the paired grid, because such labels fold to $b<\lceil m/2\rceil\leq\Lambda_m<B_m$, and if $K<4B_m$, it contains all folded labels. As in Section~\ref{sec:qet}, these pieces act on disjoint sets of folded labels and preserve the label and order registers, so the encoded block is a direct sum whose error is the largest error among the pieces. This implements the block encoding in Eq.~\eqref{eq:ideal-input-block} with approximation error $\|\widehat\Gamma_s-\Gamma_s\|_2\leq\eta_\Gamma$. The Hadamard gates on the two selector qubits of Eq.~\eqref{eq:main-polynomial-block} surround the whole sequence, so all regions share them as well as the signal selector, the predicate, and the flag.

\emph{Gate count.} To count gates, we use decompositions whose cost is linear in the number of controls and that borrow a constant number of qubits~\cite{Khattar2025}. Preparing the selector of $B_\nu$ costs $O(n)$ gates and each of its $\nu$ selected Pauli operators costs $O(\log n)$, so a signal call costs $O(n\log n)$. Each phase step is a uniformly controlled rotation over the $O(r+1)$ combinations of $s$, the polynomial part, and $\rho$, which costs $O(r+1)$ gates~\cite{Mottonen2004}. Setting and clearing the predicate costs $O(n)$. A degree-$d_\Gamma$ sequence therefore costs $O(d_\Gamma[n\log n+r+1]+n)$ gates per interval. By summing over the $O(n)$ intervals and adding $O(B_m(r+1)+n)$ for the endpoint rotation, we obtain
\begin{equation}
    C_\Gamma=O_m\bigl((d_\Gamma+1)[n^2\log n+n(r+1)]\bigr),
\end{equation}
which is shown in the first line of Eq.~\eqref{eq:component-gate-costs}. A piecewise-QSVT circuit with one signal shared by all intervals~\cite{OBrien2025} would remove the $n^2\log n$ signal cost at the price of $O(n)$ further qubits. However, we do not use it in this work.

\emph{Classical precomputation.} All angles, polynomial coefficients, and QET phases are computed classically before execution. On each dyadic interval, we construct the Chebyshev interpolant from its $d_\Gamma+1$ target points. At each point, we determine the scaled displacement using the fixed-point iteration in Eq.~\eqref{eq:general-input-contraction}, starting from $t=0$. The error decreases by at least a factor of two per iteration. Then Eq.~\eqref{eq:general-input-complex-weight} gives the corresponding weight correction $\xi_\zeta(b)$. 

\emph{Endpoint labels.} At each paired grid label $\ell$ whose folded label satisfies $b<B_m$, we write the angle as $\theta=[\pi(\ell+\zeta_m)+t]/K$, following Eq.~\eqref{eq:main-general-displacement}, and solve Eq.~\eqref{eq:main-general-angle} for $t$ directly. As a function of $t$, the left-hand side of Eq.~\eqref{eq:main-general-angle} differs from the phase $\Phi_K^{(m)}(\theta)$ of Eq.~\eqref{eq:main-phase-representation} only by a constant. Its derivative is therefore $\Phi_K^{(m)\prime}(\theta)/K=1/F_{m,K}(\theta)$, which is positive because $F_{m,K}$ is a sum of positive terms (Eq.~\eqref{eq:normalization-positive-expansion}). The equation thus has exactly one root, at which $t=-\psi_m$. Since $\psi_m$ is a sum of $m-1$ arctangents, each of modulus below $\pi/2$, the root lies in $|t|<(m-1)\pi/2$. 

Starting from this interval, we apply the bisection method that halves the search interval at each step. Thus, obtaining $p$ precision bits of $t$ requires $p+O(\log m)=O_m(p)$ steps. We evaluate the fixed-degree polynomial $\mathsf P_{m-1}$ once for each step. All roots lie at angles in $[\pi/K,\pi-\pi/K]$, because $\theta_0\geq\vartheta_{m,K}>\pi/K$ and the nodes are symmetric about $\pi/2$. Every term of Eq.~\eqref{eq:normalization-positive-expansion} decreases as $\sin\theta$ grows, so on this range the derivative is at least $1/F_{m,K}(\pi/K)$. By Eqs.~\eqref{eq:normalization-positive-expansion} and~\eqref{eq:normalization-coefficients},
\begin{equation}
F_{m,K}\left(\frac{\pi}{K}\right)=\sum_{q=0}^{m-1}\frac{c_{m,q}}{\sin^{2q}(\pi/K)\prod_{a=1}^{q}(K^2-a^2)},
\end{equation}
where $c_{m,q} = \frac{(m-1+q)!}{(m-1-q)!}\binom{2q}{q}4^{-q}$. For $K\geq m+1$ and $a\leq q\leq m-1$, we have $K^2-a^2\geq K^2-(m-1)^2\geq4mK^2/(m+1)^2$. Moreover, using Jordan's inequality $\sin x \geq 2x/\pi$ for $x \in [0,\pi/2]$, we obtain $\sin(\pi/K)\geq2/K$. Each denominator is therefore at least $[16m/(m+1)^2]^q$, so that 
\begin{equation}
F_{m,K}\left(\frac{\pi}{K}\right)\leq\sum_{q=0}^{m-1}c_{m,q}\left[\frac{(m+1)^2}{16m}\right]^q =: C_{F,m},
\end{equation}
which is independent of $K$. The derivative of the left-hand side of Eq.~\eqref{eq:main-general-angle} is thus at least $1/C_{F,m}$ on $[\pi/K,\pi-\pi/K]$. An error in evaluating this left-hand side therefore enlarges the error of the root by at most the factor $C_{F,m}$. At a node label $\ell_j$, the root is the scaled displacement $t_j=K\delta_j$, and the weight factor follows as $\xi_j=F_{m,K}(\theta_j)^{1/2}$.

\emph{Precision and classical cost.} In both cases, the target values of the interpolants and the fixed rotations obey bounds that do not depend on $K$. Computing every coefficient and angle to $O_m(n+\log[(r+1)(d_\Gamma+1)/\eta_\Gamma])$ bits of precision therefore keeps the total coefficient error below $\eta_\Gamma/4$. Here, the $O(n)$ bits resolve angles of order $1/K=2^{-n}$, while the logarithmic term accounts for summing rounding errors over the $(r+1)(d_\Gamma+1)$ polynomial coefficients. For each of the $r+1$ Taylor orders, we need $d_\Gamma+1$ coefficients on each of the $O(n)$ dyadic intervals and one angle for each of the $B_m$ endpoint labels. This total of $O((r+1)(nd_\Gamma+B_m))$ numbers takes time polynomial in $n$, $r$, and $\log(1/\eta_\Gamma)$ to compute. The QET phase sequences can be computed with the methods of~\cite{Haah2019,Dong2021}.

\section{\texorpdfstring{Fourier block and frequency powers (Section~\ref{sec:fourier-block})}{Fourier block and frequency powers}}
\label{app:fourier}

This appendix constructs the pairing unitaries $B_{\mathrm{in}}(s)$ and $B_{\mathrm{out}}(s)$, verifies that $U_E(s)$ in Eq.~\eqref{eq:controlled-fourier-stage} encodes $E_s/(K-1)^s$, and shows that the shared frequency sequence reproduces every Taylor order. The pairing circuits act on the $n+1$ qubits $\mathrm{br}\otimes\mathrm{sys}$, with the branch qubit as the most significant bit of the joint label.

\subsection{Pairing circuits and sampled derivatives}
\label{app:fourier-pairing}

The input pairing $B_{\mathrm{in}}(s)$ applies a Hadamard gate and the phase $Z^{\sigma_s}$ to the branch qubit and then, controlled on $\mathrm{br}=1$, maps the system label $\ell$ to $\operatorname{ref}_m(\ell)-K$, which is the bitwise complement $K-1-\ell$ for even $m$ and the negation $K-\ell$ modulo $K$ for odd $m$. Since the paired labels satisfy $\ell_j\geq1$ for odd $m$, this prepares the state of Eq.~\eqref{eq:Bin-action},
\begin{equation}
    \ket{\psi_\ell}=\frac{\ket\ell+(-1)^{\sigma_s}\ket{\operatorname{ref}_m(\ell)}}{\sqrt2}.
\end{equation}
The output pairing $B_{\mathrm{out}}(s)$ reverses the frequency pairing in four steps. Controlled on $\mathrm{br}=1$, it first negates the system label modulo $K$, which maps the joint label $2K-p$ to $\ket1_{\mathrm{br}}\ket p_{\mathrm{sys}}$ and leaves the label $K$, which is paired with itself, at $\ket1_{\mathrm{br}}\ket0_{\mathrm{sys}}$. The diagonal phase $\diag(e^{\ii\phi_{m,p}},e^{-\ii\phi_{m,p}})$ on the branch qubit, controlled by the frequency label, then removes the half-grid phases. Since $\phi_{m,p}$ is linear in $p$, this phase operator is a product of $n$ singly controlled rotations. For $p\neq0$, a Hadamard gate followed by $X^{\sigma_s}$ on the branch qubit maps the frequency pair of Eq.~\eqref{eq:main-output-pairing} to $\ii^{\sigma_s}\ket0_{\mathrm{br}}\ket p_{\mathrm{sys}}$ and the orthogonal pair to the branch state $\ket1_{\mathrm{br}}$, while the labels with $p=0$ are left unchanged. Finally, the order-dependent phase $(-\ii)^{\sigma_s}\chi_{m,s}$ completes Eq.~\eqref{eq:main-output-pairing}. Both pairings use $O(n)$ gates.

For $1\le p<K$, define the frequency-pair state
\begin{equation}
    \lvert Y_{s,p}\rangle \coloneqq
    \frac{\ii^{\sigma_s}}{\sqrt{2}} \left(e^{-\ii\phi_{m,p}}\lvert p\rangle +(-1)^{\sigma_s}e^{\ii\phi_{m,p}}\lvert 2K-p\rangle \right),
\end{equation}
where $\phi_{m,p}=\pi p/(2K)$ for even $m$ and $\phi_{m,p}=0$ for odd $m$. This is the frequency-pair state selected by $B_{\mathrm{out}}(s)$ in Eq.~\eqref{eq:main-output-pairing}. The QFT in Eq.~\eqref{eq:qft-convention} then produces the real sampling amplitudes
\begin{equation}
    \bra{Y_{s,p}}\mathcal F\ket{\psi_\ell}
    =\sqrt{\frac2K}\times
    \begin{cases}
        \cos(p\widetilde\theta_\ell), & \sigma_s=0,\\
        \sin(p\widetilde\theta_\ell), & \sigma_s=1.
    \end{cases}
    \label{eq:real-transform-from-qft}
\end{equation}
For odd $m$, this follows directly from $\operatorname{ref}_m(\ell)=2K-\ell$. For even $m$, the reflection $\operatorname{ref}_m(\ell)=2K-1-\ell$ provides opposite phases to the Fourier amplitudes at $p$ and $2K-p$. The phase factors in $\lvert Y_{s,p}\rangle$ cancel those phases, producing either cosine or sine modes. Differentiating the basis functions in Eqs.~\eqref{eq:cosine-basis} and~\eqref{eq:sine-basis} $s$ times multiplies each frequency-$p$ mode by $p^s$, switches between cosine and sine according to $\sigma_s$, and contributes the sign $\chi_{m,s}$ from Eq.~\eqref{eq:derivative-sign}. Thus, $B_{\mathrm{out}}{(s)}$ selects each frequency pair with the correct phase and sign, while $R_p{(s)}$ supplies the factor $[p/(K-1)]^s$. Consequently, their composition $U_E(s) = R_p(s) B_{\mathrm{out}}{(s)} \mathcal{F} B_{\mathrm{in}}{(s)}$ implements the block $E_s/(K-1)^s$ on rows $1\le p<K$.

The endpoint labels $p=0$ and $p=K$ behave as follows. At $p=0$, the contribution vanishes for every $s\geq 1$ because the frequency factor $\bigl[p/(K-1)\bigr]^s$ is zero. For even $m$ and $s=0$, the constant-mode amplitude is $(\pi/K)^{1/2}e_0=K^{-1/2}$, matching the QFT output $\bra0\mathcal F\ket{\psi_\ell}$. At $p=K$, the amplitude remains in the branch state $\lvert 1\rangle_{\mathrm{br}}$, so it is excluded from the selected output block.

\subsection{Shared frequency sequence}
\label{app:frequency-polynomials}
For each Taylor order $s$, the circuit uses quantum eigenvalue transformation (QET) to apply the polynomial $x^s$ to the frequency signal $U=U_{\mathrm{freq}}$. For $s\geq 1$, this uses a sequence of $s$ signal calls and $s+1$ rotations with a phase list $(\phi_0^{(s)}, \ldots, \phi_s^{(s)})$:
\begin{equation}
    V_s=D(\phi_0^{(s)})\,U D(\phi_1^{(s)})\cdots U D(\phi_s^{(s)}),
\end{equation}
where $D(\phi)=e^{\ii\phi(2\Pi-I)}$ and $\Pi$ is the projector onto the zero state of the frequency selector. For $s=0$, set $V_0=I$. 

To use the same signal calls for every Taylor order, we pad each phase list with zeros so that all lists have length $r+1$. For $s=0$, every phase is zero. The shared circuit has the same alternating pattern of phase rotations and calls to $U$ as $V_r$, with the order register selecting the padded phase appropriate to each value of $s$.

The circuit contains $r$ calls to $U$. The first call to act on the state is controlled by the least significant bit of the order register and is applied only when $s\equiv r\pmod 2$; the remaining $r-1$ calls are unconditional. When $s=r$, the circuit is exactly $V_r$. When $s<r$, all phase rotations after position $s$ are identities. The remaining calls therefore combine into a power of $U$; the control ensures this power is even, so it cancels because $U_{\mathrm{freq}}^2=I$. The circuit consequently reduces to $V_s$.

This cancellation holds on the full register, including the selector workspace, and also when the order register is in a superposition, since the circuit preserves the order register. Padding the negated phase list in the same way also preserves the real-part extraction. Finally, because $x^s$ has definite parity, the even and odd parts do not need to be combined. Thus each order branch produces the block in Eq.~\eqref{eq:main-frequency-flag}, while the $s=0$ branch acts as the identity.

Each of the $r$ signal-query positions requires $O(n\log n)$ gates. This is because $U_{\mathrm{freq}}$ selects among $n+1$ Pauli terms using a selector register of $\lceil\log_2(n+1)\rceil$ qubits. At each of the $r+1$ phase positions, the circuit applies a uniformly controlled rotation selected by the order register. Since there are fewer than $2(r+1)$ possible order values, each such position costs $O(r)$ gates~\cite{Mottonen2004}. The reflection $2\Pi-I$, about the zero state of the frequency selector, adds $O(\log n)$ gates per position, or $O(r\log n)$ gates overall. Combining these costs gives
\begin{equation}
    C_{\mathrm{freq}}=O\bigl(rn\log n+r^2\bigr),
\end{equation}
the second line of Eq.~\eqref{eq:component-gate-costs}. By comparison, implementing a separate order-controlled sequence for each Taylor order would require $r(r+1)/2$ query positions. When $r=0$, the frequency circuit is unnecessary and is omitted.

\section{\texorpdfstring{Weighted Helmert basis change (Section~\ref{sec:helmert-circuit})}{Weighted Helmert basis change}}
\label{app:weighted-contrasts}

This appendix derives the weighted contrasts of Section~\ref{sec:helmert-circuit} from polynomial connection formulas, shows that the two-pass tree of rotations produces those contrasts, and specifies how to complete the transformation into a unitary. It also describes how to implement the rotation layers using polynomial approximations and presents a gate-count estimate. We recall from Eqs.~\eqref{eq:contrast-weight} and~\eqref{eq:prefix-mass} that $\mu_j$ is the weight assigned to degree $j$ at parameter $\lambda$, and $M_k$ is the total weight of the preceding degrees with the same parity as $k$.

\subsection{Conversion coefficients and weights}
\label{app:helmert-weights}

Each conversion step expresses $G_k^{(\lambda+2)}$ in the basis $\{G_j^{(\lambda)}\}$. Since $G_k^{(\lambda)}(\theta)=\sin^\lambda\theta\,p_k^{(\lambda)}(\cos\theta)$, increasing the parameter from $\lambda$ to $\lambda+2$ introduces an extra factor of $\sin^2\theta=1-x^2$, where $x=\cos\theta$. For integer $\lambda\ge 1$, the required expansion is
\begin{equation}
    \begin{aligned}
        (1-x^2)\,C_k^{(\lambda+2)}={}&\frac{2\lambda+1}{2\lambda(\lambda+1)}\sum_j(j+\lambda)\,C_j^{(\lambda)}\\
        &-\frac{(k+1)(k+2)}{4\lambda(\lambda+1)}\,C_{k+2}^{(\lambda)},
    \end{aligned}
    \label{eq:weighted-conversion}
\end{equation}
where the sum runs over $j\leq k$ with $j\equiv k\pmod2$. To obtain this, we take Eq.~\eqref{eq:gegenbauer-conversion} at parameter $\lambda+1$: $C_j^{(\lambda+2)}-C_{j-2}^{(\lambda+2)}=(j+\lambda+1)\,C_j^{(\lambda+1)}/(\lambda+1)$. Summing over $j=k,k-2,\ldots$ causes the left-hand side to telescope, leaving $C_k^{(\lambda+2)} =(\lambda+1)^{-1}\sum_{j=k,k-2,\dots} (j+\lambda+1)\,C_j^{(\lambda+1)}$. We then multiply both sides by $1-x^2$ and replace each term $(1-x^2)\,C_j^{(\lambda+1)}$ by the following identity~\cite{DLMF}
\begin{equation}
    \begin{aligned}
        (1-x^2)\,C_j^{(\lambda+1)}={}&\frac{(j+2\lambda)(j+2\lambda+1)}{4\lambda(j+\lambda+1)}\,C_j^{(\lambda)}\\
        &-\frac{(j+1)(j+2)}{4\lambda(j+\lambda+1)}\,C_{j+2}^{(\lambda)}.
    \end{aligned}
\end{equation}
The resulting coefficients provide the identity in Eq.~\eqref{eq:weighted-conversion}. For $\lambda=0$, using the Chebyshev limit in Eq.~\eqref{eq:chebyshev-limit}, we obtain $(1-x^2)C_k^{(2)}=\sum'_jT_j-(k+1)T_{k+2}/2$. The sum is over the same degrees as before, except that the prime means that the $j=0$ term is counted with half weight.

Multiplying Eq.~\eqref{eq:weighted-conversion} by $\sin^\lambda\theta$, substituting $x=\cos\theta$, and normalizing gives the coefficient vector $\boldsymbol v_k$ of $G_k^{(\lambda+2)}$ in the basis $\{G_j^{(\lambda)}\}$. For $\lambda\geq1$, the coefficients for degrees $j\leq k$ are positive and proportional to $(j+\lambda)(h_j^{(\lambda)})^{1/2}$. By Eq.~\eqref{eq:gegenbauer-norm}, these are proportional to $\sqrt{\mu_j}$, where $\mu_j$ is defined in Eq.~\eqref{eq:contrast-weight}. For $\lambda=0$, the Chebyshev limiting case gives coefficients proportional to $(h_j^{(0)})^{1/2}$, except that the coefficient for $j=0$ is halved. Since $h_0^{(0)}=\pi$ and $h_j^{(0)}=\pi/2$ for $j\geq1$, the corresponding weights are $\mu_0=1/2$ and $\mu_j=1$ for $j\geq1$. In both cases, the coefficient at degree $k+2$ is negative. Thus the coefficient vector has the form
\begin{equation}
    \boldsymbol v_k=a_k\ket{w_{k+2}}-c_k\ket{k+2},
    \qquad
    a_k,c_k>0,
\end{equation}
where $|w_{k+2}\rangle$ is the weighted mean defined in Eq.~\eqref{eq:prefix-mass}. The values of $a_k$ and $c_k$ are fixed by the orthonormality conditions of $G_k^{(\lambda+2)}$. 

Since $\ket{w_{k+2}}$ is a unit vector orthogonal to $\ket{k+2}$, the unit norm of $\boldsymbol v_k$ implies $a_k^2+c_k^2=1$. Now let $k'>k$ be a later degree of the same parity, so that $\boldsymbol v_{k'}=a_{k'}\ket{w_{k'+2}}-c_{k'}\ket{k'+2}$ with $k'+2\geq k+4$. Because $\boldsymbol v_k$ has no component at degree $k'+2$, the overlap $\langle\boldsymbol v_{k'}\vert\boldsymbol v_k\rangle$ equals $a_{k'}\langle w_{k'+2}\vert\boldsymbol v_k\rangle$, which is proportional to $a_k\sqrt{M_{k+2}}-c_k\sqrt{\mu_{k+2}}$. Setting this overlap to zero and using $M_{k+4}=M_{k+2}+\mu_{k+2}$, we obtain
\begin{equation}
    a_k=\sqrt{\frac{\mu_{k+2}}{M_{k+4}}},
    \qquad
    c_k=\sqrt{\frac{M_{k+2}}{M_{k+4}}},
\end{equation}
so that $\boldsymbol v_k=-\ket{h_{k+2}}$, following the expression of the weighted contrast of Eq.~\eqref{eq:main-weighted-contrast}. Since $\boldsymbol v_k$ is the coefficient vector of $G_k^{(\lambda+2)}$ in the basis $\{G_j^{(\lambda)}\}$, this identity gives the conversion in Eq.~\eqref{eq:main-angular-contrast}. The cumulative weights in Eq.~\eqref{eq:contrast-prefix-mass} follow by telescoping: the product increases by $\mu_k$ from degree $k$ to $k+2$, and its values $M_2=\mu_0$ and $M_3=\mu_1$ start the two parity classes. Every block weight in the tree factorization is therefore a difference of two values of a polynomial of degree $2\lambda+1$.

\subsection{Tree factorization and unitary completion}
\label{app:helmert-tree}

We now show that the two passes of Section~\ref{sec:helmert-circuit} produce these contrasts exactly. 

For a set $S$ of modes, let $\mu_S$ denote the total weight and $\ket{w_S}$ denote the normalized weighted mean of the modes in $S$, defined analogously to Eq.~\eqref{eq:prefix-mass}. For disjoint sets $S$ and $S'$, define the normalized contrast of $S'$ relative to $S$ by
\begin{equation}
    \ket{c_{S'|S}}=\sqrt{\frac{\mu_S}{\mu_{S\cup S'}}}\ket{w_{S'}}-\sqrt{\frac{\mu_{S'}}{\mu_{S\cup S'}}}\ket{w_S}.
    \label{eq:helmert-set-contrast}
\end{equation}
In particular, the contrast $\ket{h_k}$ defined in Eq.~\eqref{eq:main-weighted-contrast} is the contrast of mode $k$ with the set $P_k$ of preceding modes; equivalently, $\ket{h_k}=\ket{c_{\{k\} | P_k}}$. After each layer of rotations in either pass, the amplitude at the smallest label of each block equals the overlap $\langle v | \psi\rangle$ between the input state $\ket{\psi}$ and some vector $\ket{v}$. We describe each layer by identifying these vectors.

The splitting pass first builds weighted means and contrasts within each block. For a block $A=L\cup R$, the first labels of $L$ and $R$ store the overlaps of the input state $\ket{\psi}$ with their weighted means, $\ket{w_L}$ and $\ket{w_R}$. The rotation in Eq.~\eqref{eq:main-splitting-rotation} replaces these with overlaps against the weighted mean of the whole block, $\ket{w_A}$, and the contrast $\ket{c_{R\mid L}}$. Repeating this from single modes upward leaves the mean of the full parity group at its first label and a contrast at the first label of each right half.

The correction pass then moves back down the tree to put each contrast into the form required by the weighted Helmert basis. Consider a block $A=L\cup R$ with preceding modes $P$. The first label of $A$ stores the overlap with $\ket{c_{A\mid P}}$, or with the group mean if $P$ is empty, while the first label of $R$ stores the overlap with $\ket{c_{R\mid L}}$. The desired overlaps are instead those with $\ket{c_{L\mid P}}$ and $\ket{c_{R\mid P\cup L}}$. These two pairs of vectors are orthonormal bases for the same two-dimensional subspace, and their overlaps determine a rotation by the angle $\vartheta$ in Eq.~\eqref{eq:main-helmert-angles}. Thus, applying $R(-\vartheta)$ converts the stored pair into the desired one. When $P$ is empty, $\vartheta=0$, so the pair already has the required form.

As the correction pass reaches single modes, the label for mode $k$ stores the overlap with $\ket{h_k}$, the contrast of that mode against its preceding modes. This establishes that the two passes together implement the weighted Helmert transform in Eq.~\eqref{eq:helmert-tree-factorization}.

The labels $p<\lambda$ do not correspond to actual polynomial degrees in the basis at parameter $\lambda$. To define the unitary completion, temporarily assign each such inactive label the same positive weight $\omega$. This makes every block weight positive and all rotation angles well-defined. Then take the limit as $\omega$ approaches zero from above. Since the angle formulas are bounded ratios of polynomials in $\omega$, they have limits; when both numerator and denominator vanish at zero, the limit is determined by their lowest-order nonzero powers.

In this limit, each active label $p\geq\lambda+2$ stores the weighted contrast of the active modes described by Eq.~\eqref{eq:main-weighted-contrast}. The remaining labels store weighted means or combinations involving inactive labels. These do not affect the completed transform: they are inactive at the next parameter step and therefore cannot contribute to a valid output of $L$. There are fewer than $m/2$ inactive positions in each parity group, so the affected blocks occur only at small block addresses, below $C_m$, where the circuit implements them using fixed rotations.

\subsection{Polynomial implementation of the rotation layers}
\label{app:helmert-polynomials}

Each rotation layer is determined by the total weights of three groups of modes: the modes before the current block, denoted $P$, and the block's left and right halves, $L$ and $R$. These weights are sums of the individual mode weights $\mu_j$. To calculate the sums, write a system label as $p=2a+\sigma$, where $\sigma = p_0$ is the parity bit and $a=(p_{n-1}\cdots p_1)_2$ indicates the position within the parity group. At parameter $\lambda$, this label represents degree $k=2a+\sigma-\lambda$ when $k\geq 0$; labels with $k<0$ are inactive. The positions before $a$ correspond to the degrees in $\mathcal I_k$, whose total weight is the cumulative weight $M_k$ from Eq.~\eqref{eq:prefix-mass}. This weight is zero when $k\leq 1$.

Define the rescaled cumulative weight $E_\lambda(a,\sigma):=2(2\lambda+1)M_{2a+\sigma-\lambda}$. The prefactor does not affect the rotation angles, because those depend only on ratios of weights. Substituting $k=2a+\sigma-\lambda$ into Eq.~\eqref{eq:contrast-prefix-mass} yields
\begin{equation}
    E_\lambda(a,\sigma)=
    \begin{cases}
        \prod_{v=1}^{2\lambda+1}(2a+\sigma-\lambda-2+v), & k\geq2,\\
        0, & k\leq1.
    \end{cases}
    \label{eq:helmert-prefix-product}
\end{equation}
For $\lambda=0$, this simplifies to $E_0(a,\sigma)=2a+\sigma-1$ for $a\geq1$; this formula incorporates the half-weight assigned to the constant mode.

In layer $l$, each block has length $B=2^{l+1}$. Let $F(b)=E_\lambda(Bb,\sigma)$ be the cumulative weight before the block at address $b$, that is, the weight $\mu_P=F(b)$ of its predecessors $P$. Then the weights of the two halves of the block are
\begin{equation}
    \mu_L=F\bigl(b+\tfrac12\bigr)-F(b),
    \quad
    \mu_R=F(b+1)-F\bigl(b+\tfrac12\bigr).
    \label{eq:helmert-local-masses}
\end{equation}
In other words, differences of cumulative weights provide the weights of each block segment.

The rotation angles depend on four quantities
\begin{equation}
    \begin{aligned}
        \cos\varphi&=\sqrt{\frac{\mu_L}{\mu_A}},
        &\cos\vartheta&=\sqrt{\frac{\mu_L\,\mu_{P\cup A}}{\mu_A\,\mu_{P\cup L}}},\\
        \sin\varphi&=\sqrt{\frac{\mu_R}{\mu_A}},
        &\sin\vartheta&=\sqrt{\frac{\mu_P\,\mu_R}{\mu_A\,\mu_{P\cup L}}},
    \end{aligned}
    \label{eq:helmert-four-functions}
\end{equation}
where $\mu_A=\mu_L+\mu_R$, $\mu_{P\cup L}=\mu_P+\mu_L$, and $\mu_{P\cup A}=\mu_P+\mu_A$. If one of these functions is constant\textemdash for example, the bulk splitting functions when $\lambda=0$\textemdash the circuit can use a fixed rotation rather than compute the angle from the address.

For each dyadic interval $I_\nu$, write an address as $b=2^\nu+v$, where $v$ is the offset within the interval and $0\le v<2^\nu$. The circuit encodes that offset as an angle
\begin{equation}
    \theta_{\nu,v}=0.9+0.6\frac{v}{2^\nu-1},
    \label{eq:angle-encoding}
\end{equation}
so the angle ranges from $0.9$ to $1.5$ radians. A single signal qubit $g$ is enough to encode the value $\cos\theta_{\nu,v}$. With Pauli-Y rotation $R_y(\theta)=e^{-\ii\theta Y/2}$, the circuit applies
\begin{equation}
    \begin{aligned}
        U_{\nu,v}^{\mathrm H}&=R_y(\theta_{\nu,v})\,Z\,R_y(-\theta_{\nu,v})\\
        &=\cos\theta_{\nu,v}\,Z+\sin\theta_{\nu,v}\,X,
    \end{aligned}
    \label{eq:helmert-signal-reflection}
\end{equation}
where the angles are controlled by the offset bits. The selected block with $g=0$ represents the diagonal signal operator $B_\nu^{\mathrm H}=\diag(\cos\theta_{\nu,v})$. Because $\theta_{\nu,v}\in[0.9,1.5]$, the diagonal entries lie in the fixed interval $J=[\cos1.5,\cos0.9]\subset(0,1)$ regardless of the interval, layer, or system size $K$. The rotation angle can be built from the bits $v_a$ of $v$, since $R_y(\theta_{\nu,v})=R_y(0.9)\prod_aR_y(0.6\cdot2^a/(2^\nu-1))^{v_a}$. Each bit controls a rotation by $0.6\,2^a/(2^\nu-1)$, in addition to a fixed rotation by $0.9$. Thus, one signal call requires $O(\nu)$ gates. For addresses outside the chosen interval, the circuit turns off the middle $Z$ operation, so the two rotations cancel and the circuit acts as the identity.

On the signal register, the QET circuit needs to implement functions of the form $f_\nu(x):=f(b^{\mathrm H}_\nu(x))$, where $f$ is one of the functions listed in Eq.~\eqref{eq:helmert-four-functions} and 
\begin{equation}
    b^{\mathrm H}_\nu(x)=2^\nu + \frac{2^\nu-1}{0.6} (\arccos x-0.9).
    \label{eq:inverse-angle-encoding}
\end{equation}
This formula reverses the angle encoding: at the encoded value $x=\cos\theta_{\nu,v}$, its value is $b^{\mathrm H}_{\nu}(x)=2^\nu+v$. Because $\arccos x$ has singular behavior at $x=\pm1$, the functions $f_\nu$ are not approximated over the entire interval $[-1,1]$. Instead, the circuit approximates each one only on the interval $J$, where the signal values actually lie. The approximating polynomial must still be bounded on $[-1,1]$.

\begin{lemma}[Helmert rotation polynomials]
\label{lem:helmert-approximation}
Fix $m\geq2$ and $0<\eta_L\leq10^{-2}$. For every function in Eq.~\eqref{eq:helmert-four-functions}, every layer and parity, and every dyadic interval with $2^\nu\geq C_m$ for $C_m \geq \frac{2}{3}(m-1)$, there is a real polynomial $p$ of degree
\begin{equation}
    d_L=O\Bigl(m^2\log\frac m{\eta_L}\Bigr),
    \label{eq:helmert-degree-bound}
\end{equation}
uniformly in the layer, parity, interval, and $K$, such that
\begin{equation}
    \|p-f_\nu\|_J\leq\eta_L/2,
    \qquad
    \|p\|_{[-1,1]}\leq1-\eta_L/4,
    \label{eq:helmert-bounded-approximation}
\end{equation}
where $\|\cdot\|_I$ denotes the supremum norm on $I$.
\end{lemma}

\begin{proof}
We begin by bounding the target on a complex neighborhood of the interval $J=[\cos(1.5),\cos(0.9)]$. Define three quantities: $D_m=2m-3$, $\tau_m=1/(32D_m)$, and $\delta_m=3\tau_m/40$. Since $D_m\geq 1$, we have $\tau_m\leq 1/32$ and $\delta_m\leq 3/1280<2.4\times10^{-3}$. Let $
\mathcal N = \{z\in\mathbb C:\operatorname{dist}(z,J)<\delta_m\}
$. For every $z\in\mathcal N$, it follows that $|z|<\cos0.9+\delta_m<3/4$. Hence $\arccos$ is analytic there, and $|\arccos'(z)|=|1-z^2|^{-1/2}<(1-9/16)^{-1/2}<2$. For any $x\in J$ such that $|z-x|<\delta_m$, we have $|b^{\mathrm H}_\nu(z)-b^{\mathrm H}_\nu(x)|\leq\tfrac{10}3\,2^\nu\delta_m=2^\nu\tau_m/4$ using Eq.~\eqref{eq:inverse-angle-encoding}. Since $b^{\mathrm H}_\nu(x)\in[2^\nu,2^{\nu+1}-1]$, the address $b=b^{\mathrm H}_\nu(z)$ has imaginary part $|\operatorname{Im} b| \leq 2^\nu\tau_m/4$ and real part $u = \operatorname{Re} b \geq2^\nu(1-\tau_m/4)\geq3\cdot2^\nu/4$.

By Eq.~\eqref{eq:helmert-prefix-product}, $F(b)=\prod_{v=1}^{d_\lambda}y_v$ with $d_\lambda=2\lambda+1$ and $y_v=2Bb+\sigma-\lambda-2+v$. Since $F(b+\tfrac j2)=\prod_v(y_v+jB)$, the weights in Eq.~\eqref{eq:helmert-local-masses} are built from products of the factors $y_v+jB$ with $j\in\{0,1,2\}$. Each Helmert stage changes $\lambda$ to $\lambda+2$ with $\lambda+2\leq m$, so $\lambda\leq m-2$. These products therefore have at most $d_\lambda\leq2m-3=D_m$ factors. Moreover, the constant term $\sigma-\lambda-2+v$ of $y_v$ is at least $-\lambda-1\geq1-m$, with the minimum at $v=1$ and $\sigma=0$. Since $u=\operatorname{Re}b$, every factor satisfies $\operatorname{Re}(y_v+jB)\geq2Bu+1-m$. This is at least $Bu>0$ whenever $Bu\geq m-1$. Because $B=2^{l+1}\geq2$ and $u\geq\tfrac34\,2^\nu$, this condition holds whenever $2^\nu\geq\tfrac23(m-1)$, which is given by the lemma's hypothesis. As $\operatorname{Im}(y_v+jB)=2B\operatorname{Im}b$, it follows that
\begin{equation}
\frac{|\operatorname{Im}(y_v+jB)|}{\operatorname{Re}(y_v+jB)}\leq\frac{2|\operatorname{Im}b|}{u}\leq\frac{2\tau_m}3=\frac1{48D_m}.
\end{equation}

Each factor equals $a(1+\ii\epsilon)$, where $a=\operatorname{Re}(y_v+jB)>0$ is its value at the real address $u$ and $\epsilon=2B\operatorname{Im}b/a$. Since $|\epsilon|$ is the ratio bounded above, $|\epsilon|\leq1/(48D_m)$. Let $\Pi$ be a product of at most $D_m$ such factors $a_i(1+\ii\epsilon_i)$ with $a_i>0$. Its argument has magnitude at most $\sum_i |\epsilon_i| \leq 1/48$, and its modulus satisfies
\begin{equation}
\begin{aligned}
    \prod_i a_i \le|\Pi| &\le \left(\prod_i a_i\right) \exp\!\left(\frac12\sum_i\epsilon_i^2\right) \\
    &\le \left(\prod_i a_i\right) e^{1/(4608D_m)}.
\end{aligned}
\end{equation}
Therefore, $\operatorname{Re}\Pi\ge \left(\prod_i a_i\right)\cos(1/48)$.

By definition of $F(b)$, the cumulative weights can each be written as a single product: $\mu_P =\prod_{v=1}^{d_\lambda}y_v$, $\mu_{P\cup L} =\prod_{v=1}^{d_\lambda}(y_v+B)$, and $\mu_{P\cup A} =\prod_{v=1}^{d_\lambda}(y_v+2B)$. The weight of $L$ is the increase in cumulative weight when $L$ is added to $P$, so $\mu_L=\mu_{P\cup L}-\mu_P$. Expanding this difference by replacing the factors one at a time yields
\begin{equation}
    \mu_L(b)=B\sum_{k=1}^{d_\lambda}\prod_{v<k}(y_v+B)\prod_{v>k}y_v.
    \label{eq:helmert-positive-products}
\end{equation}
Similarly, $\mu_R$ is obtained by replacing each $y_v$ in this sum with $y_v+B$. Every weight $\mu\in\{\mu_P,\mu_L,\mu_R,\mu_A,\mu_{P\cup L},\mu_{P\cup A}\}$ is therefore a sum, with positive coefficients, of products of at most $D_m$ factors $y_v+jB$. Applying $\operatorname{Re}\Pi\geq(\prod_ia_i)\cos(1/48)$ to each term gives $\operatorname{Re}\mu(b)\geq\cos(1/48)\,\mu(u)>0$. Meanwhile, the triangle inequality and the bound $|\Pi|\leq(\prod_ia_i)\,e^{1/(4608D_m)}$ imply $|\mu(b)|\leq e^{1/(4608D_m)}\mu(u)$.

The ratios under the square roots in Eq.~\eqref{eq:helmert-four-functions}, as functions of $z$ through $b=b^{\mathrm H}_\nu(z)$, are therefore analytic and nonzero on $\mathcal N$, because every weight has positive real part. Because $\mathcal N$ is simply connected, each ratio has an analytic square root there, chosen to agree with the positive square root on the real axis. Define  $\kappa_m\coloneqq e^{1/(4608D_m)}/\cos(1/48)$, which satisfies $\kappa_m<1.0005$. For a splitting ratio, such as $\mu_L/\mu_A$, the modulus bound on the numerator and the real-part bound on the denominator imply $\left|\frac{\mu_L(b)}{\mu_A(b)}\right| \le \kappa_m\,\frac{\mu_L(u)}{\mu_A(u)}$. Similarly, each correction ratio, such as $\mu_L\mu_{P\cup A}/(\mu_A\mu_{P\cup L})$, has modulus at most $\kappa_m^2$ times its value at $u$. For real addresses, all weights are positive, and each cosine--sine pair has squared values summing to one. Thus, these ratios lie in $(0,1)$ on the real axis, which implies $|f_\nu|\le\kappa_m$ on $\mathcal N$ and $0<f_\nu<1$ on $\mathcal N\cap\mathbb R$.

We now apply the bounded approximation theorem for multiple local Taylor series~\cite[Theorem~68, arXiv version]{Gilyen2019}. It combines Taylor series centered along a chain of intervals, where only neighboring intervals may overlap, into a single polynomial that approximates the function throughout the chain and remains bounded on $[-1,1]$ by the function's maximum near that chain.

To cover $J$, choose $N_0=O(\delta_m^{-1})$ centers $x_j$, spaced $3r_0/2$ apart, where $r_0=\delta_m/8$. The intervals $[x_j - r_0, x_j + r_0]$ overlap with their immediate neighbors by $r_0/2$, and non-neighboring intervals do not overlap. At each center, the theorem also allows an extension of width $r_0/2$, so it uses $f_\nu$ on the larger interval $[x_j - 3r_0/2, x_j + 3r_0/2]$. On the larger interval, $|x-x_j|\le 3r_0/2$. Thus the $k$th term in the expansion $f_\nu(x) = \sum_{k=0}^{\infty} c_k (x-x_j)^k$ is bounded in magnitude by $|c_k|(3r_0/2)^k$. The theorem requires $ \sum_k |c_k|(3r_0/2)^k \leq B_0$ for some bound $B_0$. To obtain such a bound, we use the analyticity of $f_{\nu}$: a complex disk of radius $\delta_m/2$ around $x_j$ lies in the region where $|f_\nu|\le\kappa_m$. By Cauchy's estimate, $
|c_k|\le \kappa_m\left(2 / \delta_m\right)^k$, therefore, $
\sum_k |c_k|\left(3r_0/2\right)^k
\le \kappa_m\sum_k\left(3r_0 / \delta_m\right)^k
$. Since $r_0=\delta_m/8$, the geometric-series ratio is $3/8$, so the sum is at most $\frac{\kappa_m}{1-3/8}<2$. Hence one may take $B_0=2$.

The extended intervals also lie inside the neighborhood where $0<f_\nu<1$ for real inputs. Set $\epsilon_0=\min\{\eta_L/4,1/(32 N_0)\}$. This choice satisfies the theorem's requirement $\epsilon_0\leq 1/(2 B_0 N_0)$. The theorem therefore gives a polynomial $\widetilde p$ such that $|\widetilde p-f_\nu|\leq\epsilon_0$ on $J$ and $|\widetilde p|\leq1$ on $[-1,1]$. As the extension width and the overlap both equal $r_0/2$, the polynomial's degree is $O\!\left(\frac{N_0}{r_0}\log\frac{8N_0}{\epsilon_0}\right)$. 

If $\widetilde p$ has complex coefficients, replace it by the polynomial formed from the real parts of its coefficients. This preserves both bounds for real inputs, since $f_\nu$ is real. Now define $p=(1-\eta_L/4)\operatorname{Re}\widetilde p$. Then $|p|\leq1-\eta_L/4$ on $[-1,1]$. On $J$, since $|f_\nu|<1$, $|p-f_\nu|\leq\epsilon_0+\eta_L/4\leq\eta_L/2$, which proves Eq.~\eqref{eq:helmert-bounded-approximation}. Finally, $\delta_m^{-1}=O(m)$, so $N_0=O(m)$, $r_0^{-1}=O(m)$, and $\log(8N_0/\epsilon_0)=O(\log(m/\eta_L))$. Substituting these estimates yields the degree bound in Eq.~\eqref{eq:helmert-degree-bound}.
\end{proof}

The proof uses the cutoff only through the condition $2^\nu\geq\frac23(m-1)$. Any power of two $C_m\geq\max\{2,\frac23(m-1)\}$ therefore suffices, where the lower bound $2$ keeps the angle encoding $\theta_{\nu,v}$ in Eq.~\eqref{eq:angle-encoding} well defined. Since fixed rotations are cheaper than QET sequences on short intervals, we use the smallest power of two no smaller than $16(m+1)$ in our implementation.

At fixed $m$, Lemma~\ref{lem:helmert-approximation} shows that the polynomial degree grows only logarithmically with the target error, i.e., $d_L=O_m(\log(1/\eta_L))$, as stated in Eq.~\eqref{eq:helmert-polynomial-degree}.

The reason is that the rotation functions have Taylor series that converge geometrically, so truncating each series after $O_m(\log(1/\eta_L))$ terms achieves the required accuracy. The series coefficients can be obtained from the fixed-degree weight formulas and the expansion of $\arccos$. To compute them stably, the construction uses Eq.~\eqref{eq:helmert-positive-products} to scale out powers of $B2^\nu$, avoiding cancellation when subtracting nearby weights. The resulting classical computation takes time polynomial in $n$ and $\log(1/\eta_L)$.

To implement the polynomials $P_C$ and $P_S$ with quantum eigenvalue transformation (QET), the circuit uses the projector onto $g=0$. At each QET phase step, the phase rotation on the real-part-extraction selector depends on three choices: whether the circuit is implementing $P_C$ or $P_S$, the parity $\sigma$, and whether it is using the even or odd part of the polynomial. Thus, eight possible phase angles are needed per step.

As in the input-correction construction in Appendix~\ref{app:input-circuit}, the even- and odd-part sequences share their signal calls. One additional signal call, controlled by the even/odd selector, is inserted at the middle $Z$ operation in Eq.~\eqref{eq:helmert-signal-reflection}. The Hadamard gates on the two selector qubits from Eq.~\eqref{eq:main-polynomial-block} surround all address regions in a layer. Consequently, the selected blocks are $P_C(B^{\mathrm H}_{\nu})/2$ and $P_S(B^{\mathrm H}_{\nu})/2$, as used in Eq.~\eqref{eq:qet-helmert-rotation-block}.

\subsection{Gate count}
\label{app:helmert-resources}

In layer $l$, the block address contains $n-2-l$ bits. Therefore, the addresses satisfying $b \ge C_m$ can be grouped into $O(n)$ dyadic intervals. For an interval with exponent $\nu$, processing a degree-$d_L$ sequence takes $O(d_L(\nu+1))$ gates. Each signal call costs $O(\nu)$ gates, and each phase step is a uniformly controlled rotation with eight angles. Setting and clearing the interval predicate adds $O(n)$ gates. Summing across the intervals gives a cost of $O((d_L+1)n^2)$ per layer. Since one parameter step has $2n-3$ layers, its total cost is $O((d_L+1)n^3)$. The fixed rotations at addresses $b<C_m$ add $O(C_m+n)$ gates per layer. Amplifying each layer uses seven calls to its padded block. Repeating this for the $\lfloor m/2\rfloor$ parameter steps changes the cost only by factors that depend on $m$. Thus, the basis-change cost is
\begin{equation}
    C_{\mathcal L}=O_m\bigl((d_L+1)n^3\bigr),
\end{equation}
which is the third line of Eq.~\eqref{eq:component-gate-costs}. The basis change requires only the signal qubit $g$ and a constant number of selector, predicate, and padding qubits; these qubits are reused across all layers.

\section{\texorpdfstring{Amplification errors (Sections~\ref{sec:helmert-circuit} and~\ref{sec:circuit-assembly})}{Amplification errors}}
\label{app:amplification}

Both the basis-change layers and the full transform are amplified by the circuit~\eqref{eq:main-amplified-circuit} from block encodings that are only approximately unitary. The following lemma bounds the resulting output error, including leakage outside the successful subspace. We then apply it to both cases. As defined in Section~\ref{sec:amplification-primitive}, $\beta_t=1/\sin[\pi/(4t+2)]$.

\begin{lemma}[Robust amplification]
\label{lem:robust-oaa}
Let $W$ be a unitary and let $J_{\mathrm{in}}$ and $J_{\mathrm{out}}$ be isometries onto its input and successful output subspaces such that $J_{\mathrm{out}}^\dagger WJ_{\mathrm{in}}=\widehat V/\beta_t$ with $t\geq1$, where $\|\widehat V-V\|_2\leq\Delta\leq1/10$ for a unitary $V$. With the reflections $S_{\mathrm{in}}=I-2J_{\mathrm{in}}J_{\mathrm{in}}^\dagger$ and $S_{\mathrm{out}}=I-2J_{\mathrm{out}}J_{\mathrm{out}}^\dagger$, the circuit $\mathcal U_t=(-WS_{\mathrm{in}}W^\dagger S_{\mathrm{out}})^tW$ satisfies
\begin{equation}
    \|\mathcal U_tJ_{\mathrm{in}}-J_{\mathrm{out}}V\|_2\leq4\Delta.
    \label{eq:amplification-coherent-bound}
\end{equation}
\end{lemma}

\begin{proof}
Write $\widehat V=V'H$ with $V'$ unitary and $H=(\widehat V^\dagger\widehat V)^{1/2}$. The singular values of $\widehat V$ lie in $[1-\Delta,1+\Delta]$, so $\|H-I\|_2\leq\Delta$ and $\|V'-V\|_2\leq\|V'-\widehat V\|_2+\Delta\leq2\Delta$. On each singular vector of $\widehat V$ with singular value $z$, the two-dimensional analysis of oblivious amplitude amplification~\cite{Berry2015,Gilyen2019} shows that the amplified amplitude is
\begin{equation}
    g_t(z)=\sin\bigl[(2t+1)\arcsin(z/\beta_t)\bigr].
    \label{eq:amplified-amplitude}
\end{equation}
Consequently $J_{\mathrm{out}}^\dagger\mathcal U_tJ_{\mathrm{in}}=V'g_t(H)$. Because $\mathcal U_tJ_{\mathrm{in}}$ and $J_{\mathrm{out}}V'$ are isometries, their squared difference satisfies
\begin{equation*}
    \bigl(\mathcal U_tJ_{\mathrm{in}}-J_{\mathrm{out}}V'\bigr)^\dagger\bigl(\mathcal U_tJ_{\mathrm{in}}-J_{\mathrm{out}}V'\bigr)=2\bigl[I-g_t(H)\bigr].
\end{equation*}
Write $g_t(z)=\cos h_t(z)$, where $h_t(z)=(2t+1)\arcsin(z/\beta_t)-\pi/2$. Since $h_t(1)=0$, we can bound the operator norm as
\begin{equation}
    \begin{aligned}
        \|\mathcal U_tJ_{\mathrm{in}}-J_{\mathrm{out}}V'\|_2
        &= 2 \max_z |\sin(h_t(z)/2)| \\
        &\leq \max_z|h_t(z)|.
    \end{aligned}
\end{equation}
For $z\in[0.9,1.1]$, the bounds $\beta_t\geq2$ and $(2t+1)/\beta_t\leq\pi/2$ imply
\begin{equation}
    |h_t'(z)|=\frac{2t+1}{\sqrt{\beta_t^2-z^2}}\leq\frac{\pi/2}{\sqrt{1-0.55^2}}<1.89,
\end{equation}
so $|h_t(z)|\leq1.89\Delta$ on the singular values of $\widehat V$. Hence $\|\mathcal U_tJ_{\mathrm{in}}-J_{\mathrm{out}}V'\|_2\leq1.89\Delta$. Since $J_{\mathrm{out}}$ is an isometry, it follows that
\begin{equation}
    \begin{aligned}
        \|\mathcal U_tJ_{\mathrm{in}}-J_{\mathrm{out}}V\|
        &\leq\|\mathcal U_tJ_{\mathrm{in}}-J_{\mathrm{out}}V'\| + \|V'-V\| \\
        &\leq1.89\Delta+2\Delta<4\Delta,
    \end{aligned}
\end{equation}
which proves the inequality of Eq.~\eqref{eq:amplification-coherent-bound}.
\end{proof}
\subsection{Basis-change layers}
\label{app:helmert-amplification}

Each layer of the basis change is amplified by Eq.~\eqref{eq:main-amplified-circuit} with $t=3$, applied to the layer block padded to $\beta_3$, and both its input and successful subspaces require the layer workspace to be zero. At an address treated by QET, the encoded operator $\widehat R=P_CI\pm\ii P_SY$ of Eq.~\eqref{eq:qet-helmert-rotation-block} differs from the exact rotation $\cos\alpha\,I\pm\ii\sin\alpha\,Y$ by at most $2\eta_L$, where $\eta_L$ bounds the errors of the four angle functions as in Section~\ref{sec:complexity-analysis}; the fixed-rotation and idle addresses are exact. Since the layer block is a direct sum over addresses, Lemma~\ref{lem:robust-oaa} with $\Delta=2\eta_L$ shows that, for $\eta_L\leq10^{-2}$, the amplified layer $\mathcal U_l$ and its ideal rotation layer $R_l$ satisfy
\begin{equation}
    \|\mathcal U_l(\ket0\otimes I)-\ket0\otimes R_l\|_2\leq8\eta_L.
    \label{eq:helmert-layer-error}
\end{equation}
This is within the per-layer bound $10\eta_L$ used in Section~\ref{sec:complexity-analysis}. These errors add over the $L_{\mathrm{tot}}=\lfloor m/2\rfloor(2n-3)$ layers. With $\mathcal J=\ket0\otimes I$, compare the implemented product of layers, $U_{\mathcal L}$, with the ideal product by replacing one layer at a time. For each term in this sum, the layers before the changed layer are ideal and therefore leave the workspace in the zero state; the layers after it are unitary and cannot increase the error. Thus,
\begin{equation}
    \|U_{\mathcal L}\mathcal J-\mathcal J\mathcal L\|_2\leq\sum_{l=1}^{L_{\mathrm{tot}}}\|\mathcal U_l\mathcal J-\mathcal JR_l\|_2\leq10L_{\mathrm{tot}}\eta_L.
    \label{eq:helmert-accumulated-error}
\end{equation}
This argument also applies when all layers use the same workspace, and it accounts for any amplitude that remains in nonzero workspace states. Therefore, the choice $\eta_L \leq \varepsilon_L/(10L_{\mathrm{tot}})$ from Section~\ref{sec:complexity-analysis} establishes Eq.~\eqref{eq:helmert-basis-error}.

\subsection{Encoded transform}
\label{app:transform-amplification}

For the full transform, take $W$ to be the padded block $W_t$ from Section~\ref{sec:full-circuit}. The isometries $J_{\mathrm{in}}$ and $J_{\mathrm{out}}$ embed the valid labels $\ell_j$ and $k+m$, respectively, with all auxiliary registers set to zero. Let $\widehat V$ be the unnormalized encoded transform $\widehat Q$, which satisfies $\|\widehat Q-Q_N^{(m)}\|_2\leq\Delta$ by Eq.~\eqref{eq:qet-transform-error}. Applying Lemma~\ref{lem:robust-oaa} with $V=Q_N^{(m)}$ shows that, for every $t\geq 1$ and $\Delta\leq 1/10$, the output error is at most $4\Delta$, uniformly over all normalized valid inputs. This bound applies to the entire output vector, so it also accounts for amplitude leaking into unused labels or nonzero auxiliary states.

Finally, implementation errors in the $2t+1$ block calls and $2t$ reflections add to this error by the triangle inequality, yielding Eq.~\eqref{eq:qet-general-total-error}. Here, $\varepsilon_W$ bounds the operator-norm difference between the implemented block and the ideal padded block, viewed as full unitaries. To meet this error budget, each of the $G$ elementary operations in one block---including operations repeated during basis-change amplification---should be implemented to error at most $\varepsilon_W/G$.

\section{\texorpdfstring{Ultraspherical spectral solver (Section~\ref{sec:us-solver})}{Ultraspherical spectral solver}}
\label{app:us-proofs}

This appendix provides supporting results and calculations for Section~\ref{sec:us-solver}, organized as follows. Appendix~\ref{app:us-identities} establishes the identities of Section~\ref{sub:us-discretization}, Appendices~\ref{app:us-bc} and~\ref{app:us-conditioning} prove Propositions~\ref{prop:us-bc} and~\ref{prop:us-conditioning}, and Appendix~\ref{app:us-certificate} shows a way to compute an explicit bound on $\norm{\mathcal A_K^{-1}}_2$. Appendix~\ref{app:us-natural} discusses the reformulations for natural boundary conditions used in Section~\ref{sub:us-conditioning}, Appendix~\ref{app:us-encoding} constructs the block encoding and proves Proposition~\ref{prop:us-inverse}, and Appendix~\ref{app:us-error} examines the discretization error. 

Throughout, $\ell^2=\ell^2(\mathbb N)$ contains coefficient vectors indexed from zero, $\Pi_{<K}$ projects onto their first $K$ entries, and $\norm{\cdot}=\norm{\cdot}_2$ is the Euclidean norm or the induced operator norm. For functions, $\mathcal T_Ng$ keeps the terms of degree less than $N$ in the orthonormal expansion of $g$ in the relevant basis $p^{(\lambda)}$, and $\norm{u}_{C^j}\coloneqq\max_{i\leq j}\norm{\partial_x^iu}_\infty$.

\subsection{Differentiation, conversion, and multiplication}
\label{app:us-identities}

The derivative identity~\eqref{eq:us-derivative} can be obtained by differentiating the normalized ultraspherical polynomials. For $j\geq1$ and $k\geq1$, write $p_k^{(j)}=C_k^{(j)}/(h_k^{(j)})^{1/2}$. By Eqs.~\eqref{eq:gegenbauer-first-derivative} and~\eqref{eq:gegenbauer-norm},
\begin{equation}
    \frac{\dd p_k^{(j)}}{\dd x}=2j\sqrt{\frac{h_{k-1}^{(j+1)}}{h_k^{(j)}}}\,p_{k-1}^{(j+1)}=\sqrt{k(k+2j)}\,p_{k-1}^{(j+1)}.
    \label{eq:us-single-derivative}
\end{equation}
At $j=0$, Eq.~\eqref{eq:chebyshev-first-derivative} with $h_k^{(0)}=h_{k-1}^{(1)}=\pi/2$ implies a similar formula, $\dd p_k^{(0)}/\dd x=k\,p_{k-1}^{(1)}$. Differentiating $p_k^{(0)}$ a total of $\lambda$ times passes through $p_{k-j}^{(j)}$ for $j=0,\ldots,\lambda-1$, and the factor of step $j$ is $\sqrt{(k-j)(k+j)}=\sqrt{k^2-j^2}$.

The conversion coefficients follow from Eq.~\eqref{eq:gegenbauer-conversion} in the same way. For $\lambda\geq1$, we write each $C_j^{(\nu)}$ in Eq.~\eqref{eq:gegenbauer-conversion} as $(h_j^{(\nu)})^{1/2}p_j^{(\nu)}$ and then normalize by $(h_k^{(\lambda)})^{1/2}$. The resulting coefficients are $s_k^{(\lambda)}=\frac{\lambda}{k+\lambda}(h_k^{(\lambda+1)}/h_k^{(\lambda)})^{1/2}$ and $t_k^{(\lambda)}=\frac{\lambda}{k+\lambda}(h_{k-2}^{(\lambda+1)}/h_k^{(\lambda)})^{1/2}$. From Eq.~\eqref{eq:gegenbauer-norm}, we compute the norm ratios
\begin{equation}
    \begin{aligned}
        \frac{h_k^{(\lambda+1)}}{h_k^{(\lambda)}}&=\frac{(k+2\lambda)(k+2\lambda+1)(k+\lambda)}{4\lambda^2(k+\lambda+1)},\\
        \frac{h_{k-2}^{(\lambda+1)}}{h_k^{(\lambda)}}&=\frac{k(k-1)(k+\lambda)}{4\lambda^2(k+\lambda-1)},
    \end{aligned}
    \label{eq:us-norm-ratios}
\end{equation}
which demonstrate the formulas in Eq.~\eqref{eq:us-conversion-entries}. At $\lambda=0$, the ultraspherical polynomials are related to the Chebyshev polynomials by $T_0=C_0^{(1)}$, $T_1=C_1^{(1)}/2$, and $T_k=(C_k^{(1)}-C_{k-2}^{(1)})/2$ for $k\geq2$. Using the normalization constants $h_0^{(0)}=\pi$ and $h_k^{(0)}=h_k^{(1)}=\pi/2$ for $k\ge1$, we obtain $s_0^{(0)}=1/\sqrt{2}$, $s_k^{(0)}=1/2$ for $k\ge1$, and $t_k^{(0)}=1/2$ for $k\ge2$.

Every conversion is contractive: if $\boldsymbol z$ contains the coefficients of a polynomial $g$ of degree less than $N_\lambda$ in the basis $p^{(\lambda)}$, then $\widehat S_{\lambda\to\mu}\boldsymbol z$ contains its first $N_\mu$ coefficients in the basis $p^{(\mu)}$. By orthonormality, 
\begin{equation}
    \norm{\widehat S_{\lambda\to\mu}\boldsymbol z}^2\leq\int_{-1}^1w_\mu g^2\,\dd x\leq\int_{-1}^1w_\lambda g^2\,\dd x=\norm{\boldsymbol z}^2.
\end{equation}

For the multiplication matrix, let $N=N_\lambda$ and let $\boldsymbol p(x)=(p_0^{(\lambda)}(x),\ldots,p_{N-1}^{(\lambda)}(x))^\T$ contain the first $N$ orthonormal polynomials. The three-term recurrence for these polynomials implies 
\begin{equation}
    J_N^{(\lambda)}\boldsymbol p(x)=x\boldsymbol p(x)-c_N\,p_N^{(\lambda)}(x)\,\boldsymbol e_{N-1},
\end{equation}
where $c_N>0$, and $\boldsymbol e_{N-1}$ is the last coordinate vector. At any zero $x_j$ of $p_N^{(\lambda)}$, the final term is zero. Thus $\boldsymbol p(x_j)$ is an eigenvector of $J_N^{(\lambda)}$ with eigenvalue $x_j$. The columns of $Q_N^{(\lambda)}$ are the vectors $\sqrt{w_j}\, \boldsymbol p(x_j)$. They are orthonormal, so $Q_N^{(\lambda)}$ diagonalizes $J_N^{(\lambda)}$. This makes the definition of $a(J_N^{(\lambda)})$ in Eq.~\eqref{eq:us-multiplication} a spectral definition when $a$ is applied to each eigenvalue $x_j$. The entries of $a(J_N^{(\lambda)})$ are $\sum_jw_jp_i^{(\lambda)}(x_j)a(x_j)p_k^{(\lambda)}(x_j)$. These are Gauss-quadrature sums that equal the corresponding integrals exactly whenever $a$ is a polynomial and $i+k+\deg a\leq2N-1$. Operationally, $Q_N^{(\lambda)\dagger}$ converts the coefficients of $g$ into its values at the nodes $x_j$, scaled by $\sqrt{w_j}$. Multiplication by $a(x_j)$ then acts pointwise on those samples. Applying $Q_N^{(\lambda)}$ converts the result back into coefficients of the polynomial interpolant of $ag$.

\subsection{Boundary functionals}
\label{app:us-bc}

\begin{proof}[Proof of Proposition~\ref{prop:us-bc}]
We first bound the entries of the rescaling matrix. For $k\geq q$, substituting Eq.~\eqref{eq:us-derivative} into Eq.~\eqref{eq:us-R} yields $R_{kk}=\prod_{l<q}(k^2-l^2)^{-1/2}$. Since $l\leq q-1\leq(q-1)k/q$ for $l<q$, each factor satisfies $(2q-1)k^2/q^2\leq k^2-l^2\leq k^2$. Then, $k^{-q}\leq R_{kk}\leq c_qk^{-q}$ with $c_q=[q^2/(2q-1)]^{q/2}$.

(i)~Section~\ref{sub:us-conditioning} uses Markov brothers' inequality to show that, for a functional $B$ bounded on $C^j([-1,1])$, $|\mathcal B(p_k^{(0)})|=O(k^{2j})$. After rescaling, this gives $|r_k|=O(k^{2j-q})$. The sequence $(r_k)$ is square-summable if $4j-2q<-1$, which, for integer $j$ and $q$, is equivalent to $j<q/2$. Endpoint derivatives of order $j$ are bounded functionals on $C^j([-1,1])$. Likewise, the functional $u\mapsto\int_{-1}^{1}\varpi u\,dx$ is bounded on $C^0([-1,1])$ whenever $\varpi$ is integrable.

(ii)~Write $x_0=\cos\theta_0$ with $0<\theta_0<\pi$. Since $T_k(\cos\theta)=\cos k\theta$ and $\partial_x=-(\sin\theta)^{-1}\partial_\theta$, repeated differentiation yields $|\partial_x^jT_k(x_0)|\le Ck^j$, where $C$ does not depend on $k$. Therefore, if $|r_k|=O(k^{j-q})$, then $(r_k)$ is square-summable when $j<q$.

For the final statement, let $\mathcal Bu=u^{(j)}(\pm1)$ and suppose $k\geq q$. Using $p_k^{(0)}= \sqrt{2/\pi} T_k$, endpoint derivatives $\bigl|T_k^{(j)}(\pm1)\bigr|=\prod_{l=0}^{j-1}\frac{k^2-l^2}{2l+1}$, together with the diagonal entries $R_{kk}=\prod_{l<q}(k^2-l^2)^{-1/2}$, the magnitude of $r_k=R_{kk}\,\mathcal B(p_k^{(0)})$ is given by
\begin{equation}
    |r_k|=\sqrt{\frac2\pi} \left( \prod_{l=0}^{j-1}\frac{\sqrt{k^2-l^2}}{2l+1}\right) \left( \prod_{l=j}^{q-1}\frac1{\sqrt{k^2-l^2}} \right).
    \label{eq:us-row-entries}
\end{equation}
Thus, for positive constants $c_-$ and $c_+$ depending only on $q$ and $j$, $c_-k^{2j-q}\leq|r_k|\leq c_+k^{2j-q}$. When $j\geq q/2$, the partial sums of $\sum_kk^{4j-2q}$ up to $K$ grow as $K^{4j-2q+1}$. Consequently, $\norm{\Pi_{<K}\boldsymbol r}=\Theta(K^{2j-q+1/2})$.
\end{proof}

\subsection{Uniform conditioning}
\label{app:us-conditioning}

\begin{proof}[Proof of Proposition~\ref{prop:us-conditioning}]
\emph{Limit operator.} Let $\widehat D_\lambda^\infty$, $\widehat S_{\lambda\to\mu}^\infty$, and $M_\lambda^\infty[a]$ denote the infinite matrices of differentiation, conversion, and multiplication by $a$ in the orthonormal bases. $\widehat S_{\lambda\to\mu}^\infty$ and $M_\lambda^\infty[a]$ represent the inclusion $L^2(w_\lambda)\subset L^2(w_\mu)$ and multiplication by $a$ on $L^2(w_\lambda)$, so their norms are at most one and $\bar a$, respectively. In the register layout of Section~\ref{sub:us-discretization}, the operator $\mathcal A_\infty$ on $\ell^2$ has the rows $\bra{\boldsymbol r_i}$ for $i<q$, and for $\boldsymbol v\in\ell^2$, the entry $(\mathcal A_\infty\boldsymbol v)_{q+i}$ is the coefficient of degree $i$, in the parameter-$q$ basis, of $\widehat D_q^\infty R\boldsymbol v+\sum_\lambda\widehat S_{\lambda\to q}^\infty M_\lambda^\infty[a_\lambda]\widehat D_\lambda^\infty R\boldsymbol v$. Thus $\mathcal A_\infty=I+\mathcal K$ with $\mathcal K=-\Pi_{<q}+\sum_i\ket i\bra{\boldsymbol r_i}+\mathcal C$, where $\mathcal C\coloneqq\sum_{i\geq0}{\ket{q+i}}\bra i\sum_{\lambda<q}\widehat S_{\lambda\to q}^\infty M_\lambda^\infty[a_\lambda]\widehat D_\lambda^\infty R$ collects the lower-order terms. The first term, $-\Pi_{<q}$, is a finite-rank projection and is therefore bounded. The second term, $\sum_i |i\rangle\langle \boldsymbol r_i|$, also has finite rank and is bounded because $\boldsymbol r_i\in\ell^2$. The last term, $\mathcal C$, is compact: for $\lambda<q$, the shifted diagonal operator $\widehat{D}^{\infty}_{\lambda}R$ is compact because its entries $d^{(\lambda)}_k R_{kk}$ tend to zero, and the conversion and multiplication operators applied afterward are bounded.

\emph{Boundary functionals on smooth functions.} For a boundary functional $\mathcal{B}_i$ and a function $g\in C^q([-1,1])$ with orthonormal Chebyshev coefficients $\widehat{ \boldsymbol g}$, we claim that $\mathcal B_i g=\langle\boldsymbol r_i,R^{-1}\widehat{\boldsymbol g}\rangle$. This identity holds for polynomials, where the coefficient sum is finite. By Eq.~\eqref{eq:us-derivative}, the entries $d_k^{(q)}\widehat{g}_k$ of $R^{-1}\widehat{\boldsymbol g}$ for $k\ge q$ are the coefficients of $\frac{\mathrm{d}^q g}{\mathrm{d}x^q}$ in the basis $p^{(q)}$. Since $p^{(q)}$ is orthonormal in $L^2(w_q)$, $p^{(0)}$ is orthonormal in $L^2(w_0)$, and $\int_{-1}^1 w_\lambda\,dx=h_0^{(\lambda)}$, it follows that
\begin{equation}
\begin{aligned}
    \norm{R^{-1}\widehat{\boldsymbol g}}^2
    &= \sum_{k<q}|\widehat g_k|^2+\norm{g^{(q)}}^2_{L^2(w_q)} \\
    &\leq (h_0^{(0)}+h_0^{(q)})\norm g^2_{C^q}.
\end{aligned}
\end{equation}
Consequently, both sides of the identity depend continuously on $g$ in the $C^q$ norm. The left-hand side is continuous because $\mathcal{B}_i$ is bounded on $C^q([-1,1])$, and the right-hand side is continuous because $ |\langle \boldsymbol r_i,R^{-1}\widehat{\boldsymbol g}\rangle|
\le \|\boldsymbol r_i\|_2\,\|R^{-1}\widehat{\boldsymbol g}\|_2 $ with $\boldsymbol r_i \in\ell^2$. Therefore, approximating $g$ in the $C^q$ norm by polynomials, for instance, by integrating a uniform polynomial approximation of the $q$th derivative of $g$ for $q$ times, extends the identity to all $g\in C^q([-1,1])$.

\emph{Injectivity.} 
Assume $\mathcal A_\infty \boldsymbol v=0$, and define 
\begin{equation}
    u \coloneqq \sum_kR_{kk}v_kp_k^{(0)}\in L^2(w_0).
    \label{eq:preconditioned-solution-expansion}
\end{equation}
For each $\lambda\le q$, the series $\sum_kd_k^{(\lambda)}R_{kk}v_kp_{k-\lambda}^{(\lambda)}$ converges in $L^2(w_\lambda)$ and represents the $\lambda$th distributional derivative $u^{(\lambda)}$. The interior rows of $ \mathcal A_\infty \boldsymbol v=0$ imply that
\begin{equation}
  u^{(q)}+\sum_\lambda a_\lambda u^{(\lambda)}=0
  \quad\text{in }L^2(w_q).
\end{equation}
On every compact subinterval $[-1+\eta,1-\eta]$, each weight $w_\lambda(x)=(1-x^2)^{\lambda-1/2}$ is bounded above and below by positive constants, so convergence in $L^2(w_\lambda)$ implies convergence in $L^2(-1+\eta,1-\eta)$. Hence $u$ has weak derivatives $u^{(\lambda)}\in L^2_{\mathrm{loc}}(-1,1)$ for all $\lambda\leq q$ and satisfies the regular homogeneous equation almost everywhere on $(-1,1)$, so $u$ is a strong solution there. Since the equation is regular with continuous coefficients, this strong solution is classical, so $u \in C^q([-1,1])$ is a combination of fundamental solutions $y_0,\ldots,y_{q-1}$. By Eq.~\eqref{eq:preconditioned-solution-expansion}, $u$ is represented by the coefficient vector $R\boldsymbol v$. The previous identity $\mathcal B_i g=\langle\boldsymbol r_i,R^{-1}\widehat{\boldsymbol g}\rangle$ shows that $\mathcal B_iu=\langle\boldsymbol r_i,\boldsymbol v\rangle=0$ for every $i$. By uniqueness for the boundary-value problem, $u=0$, which in turn implies $\boldsymbol v=0$.

\emph{Invertibility.} Since $\mathcal K$ is compact and $I+\mathcal K$ is injective, the Fredholm alternative implies that $\mathcal A_\infty$ is invertible.

\emph{Norm convergence.} 
Let $\mathcal A_K \oplus I$ denote $\mathcal A_K$ extended by the identity on labels $k \ge K$. The difference from the limiting operator can be written as
\begin{equation}
    \mathcal A_K \oplus I - \mathcal A_\infty = (\Pi_{<K} \mathcal K \Pi_{<K} - \mathcal K)+E_K.
\end{equation}
The first term tends to zero in norm because $\mathcal K$ is compact and $\Pi_{<K}$ converges strongly to the identity. The remaining term $E_K$, which accounts for the finite Gauss multiplications and truncations in $\mathcal A_K$, is the difference between two ways of incorporating the coefficient-dependent part of the operator,
\begin{equation}
    E_K\coloneqq(C_K\oplus0)-\Pi_{<K}\mathcal C\,\Pi_{<K},
\end{equation}
where $C_K\oplus 0$ is the finite matrix $C_K$ extended by zero for $k \geq K$.

To show that $\|E_K\|\to 0$, we compare $C_K\oplus 0$ and $\Pi_{<K}\mathcal C\Pi_{<K}$ term by term, viewing each finite matrix as an operator on $\ell^2$ that is zero outside its block.
The finite matrices in $C_K$ are truncations of the corresponding infinite operators. In particular, $\widehat D_\lambda R=\widehat D^\infty_\lambda R\,\Pi_{<K}$ and $\widehat S_{\lambda\to\mu}=\Pi_{<N_\mu}\widehat S^\infty_{\lambda\to\mu}\Pi_{<N_\lambda}$. The first identity holds because $\widehat D^\infty_\lambda R$ maps the columns with $k<K$ to degrees below $N_\lambda$. The second follows from how the finite conversion matrix is defined. In the infinite operator $\mathcal{C}$, multiplication by a function is consistent across the nested function spaces $L^2(w_\lambda)\subset L^2(w_{\mu^\star})\subset L^2(w_q)$. Thus the different routes through the spaces agree: 
\begin{equation}
    \widehat S^\infty_{\lambda\to q}M^\infty_\lambda[a]=\widehat S^\infty_{\mu^\star\to q}M^\infty_{\mu^\star}[a]\widehat S^\infty_{\lambda\to\mu^\star}.
\end{equation}
After these identifications, the only discrepancy comes from using Gauss multiplication on a finite set of coefficients instead of exact multiplication in the infinite space. For each $\lambda$, that difference is $\Delta M_\lambda\coloneqq(M_{\mu^\star}[a_\lambda]\oplus0)-M^\infty_{\mu^\star}[a_\lambda]$. The zero extension also captures the truncation to the first $N_{\mu^\star}$ coefficients. Putting the pieces together gives
\begin{equation}
    E_K=\sum_{i\geq0}{\ket{q+i}}\bra i\,\Pi_{<N_q}\sum_{\lambda<q}E_K^{(\lambda)},
\end{equation}
where $E_K^{(\lambda)}\coloneqq\widehat S^\infty_{\mu^\star\to q}\,\Delta M_\lambda\,\widehat S^\infty_{\lambda\to\mu^\star}\widehat D^\infty_\lambda R\,\Pi_{<K}$. The shift $|q+i\rangle\langle i|$ places the output at labels $q+i$; $\Pi_{<N_q}$ keeps precisely those labels for which $i<N_q$. Since neither the shift nor the projection $\Pi_{<N_q}$ can increase an operator's norm, applying the triangle inequality therefore yields $\norm{E_K}\leq\sum_\lambda\norm{E_K^{(\lambda)}}$. This reduces the task of bounding the full error $E_K$ to bounding each individual contribution.

We start by approximating the coefficient functions by polynomials. Fix any $\varepsilon>0$. For each $\lambda$, choose a polynomial $p_\lambda$ of degree at most $d$ such that $ \norm{a_\lambda-p_\lambda}_\infty\le \varepsilon$. Replacing $a_\lambda$ by $p_\lambda$ changes the multiplication-error operator $\Delta M_\lambda$ by at most $2\varepsilon$. It changes $E_K^{(\lambda)}$ by at most $2\varepsilon\rho_\lambda$ because the conversion operators are contractions and $\|\widehat D_\lambda^\infty R\|\le \rho_\lambda$.

For a polynomial coefficient, the multiplication error disappears on low-degree inputs. With $p_\lambda$ in place of $a_\lambda$, $\Delta M_\lambda$ is zero on every input of degree $k<N_{\mu^\star}-d$. The reason is that multiplying an input polynomial of degree $k$ by $p_\lambda$, whose degree is at most $d$, gives a polynomial of degree less than $N_{\mu^\star}$. The product therefore remains within the first $N_{\mu^\star}$ coefficients, and the Gauss quadrature integrates the relevant products exactly. Consequently, since the conversion $\widehat S_{\lambda\to\mu^\star}^{\infty}$ does not increase degree, $E_K^{(\lambda)}$ vanishes on columns $k<N_{\mu^\star}-d+\lambda$. On the remaining columns, the diagonal entries $d_k^{(\lambda)}R_{kk}$ of $\widehat D_\lambda^\infty R$ are $O(K^{\lambda-q})$. Since $\norm{\Delta M_\lambda}\leq2\norm{p_\lambda}_\infty$, it follows that $\norm{E_K^{(\lambda)}}=O(K^{\lambda-q})$. This tends to zero as $K\to\infty$, because $\lambda<q$. Summing the approximation errors gives $\limsup_{K\to\infty}\norm{E_K}\leq2\varepsilon\sum_\lambda\rho_\lambda$. Since $\varepsilon$ is arbitrary, $\norm{E_K} \rightarrow 0$, hence $\norm{\mathcal A_K\oplus I-\mathcal A_\infty}\to0$.

\emph{Condition numbers.} Let $\mathcal R_K\coloneqq\mathcal A_K\oplus I-\mathcal A_\infty$ and $\Delta_K\coloneqq\norm{\mathcal R_K}$, so that $\mathcal A_K\oplus I=\mathcal A_\infty(I+\mathcal A_\infty^{-1}\mathcal R_K)$. Since $\Delta_K\to0$, the condition $\Delta_K\norm{\mathcal A_\infty^{-1}}<1$ holds for all sufficiently large $K$. Then $\norm{\mathcal A_\infty^{-1}\mathcal R_K}\leq\norm{\mathcal A_\infty^{-1}}\Delta_K<1$. The Neumann series $(I+\mathcal A_\infty^{-1}\mathcal R_K)^{-1}=\sum_{j\geq0}(-\mathcal A_\infty^{-1}\mathcal R_K)^j$ therefore converges, and its norm is bounded by a geometric series:
\begin{equation}
    \bigl\|\bigl(I+\mathcal{A}_\infty^{-1} \mathcal{R}_K\bigr)^{-1}\bigr\| 
    \le \frac{1}{1-\Delta_K\|\mathcal{A}_\infty^{-1}\|}.
\end{equation}
Hence $\mathcal A_K\oplus I$ is invertible. Since it is block diagonal, $\mathcal A_K$ is also invertible such that $(\mathcal A_K\oplus I)^{-1}=\mathcal A_K^{-1}\oplus I$. It follows that
\begin{equation}
    \norm{\mathcal{A}_K^{-1}} 
    \le \norm{(\mathcal{A}_K\oplus I)^{-1}}
    \le \frac{\|\mathcal{A}_\infty^{-1}\|} {1-\Delta_K\|\mathcal{A}_\infty^{-1}\|}.
\end{equation}
Similarly, $\norm{\mathcal A_K}\leq\norm{\mathcal A_K\oplus I}\leq\norm{\mathcal A_\infty}+\Delta_K$. As $K\to\infty$, the product of these bounds converges to $\kappa_2(\mathcal A_\infty)$. Therefore $\limsup_{K\to\infty}\kappa_2(\mathcal A_K)\leq\kappa_2(\mathcal A_\infty)$, which gives Eq.~\eqref{eq:us-uniform} for every $\gamma>1$.
\end{proof}

\subsection{\texorpdfstring{Certified bound on the inverse}{Certified bound on the inverse}}
\label{app:us-certificate}

Proposition~\ref{prop:us-inverse} requires a bound $\bar\kappa\geq\alpha\norm{\mathcal A_K^{-1}}_2$ before the circuit is built. Although Proposition~\ref{prop:us-conditioning} guarantees that such a bound exists for large $K$, it does not provide an explicit value.  We obtain one by performing a single classical computation at a small, fixed resolution, following the approach of validated Chebyshev methods for linear ordinary differential equations~\cite{Brehard2018}. The formulas in Section~\ref{sub:us-discretization} define $A_K$ for every integer $K>q$, so we use them even when $K$ is not a power of two.

The computation uses polynomial approximations of the coefficients. For each $\lambda<q$ with $a_\lambda\not\equiv0$, let $p_\lambda$ be a polynomial of degree at most $d_\lambda$ with $\norm{a_\lambda-p_\lambda}_\infty\leq\varepsilon_\lambda$. We take $p_\lambda=a_\lambda$ and $\varepsilon_\lambda=0$ when $a_\lambda$ is a polynomial, and the Chebyshev interpolant of degree $d_\lambda$ with $\varepsilon_\lambda=4M_\lambda\varrho^{-d_\lambda}/(\varrho-1)$ when $a_\lambda$ is analytic and bounded by $M_\lambda$ on a neighborhood of the Bernstein ellipse $E_\varrho$~\cite[Theorem~8.2]{Trefethen2013}. Define $\mathcal A_K[p]$ as the matrix formed using the polynomial coefficients $p_\lambda$. In all sums over $\lambda$, include only indices with $\lambda<q$ and $a_\lambda\not\equiv 0$. For each such $\lambda$, define
\begin{equation}
    \rho_\lambda(k_0)\coloneqq\max_{k\geq\max\{k_0,0\}}d_k^{(\lambda)}R_{kk}.
\end{equation}
Then, $\rho_\lambda(0)=\rho_\lambda$ and $\rho_\lambda(k_0)=\prod_{j=\lambda}^{q-1}(k_0^2-j^2)^{-1/2}$ for $k_0\geq q$.

Choose a cutoff $K_1$ with $K_1\ge q$, fixed independently of $K$. Increasing $K_1$ makes the omitted couplings smaller, but also increases the computation needed for the part below the cutoff. For indices $k\geq K_1$, the lower-order terms have norm at most the small quantity $\Delta$ defined below, so $\mathcal A_\infty$ is close to the identity there on this tail of the indices. The main computation only needs to focus on the leading block. Split the index set into two parts: $k<K_1$ and $k\ge K_1$. In this split, $\mathcal A_{11}, \mathcal A_{12}, \mathcal A_{21}, \mathcal A_{22}$ denote the corresponding four blocks of an operator $\mathcal A$. Define
\begin{equation}
    K_2\coloneqq K_1+q+\max\{0,\max_\lambda(d_\lambda-\lambda)\}
\end{equation}
and $K_3\coloneqq2K_2-K_1$. The columns with indices below $K_1$ can couple only to indices below $K_2$, except for the boundary rows. At resolution $K_3$, these couplings are not distorted by truncating the matrix. Both claims are established by the proof below.

From $\mathcal A_{K_3}[p]$, extract three blocks
\begin{equation}
\begin{aligned}
    G:& [0,K_1)\times[0,K_1),\\
  B_{12}:& [0,K_1)\times[K_1,K_2),\\
  B_{21}:& [K_1,K_2)\times[0,K_1).
\end{aligned}
\end{equation}
In other words, $G$ is the leading square block, while $B_{12}$ and $B_{21}$ describe the couplings between the leading indices and the next band of indices. Any remaining coupling is bounded using the tail norm $r_{\mathrm{tail}}^2 \coloneqq \sum_{i<q} \norm{(I-\Pi_{<K_2})\boldsymbol r_i}_2^2$ and the tail error bound $\Delta\coloneqq\sum_\lambda(\bar a_\lambda+\varepsilon_\lambda)\,\rho_\lambda(K_1)$. The tail norm measures how much of the boundary rows $\boldsymbol r_i$ lies beyond the cutoff $K_2$, while $\Delta$ bounds the contribution of the lower-order terms on the tail $k\ge K_1$. Using these bounds, we define
\begin{equation}
    b_{12}\coloneqq \frac{(\norm{B_{12}}_2^2 + r_{\mathrm{tail}}^2)^{1/2}}{1-\Delta},
    \qquad
    b_{21}\coloneqq\frac{\norm{B_{21}}_2}{1-\Delta},
    \label{eq:us-cert-quantities}
\end{equation}
and $\underline\sigma\coloneqq\sigma_{\min}(G-B_{12}B_{21})-\Delta(1-\Delta)\,b_{12}b_{21}$.

\begin{lemma}[Certified inverse bound]
\label{lem:us-certificate}
Assume that every rescaled boundary row lies in $\ell^2(\mathbb N)$, that $\Delta<1$, and that $\underline\sigma>0$. Define
\begin{equation}
    \kappa_p\coloneqq \frac1{\underline\sigma}\left\lVert\begin{bmatrix}1&b_{12}\\b_{21}&\underline\sigma/(1-\Delta)+b_{12}b_{21}\end{bmatrix}\right\rVert_2
\end{equation}
and $e\coloneqq\sum_\lambda\varepsilon_\lambda\rho_\lambda$. If $e\kappa_p<1$, then $\mathcal A_\infty$ is invertible. Moreover, for every $K\geq K_3$,
\begin{equation}
    \norm{\mathcal A_K^{-1}}_2\leq\kappa_{\mathrm{cert}}\coloneqq\frac{\kappa_p}{1-e\kappa_p}.
    \label{eq:us-certificate}
\end{equation}
The same bound holds for $\norm{\mathcal A_\infty^{-1}}_2$. In particular, $\bar\kappa\coloneqq\alpha\kappa_{\mathrm{cert}}$ satisfies the assumption of Proposition~\ref{prop:us-inverse} for every $K\geq K_3$.
\end{lemma}

\begin{proof}
The proof has two parts. We first bound the inverse for the operators built from the polynomials $p_\lambda$, and only in the last step do we account for the difference between $a_\lambda$ and $p_\lambda$. Let $\mathcal A_\infty[p]$ denote the operator $\mathcal A_\infty=I+\mathcal K$ defined in the first step of the proof in Appendix~\ref{app:us-conditioning}, with every $a_\lambda$ replaced by $p_\lambda$. Until the last step, $\mathcal A$ stands for either $\mathcal A_K[p]\oplus I$ with some $K\geq K_3$ or $\mathcal A_\infty[p]$. For now, we omit the argument bracket for brevity. The operator $\mathcal A_\infty$ is bounded on $\ell^2$, because the rescaled boundary rows lie in $\ell^2$ by assumption and the lower-order terms have norm at most $\sum_\lambda\norm{p_\lambda}_\infty\rho_\lambda$. We can treat the two cases together, since the argument below uses only entries that the two operators share and norm bounds that hold for both. For the finite-dimensional case, suppose $\mathcal A_K\oplus I$ is invertible. Since it is block diagonal,
$(\mathcal A_K\oplus I)^{-1}=\mathcal A_K^{-1}\oplus I$,
and hence $\|(\mathcal A_K\oplus I)^{-1}\|_2\ge \|\mathcal A_K^{-1}\|_2$. Thus, it suffices to bound $\|(\mathcal A_K\oplus I)^{-1}\|_2$.

As discussed before the lemma, we split the labels into $k<K_1$ and $k\geq K_1$ and write
\begin{equation}
    \mathcal A=\begin{bmatrix}\mathcal A_{11}&\mathcal A_{12}\\ \mathcal A_{21}&\mathcal A_{22}\end{bmatrix}.
\end{equation}
We will show that $\mathcal A_{22}$ is close to the identity and that the other three blocks are given by the computed matrices $G$, $B_{12}$, and $B_{21}$, up to the tails of the boundary rows. Eliminating $\mathcal A_{22}$ then leaves a $K_1\times K_1$ Schur complement, whose smallest singular value can be bounded from below by the computed quantities.

\emph{Degree count.} We first determine which entries of $\mathcal A_\infty$ can be nonzero. In $\mathcal A_\infty$, the term of order $\lambda$ maps column $k$ first to degree $k-\lambda$ through $\widehat D_{\lambda}^{\infty}$. By Eq.~\eqref{eq:us-conversion}, each of the $q-\lambda$ conversion steps either preserves the degree or lowers it by two, while multiplication by $p_\lambda$ changes the degree by at most $d_\lambda$. The resulting degree $i$ therefore lies between $k-\lambda-d_\lambda-2(q-\lambda)$ and $k-\lambda+d_\lambda$. Since interior degree $i$ is stored at label $q+i$, the term can map column $k$ only to labels $\ell$ with
\begin{equation}
    k-(q+d_\lambda-\lambda)\leq\ell\leq k+(q+d_\lambda-\lambda).
    \label{eq:us-cert-band}
\end{equation}

The cutoff $K_2$ was chosen to cover this band. By its definition, every $\lambda$ satisfies $q+d_\lambda-\lambda\leq K_2-K_1$. Each lower-order term therefore maps column $k$ to labels that differ from $k$ by at most $K_2-K_1$, and so does the identity part $I-\Pi_{<q}$, which maps a column $k\geq q$ to the label $k$ itself. For a column $k<K_1$, this gives $\ell\leq K_1-1+(K_2-K_1)=K_2-1$. For a column $k\geq K_2$, it gives $\ell\geq K_2-(K_2-K_1)=K_1$. The boundary rows are the only exception, since they can have nonzero entries in every column. Their labels $0,\ldots,q-1$ all lie below $K_1$, because $K_1\geq q$. In summary, apart from the boundary rows, the columns $k<K_1$ reach only labels below $K_2$, and the columns $k\geq K_2$ reach only labels of at least $K_1$.

\emph{Finite resolution.} We now compare $\mathcal A_K\oplus I$, for a fixed $K\geq K_3$, with $\mathcal A_\infty$. As in the proof of Proposition~\ref{prop:us-conditioning}, we route each lower-order term of $\mathcal A_\infty$ through the parameter-$\mu^\star$ basis, using $\widehat S^\infty_{\lambda\to q}M^\infty_\lambda[p_\lambda]=\widehat S^\infty_{\mu^\star\to q}M^\infty_{\mu^\star}[p_\lambda]\widehat S^\infty_{\lambda\to\mu^\star}$. The finite and infinite terms therefore use the same factors, except that the finite derivative and conversion operators are truncated. In particular, $\widehat D_\lambda R=\widehat D^\infty_\lambda R\,\Pi_{<K}$, and the finite conversions retain only the first $N_{\mu^\star}$ or $N_q$ coefficients. As a result, the two operators can differ for only three reasons. First, for columns $k\geq K$, $\mathcal A_K \oplus I$ has only the diagonal entry $1$, whereas $\mathcal A_\infty[p]$ also has boundary-row and lower-order entries. Second, the finite conversion $\widehat S_{\mu^\star\to q}$ discards the rows $\ell\geq K$ in $\mathcal A_\infty$, while $\mathcal A_K\oplus I$ has identity rows there. Third, within the leading $K\times K$ block, the finite operator uses the Gauss multiplication matrix $M_{\mu^\star}[p_\lambda]$ on the first $N_{\mu^\star}$ coefficients rather than the exact infinite multiplication operator.

The difference in the third case is $\Delta M_\lambda=(M_{\mu^\star}[p_\lambda]\oplus0)-M^\infty_{\mu^\star}[p_\lambda]$ from the proof of Proposition~\ref{prop:us-conditioning}. This also accounts for the truncation of $\widehat S_{\lambda\to\mu^\star}$ to $N_{\mu^\star}$ coefficients. The difference vanishes on inputs of degree below $N_{\mu^\star}-d_\lambda$: multiplying such an input by $p_\lambda$ keeps the result below degree $N_{\mu^\star}$, so no terms are truncated and the Gauss rule is exact. Since both multiplication matrices are symmetric, $\Delta M_\lambda$ also produces no outputs below that degree threshold.

Using the band in Eq.~\eqref{eq:us-cert-band}, together with $q+d_\lambda-\lambda\leq K_2-K_1$, we can locate where the differences can occur. In the first case, the column satisfies $k\geq K$, so the lower-order entries lie in the rows $\ell\geq k-(K_2-K_1)\geq K-(K_2-K_1)$. In the second case, the row satisfies $\ell\geq K$, and the relevant columns are $k\geq\ell-(K_2-K_1)\geq K-(K_2-K_1)$. In the third case, the affected input and output labels satisfy $k\ge K-\mu^\star-d_\lambda+\lambda$ and $\ell\ge K-q+\mu^\star-d_\lambda$. Since $\lambda\le\mu^\star<q$, both bounds are at least $K-(q+d_\lambda-\lambda)\ge K-(K_2-K_1)$.

Thus, in every case, any changed entry outside the boundary rows has both its row and column labels at least $K-(K_2-K_1)$. This quantity is at least $K_2$, because $K\ge K_3=2K_2-K_1$. Therefore, $\mathcal A_K\oplus I$ and $\mathcal A_\infty$ agree in all rows and columns with labels below $K_2$, except that the finite boundary rows stop at column $K-1$, whereas the infinite boundary rows continue beyond $K$. This exception concerns only columns $k\ge K>K_2$. In particular, the blocks $G$, $B_{12}$, and $B_{21}$ are the same in $\mathcal A_{K_3}$, in $\mathcal A_K\oplus I$ for every $K\ge K_3$, and in $\mathcal A_\infty$.

\emph{Off-diagonal blocks.} We can now examine the three blocks of $\mathcal A$ that involve the first $K_1$ labels. By the previous step, these blocks are identical for $\mathcal A_K \oplus I$ and $\mathcal A_\infty$, except for the boundary-row entries in the columns $k\geq K$. 

The leading block is $\mathcal A_{11}=G$. The block $\mathcal A_{21}$ contains rows $\ell\geq K_1$ and columns $k<K_1$. Its rows with $K_1\leq\ell<K_2$ form $B_{21}$; all rows with $\ell\geq K_2$ are zero by the degree count. Thus $\mathcal A_{21}=\begin{bmatrix}B_{21}\\0\end{bmatrix}$. The block $\mathcal A_{12}$ consists of rows $\ell<K_1$ and columns $k\geq K_1$. Its columns with $K_1\le k<K_2$ form $B_{12}$. For the remaining columns, $k\ge K_2$, the degree count shows that only the boundary rows can be nonzero. Call this remaining part $T$. Then $\mathcal A_{12}=\begin{bmatrix}B_{12}&T\end{bmatrix}$. The only nonzero rows of $T$ are the $q$ boundary rows, whose entries are $r_{ik}$ for $k\ge K_2$. In $\mathcal A_K\oplus I$, entries with $k\ge K$ are set to zero. In either case, $\|T\|_2$ is at most its Frobenius norm, which is at most $r_{\mathrm{tail}}$.

These block forms imply three useful facts. First, $T$ multiplies the zero rows of $\mathcal A_{21}$, so $\mathcal A_{12} \mathcal A_{21} = B_{12} B_{21}$. Second, write a vector $\boldsymbol v$ in parts matching the columns of $\mathcal A_{12}$, as $\boldsymbol v=(\boldsymbol v_a, \boldsymbol v_b)$. By the triangle inequality and Cauchy--Schwarz,
\begin{equation}
    \begin{aligned}
        \norm{B_{12}\boldsymbol v_a+T\boldsymbol v_b}_2
        &\leq\norm{B_{12}}_2\norm{\boldsymbol v_a}_2+r_{\mathrm{tail}}\norm{\boldsymbol v_b}_2\\
        &\leq\bigl(\norm{B_{12}}_2^2+r_{\mathrm{tail}}^2\bigr)^{1/2}\norm{\boldsymbol v}_2.
    \end{aligned}
\end{equation}
By the definition of $b_{12}$ in Eq.~\eqref{eq:us-cert-quantities}, this means $\norm{\mathcal A_{12}}_2\leq(1-\Delta)b_{12}$. Third, the zero rows of $\mathcal A_{21}$ do not contribute to its norm, so $\norm{\mathcal A_{21}}_2=\norm{B_{21}}_2=(1-\Delta)b_{21}$.

\emph{Diagonal block.} 
The block $\mathcal{A}_{22}$ contains only rows and columns whose labels are at least $K_1$. Because $K_1\ge q$, the boundary-condition rows and the boundary projection do not affect this block, so its identity contribution is exactly $I$. Therefore, $\mathcal{A}_{22}-I$ consists only of the lower-order differential terms.

For each term of order $\lambda$, the rescaled derivative has norm at most $\rho_\lambda(K_1)$ on columns with labels $k\ge K_1$. The basis conversions are contractions, so they do not increase this bound. The multiplication operator has norm at most $\|p_\lambda\|_\infty$, which is no greater than $\bar a_\lambda+\varepsilon_\lambda$. Shifting labels and truncating the matrix also cannot increase the norm. Adding the bounds for all lower-order terms gives
\begin{equation}
    \norm{\mathcal A_{22}-I}_2\leq\sum_\lambda(\bar a_\lambda+\varepsilon_\lambda)\,\rho_\lambda(K_1)=\Delta<1.
\end{equation}
Thus $\mathcal{A}_{22}$ is the identity plus a perturbation of norm less than one, so it is invertible by the convergent Neumann series. 
\begin{equation}
    \mathcal A_{22}^{-1}=\sum_{j\geq0}(I-\mathcal A_{22})^j=I+\sum_{j\geq1}(I-\mathcal A_{22})^j.
\end{equation}
Moreover, since $\norm{(I - \mathcal{A}_{22})^j}_2 \leq \Delta^j$, it follows that $\norm{\mathcal{A}_{22}^{-1}}_2 \leq 1/(1-\Delta)$ and $\norm{\mathcal A_{22}^{-1}-I}_2\leq\Delta/(1-\Delta)$.

\emph{Schur complement.} The bounds above let us eliminate the block $\mathcal A_{22}$ using the Schur complement. Define $X\coloneqq\mathcal A_{12}\mathcal A_{22}^{-1}$ and $Y\coloneqq\mathcal A_{22}^{-1}\mathcal A_{21}$. Then $\norm X_2\leq\norm{\mathcal A_{12}}_2\norm{\mathcal A_{22}^{-1}}_2\leq(1-\Delta)b_{12}/(1-\Delta)=b_{12}$ and, in the same way, $\norm Y_2\leq b_{21}$. The Schur complement of $\mathcal A_{22}$ is 
\begin{equation}
    S\coloneqq \mathcal A_{11}-\mathcal A_{12}\mathcal A_{22}^{-1}\mathcal A_{21}.
\end{equation}
Using $\mathcal A_{11}=G$, $\mathcal A_{12}\mathcal A_{21}=B_{12}B_{21}$, and writing $\mathcal A_{22}^{-1}=I+(\mathcal A_{22}^{-1}-I)$, we obtain
\begin{equation}
    S=G-B_{12}B_{21}-\mathcal A_{12}(\mathcal A_{22}^{-1}-I)\mathcal A_{21}.
\end{equation}
The final term has norm at most $\norm{\mathcal A_{12}}_2\norm{\mathcal A_{22}^{-1}-I}_2\norm{\mathcal A_{21}}_2 \leq \Delta(1-\Delta)b_{12}b_{21}$. By Weyl's inequality, adding a matrix of norm at most $\delta$ changes each singular value by at most $\delta$. Therefore,
\begin{equation}
    \begin{aligned}
        \sigma_{\min}(S)&\geq\sigma_{\min}(G-B_{12}B_{21})-\Delta(1-\Delta)b_{12}b_{21}\\
        &=\underline\sigma>0.
    \end{aligned}
\end{equation}
In particular, $S$ is invertible, and $\norm{S^{-1}}_2=1/\sigma_{\min}(S)\leq1/\underline\sigma$. Since $\mathcal A_{22}$ and $S$ are both invertible, so is $\mathcal A$. Indeed,
\begin{equation}
    \mathcal A=\begin{bmatrix}I&X\\0&I\end{bmatrix}\begin{bmatrix}S&0\\0&\mathcal A_{22}\end{bmatrix}\begin{bmatrix}I&0\\Y&I\end{bmatrix},
\end{equation}
and hence
\begin{equation}
    \mathcal A^{-1}=\begin{bmatrix}S^{-1}&-S^{-1}X\\-YS^{-1}&\mathcal A_{22}^{-1}+YS^{-1}X\end{bmatrix}.
\end{equation}
The four blocks of $\mathcal A^{-1}$ satisfy
\begin{equation}
\begin{aligned}
\|S^{-1}\|_2\le\frac1{\underline\sigma}, \quad
\|S^{-1}X\|_2\le\frac{b_{12}}{\underline\sigma}, \quad 
\|YS^{-1}\|_2\le\frac{b_{21}}{\underline\sigma},
\end{aligned}
\end{equation}
and $\norm{\mathcal A_{22}^{-1}+YS^{-1}X}_2\leq1/(1-\Delta)+b_{12}b_{21}/\underline\sigma$. Using these bounds, we can show that, for any block vector $\boldsymbol v=(\boldsymbol v_1,\boldsymbol v_2)$, $\norm{\mathcal A^{-1}\boldsymbol v}_2\leq\kappa_p\norm{\boldsymbol v}_2$, where $\kappa_p$ is the constant defined from the corresponding $2\times2$ matrix in the lemma's statement. Consequently, $\norm{\mathcal A^{-1}}_2\leq\kappa_p$. This establishes the inverse-norm bound when the coefficients are polynomials.

\emph{General coefficients.} Replacing the polynomial approximations $p_\lambda$ with the true coefficients $a_\lambda$ changes the operator only in its multiplication steps. Since multiplication is linear in the coefficient, the difference $\mathcal R=\mathcal{A}[a]-\mathcal{A}[p]$ is the sum of the lower-order terms in which $a_\lambda$ is replaced by $a_\lambda-p_\lambda$. Each such term has norm at most $\varepsilon_\lambda\rho_\lambda$: multiplication by the coefficient difference has norm at most $\varepsilon_\lambda$, the basis conversions are contractions, and the rescaled derivative has norm at most $\rho_\lambda$. Therefore, $\norm{\mathcal R}_2\leq\sum_\lambda\varepsilon_\lambda\rho_\lambda=e$.

We now write $\mathcal A[a]=\mathcal A[p]\,(I+\mathcal A[p]^{-1}\mathcal R)$. The second factor is a perturbation of the identity whose difference from the identity has norm at most $\kappa_p e$. Because $e\kappa_p<1$, the Neumann-series theorem guarantees that this factor is invertible, and $\norm{(I+\mathcal A[p]^{-1}\mathcal R)^{-1}}_2 \leq 1/(1- e\kappa_p)$.  It follows that $\mathcal{A}[a]$ is invertible and $ \norm{\mathcal{A}[a]^{-1}}_2 \leq \frac{\kappa_p}{1-e\kappa_p} =\kappa_{\mathrm{cert}}$. This result applies both to $\mathcal{A}_\infty$ and to $\mathcal{A}_K$ for every $K\geq K_3$. Therefore, $\bar{\kappa}=\alpha\kappa_{\mathrm{cert}}$ is a valid upper bound on $\alpha\|\mathcal{A}_K^{-1}\|_2$ for all $K\geq K_3$, which is assumed by Proposition~\ref{prop:us-inverse}.
\end{proof}

The certification procedure verifies three conditions in Lemma~\ref{lem:us-certificate} using quantities computed at a fixed matrix size $K_3$: $\Delta<1$, $\underline\sigma>0$, and $e\kappa_p<1$. For polynomial coefficients, $e=0$, so the last condition holds automatically. We assemble $\mathcal A_{K_3}[p]$, extract the blocks $G$, $B_{12}$, and $B_{21}$, compute the norms $\norm{B_{12}}_2$ and $\norm{B_{21}}_2$, and bound the tail terms $r_{\mathrm{tail}}$ and $\Delta$. We then use a singular value decomposition of the $K_1\times K_1$ matrix $G-B_{12}B_{21}$ to find its smallest singular value. These calculations require $O(K_3^3)$ classical operations.

Rounding errors do not undermine the certification: their effects can be bounded and added to the computed singular value and norms using Weyl's inequality and the triangle inequality. The test requires no prior assumption that the differential equation has a unique solution; passing the test itself proves that the limiting operator $\mathcal A_\infty$ is invertible. Conversely, if $\mathcal A_\infty$ is invertible, the test succeeds for sufficiently large $K_1$, provided the coefficient-approximation error $e$ is small enough.

The test works because the Schur complement of $\mathcal A_\infty$ has smallest singular value at least $1/\norm{\mathcal A_\infty^{-1}}_2$, and its difference from $G-B_{12}B_{21}$ can be bounded. As $K_1$ increases, the relevant tail and block quantities shrink, making the certification more likely to succeed. Table~\ref{tab:us-certificate} reports the certified bounds for the four examples in Fig.~\ref{fig:us-checks}(a). Using $K_1=32$, the certified bounds are at most $2.1\%$ above the largest directly computed values of $\alpha \norm{\mathcal{A}_K^{-1}}_2$ for $K_3\le K\le4096$. The bounds are less tight for the first- and third-order examples because their boundary rows decay relatively slowly, like $1/k$.

\begin{table}[t]
\centering
\caption{\label{tab:us-certificate}Certified bounds $\bar\kappa=\alpha\kappa_{\mathrm{cert}}$ from Lemma~\ref{lem:us-certificate} for the four problems of Fig.~\ref{fig:us-checks}(a). The last column gives the largest computed value of $\alpha\norm{\mathcal A_K^{-1}}_2$ for $K_3\leq K\leq4096$.}
\setlength{\tabcolsep}{6pt}
\begin{tabular}{@{}rrrrrrr@{}}
\toprule
 & & \multicolumn{2}{c}{$K_1=16$} & \multicolumn{2}{c}{$K_1=32$} & \\
\cmidrule(lr){3-4}\cmidrule(lr){5-6}
$q$ & $\alpha$ & $K_3$ & $\bar\kappa$ & $K_3$ & $\bar\kappa$ & Computed\\
\midrule
1 & $8.17$ & $20$ & $116.0$ & $36$ & $107.0$ & $105.2$\\
2 & $11.27$ & $52$ & $14.94$ & $68$ & $14.90$ & $14.89$\\
3 & $24.44$ & $54$ & $37.21$ & $70$ & $36.19$ & $35.47$\\
4 & $82.48$ & $56$ & $166.49$ & $72$ & $166.48$ & $166.48$\\
\bottomrule
\end{tabular}
\end{table}

\subsection{Natural conditions and block systems}
\label{app:us-natural}

High-order endpoint conditions can make the discretized matrix $\mathcal A_K$ ill-conditioned. In particular, if $\mathcal Bu=u^{(j)}(\pm1)$ with $q/2\le j<q$, Proposition~\ref{prop:us-bc} implies that the associated rescaled boundary row grows in norm as $\norm{\Pi_{<K}\boldsymbol r}_2=\Theta\!\left(K^{2j-q+1/2}\right)$. Thus, $\|\mathcal A_K\|_2$ grows at least at this rate. At the same time, a fixed column, such as $\mathcal A_K\boldsymbol e_q$, has norm bounded independently of $K$, so $\norm{\mathcal A_K^{-1}}_2 \geq 1/\norm{\mathcal A_K\boldsymbol e_q}_2$ is bounded below. It follows that the condition number grows at least as $K^{2j-q+1/2}$. For example, in the Neumann example in Fig.~\ref{fig:us-checks}, the condition number grows approximately linearly.

A way to avoid this growth is to rewrite the equation as a first-order system. Introduce $U_j=u^{(j)}$ for $j<q$, so that $U_j'-U_{j+1}=0$ for $j<q-1$. The reformulated equation is $U_{q-1}'+\sum_\lambda a_\lambda U_\lambda=f$. Each component can then be discretized using the first-order operator, with the derivative matrix $\widehat D_1$, the rescaling $R^{(1)}$, where $R^{(1)}_{00}=1$ and $R^{(1)}_{kk}=1/k$ for $k\ge1$, and the lower-order terms $\widehat S_{0\to1}M_0[a_{\lambda}]$. The coupling term $-U_{j+1}$ is represented by $M_0[-1]=-I$.

Under this reformulation, each boundary condition becomes a point value, an interior value, an integral of a single component, or a linear combination of such quantities. The rescaled boundary rows therefore satisfy Proposition~\ref{prop:us-bc} for a first-order problem. The coupling terms are compact because their columns decay as $1/k$, and uniqueness of the reformulated system is equivalent to uniqueness of the original problem. Consequently, the argument in Proposition~\ref{prop:us-conditioning} gives uniform conditioning for the resulting $qK\times qK$ system, which uses only $Q^{(0)}_K$ and $Q^{(1)}_{N_1}$.

If $p\ge2$ divides $q$, one can instead use a block formulation with fewer unknowns. Set $U_l=u^{(lp)}$ for $l<q/p$. Then $U_l^{(p)}=U_{l+1}$ for $l<q/p-1$, and the equation is
\begin{equation}
    U_{q/p-1}^{(p)}+\sum_\lambda a_\lambda\,U_{\lfloor\lambda/p\rfloor}^{(\lambda\bmod p)}=f.
\end{equation}
Each block uses the parameter-$p$ discretization and its corresponding rescaling. A boundary condition on $u^{(j)}(\pm1)$, with $j=lp+i$ and $0\le i<p$, becomes a condition on $U_l^{(i)}(\pm1)$; it satisfies the conditioning criterion when $i<p/2$. For example, in the simply supported beam problem, choosing $p=2$ turns the conditions on $u$ and $u''$ into Dirichlet conditions on $U_0$ and $U_1$. Both blocks use $Q^{(2)}_{N_2}$, and the reformulation avoids the poor conditioning of the direct high-order discretization.

\subsection{Block encoding and the preconditioned inverse}
\label{app:us-encoding}
We prepare the boundary-row state $V_i$ using a dyadic envelope. Let $r=r_i$ be a nonzero rescaled boundary row, with entries $r_k=R_{kk}\mathcal B_i(p_k^{(0)})$. Suppose that, for every $k\ge k_0$, $|r_k|\leq Ck^{-\beta}$ for some constants $C>0$, $\beta>\tfrac12$, and $k_0=2^{s_0}$ that do not depend on $K$. Define an envelope $g_k$ by setting $g_k=|r_k|$ for the finitely many indices $k<k_0$, and setting $g_k=C2^{-s\beta}$ on each dyadic interval $2^s\le k<2^{s+1}$, for $s_0\le s<n$, where $K=2^n$. Because $k\ge 2^s$ on that interval, $g_k\ge |r_k|$. Moreover, the squared norm of the envelope is bounded uniformly in $K$:
\begin{equation}
    \gamma_K^2\coloneqq \sum_{k<K}g_k^2 \leq  \sum_{k<k_0}|r_k|^2+\frac{C^2\,2^{s_0(1-2\beta)}}{1-2^{1-2\beta}} := \gamma_\infty^2.
    \label{eq:us-envelope-normalization}
\end{equation}
The condition $\beta>\tfrac12$ ensures that the dyadic contributions form a convergent geometric series. If $K<k_0$, the state can instead be prepared directly, since there are only finitely many such cases.

An envelope state can be prepared without storing all $K$ entries in a table. Prepare a selector over the exceptional labels and dyadic intervals, assigning probabilities proportional to their contributions to the squared envelope norm. For an interval $2^s \le k < 2^{s+1}$, set bit $s$ of the degree register and apply Hadamard gates to the $s$ lower bits to create a uniform superposition over that interval. The degree label identifies its interval through its most significant nonzero bit, allowing the selector to be uncomputed; the exceptional labels are handled separately. This prepares the degree register in the state $\gamma_K^{-1}\sum_{k<K} g_k|k\rangle$. A flag rotation with signed amplitude $r_k/g_k$, computed and uncomputed by reversible arithmetic and defined to be zero when $g_k=0$, then gives a unitary $U_{\mathrm{env}}$ such that
\begin{equation}
    U_{\mathrm{env}}\ket{0^{a_r}}\ket0 
    = \ket{0^{a_r}}\frac1{\gamma_K}\sum_{k<K}r_k \ket k + \ket{\mathrm{fail}},
    \label{eq:us-envelope-preparation}
\end{equation}
where $|\mathrm{fail}\rangle$ has no component with all auxiliary registers in the zero state. If the selector amplitudes and ratios $r_k/g_k$ can be computed efficiently, $U_{\mathrm{env}}$ can be implemented to error $\eta$ with gate cost polynomial in $n$ and $\log(1/\eta)$, since the selector has only $O(n+k_0)$ possible values.

The boundary term can be block-encoded using $U_{\mathrm{env}}$, first without amplifying its success probability. Let $
E=(\langle 0^{a_r}|\otimes I)U_{\mathrm{env}}(|0^{a_r}\rangle\otimes I)$, so that $E|0\rangle=\Pi_{<K}\boldsymbol r/\gamma_K$. A comparator block-encodes $J_0=|0\rangle\langle 0|$, while Pauli gates $X_i$ map the label $0$ to the boundary-condition label $i$. Applying $U_{\mathrm{env}}^\dagger$, the comparator, and $X_i$, each with its own flag register, block-encodes
\begin{equation}
X_iJ_0E^\dagger = \frac{|i\rangle\langle\Pi_{<K}\boldsymbol r|}{\gamma_K}.
\end{equation}
Thus, the $i$th boundary term can be included in the LCU using weight $\gamma_K$, which is bounded above by the $K$-independent constant $\gamma_\infty$. Alternatively, a further flag rotation with amplitude $\gamma_K/\gamma_\infty$ gives the uniform weight $\gamma_\infty$.

For an endpoint condition $\mathcal B_i u=u^{(j)}(\pm 1)$ with $j<q/2$, Eq.~\eqref{eq:us-row-entries} gives an envelope bound with $\beta=q-2j$ and $C=c_+$. We choose $k_0$ to be the smallest power of two satisfying $k_0\geq q$. The finitely many exceptional entries are evaluated directly from $R_{kk}\partial_x^j p_k^{(0)}(\pm 1)$, while the remaining entries are explicit algebraic functions of $k$. At the left endpoint, the sign of the rotation also includes the factor $(-1)^{k+j}$. The envelope has a bounded norm because $1-2\beta=1+4j-2q<0$, which, for integer $j$ and $q$, is equivalent to $j<q/2$. Since $r_j\neq 0$ and $j<q<K$, Eq.~\eqref{eq:us-envelope-preparation} gives the uniform lower bound
\begin{equation}
    \frac{\norm{\Pi_{<K}\boldsymbol r}_2^2}{\gamma_K^2}\geq\frac{|r_j|^2}{\gamma_\infty^2}>0.
    \label{eq:us-envelope-success}
\end{equation}

To prepare the normalized row $\Pi_{<K} \boldsymbol r/\norm{\Pi_{<K} \boldsymbol r}_2$, we first amplify the successful branch of $U_{\mathrm{env}}$. Because the success probability is known, a padding rotation and a constant number of amplitude-amplification rounds suffice. A final flag rotation, with amplitude $\norm{\Pi_{<K} \boldsymbol r}_2 / \norm{\boldsymbol r}_2$, then produces the preparation circuit $V_i$.

For fixed $q$ and $j$, the squared tail entries are rational functions of $k$. Partial-fraction decomposition reduces their finite and infinite sums to telescoping sums and second-order harmonic sums. Thus, the required norms and padding angles can be computed classically without enumerating all $K$ entries, and the endpoint-state preparation has cost $T_{\boldsymbol r_i}(\eta)=\mathrm{poly}_q(n,\log(1/\eta))$. The same approach applies to other boundary functionals when an efficiently evaluable envelope and ratio are available. However, square summability alone does not guarantee an efficient circuit; in general, we assume that a circuit for $V_i$ is given and denote its cost by $T_{\boldsymbol r_i}$.

\begin{proof}[Proof of Proposition~\ref{prop:us-inverse}]
\emph{Linear combination.} We construct the block encoding using the LCU method from Section~\ref{sec:block-encoding}. A selector register prepares a superposition whose amplitudes are proportional to the square roots of the term weights: $(1/\alpha)^{1/2}$, $(\norm{\boldsymbol r_i}/\alpha)^{1/2}$, and $[(3/2)^{q-\lambda}\rho_\lambda\bar a_\lambda/\alpha]^{1/2}$. The selector includes the identity term, one term for each boundary row, and one term for each nonzero lower-order coefficient $a_\lambda$. Thus, there are $1+2q$ possible terms before omitting any terms with $a_\lambda\equiv 0$. Each selected circuit implements its corresponding term divided by its assigned weight, with subnormalization one. The combined circuit therefore block-encodes $\mathcal A_K/\alpha$.

\emph{Components.} For the identity term, we use a comparator flag to check whether the label satisfies $k\ge q$. This implements the identity contribution exactly on the required labels. For a boundary term $i$, the circuit applies $V_i^\dagger$, selects the zero label and the flag marking the successful branch of $V_i$, and then maps label $0$ to label $i$ using Pauli-$X$ gates. Its encoded block is $\ket i\bra{\Pi_{<K}\boldsymbol r_i}/\norm{\boldsymbol r_i}$. 

For the lower-order term $\lambda$, the derivative, multiplication, and basis-conversion block encodings are composed using separate flags. Their subnormalization factors multiply, while their approximation errors contribute additively. The controlled rotations used for $\widehat D_\lambda R/\rho_\lambda$ and for the diagonal factors in each conversion step have amplitudes that are algebraic functions of the label. Reversible arithmetic on $O(n)$-bit integers can calculate these amplitudes to $b$ bits of precision using $\operatorname{poly}(n,b)$ gates. Each conversion step is implemented as $\diag(s_k^{(\nu)})-\tfrac12\mathrm{ADD}_{-2}\diag(2t_k^{(\nu)})$ with subnormalization $3/2$. Comparators restrict the result to the first $N_{\mu^\star}$ or $N_q$ labels, as appropriate.

\emph{Shared transforms.} The transform pair $Q_{N_{\mu^\star}}^{(\mu^\star)}$ and $Q_{N_{\mu^\star}}^{(\mu^\star)\dagger}$ is implemented using Theorem~\ref{thm:efficient-qut} when $\mu^\star\ge 2$, and using the DCT-II or DST-I circuits~\cite{Klappenecker2001} when $\mu^\star\le 1$. The same pair is needed for every lower-order term, so it is applied only once before and once after the $\lambda$-controlled diagonal operations. A flag identifies the selector values corresponding to lower-order terms. Comparators around the transform pair restrict its action to valid labels $k<N_{\mu^\star}$; when $\mu^\star\ge 2$, this is precisely the subspace on which Theorem~\ref{thm:efficient-qut} guarantees the stated error bound.

\emph{Error and cost.} Choose each component's error to be $O(\delta/q)$, and use arithmetic precision $b=O(\log(q/\delta))$. The combined circuit then block-encodes $\mathcal A_K/\alpha$ with operator-norm error $O(\delta)$, at the cost stated in Eq.~\eqref{eq:us-blockcost}.

\emph{Singular-value inversion.} Let $\mathcal A_K=W\Sigma V^\dagger$ be a singular value decomposition. By assumption, the singular values of $\mathcal{A}_K/\alpha$ lie in $[1/\bar\kappa,1]$. There exists an odd polynomial $P$ of degree~\cite{Gilyen2019,Childs2017} 
\begin{equation}
    L_\eta=O(\bar\kappa\log(\bar\kappa/\eta))
\end{equation}
such that $|P(x)|\le 1$ for $x\in[-1,1]$, and
\begin{equation}
    \left|P(x)-\frac{1}{2\bar\kappa x}\right|\le \frac{\eta}{2},
\qquad x\in[1/\bar\kappa,1].
\end{equation}
Applying quantum singular-value transformation to the adjoint block encoding of $\mathcal{A}_K^\dagger/\alpha$ implements a block encoding of $VP(\Sigma/\alpha)W^\dagger$. This operator approximates $\alpha \mathcal{A}_K^{-1}/(2\bar\kappa)$ to error at most $\eta/2$, using $L_\eta$ calls to the block encoding and $L_\eta+1$ phase rotations.

If each block-encoding call has error $\delta$, the resulting encoded polynomial changes by at most $4L_\eta\sqrt{\delta}$~\cite{Gilyen2019}. Choosing $\delta_\eta=\eta^2/(64L_\eta^2)$ makes this contribution at most $\eta/2$. Together with the polynomial-approximation error, this establishes the claimed inverse block-encoding guarantee.
\end{proof}

\subsection{Discretization error}
\label{app:us-error}

We first establish the discretization estimate in Eq.~\eqref{eq:us-discretization}. We then bound the residual $\boldsymbol\zeta_K$ using the tails of the rescaled boundary rows and the Chebyshev and ultraspherical expansions, obtaining the final bound in Eq.~\eqref{eq:us-residual}. Throughout, $u\in C^q([-1,1])$ is the exact solution of the boundary-value problem, with orthonormal Chebyshev coefficients $u_k^\star$, and $\boldsymbol u^\star=(u_0^\star,\ldots,u_{K-1}^\star)^\T$ is the coefficient vector of its truncation $\mathcal T_Ku$. The discrete solution $u_K=\sum_{k<K}u_kp_k^{(0)}$ defined in Eq.~\eqref{eq:us-expansion} has the coefficient vector $\boldsymbol u=R\boldsymbol v$, where $\boldsymbol v$ solves $\mathcal A_K\boldsymbol v=(\boldsymbol c,\widehat{\boldsymbol f})^\T$.

\emph{Error equation.} Since $\mathcal A_K=A_KR$, the definition of $\boldsymbol\zeta_K$ in Section~\ref{sub:us-scope} reads $\mathcal A_KR^{-1}\boldsymbol u^\star=(\boldsymbol c,\widehat{\boldsymbol f})^\T+\boldsymbol\zeta_K$. Subtracting $\mathcal A_K\boldsymbol v=(\boldsymbol c,\widehat{\boldsymbol f})^\T$ side by side, we obtain $\mathcal A_K(\boldsymbol v-R^{-1}\boldsymbol u^\star)=-\boldsymbol\zeta_K$. Thus, $\boldsymbol v-R^{-1}\boldsymbol u^\star=-\mathcal A_K^{-1}\boldsymbol\zeta_K$. By Eq.~\eqref{eq:us-R}, the entries of $\boldsymbol v-R^{-1}\boldsymbol u^\star=R^{-1}(\boldsymbol u-\boldsymbol u^\star)$ are $u_k-u_k^\star$ for $k<q$ and $d_k^{(q)}(u_k-u_k^\star)$ for $q\leq k<K$. By Eq.~\eqref{eq:us-derivative}, the latter are the coefficients of $u_K^{(q)}-(\mathcal T_Ku)^{(q)}$ in the orthonormal basis $p^{(q)}$ of $L^2(w_q)$, therefore
\begin{equation}
    \begin{aligned}
        \norm{\boldsymbol v-R^{-1}\boldsymbol u^\star}^2={}&\sum_{k<q}|u_k-u_k^\star|^2\\
        &+\norm{u_K^{(q)}-(\mathcal T_Ku)^{(q)}}^2_{L^2(w_q)}.
    \end{aligned}
    \label{eq:us-v-error}
\end{equation}
In particular, $\norm{u_K^{(q)}-(\mathcal T_Ku)^{(q)}}_{L^2(w_q)}\leq\norm{\boldsymbol v-R^{-1}\boldsymbol u^\star}$. The coefficient vector of $u_K-\mathcal T_Ku$ is $\boldsymbol u-\boldsymbol u^\star=R(\boldsymbol v-R^{-1}\boldsymbol u^\star)$. Since $\|R\|_2\leq 1$, its norm is at most $\norm{\boldsymbol v-R^{-1}\boldsymbol u^\star}$, and therefore
\begin{equation}
    \norm{u_K-\mathcal T_Ku}_{L^2(w_0)}\leq \norm{\boldsymbol v-R^{-1}\boldsymbol u^\star}.
\end{equation}
By Eq.~\eqref{eq:us-v-error}, the $L^2(w_q)$-norm of the difference between the $q$th derivatives is also at most $\norm{\boldsymbol v-R^{-1}\boldsymbol u^\star}$. From the system equation, 
\begin{equation}
    \norm{\boldsymbol v-R^{-1}\boldsymbol u^\star}=\norm{\mathcal A_K^{-1}\boldsymbol\zeta_K}\leq\norm{\mathcal A_K^{-1}}\norm{\boldsymbol\zeta_K}.
\end{equation}
Thus, each of the solution error and the $q$th derivative error is bounded by $\norm{\mathcal A_K^{-1}}\norm{\boldsymbol\zeta_K}$. Summing these two bounds proves Eq.~\eqref{eq:us-discretization}.

\emph{Boundary part.} The residual $\boldsymbol\zeta_K=A_K\boldsymbol u^\star-(\boldsymbol c,\widehat{\boldsymbol f})^\T$ measures how far the Chebyshev coefficients of the truncated solution $\mathcal T_Ku$ are from satisfying the discrete system~\eqref{eq:us-system}. Its first $q$ entries correspond to the boundary conditions; the remaining $N_q$ entries correspond to the interior equations. 

For the $i$th boundary condition, the residual entry is
\begin{equation}
    (\boldsymbol\zeta_K)_i=\mathcal B_i(\mathcal T_Ku)-c_i=-\mathcal B_i(u-\mathcal T_Ku)
\end{equation}
because the exact solution satisfies $\mathcal{B}_i u = c_i$. Assume, as in Proposition~\ref{prop:us-conditioning}, that each rescaled boundary row $\boldsymbol r_i$ belongs to $\ell^2$. Then, for any $g\in C^q([-1,1])$ with orthonormal Chebyshev coefficient vector $\widehat{\boldsymbol{g}}$,
\begin{equation}
    \mathcal B_ig=\langle\boldsymbol r_i,R^{-1}\widehat{\boldsymbol g}\rangle.
\end{equation}
Apply this identity to $g=u-\mathcal T_Ku$. Its coefficients vanish for $k<K$ and equal $u_k^\star$ for $k\ge K$. Defining $v_k^\star=u_k^\star/R_{kk}$, we obtain
\begin{equation}
    (\boldsymbol\zeta_K)_i=-\sum_{k\geq K}r_{ik}v^\star_k.
\end{equation}
By the Cauchy--Schwarz inequality, $|(\boldsymbol\zeta_K)_i|\leq\norm{(I-\Pi_{<K})\boldsymbol r_i}\,\norm{(I-\Pi_{<K})\boldsymbol v^\star}$. Consequently, the norm of the boundary part of the residual is bounded by $(\sum_{i<q}\norm{(I-\Pi_{<K})\boldsymbol r_i}^2)^{1/2}\,\norm{(I-\Pi_{<K})\boldsymbol v^\star}$.

\emph{Interior part.} Let $\mathcal I_N$ denote interpolation at the $N$ Gauss nodes of the relevant ultraspherical basis, and define $\tilde u_\lambda \coloneqq\mathcal T_{N_{\mu^\star}}\left( (\mathcal T_Ku)^{(\lambda)} \right)$, the truncation of the $\lambda$th derivative of $\mathcal{T}_K u$ in the parameter-$\mu^\star$ basis. The term of order $\lambda$ in Eq.~\eqref{eq:us-operator}, shown in Fig.~\ref{fig:us-flow}(c), acts on $\boldsymbol u^\star$ in three steps. First, $\widehat S_{\lambda\to\mu^\star}\widehat D_\lambda\boldsymbol u^\star$ gives the coefficients of $\tilde u_\lambda$. Second, $M_{\mu^\star}[a_\lambda]$ maps those coefficients to the coefficients of the interpolant $\mathcal I_{N_{\mu^\star}}(a_\lambda\tilde u_\lambda)$. Third, $\widehat S_{\mu^\star\to q}$ retains the first $N_q$ coefficients of this interpolant in the parameter-$q$ basis. 

The vectors $\widehat D_q\boldsymbol u^\star$ and $\widehat{\boldsymbol f}$ contain the coefficients of $(\mathcal T_Ku)^{(q)}$ and $\mathcal I_{N_q}f$, respectively. Therefore, $(\boldsymbol\zeta_K)_{q+i}$, for $0\leq i<N_q$, is the $i$th coefficient in the orthonormal basis $p^{(q)}$ of
\begin{equation*}
    (\mathcal T_Ku)^{(q)}+\sum_{\lambda<q}\mathcal I_{N_{\mu^\star}}(a_\lambda\tilde u_\lambda)-\mathcal I_{N_q}f.
\end{equation*}
Since $u$ satisfies Eq.~\eqref{eq:us-ode}, with $a_q\equiv1$,  $u^{(q)}+\sum_{\lambda<q}a_\lambda u^{(\lambda)}-f=0$. Subtracting this identity from the expression above gives the residual in terms of discretization errors. By Eq.~\eqref{eq:us-derivative}, $(\mathcal T_Ku)^{(q)}$ is exactly the truncation of $u^{(q)}$ in the parameter-$q$ basis, so their first $N_q$ coefficients agree.  Likewise, the first $N_q$ coefficients of $f$ and $\mathcal T_{N_q}f$ agree. Thus, for $0\le i<N_q$,
\begin{align}
        (\boldsymbol\zeta_K)_{q+i}={}&\sum_{\lambda<q}\int_{-1}^{1}w_q\,p_i^{(q)}\bigl[\mathcal I_{N_{\mu^\star}}(a_\lambda\tilde u_\lambda)-a_\lambda u^{(\lambda)}\bigr]\,\dd x \nonumber \\
        &-\int_{-1}^{1}w_q\,p_i^{(q)}\bigl(\mathcal I_{N_q}f-\mathcal T_{N_q}f\bigr)\,\dd x.
    \label{eq:us-interior-rows}
\end{align}
We call $\mathcal I_{N_q}f-\mathcal T_{N_q}f$ the forcing error and $\mathcal I_{N_{\mu^\star}}(a_\lambda\tilde u_\lambda)-a_\lambda u^{(\lambda)}$ the lower-order error of order~$\lambda$.

\emph{Bound on the residual.} The $N$-point Gauss rule associated with the parameter-$\lambda$ basis integrates $(\mathcal I_Ng)^2$ exactly. Since its weights are positive and sum to 
$h_0^{(\lambda)}=\int_{-1}^{1}w_\lambda(x)\,dx$, we have 
\begin{equation}
    \norm{\mathcal I_Ng}_{L^2(w_\lambda)}\leq(h_0^{(\lambda)})^{1/2}\norm g_\infty.
\end{equation}
The same bound holds for $\norm g_{L^2(w_\lambda)}$. Moreover, $\mathcal I_N$ reproduces every polynomial of degree less than $N$. Let $e_N^{(\lambda)}(g)\coloneqq\norm{g-\mathcal T_Ng}_\infty$ denote the uniform truncation error in the parameter-$\lambda$ basis.

By Bessel's inequality, the first $N_q$ coefficients of a function $h$, as in Eq.~\eqref{eq:us-interior-rows}, have norm at most $\norm h_{L^2(w_q)}$. Because $w_q\leq w_{\mu^\star}$, it is bounded by $\norm h_{L^2(w_{\mu^\star})}$. The forcing error can be written as $\mathcal I_{N_q}(f-\mathcal T_{N_q}f)$, since $\mathcal I_{N_q}$ reproduces $\mathcal T_{N_q}f$. Consequently,
\begin{equation}
    \|\mathcal I_{N_q}f - \mathcal T_{N_q}f\|_{L^2(w_q)}
\leq \sqrt{h_0^{(q)}}\,e_{N_q}^{(q)}(f).
\end{equation}
For each lower-order term $\lambda$, let $g_\lambda\coloneqq a_\lambda u^{(\lambda)}$. We decompose its error as
\begin{equation}
    \begin{aligned}
        &\mathcal I_{N_{\mu^\star}}(a_\lambda\tilde u_\lambda)-g_\lambda=\mathcal I_{N_{\mu^\star}}\bigl[a_\lambda(\tilde u_\lambda-u^{(\lambda)})\bigr]\\
        &\quad+\mathcal I_{N_{\mu^\star}}(g_\lambda-\mathcal T_{N_{\mu^\star}}g_\lambda)-(g_\lambda-\mathcal T_{N_{\mu^\star}}g_\lambda).
    \end{aligned}
    \label{eq:lower-order-error}
\end{equation}
The first term is bounded in $L^2(w_{\mu^\star})$ by $\sqrt{h_0^{(\mu^\star)}}\,\bar a_\lambda\norm{\tilde u_\lambda-u^{(\lambda)}}_\infty$. Each of the remaining two terms is bounded by $\sqrt{h_0^{(\mu^\star)}}\,e^{(\mu^\star)}_{N_{\mu^\star}}(g_\lambda)$. Thus, together, they contribute twice this truncation-error bound. Summing over $\lambda$, and adding the forcing error and the boundary part, gives
\begin{equation}
    \begin{aligned}
        \norm{\boldsymbol\zeta_K}\leq{}&\Bigl(\sum_{i<q}\norm{(I-\Pi_{<K})\boldsymbol r_i}^2\Bigr)^{1/2}\norm{(I-\Pi_{<K})\boldsymbol v^\star}\\
        &+\sqrt{h_0^{(q)}}\,e_{N_q}^{(q)}(f)\\
        &+2\sqrt{h_0^{(\mu^\star)}}\sum_{\lambda<q}e_{N_{\mu^\star}}^{(\mu^\star)}\bigl(a_\lambda u^{(\lambda)}\bigr)\\
        &+\sqrt{h_0^{(\mu^\star)}}\sum_{\lambda<q}\bar a_\lambda\norm{u^{(\lambda)}-\tilde u_\lambda}_\infty.
    \end{aligned}
    \label{eq:us-residual}
\end{equation}
The first term bounds the boundary contribution, and the second bounds the forcing error. The third term bounds the second and third terms in Eq.~\eqref{eq:lower-order-error}; this accounts for the factor of $2$. The final term bounds the first term from the same lower-order error decomposition.

Each contribution in Eq.~\eqref{eq:us-residual} is controlled by a tail of an expansion of $u$, $f$, or $a_\lambda u^{(\lambda)}$. In the boundary term, $\norm{(I-\Pi_{<K})\boldsymbol v^\star}$ is the $\ell^2$ norm of the rescaled coefficients $v_k^\star=d_k^{(q)}u_k^\star$ for $k\geq K$. The forcing and lower-order tail terms, $e_{N_q}^{(q)}(f)$ and $e_{N_{\mu^\star}}^{(\mu^\star)}(a_\lambda u^{(\lambda)})$, are uniform norms of the tails in the parameter-$q$ and parameter-$\mu^\star$ bases, respectively. In the last term, the difference $\widetilde u_\lambda-u^{(\lambda)}$ involves only coefficients $u_k^\star$ with $k\ge N_{\mu^\star}+\lambda$: truncation at degree $K$ removes coefficients with $k\ge K$, while truncation at $N_{\mu^\star}$ additionally removes contributions with $N_{\mu^\star}+\lambda\le k<K$. If $f$ and the coefficients $a_\lambda$ are analytic in a Bernstein ellipse, then $u$ is analytic there as well, and these expansion tails decay geometrically with $K$~\cite{Trefethen2013}. For functions with only finite smoothness, the decay is algebraic. The boundary-row tails decay algebraically even when the solution is analytic. For an endpoint condition on $u^{(j)}$ with $j<q/2$, Eq.~\eqref{eq:us-row-entries} implies $\norm{(I-\Pi_{<K})\boldsymbol r_i}=\Theta(K^{2j-q+1/2})$. Nevertheless, the boundary contribution in Eq.~\eqref{eq:us-residual} still decays geometrically, because this tail is multiplied by $\norm{(I-\Pi_{<K})\boldsymbol v^\star}$, which decays geometrically in the analytic case.

\section{\texorpdfstring{Darboux and Jacobi extensions (Section~\ref{sec:input-correction})}{Darboux and Jacobi extensions}}
\label{app:darboux-extensions}

This appendix shows that two parts of the input correction of Section~\ref{sec:input-correction}, namely the separated structure of the Darboux ladder and the phase representation of the Gauss nodes and weights, carry over to Jacobi polynomials. We also explain why an exact connection to Fourier modes requires half-integer parameters. No result of the main text depends on this appendix.

\emph{Darboux steps for Jacobi parameters.} For $\alpha,\beta>-1$, which we call admissible parameters, define the Jacobi weight by $w(x)=(1-x)^\alpha(1+x)^\beta$. Using the change of variable $x=\cos2t$ with $0<t<\pi/2$, define the angular functions
\begin{equation}
    v_k(t)=\sqrt{w(x)|x'(t)|} \, p_k^{(\alpha,\beta)}(x),
    \label{eq:jacobi-angular-functions}
\end{equation}
where $p_k^{(\alpha,\beta)}$ is the orthonormal Jacobi polynomial of degree $k$. The change-of-variables factor ensures that these functions are orthonormal in $L^2(0,\pi/2)$. In the variable $t$, the Jacobi differential equation becomes an eigenvalue problem for the trigonometric P\"oschl--Teller operator:
\begin{equation}
    H_{\alpha,\beta}v_k=\omega_k^2v_k,
    \qquad
    \omega_k=2k+\alpha+\beta+1,
    \label{eq:jacobi-angular}
\end{equation}
where
\begin{equation}
    H_{\alpha,\beta}=-\partial_t^2+\frac{\alpha^2-1/4}{\sin^2t}+\frac{\beta^2-1/4}{\cos^2t}.
\end{equation}
For signs $\sigma,\tau\in\{+1,-1\}$, define the seed function
\begin{equation}
    \phi_{\sigma\tau} = \sin^{1/2+\sigma\alpha}t\,\cos^{1/2+\tau\beta}t
    \label{eq:seed-function}
\end{equation}
and the associated Darboux operator
\begin{equation}
\begin{aligned}
     A_{\sigma\tau} 
     &= \frac{\partial}{\partial t} - \frac{\partial}{\partial t} \log\phi_{\sigma\tau} \\
     &= \frac{\partial}{\partial t} -\left(\frac12 + \sigma\alpha\right) \cot t + \left(\frac12 + \tau\beta\right) \tan t.
\end{aligned}
    \label{eq:darboux-operator}
\end{equation}
The seed is an eigenfunction of $H_{\alpha,\beta}$:
\begin{equation}
    H_{\alpha,\beta}\phi_{\sigma\tau} = s_{\sigma\tau}^2\phi_{\sigma\tau}, \qquad s_{\sigma\tau}=1+\sigma\alpha+\tau\beta.
\end{equation}
The Darboux operator intertwines two operators with shifted parameters: $H_{\alpha+\sigma,\beta+\tau}A_{\sigma\tau} = A_{\sigma\tau}H_{\alpha,\beta}$. Writing $v_k^{(\alpha,\beta)}$ to indicate the parameters of $v_k$, the Jacobi derivative identities~\cite[Section~18.9]{DLMF} imply
\begin{equation}
    A_{\sigma\tau}v_k^{(\alpha,\beta)}=-\sigma\, \sqrt{\omega_k^2-s_{\sigma\tau}^2} \,v_{k-(\sigma+\tau)/2}^{(\alpha+\sigma,\beta+\tau)}.
    \label{eq:darboux-step-action}
\end{equation}
This identity applies when $\alpha+\sigma>-1$, $\beta+\tau>-1$, $k-(\sigma+\tau)/2\geq0$, and $\omega_k^2>s_{\sigma\tau}^2$. Because the shifted angular function on the right has unit norm, the norm of the result of applying $A_{\sigma\tau}$ is $
\left\|A_{\sigma\tau}v_k^{(\alpha,\beta)}\right\| = \sqrt{\omega_k^2-s_{\sigma\tau}^2}$. 

In the ultraspherical case, $\alpha=\beta=q-\tfrac12$. Setting $\theta=2t$ gives $H_{\alpha,\beta}=4H_q$, and the angular functions satisfy $v_k(t)=\sqrt{2}\,G_k^{(q)}(\theta)$. Under this change of variables, the $(+,+)$ Darboux step, divided by $2$, becomes $A_q=\partial_\theta-q\cot\theta$, which is precisely the operator $A_q$ used in Section~\ref{sec:input-correction}.

\emph{Chains from a free base.} Starting from one of the four free parameter pairs $(\pm \tfrac12,\pm \tfrac12)$, a finite sequence of Darboux steps reaches any admissible half-integer pair $(\alpha,\beta)$. The number of steps depends on the target parameters but is independent of the transform dimension $N$. For the free pairs, the potential terms vanish and the operator reduces to $H_0=-\partial_t^2$, whose frequency-$\omega$ solutions are Fourier waves. Let $\mathcal D$ denote the composition of the $r_{\mathcal D}$ first-order operators in a chain from a free pair to $(\alpha,\beta)$. It can be written as
\begin{equation}
    \mathcal D = \sum_{\nu=0}^{r_{\mathcal D}} a_\nu(t)\,\partial_t^\nu,
\end{equation}
where the coefficient functions $a_\nu(t)$ are independent of the mode frequency. The intertwining relation $H_{\alpha,\beta}\mathcal D=\mathcal D H_0$~\cite{Crum1955} implies that, under the conditions stated after Eq.~\eqref{eq:darboux-step-action}, $\mathcal D$ maps a Fourier mode to a multiple of the corresponding angular function for $(\alpha,\beta)$. In particular,
\begin{equation}
    e^{-\ii\omega t}\mathcal De^{\ii\omega t}=\sum_{\nu=0}^{r_{\mathcal D}}a_\nu(t)(\ii\omega)^\nu.
    \label{eq:darboux-fourier}
\end{equation}
Thus, the correction to each Fourier mode is a polynomial in its frequency with coefficients drawn from a fixed set of functions of $t$. The number of such terms, $r_{\mathcal D}+1$, called the \emph{separation rank}, is independent of the mode index and of $N$, as in the ultraspherical construction.

\emph{Phase and quadrature.} The phase construction from Section~\ref{sec:input-correction} extends to these Darboux chains. Label the $r_{\mathcal D}$ steps by $i=1,\ldots,r_{\mathcal D}$, and denote the $i$th Darboux operator, seed, seed value, and resulting parameters by $A_i$, $\phi_i$, $s_i$, and $(\alpha_i,\beta_i)$, respectively. Let $(\alpha_0,\beta_0)$ be the parameters of the free starting problem. Choose a frequency $\omega$ such that $\omega^2>s_i^2$ for every step and such that the mode exists at every stage:
\begin{equation}
\omega=2k_i+\alpha_i+\beta_i+1,\quad k_i\in\mathbb{Z}_{\ge 0},\, i=0,\ldots,r_{\mathcal D}.
\end{equation}
Write $N=k_{r_{\mathcal D}}$ for the degree of the final mode. The right-hand side is the same for every $i$, because by Eq.~\eqref{eq:darboux-step-action}, a step $(\sigma,\tau)$ raises $\alpha+\beta$ by $\sigma+\tau$ and lowers the degree by $(\sigma+\tau)/2$; in particular, $\omega=\omega_N$ for the final parameters.

On the free starting problem, the corresponding mode is $v_{k_0}=(2/\sqrt\pi)\sin(\omega t+\chi)$, where $\chi\in\{0,\pi/2\}$ selects the boundary condition at $t=0$. Start with the complex wave $Z_0=e^{\ii(\omega t+\chi)}$, so that $\operatorname{Im}Z_0=(\sqrt{\pi}/2)v_{k_0}$, and define each successive wave by
\begin{equation}
    Z_i=\frac{A_iZ_{i-1}}{\sqrt{\omega^2-s_i^2}}.
\end{equation}
For a complex wave $Z$, define its phase current (Wronskian) by $J[Z]:=\operatorname{Im}(\overline{Z}Z')$. The initial wave has $J[Z_0]=\omega$, and the seed equation gives $J[A_iZ]=(\omega^2-s_i^2)J[Z]$. Consequently, this normalization ensures that $J[Z_i]=\omega$ at every step. 

Writing $Z_i=|Z_i|e^{\ii\Phi_i}$, we have $J[Z_i]=|Z_i|^2\Phi_i'$. Thus the phase $\Phi_i$ increases with $t$, and its increment at step $i$ is
\begin{equation}
    \Phi_i-\Phi_{i-1}=\arg\bigl(\partial_t\ln|Z_{i-1}|-\partial_t\ln\phi_i+\ii\Phi_{i-1}'\bigr).
    \label{eq:phase-transport}
\end{equation}
This generalizes the phase increments $\arg(\mathsf P_q/\mathsf P_{q-1})$ in the ultraspherical Darboux ladder of Section~\ref{sec:input-correction}.

The final wave $Z_f:=Z_{r_{\mathcal D}}=|Z_f|e^{\ii\Phi_f}$ satisfies
\begin{equation}
\operatorname{Im}Z_f=\pm\frac{\sqrt{\pi}}{2}v_N,
\end{equation}
where $v_N$ is the degree-$N$ mode for the final parameters $(\alpha,\beta)$. Its zeros $t_j$ correspond to the Gauss--Jacobi nodes $x_j=\cos(2t_j)$; at each zero, $\Phi_f(t_j)\in\pi\mathbb{Z}$.

The corresponding Gauss--Jacobi weights are given by the Christoffel numbers~\cite[Section~3.5(v)]{DLMF}
\begin{equation}
    \gamma_j=\left(\sum_{k<N}p_k^{(\alpha,\beta)}(x_j)^2\right)^{-1}.
\end{equation}

Squaring Eq.~\eqref{eq:jacobi-angular-functions} and summing over $k<N$ gives $\sum_{k<N}v_k(t)^2 = w(x)\,|x'(t)|\sum_{k<N}p_k^{(\alpha,\beta)}(x)^2$. Applying the confluent Christoffel--Darboux formula and the Jacobi derivative identities~\cite[Sections~18.2(v) and 18.9]{DLMF} yields
\begin{equation}
\sum_{k<N}v_k(t_j)^2
=\frac{v_N'(t_j)^2}{4\omega}
=\frac{\Phi_f'(t_j)}{\pi}.
\end{equation}
Therefore, for $N\ge1$, the same phase determines both the nodes and their weights:
\begin{equation}
    \Phi_f(t_j)\in\pi\mathbb{Z},
    \qquad \frac{\gamma_j}{w(x_j)|x'(t_j)|} = \frac{\pi}{\Phi_f'(t_j)}.
\end{equation}
This exact relation alone does not guarantee an efficient circuit: that also requires uniform smoothness of the corrections and bounds on the waves $Z_i$ and the derivatives of the phase.

\emph{Why the exact connection requires half-integers.} An exact connection by a finite-order differential operator is possible only when both Jacobi parameters are half-integers. Near $t=0$, the two independent solutions of $H_{\alpha,\beta}Z=\omega^2Z$ behave like $t^{1/2+\alpha}$ and $t^{1/2-\alpha}$, whereas applying a differential operator with coefficients meromorphic at $t=0$ to $e^{\ii\omega t}$ produces a meromorphic function. Compatibility with both local solution behaviors therefore requires $1/2+\alpha$ and $1/2-\alpha$ to be integers, so $\alpha\in \tfrac12+\mathbb Z$. The analogous argument at $t=\pi/2$ gives $\beta\in \tfrac12+\mathbb Z$. This restriction applies only to exact differential intertwiners and does not preclude other representations or approximate circuits.

\begin{proposition}[Finite Darboux connection]
\label{prop:darboux-obstruction}
Let $\alpha,\beta>-1$, and let $\mathcal D\neq0$ be a finite-order differential operator, independent of $\omega$, with coefficients analytic on $(0,\pi/2)$. If $\mathcal De^{\ii\omega t}$ solves $H_{\alpha,\beta}Z=\omega^2Z$ for infinitely many distinct frequencies, then $\alpha,\beta\in1/2+\mathbb Z$. Conversely, such parameters admit an intertwiner built from Darboux operators in Eq.~\eqref{eq:darboux-operator}.
\end{proposition}

\begin{proof}
If $\mathcal{D}e^{\ii\omega t}$ solves the Jacobi angular equation for infinitely many distinct frequencies, then $(H_{\alpha,\beta}\mathcal D-\mathcal DH_0)e^{\ii\omega t}$ vanishes for infinitely many values of $\omega$. This expression is $e^{\ii\omega t}$ multiplied by a polynomial in $\omega$, so that polynomial must vanish identically. Hence, $\mathcal{D}$ intertwines the two operators: $ H_{\alpha,\beta}\mathcal{D} = \mathcal{D}H_0$. If $\mathcal{D}$ has order zero, this identity forces $H_{\alpha,\beta}=H_0$. Otherwise, let $r_{\mathcal D}\geq1$ denote the order of $\mathcal D$. Comparing the highest-order derivative terms shows that the leading coefficient of $\mathcal{D}$ is a nonzero constant, denoted by $c$.

The intertwining relation also implies that $\ker \mathcal{D}$ is invariant under $H_0$. On $(0,\pi/2)$, the equation $\mathcal{D}f=0$ is a regular linear differential equation of order $r_{\mathcal{D}}$, so its solution space has dimension $r_{\mathcal{D}}$, and its elements are analytic on that interval. By the Cayley--Hamilton theorem, the restriction of $H_0$ to $\ker \mathcal{D}$ satisfies its characteristic polynomial. Consequently, every element of $\ker \mathcal{D}$ is annihilated by a linear differential operator with constant coefficients and is therefore an exponential polynomial. In particular, a basis $f_1,\ldots,f_{r_{\mathcal{D}}}$ of $\ker \mathcal{D}$ can be chosen to consist of entire functions.

For functions $g_1,\ldots,g_m$, define the Wronskian by
\begin{equation}
    W(g_1,\ldots,g_m)\coloneqq\det\bigl[g_i^{(j)}\bigr]_{0\leq j<m,\;1\leq i\leq m},
\end{equation}
whose $i$th column contains $g_i$ and its first $m-1$ derivatives.
For two functions, $W(\operatorname{Re}Z,\operatorname{Im}Z)$ is the phase current $J[Z]$ used above. The Wronskian $W(f_1,\ldots,f_{r_{\mathcal{D}}})$ has no zeros on $(0,\pi/2)$. Indeed, if it vanished at a point $t_0$, a nontrivial linear combination of the $f_i$ would solve $\mathcal Df = 0$ with zero initial values at $t_0$ and would therefore vanish identically, contradicting their linear independence.

We now establish the Wronskian representation of $\mathcal{D}$. Define the differential operator
\begin{equation}
\mathcal{L}f :=c\, \frac{W(f_1,\ldots,f_{r_{\mathcal{D}}},f)}
{W(f_1,\ldots,f_{r_{\mathcal{D}}})}.
\end{equation}Expanding the numerator determinant along its last column shows that $\mathcal{L}$ is a differential operator of order $r_{\mathcal{D}}$. The coefficient of $f^{(r_{\mathcal{D}})}$ is $c$, since its cofactor is $W(f_1,\ldots,f_{r_{\mathcal{D}}})$. Also, $\mathcal{L}f_i=0$ for every $i$, because the numerator then has two identical columns. Thus, $\mathcal{D}-\mathcal{L}$ has order at most $r_{\mathcal{D}}-1$ and annihilates all $r_{\mathcal{D}}$ linearly independent functions $f_1,\ldots,f_{r_{\mathcal{D}}}$. If $\mathcal{D}-\mathcal{L}$ were nonzero, its kernel would have dimension at most $r_{\mathcal{D}}-1$ on any subinterval where its leading coefficient is nonzero. This contradicts the fact that $f_1, \dots, f_{r_{\mathcal D}}$ are linearly independent on every subinterval. Therefore $\mathcal{D}=\mathcal{L}$.

We use the same numerator expansion to write $\mathcal{D}$ as a linear combination of $f,f',\ldots,f^{(r_{\mathcal{D}})}$. The coefficient multiplying each derivative is, up to a sign, $c$ times a cofactor of the numerator divided by $W(f_1,\ldots,f_{r_{\mathcal{D}}})$. The cofactors and the denominator are polynomials in derivatives of entire functions $f_1, \dots, f_{r_{\mathcal{D}}}$, and are therefore themselves entire. Since the denominator is not identically zero, these quotients are meromorphic. Hence the coefficients of $\mathcal{D}$ are meromorphic.

Finally, for each $\omega\ne 0$, the functions $e^{\ii\omega t}$ and $e^{-\ii\omega t}$ span the solution space of $H_0f=\omega^2f$. The intertwining relation maps this space into the two-dimensional solution space of the Jacobi angular equation with the same eigenvalue, $H_{\alpha,\beta} Z = \omega^2 Z$. The map can fail to be injective only if $\omega^2$ is an eigenvalue of the restriction of $H_0$ to $\ker \mathcal{D}$. The restriction $H_0|_{\ker\mathcal{D}}$ is a linear operator on the $r_{\mathcal{D}}$-dimensional vector space $\ker\mathcal{D}$. It has at most $r_{\mathcal{D}}$ distinct eigenvalues, so injectivity fails for at most $2r_{\mathcal{D}}$ nonzero frequencies. Therefore, for all other nonzero $\omega$, $\mathcal{D}$ maps the Fourier solution space bijectively onto the corresponding Jacobi solution space.

Fix one nonzero frequency $\omega$ outside this finite exceptional set. By surjectivity, every solution of the target equation can be written as $Z=\mathcal Df$ with $f=ae^{\ii\omega t}+be^{-\ii\omega t}$. Since $f$ is entire and the coefficients of $\mathcal D$ are meromorphic on $\mathbb C$, $Z=\sum_\nu a_\nu f^{(\nu)}$ is meromorphic on $\mathbb C$. Poles can occur only at zeros of $W(f_1,\ldots,f_{r_{\mathcal D}})$. Since $W$ has no zeros on $(0,\pi/2)$, poles on $[0,\pi/2]$ are therefore possible only at $t=0$ and $t=\pi/2$. We now apply the Frobenius method to the target equation $-Z''+V(t)Z=\omega^2Z$, where $V(t)=\frac{\alpha^2-\tfrac14}{\sin^2t}+\frac{\beta^2-\tfrac14}{\cos^2t}$. Near $t=0$, its two local solutions are governed by the indicial equation $\rho(\rho-1) = \alpha^2 - 1/4$ with roots $\rho_{\pm} = \frac{1}{2} \pm |\alpha|$. By the Frobenius method, there is a solution whose leading behavior is $t^{\rho_+}$. Since every target-equation solution must be meromorphic at $t=0$, this exponent must be an integer. The other root, $\rho_-=1-\rho_+$, is also an integer then, so $\alpha \in \frac{1}{2} + \mathbb{Z}$. Applying the same argument near $t=\pi/2$, we can also show $\beta \in \frac{1}{2} + \mathbb{Z}$. Since $\alpha,\beta>-1$, the possible values are $\alpha,\beta\in\{-\tfrac12,\tfrac12,\tfrac32,\ldots\}$. Conversely, for these parameters the chains from a free base described above build an intertwiner from the Darboux operators in Eq.~\eqref{eq:darboux-operator}.

\end{proof}

This endpoint argument is related to local monodromy~\cite{DuistermaatGrunbaum1986}, which studies how solutions behave around singular points. The same parameter restriction also follows from the basis change: the angular functions behave as $t^{\alpha+1/2}$ and $(\pi/2-t)^{\beta+1/2}$ near the endpoints. They are finite trigonometric polynomials precisely when both powers are nonnegative integers, as required for the finite matrix $L$ by Eq.~\eqref{eq:frequency-to-degree}.

For certain exceptional orthogonal-polynomial families~\cite{Odake2011,Grandati2025}\textemdash families that omit finitely many polynomial degrees\textemdash the angular functions can also be obtained from Fourier modes by a finite chain of Darboux steps, so their separation rank is again independent of the mode index and of $N$. However, these cases need extra care: the transformed functions may have regularity issues or poles in their coefficients, and the missing degrees affect the quadrature rules.

\emph{Normalization.} A fixed separation rank does not, by itself, guarantee an efficient block encoding: the normalization required to encode the separated factors may still grow with the grid size. For example, when $(\alpha,\beta)=(3/2,1/2)$, a single Darboux step from the free pair $(1/2,-1/2)$ gives
\begin{equation}
    Z_1(t)=\frac{(\ii\omega_k-\cot t)e^{\ii\omega_k t}}{\sqrt{\omega_k^2-1}},
\qquad \omega_k=2k+3.
\end{equation}
Although this expression contains the singular factor $\cot t$, its imaginary part is regular at $t=0$. Expanding the real and imaginary parts,
\begin{equation}
    \begin{aligned}
        \sqrt{\omega_k^2-1}\,\operatorname{Im} Z_1 &= \omega_k\cos(\omega_k t)-\cot t\,\sin(\omega_k t) \\
        \sqrt{\omega_k^2-1}\,\operatorname{Re} Z_1 &= -\omega_k\sin(\omega_k t)-\cot t\,\cos(\omega_k t).
    \end{aligned}
\end{equation}
As $t\to 0$, the two terms in the imaginary part both approach $\omega_k$, and their difference is $-\frac{1}{3}\omega_k(\omega_k^2-1)t^2+O(t^4)$. This $t^2$ behavior agrees with the leading power $t^{\alpha+1/2}$ of the angular function, since $\operatorname{Im}Z_1=-(\sqrt{\pi}/2)v_k$. We refer to the cancellation between the two terms as endpoint cancellation. The real part has no corresponding cancellation; it behaves as $-1/(t\sqrt{\omega_k^2-1})$.

In the separated form of Eq.~\eqref{eq:darboux-fourier}, $Z_1$ is computed from the two terms $\ii\omega_k e^{\ii\omega_k t}$ and $-\cot t\,e^{\ii\omega_k t}$. A circuit that implements this form as a linear combination, say via LCU, must be able to encode $\cot t$ separately. The subnormalization in a block encoding is at least as large as the norm of the encoded matrix, and the subnormalizations of the terms in a linear combination of unitaries add. On an $N$-point Gauss grid, the node nearest $t=0$ is of order $1/N$, so $\max_j|\cot t_j|$ is of order $N$. Thus, encoding $\cot t$ separately requires a subnormalization of order $N$, even though the angular function itself is regular. Because endpoint cancellation occurs only after the two terms are combined, it cannot reduce the subnormalization of this separated implementation.

The Darboux chain avoids this difficulty because it maps normalized modes to normalized modes. By Eq.~\eqref{eq:darboux-step-action}, each step multiplies a mode's norm by $\sqrt{\omega^2-s_i^2}$, provided the stated conditions hold. Consequently, if $\omega^2>s_i^2$ at every step, then the Darboux operator $\mathcal D$, divided by $\prod_i(\omega^2-s_i^2)^{1/2}$, maps the normalized free mode $v_{k_0}$ to the normalized final mode $\pm v_N$. On these modes, the operator $\mathcal D\prod_i(H_0-s_i^2)^{-1/2}$ is therefore an isometry. In operator form, this gives the formal identity $\mathcal D^\dagger\mathcal D=\prod_i(H_0-s_i^2)$, where each seed is included with its multiplicity. This identity is formal in the sense that boundary terms from integration by parts are ignored. For the single-step example above,
$(-\partial_t-\cot t)(\partial_t-\cot t) = -\partial_t^2-1=H_0-1$.
A circuit that implemented this isometry as a single operator, rather than encoding its factors separately, could have a subnormalization of order one, allowing endpoint cancellation to occur within the operator.

\bibliography{references}

\end{document}